\documentstyle[11pt]{article}
\def\thefootnote{\fnsymbol{footnote}}

\def\lrarrow{\leftrightarrow}
\def\rarrow{\rightarrow}
\def\larrow{\leftarrow}
\def\tilD{\tilde{D}}
\def\tilDp{\tilde{D}^\prime}
\def\bs{\indent\indent}
\def\be {\begin{eqnarray}}
\def\ee {\end{eqnarray}}
\def\bea {\begin{eqnarray}}
\def\eea {\end{eqnarray}}
\def\ben {\begin{eqnarray}}
\def\een {\end{eqnarray}}
\def\bitem{\begin{itemize}}
\def\eitem{\end{itemize}}
\def\etap{\eta^\prime}
\def\A{{\cal A}}
\def\D{{\cal D}}
\def\I{{\cal I}}
\def\L{{\cal L}}
\def\V{{\cal V}}
\def\ri{\right}
\def\le{\left}
\def\boxB{\mbox{\tiny B}}
\def\barB{\overline{B}}
\def\ie{{\it i.e}}
\def\viz{{\it viz}}
\def\eg{{\it e.g.}}

\def\del{\partial}
\def\M{{\cal M}}

\def\gB {g_A}

\def\Nbar{{\overline {N}}}
\def\barN{{\overline {N}}}
\def\Bbar{{\overline {B}}}

\def\barK{{\overline {K}}}

\def\pivec{{\vec \pi}}

\def\tauvec{{\vec \tau}}

\def\dmu{{\partial_\mu}}

\def\Tr{{\mbox{Tr}}}

\def\prl {Phys. Rev. Lett.}
\def\pl {Phys. Lett.}
\def\pr {Phys. Rev.}
\def\np {Nucl. Phys.}

\begin{document}

\begin{titlepage}
\begin{center}
\begin{flushright}
{\bf SPhT T93/110}
\end{flushright}
\hfill\today
\vskip 3.0cm
{\Large\bf CHESHIRE CAT HADRONS}\footnote{
Lectures given at Latin American School of Physics 93 (ELAF '93),
5-16 July 1993, Mar del Plata, Argentina.}
\vskip 3cm
{\large  Mannque Rho}\\
{\it Service de Physique Th\'{e}orique, CEA Saclay}\\
{\it 91191 Gif-sur-Yvette, France}
\vskip 5cm
{\bf ABSTRACT}
\begin{quotation}
This article is based on a series of lectures given at ELAF 93
on the description of low-energy hadronic
systems {\it in and out} of hadronic medium.  The focus is put
on identifying, with the
help of a Cheshire Cat philosophy, the
effective degrees of freedom relevant for the strong interactions
from a certain number of generic symmetry properties of
QCD. The matters treated are the ground-state and excited-state properties
of light- and heavy-quark baryons and applications to nuclei and nuclear
matter under normal as well as extreme conditions.
\end{quotation}
\end{center}
\end{titlepage}
\tableofcontents
\newpage
\renewcommand{\thefootnote}{\#\arabic{footnote}}
\centerline{\large\bf INTRODUCTION}
\indent \indent

In this series of lectures, I would like to describe how far one can go in
understanding hadronic properties, both elementary and many-body, at low energy
based principally on generic features of the fundamental theory, QCD.
For this I will use a specific simplified model, the chiral bag model, but
the results should not crucially depend upon its characteristics. In fact,
what matters is the generic consequence of symmetries considered, not the
dynamical details of the model. If one wishes, once the generic structure is
extracted, one can simply forget about the model without losing relevant
physics.

This way of approaching hadron and nuclear physics is in line with the
effective theory approach to other fields of physics. As emphasized
since some time by Nambu \cite{nambu}, one generic
effective theory, Ginzburg-Landau
or Gell-Mann-L\'{e}vy Lagrangian, is able to capture the essence of
the physics of superconductivity,
$^3$He superfluidity, nuclear pairing, low-energy QCD and standard model
with appropriately rescaled length parameters such as gap and condensate
parameters. It is an intriguing fact of Nature that a sigma model
so aptly describes the generic structure
of {\it all} these interactions. Recently this point is given a more rigorous
basis by Polchinski, Weinberg and others
\cite{polchinski,wein93} in terms of effective actions. Effective actions
play an important role in two ways: One, when fundamental theory is not known
above a certain energy scale as in the case of the standard model;
two, when fundamental theory is known but the solution is unknown
as in the case of QCD at low energy. In both cases, symmetries of the
physical particles that one is trying to describe allow one to write the
effective actions. In this lecture we are concerned with the latter.

The aim of this lecture is to show how much of low-energy properties
of elementary hadrons and many-hadron systems can be understood with
a set of effective degrees of freedom consistent with chiral symmetry of
light quarks and heavy-quark symmetry of massive quarks. In doing this,
I will resort to the Cheshire Cat philosophy \cite{cheshire}. Now
while exact in two dimensions, the Cheshire cat can only be approximate in
four dimensions. Clearly it would be meaningless to try to be too
quantitative. Furthermore this cannot be a substitute for a real QCD
calculation. But I feel that one can learn a lot here as one does in
other areas of physics \footnote{ As sketched in \cite{history}, a
minority of nuclear physicists, to which we belonged, argued early on
that understanding of the structure of the nucleon should come mostly
from chiral symmetry of QCD -- in particular the pion cloud -- rather
than asymptotic freedom and confinement. This led to the description of
the nucleon as a ``little bag" in which quarks and gluons are confined,
surrounded by a big pion cloud \cite{br79}. It was proposed that nuclear
dynamics at low energy in which nonperturbative aspect of QCD dominates
be mainly attributed to
the interactions involving pion and associated heavier meson degrees of
freedom. Although the initial crude model is by now largely superseded
by more sophisticated topological as well as nontopological models,
the physical picture proposed then with the ``little bag" model appears
to have survived more or less intact
as we get to understand better the nucleon structure through
lattice calculations and QCD-based models. What follows is an attempt
to give a more modern and certainly personal account of what we have
learned since the notion of the ``little bag" was put forward.}.

This lecture consists of three broad sections. In the first, I use
the chiral bag model and the Cheshire Cat approach to ``guess" an
effective Lagrangian for hadrons. The notion of induced gauge fields
as Berry potentials on the strong interaction sector is introduced.
In the second, I apply a simple effective theory to describe the ground-state
and excited-state properties of both light and heavy baryons.
In the last, some properties of nuclei and nuclear matter are treated in
terms of effective chiral Lagrangians motivated in the two preceding
lectures.

The nature of this note is strongly influenced by my collaborations or
discussions with many people, in particular with Gerry Brown, Andy Jackson,
Kuniharu Kubodera, Hyun Kyu Lee, Dong-Pil Min, Holger Bech Nielsen, Maciej
Nowak, Tae-Sun Park, Dan Olof Riska, Norberto Scoccola, Andreas Wirzba,
Koichi Yamawaki and Ismail Zahed.
\vskip 1cm
\part{``Guessing" an Effective Theory}
\renewcommand{\theequation}{I.\arabic{equation}}
\section{The Cheshire Cat Principle (CCP)}
\subsection{Toy Model in (1+1) Dimensions}
\setcounter{footnote}{0}
\indent

In two dimensions the Cheshire Cat principle (CCP in short from now on)
can be formulated exactly. Let us discuss it to have a general idea involved.
Later we will make a leap to reality in four dimensions by making a guess.
We shall use for definiteness the chiral bag picture\cite{brcom88}.
The result is generic and should not sensitively depend
upon details of the models. Thus other models of similar structure containing
the same symmetries  could be equally used. We will leave the construction
to the aficionados of their own favorite models.

In the spirit of a chiral bag, consider a massless free single-flavored
fermion $\psi$ confined
in a region (say, inside) of ``volume" $V$ coupled on the surface $\del V$
to a massless free boson $\phi$ living in a region of ``volume" $\tilde{V}$
(say, outside). Of course in one-space dimension, the ``volume" is just
a segment of a line but we will use this symbol in analogy to higher
dimensions. We will assume that the action is invariant under global chiral
rotations and parity. Interactions invariant under these symmetries can be
introduced without changing the physics, so the simple system that we consider
captures all the essential points of the construction. The action contains
three terms\footnote{Our convention is as follows. The metric is
$g_{\mu\nu}={\rm diag} (1,-1)$ with Lorentz indices $\mu,\nu=0,1$ and the
$\gamma$ matrices are in Weyl representation, $\gamma_0=\gamma^0=\sigma_1$,
$\gamma_1=-\gamma^1=-i\sigma_2$, $\gamma_5=\gamma^5=\sigma_3$ with the
usual Pauli matrices $\sigma_i$.}

\ben
S=S_V + S_{\tilde{V}} +S_{\del V} \label{action}
\een
where
\ben
S_V &=& \int_{V} d^2x \bar{\psi}i\gamma^\mu\del_\mu \psi +\cdots,\label{Sin}\\
S_{\tilde{V}} &=& \int_{\tilde{V}} d^2x \frac{1}{2} (\del_\mu \phi)^2 +\cdots
\label{Sout}
\een
and $S_{\del V}$ is the boundary term which we will specify shortly.
Here the ellipsis stands for other terms such as interactions, fermion masses
etc. consistent with the assumed symmetries of the system on which we will
comment later. For instance, there can be a coupling to a $U(1)$ gauge field
in (\ref{Sin}) and a boson mass term in (\ref{Sout}) as would be needed in
Schwinger model. Without loss of generality, we will simply ignore what is in
the ellipsis unless we absolutely need it.
Now the boundary term is essential in making the connection
between the two regions. Its structure depends upon the physics ingredients
we choose to incorporate. We will assume that chiral symmetry holds on the
boundary even if it is, as in Nature, broken both inside and outside by
fermion mass terms. As long as the symmetry
breaking is gentle, this should be a good approximation since the surface
term is a $\delta$ function. We should also assume that the boundary term
does not break the discrete symmetries ${\cal P}$, ${\cal C}$ and ${\cal T}$.
Finally we demand that it give no decoupled boundary conditions, that is
to say, boundary conditions that involve only $\psi$ or $\phi$ fields.
These three conditions are sufficient in (1+1) dimensions to give a unique
term\cite{cheshire86}
\ben
S_{\del V}=\int_{\del V} d\Sigma^\mu \left\{\frac{1}{2}n_\mu \bar{\psi}
e^{i\gamma_5\phi/f} \psi\right\}
\een
with the $\phi$ ``decay constant" $f=1/\sqrt{4\pi}$ where $d\Sigma^\mu$ is an
area element with $n^\mu$ the normal vector, {\ie}, $n^2=-1$ and
picked outward-normal. As we will mention
later, we cannot establish the same unique relation in (3+1) dimensions
but we will continue using this simple structure even when there is no
rigorous justification in higher dimensions.

At classical level, the action (\ref{action}) gives rise to the bag
``confinement," namely that inside the bag the fermion (which we shall
call ``quark" from now on) obeys
\ben
i\gamma^\mu\del_\mu\psi =0
\een
while the boson (which we will call ``pion") satisfies
\ben
\del^2 \phi =0
\een
subject to the boundary conditions on $\del V$
\ben
in^\mu \gamma_\mu \psi &=&- e^{i\gamma_5 \phi/f}\psi, \label{conf}\\
n^\mu \del_\mu \phi &=& f^{-1}\bar{\psi}(\frac{1}{2} n^\mu\gamma_\mu
\gamma_5) \psi.\label{conf1}
\een
Equation (\ref{conf}) is the familiar ``MIT confinement condition" which
is simply a statement of the classical conservation of the vector current
$\del_\mu j^\mu=0$ or $\bar{\psi}\frac{1}{2}n^\mu\gamma_\mu \psi=0$ at
the surface while Eq. (\ref{conf1}) is just the statement of the conserved
axial vector current $\del_\mu j_5^\mu=0$ (ignoring the possible
explicit mass of the quark and the pion at the surface)\footnote{Our definition
of the currents is as follows: $j^\mu =\bar{\psi}\frac{1}{2}\gamma^\mu\psi$,
$j_5^\mu=\bar{\psi}\frac{1}{2}\gamma^\mu\gamma_5\psi$.}. The crucial point
of our argument is that these classical observations are invalidated by
quantum mechanical effects. In particular while the axial current continues
to be conserved, the vector current fails to do so due to quantum anomaly.

There are several ways of seeing that something is amiss with the classical
conservation law, with slightly different interpretations. The easiest way
is as follows. For definiteness, let us imagine that the quark is ``confined"
in the space $-\infty \leq r \leq R$ with a boundary at $r=R$. Now
the vector current $j_\mu =\bar{\psi}\gamma_\mu \psi$ is conserved inside
the ``bag"
\ben
\del^\mu j_\mu=0, \ \ \ \ \ r<R.
\een
If one integrates this from $-\infty$ to $R$ in $r$, one gets the
time-rate change of the fermion ({\ie}, quark) charge
\ben
\dot{Q}\equiv \frac{d}{dt} Q=2\int_{-\infty}^R dr\del_0 j_0=2\int_{-\infty}
^R dr \del_1 j_1=2j_1 (R)\label{vanomaly}
\een
which is just
\ben
\bar{\psi}n^\mu \gamma_\mu \psi, \ \ \ \ r=R.
\een
This vanishes classically as we mentioned above. But this is not correct
quantum mechanically because it is not well-defined locally in time.
In other words, $\psi^\dagger (t) \psi (t+\epsilon)$
goes like $\epsilon^{-1}$ and so is singular as $\epsilon\rightarrow 0$.
To circumvent this difficulty which is related to vacuum fluctuation,
we regulate the bilinear product by point-splitting in time
\ben
j_1 (t)=\bar{\psi} (t)\frac{1}{2}\gamma_1 \psi (t+\epsilon), \ \ \ \ \
\epsilon\rightarrow 0.
\een
Now using the boundary condition
\ben
i\gamma_1 \psi (t+\epsilon)&=& e^{i\gamma_5 \phi (t+\epsilon)/f}
\psi (t+\epsilon)\nonumber\\
&\approx& e^{i\gamma_5 \phi (t)/f} [1+i\epsilon \gamma_5
\dot{\phi} (t)/f] e^{\frac{1}{2}[\phi(t),\phi(t+\epsilon)]} \psi (t+\epsilon),
 \ \ \ \ r=R
\een
and the commutation relation
\ben
[\phi (t), \phi (t+\epsilon)]=i\ {\rm sign}\ \epsilon,
\een
we obtain
\ben
j_1 (t)=-\frac{i}{4f}\epsilon\dot{\phi} (t) \psi^\dagger (t)\psi (t+\epsilon)
=\frac{1}{4\pi f} \dot{\phi} (t) + O(\epsilon), \ \ \ \ r=R \label{flow}
\een
where we have used $\psi^\dagger (t)\psi (t+\epsilon)=\frac{i}{\pi\epsilon}+
[{\rm regular}]$.
Thus quarks can flow out or in if the pion field varies in time.
But by fiat,
we declared that there be no quarks outside, so the question is what
happens to the quarks when they leak out? They cannot simply disappear into
nowhere if we impose that the fermion (quark) number is conserved. To
understand what happens, rewrite (\ref{flow}) using the surface tangent
\ben
t^\mu=\epsilon^{\mu\nu} n_\nu.
\een
We have
\ben
t\cdot\del \phi=\frac{1}{2f}\bar{\psi} n\cdot \gamma\psi
=\frac{1}{2f}\bar{\psi}t\cdot \gamma\gamma_5 \psi, \ \ \ \ r=R
\label{bosoniz2}
\een
where we have used the relation $\bar{\psi}\gamma_\mu\gamma_5\psi=
\epsilon_{\mu\nu}\bar{\psi}\gamma^\nu \psi$ valid in two dimensions.
Equation (\ref{bosoniz2}) together with (\ref{conf1}) is nothing but
the bosonization relation
\ben
\del_\mu \phi= f^{-1}\bar{\psi} (\frac{1}{2}\gamma_\mu\gamma_5)\psi
\label{bosonization}
\een
at the point $r=R$ and time $t$. As is well known, this is a striking
feature of (1+1) dimensional fields that makes one-to-one correspondence
between fermions and bosons\cite{colemanmand}.
In fact, one can even write
the fermion field in terms of the boson field as
\ben
\psi(x)={\rm exp}\left\{-\frac{i}{2f}\int_{x_0}^x d\xi [\pi (\xi)+\gamma_5
\phi^\prime (\xi)]\right\}\psi (x_0)
\een
where $\pi (\xi)$ is the conjugate field to $\phi (\xi)$ and
$\phi^\prime (\xi) =\frac{d}{d\xi}\phi (\xi)$.

Equation (\ref{vanomaly}) with (\ref{flow}) is the vector anomaly, {\ie},
quantum anomaly in the vector current: The vector current is not conserved
due to quantum effects. What it says is that the vector charge which in this
case is equivalent to the fermion (quark) number inside the bag is not
conserved. Physically what happens is that the amount of fermion number
$\Delta Q$ corresponding to $\Delta t\dot{\phi}/\pi f$ is pushed into the
Dirac sea at the bag boundary and so is lost from inside and gets accumulated
at the pion side of the bag wall\footnote{One can see this more clearly by
rewriting the change in the quark charge in the bag as
$$\Delta Q_V=\frac{1}{2}\int_V \langle 0|[\psi^\dagger (x),\psi (x)]|0\rangle
-1=- \frac{1}{2}\left\{\sum_{E_n>0} 1 -\sum_{E_n<0}\right\}$$.}.
This accumulated baryon charge must be carried by something residing in the
meson sector. Since there is nothing but ``pions" outside, it must be be
the pion field that carries the leaked charge. This means that the pion
field supports a soliton. This is rather simple to verify in (1+1) dimensions
using arguments given in the work of Coleman and Mandelstam \cite{colemanmand},
referred to above.
This can also be shown to be the case in (3+1) dimensions.
In the present model, we find that
one unit of fermion charge $Q=1$ is partitioned as
\ben
Q&=& 1=Q_V + Q_{\tilde{V}},\\
Q_{\tilde{V}} &=& \theta/\pi,\nonumber\\
Q_V &=& 1-\theta/\pi\nonumber
\een
with
\ben
\theta=\phi (R)/f.
\een
We thus learn that the quark charge is partitioned into the bag and
outside of the bag, without however any dependence of the total
on the size or location
of the bag boundary. In (1+1)-dimensional case, one can calculate other
physical quantities such as the energy, response functions and more
generally, partition functions and show that the physics does not depend
upon the presence of the bag\cite{perry}.
We could work with quarks alone, or pions alone or
any mixture of the two. If one works with the quarks alone, we have to
do a proper quantum treatment to obtain something which one can obtain
in mean-field order with the pions alone. In some situations, the hybrid
description is more economical than the pure ones. {\it The complete
independence of the physics on the bag -- size or location-- is called
the ``Cheshire Cat Principle" (CCP) with the corresponding mechanism referred
to as ``Cheshire Cat Mechanism" (CCM).} Of course the CCP holds exactly in
(1+1) dimensions because of the exact bosonization rule. There is no exact
CCP in higher dimensions since fermion theories cannot be bosonized
exactly in higher dimensions but we will see later that a fairly strong case of
CCP can be made for (3+1) dimensional models. Topological quantities like
the fermion (quark or baryon) charge satisfy an exact CCP in (3+1) dimensions
while nontopological
observables such as masses, static properties and also some nonstatic
properties satisfy it approximately but rather well.

\subsubsection{\it ``Color" anomaly}
\bs
So far we have been treating the quark as ``colorless", in other
words, in the absence of a gauge field. Let us consider that the quark
carries a $U(1)$ charge $e$, coupling to a $U(1)$ gauge field $A^\mu$.
We will still continue working with a single-flavored quark, treating
the multi-flavored case later.
Then inside the bag, we have essentially the Schwinger model, namely,
(1+1)-dimensional QED\cite{schwinger}.
It is well established that the charge is confined in the Schwinger model,
so there are no charged particles in the spectrum. If now our ``leaking"
quark carries the color (electric) charge, this will at first sight pose
a problem since
the anomaly obtained above says that there will be a leakage of the charge
by the rate
\ben
\dot{Q^c}=\frac{e}{2\pi}\dot{\phi}/f  .
\een
This means that the charge accumulated on the surface will be
\ben
\Delta Q^c=\frac{e}{2\pi}\phi (t,R)/f.
\een
Unless this is compensated, we will have a violation of charge
conservation, or breaking of gauge invariance. This is unacceptable.
Therefore we are invited to
introduce a boundary condition that compensates the induced charge, {\ie},
by adding a boundary term
\ben
\delta S_{\del V}=-\int_{\del V} d\Sigma \frac{e}{2\pi}\epsilon^{\mu\nu}
n_\nu A_\mu \frac{\phi}{f}
\een
with $n^\mu=g^{\mu\nu}n_\nu$.
The action is now of the form \cite{nwirzba}
\ben
S=S_V+S_{\tilde{V}}+S_{\del V}\label{saction}
\een
with
\ben
S_V&=& \int_V d^2x \left\{\bar{\psi} (x)\left[i\del_\mu-eA_\mu\right]\gamma^\mu
\psi (x) -\frac{1}{4}F_{\mu\nu}F^{\mu\nu}\right\},\\
S_{\tilde{V}}&=& \int_{\tilde{V}} d^2x \left\{\frac{1}{2}(\del_\mu \phi (x))^2
-\frac{1}{2}m_{\phi}^2\phi (x)^2\right\},\\
S_{\del V}&=& \int_{\del V} d\Sigma \left\{\frac{1}{2}\bar{\psi}
e^{i\gamma^5\phi/f}\psi -\frac{e}{2\pi}\epsilon^{\mu\nu} n_\mu A_\nu\frac{\phi}
{f} \right\}.
\een
We have included the mass of the $\phi$ field, the reason of which will
become clear shortly.

In this example, we are having the ``pion" field (more precisely the soliton
component of it) carry the color charge. In
general, though, the field that carries the color could be different from
the field that carries the soliton. In fact, in (3+1) dimensions, it is
the $\eta^\prime$ field that will be coupled to the gauge field (although
$\eta^\prime$ is colorless) while it is the pion field that supports a soliton.
But in (1+1) dimensions, the soliton is lodged in the $U(1)$ flavor sector
(whereas in (3+1) dimensions, it is in $SU(n_f)$ with $n_f\geq 2$).
For simplicity, we will continue our discussion
with this Lagrangian.

We now illustrate how one can calculate the mass of the $\phi$ field using the
action (\ref{saction}) and the CCM. This exercise will help understand a
similar calculation of the mass of the $\eta^\prime$ in (3+1) dimensions. The
additional ingredient needed for this exercise is the $U_A (1)$
(Adler-Bell-Jackiw) anomaly \cite{ABJ} which in (3+1) dimensions reads
\ben
\del_\mu j_5^\mu=-\frac{e^2}{32\pi^2} \epsilon_{\mu\nu\rho\sigma} F^{\mu\nu}
F^{\rho\sigma}
\een
and in (1+1) dimensions takes the form
\ben
\del_\mu j_5^\mu=-\frac{e}{4\pi}\epsilon_{\mu\nu}F^{\mu\nu}
=-\frac{1}{2\pi} eE.\label{abj1}
\een
Instead of the configuration with the quarks confined in $r<R$, consider
a small bag ``inserted" into the static field configuration of the scalar
field $\phi$ which we will now call $\etap$ in anticipation of the
(3+1)-dimensional case we will consider later. Let the bag be put in
$-\frac{R}{2}\leq r \leq \frac{R}{2}$. The CCP states that
the physics should not depend on the size $R$, hence we can take it to be
as small as we wish. If it is small, then the charge accumulated at the
two boundaries will be $\pm \frac{e}{2\pi}\phi/f$. This means that
the electric field generated inside the bag is
\ben
E=F_{01}=\frac{e}{2\pi} (\phi/f).
\een
Therefore from the $U_A (1)$ anomaly (\ref{abj1}), the axial charge created
or destroyed (depending on the sign of the scalar field) in the static bag is
\ben
2\int_V dr \del_\mu j_5^\mu=2(-j_1^5|_{\frac{R}{2}} + j_1^5|_{-\frac{R}{2}})
=-\frac{e^2}{2\pi^2} (\phi/f)R.
\een
{}From the bosonization condition (\ref{bosonization}), we have
\ben
2f\left(\del_1\phi|_{\frac{R}{2}}-\del_1\phi|_{-\frac{R}{2}}\right)
=\frac{e^2}{2\pi^2}(\phi/f)R.
\een
Now taking $R$ to be infinitesimal by the CCP, we have, on cancelling $R$
from both sides,
\ben
\del^2 \phi-\frac{e^2}{4 \pi^2 f^2}\phi=0
\een
which then gives the mass, with $f^{-2}=4\pi$,
\ben
m_{\phi}^2=\frac{e^2}{\pi},\label{schmass}
\een
the well-known scalar mass in the Schwinger model which can also be obtained
by bosonizing the $(QED)_2$. A completely parallel reasoning will be used later
to calculate the $\eta^\prime$ mass in (3+1) dimensions.
\subsubsection{\it The Wess-Zumino-Witten term}
\bs
Let us consider the case of more than one flavor. For simplicity, we take
two flavors. We shall see later that in (3+1) dimensions, we have
an extra topological term called Wess-Zumino-Witten term (WZW term in short,
alternatively referred to as Wess-Zumino term) if we have more than two
flavors. In (1+1) dimensions, such an extra term arises already for two
flavors\cite{wittencomath}.
The relevant flavor symmetry is then $U(2)\times U(2)$. The $U_V(1)$
symmetry associated with the fermion number does not concern us, so we may
drop it, leaving the symmetry $U_A(1)\times SU(2)\times SU(2)$.
The scalar field discussed above corresponds to the $U_A (1)$ field, so
we will not consider it anymore and instead focus on scalars living
in $SU(2)_V$ to which $SU(2)\times SU(2)$ breaks down .
The scalars will be called $\pi^a$ with $a=1,2,3$. As stated before,
the soliton lives in the $U(1)$ sector, so the $SU(2)$ scalars have nothing
to do with the ``baryon" structure but its nonabelian nature brings in
an extra ingredient associated with the Wess-Zumino-Witten term which we
will now show emerges from the boundary condition. Let the scalar multiplets be
summarized in the form
\ben
U=e^{i\vec{\tau}\cdot \vec{\pi}/f}
\een
with $f=1/\sqrt{4\pi}$.

As before the action consists of three terms
\ben
S=S_V + S_{\tilde{V}} +S_{\del V}. \label{actionp}
\een
The bag action is given as before by
\ben
S_V &=& \int_{V} d^2x \bar{\psi}i\gamma^\mu\del_\mu \psi +\cdots,\label{Sin2}
\een
where the fermion field is a doublet $\psi^T= (\psi_1,\psi_2)$.
A straightforward generalization of the abelian boundary condition consistent
with chiral invariance suggests the boundary term of the form
\ben
S_{\del V}=\int_{\del V} d\Sigma^\mu \left\{\frac{1}{2}n_\mu\bar{\psi}
U^{\gamma_5}\psi\right\}\label{Son2}
\een
with
\ben
U^{\gamma_5}=e^{i\gamma_5 \vec{\tau}\cdot \vec{\pi}/f}.
\een

As is very well-known by now,
the meson sector contains an extra term, the WZW term, which plays an
important role
in multi-flavor systems even for noninteracting bosons. This will be
derived below using the Cheshire Cat mechanism. (For a standard derivation,
see \cite{wittencomath}).  The outside
(meson sector) contribution to the action takes the form
\ben
S_{\tilde{V}}=\int_{\tilde{V}} d^2 x\frac{f^2}{4}\Tr (\del_\mu U \del^\mu
U^\dagger)+\cdots+ S_{WZW}. \label{Sout2}
\een
As seen, the main new ingredient in this action is the presence of the extra
term $S_{WZW}$ in the boson sector. This is the WZW term
which in the (1+1)-dimensional system we are interested in
takes the form
\ben
S^{\tilde{V}}_{WZW}=\frac{2\pi n}{24\pi^2}\int_{\tilde{V}\times
[0,1]} d^3x \epsilon^{ijk}\, \Tr(L_i L_j L_k)\label{wzwout}
\een
with an integer $n$ corresponding to the number of ``colors",
where $L_\mu=g^\dagger \del_\mu g$ with $g$ ``homotopically" extended as
$g(s=0,x)=1$, $g(s=1,x)=U(x)$. The action (\ref{wzwout}) is defined
in the partial space $\tilde{V}$. As we will see later, the WZW
term plays a crucial role in the skyrmion physics in (3+1) dimensions, so
the question we wish to address here is:
How does the Cheshire Cat manifest itself for the WZW term?
To answer this question, we imagine shrinking the bag to a point and ask
what remains from what corresponds to the WZW term coming from the space $V$
in which quarks live.
The obvious answer would be that the bag gives rise to the portion of the
WZW term complement to (\ref{wzwout}) so as to yield the total defined
in the full space. This is of course the right answer but it is not at
all obvious that the chiral bag theory as formulated does this properly.
For instance, consider the variation of the action of the meson sector
(\ref{wzwout})
\ben
\delta S^{\tilde{V}}_{WZW} &=& 3 \frac{n}{12\pi}\int_{\tilde{V}}
d^2 x \epsilon^{\mu\nu}\Tr ((g^\dagger\delta g)L_\mu L_\nu)\nonumber\\
&+& 3\frac{n}{12\pi}
\int_0^1 ds \int_{\del {\tilde{V}}} d\Sigma_\mu
\epsilon^{\mu\nu\alpha} \Tr ((g^\dagger\delta g) L_\nu L_\alpha).
\een
The second term depends on both the homotopy
extension and the surface. By the CCP, such a term is unphysical and
must be cancelled by a complementary contribution from the bag. Thus
the consistency of the model requires that the cancellation occur through
the boundary condition. We will now show this is exactly what one obtains
within the model so defined \cite{bagwz}.

Consider the action for the quark bag with the quark satisfying the
Dirac equation
\ben
i\not\!{\del} \psi=0\label{dirac2}
\een
subject to the boundary condition
\ben
(e^{i\theta(\beta,t)\gamma_5}-\not\!{n})\psi(\beta,t)=0. \label{bc2}
\een
where the bag wall is put at $x=\beta$ and $\theta\equiv \vec{\tau}\cdot
\vec{\pi}/f$. We will use the Weyl representation
for the $\gamma$ matrices as before. We will move the bag wall such that
the bag will shrink to zero radius. The general solution for (\ref{dirac2})
with (\ref{bc2}) is
\ben
\psi(x,t)={\rm exp} \left\{-i[\frac{1}{2}(1+\gamma_5)\theta_-
+\frac{1}{2}(1-\gamma_5)\theta_+]\right\}\psi_0 (x,t) \label{sol}
\een
where
$\theta_{\pm}=\theta (x\pm t)$ and $\psi_0$ is the solution for $\theta=0$
(for the MIT bag). The boundary condition is satisfied if
\ben
\theta_- (\beta,t)-\theta_+ (\beta,t)=\theta (\beta,t).
\een
Since we are dealing with a time-dependent boundary condition
\footnote{An alternative way is to make a local chiral rotation on the
quark field so as
to make the boundary condition time-independent and generate an induced
gauge potential (known as Berry potential or connection). This elegant
procedure is discussed below.}, it is more
convenient to use the path-integral method instead of a canonical formalism.
Let $\not\!{\del}_\theta$ denote the Euclidean Dirac
operator\footnote{For a short dictionary for translating quantities
in Minkowski metric to those in Euclidean metric and vice-versa, see
Ref.\cite{dictionary}.}
associated with the boundary condition (\ref{bc2}). Then integrating
the quark field in the bag action, we have
\ben
e^{-S_F}={\rm det}\left(\not\!{\del}_\theta /\not\!{\del}_0\right)
\een
or
\ben
S_F=-{\rm tr}\ {\rm ln} (\not\!{\del}_\theta \not\!{\del}_0^{-1})=
-\int_0^1 ds\, {\rm tr} (\del_s \not\!{\del}_\theta \not\!{\del}_\theta^{-1})
\label{sf}
\een
where we denote by $S_F$ the effective Euclidean action for the
bag. On the right-hand side of Eq.(\ref{sf}),
$\theta$ is homotopically extended as $\theta (x,t)\rightarrow
\theta (x,t,s)$ such that $\theta (x,t,s=1)=\theta (x,t)$ and
$\theta (x,t,s=0)=0$. The trace tr goes over the ``color",
the space-time, $\gamma$ matrices
and the flavor ($\tau$) matrices. Since the $\theta$ dependence appears only
in the boundary condition, we can use the latter to rewrite
\ben
S_F=-\int_0^1 ds\, {\rm tr}\ \left(\Delta_B \del_s[e^{i\theta\gamma_5}-
\not\!{n}] \not\!{\del}_\theta^{-1}\right)
\een
where $\Delta_B$ is the surface delta function.
This expression has a meaning only when it is suitably regularized. We use the
point-splitting regularization
\ben
S_F=-\lim_{\epsilon\rightarrow 0_+} \int_{V \times [0,1]} ds d^2x
\Delta_B {\rm tr}\ (\del_s e^{i\gamma_5 \theta}
G_ \theta (x,t+\epsilon;x,t-\epsilon))
\een
with $G$ the Dirac propagator inside the chiral bag
which can be written explicitly using the time-dependent solution (\ref{sol})
\ben
G_\theta (x,t;x^\prime,t^\prime) &=& \left(\begin{array}{cc}
e^{-i\theta_- (x-t)} & 0 \\ 0 & e^{-i\theta_+ (x+t)} \end{array}\right)
\nonumber\\
& & G_0 (x,t;x^\prime,t^\prime)\left(\begin{array}{cc} e^{i\theta_+ (x^\prime
+t^\prime)} & 0 \\ 0 & e^{i\theta_- (x^\prime-t^\prime)} \end{array}\right)
\een
where $G_0$ is the bagged-quark MIT propagator which has been computed
by multiple-reflection method \cite{hansjaffe}
\ben
\lim_{x,x^\prime\rightarrow \beta} G_0 (x,t+\epsilon;x^\prime,t-\epsilon)=
\frac{1}{4}(1+\gamma_\beta^1)(\gamma^0/4\pi\epsilon)(3-\gamma_\beta^1) +O(1)
\een
with $\gamma_\beta^1=\gamma^1\cdot n_\beta$ and $n_\beta=n_\pm =\pm 1$.
The WZW term arises from the imaginary part of the action,
so calculating ${\rm Im}\ S_F$ to leading order in the derivative
expansion, we get
\ben
{\rm Im}\, S_F=i\frac{2\pi n}{24\pi^2}\int_0^1 ds\int_V dx dt \Tr\,
(\del_s\theta\del_t \theta \del_x \theta) + \cdots
\een
where the trace Tr now runs only over the flavor indices and the ellipsis
stands for higher derivative terms. Invoking chiral symmetry and going to
the Minkowski space, we finally obtain
\ben
S_{WZW}^V=-i\frac{2\pi n}{24\pi^2}\int_{V\times [0,1]} d^3x
\epsilon^{\mu\nu\alpha}\Tr\, (L_\mu L_\nu L_\alpha)
\een
which is exactly the complement to the outside contribution (\ref{wzwout}).
Notice that there is no surface-dependent term whose presence would have
signaled the breakdown of the CCP.

So far our discussion has been a bit nonrigorous. A lot more rigorous
derivation was given by Falomir, M.A. Muschietti and E.M. Santangelo
\cite{bagwz} who computed the path integral
\ben
Z=\int {\cal D}\bar{\psi} {\cal D}\psi e^{-S_E}
\een
where
\ben
S_E=\int_V d^2x \bar{\psi}e^{\gamma_5 \theta}i\not\!{\del} e^{\gamma_5\theta}
\psi +\frac{1}{2}\int_{\del V} d^1 x \bar{\psi} (1+i\not\!{n})\psi.\label{se}
\een
A careful analysis showed indeed that there is no surface term in the
effective action, corroborating the above derivation: the model is fully
consistent with the CCP. Note that in (\ref{se}), the field
$\theta$ is defined in the interior of the volume $V$ whereas in the above
consideration all the {\it action} occurred on the surface. The physics
is equivalent, however, since our argument was based on ``shrinking" the bag
whereas in the formulation of these authors, the quarks are being plainly
integrated out from the bag. (The real part of the effective action gives
$S_V=\frac{f^2}{4}\int_V d^2x \Tr\ (\del_\mu U \del^\mu U^\dagger)+
{\rm higher\ derivative\ terms}$, just to complement the meson sector.)

To summarize the point in case the readers did not find the above
discussion illuminating, the Cheshire Cat argument provides a ``derivation"
of the WZW term in (1+1) dimensions. This is certainly not a very elegant
derivation but ties the anomaly issue naturally with non-anomalous
phenomena within the model. The same argument holds for (3+1) dimensions
as shown by Chen et al. \cite{bagwz}.
\subsubsection{\it A bit more mathematics of the Wess-Zumino-Witten term}
\bs
Here we give a slightly more rigorous explanation \cite{balachandran}
of the WZW term we have
derived. The argument can be easily extended to higher dimensions but the
situation in (1+1) dimensions is vastly simpler.

Consider a field $g$ valued in the group $SU(N)$ with $N\geq 2$
(in the case of two dimensions) which we can represent by $N\times N$
unitary matrices with unit determinant. The configuration space $Q$
is given by fields in one spatial dimension. If we impose the boundary
condition that $g\rightarrow 1$ at $x=\pm\infty$, so that at a fixed time,
the space is a circle $S^1$, then $Q$ is just the set of maps
\ben
Q: S^1 &\rightarrow& SU(N)\nonumber\\
x^1 &\rightarrow& g(x^1).
\een
We will consider the WZW term in the full space $\Omega=V\cup \tilde{V}$.
Now let $D$ be a disc in $Q$ parametrized by $x^0$ corresponding to
time and an $s$ with $0\leq s \leq 1$ such that the boundary $D$ is just the
space-time $\Omega$ (we are implicitly understanding that time is included
therein).
The field is then $g(x^0,x^1,s)$ chosen so that $g(x^0,x^1,0)=1$ and
$g(x^0,x^1.1)=g(x^1)=U$. Then the WZW term in Euclidean space is
\ben
S_{WZW}=\frac{2\pi n}{24\pi^2}\int_{D[\del D=\Omega\times [0,1]]}
\epsilon_{ijk}\ \Tr (g^\dagger\del_i g g^\dagger\del_j g g^\dagger\del_k
g).
\een
This action has the properties that the quantity $n$ is quantized to an
integer value and that the action is invariant under the infinitesimal
variation in the interior of the disc $D$. One can see the latter in various
ways. One way is to note that $S_{WZW}/(2\pi n)$ is just the charge of
the winding number current
\ben
J_\mu=\frac{1}{24\pi^2}\epsilon_{\mu\nu\alpha\beta}\Tr [g^\dagger
\del^\nu g g^\dagger\del^\alpha g g^\dagger\del^\beta g],
\een
namely
\ben
S_{WZW}/(2\pi n)=\int d^3x J_0 (x)
\een
which is just the conserved winding number. Another way of seeing this is
to note that $S_{WZW}$ is just an integral of a closed three form.
Now to show that $n$ is quantized, it suffices to notice that
$D\cup \bar{D}=S^2\times S^1$ (where $\bar{D}$ is the disc complement to
$D$) and that
\ben
\frac{2\pi n}{24\pi^2}\int_{S^2\times S^1}\ \Tr [g^\dagger dg]^3=2\pi \times
{\rm integer}.
\een
This proves the assertion. Similar quantization of the coefficient of the
WZW term in four dimensions will turn out to be crucial for skyrmion physics
later.

\subsection{Going to (3+1) Dimensions}
\bs
Since we cannot bosonize completely even a free Dirac theory in four dimensions
(not to mention highly nonlinear QCD), we will essentially follow step-by-step
the (1+1)-dimensional reasoning and construct in complete parallel
a simple, yet hopefully sufficiently realistic
chiral bag model which will then be {\it partially} bosonized.
The roles of CCP and bosonization will then be reversed here: We will
invoke CCP in bosonizing the chiral bag. Here again the boundary conditions
play a key role.
Ultimately we will motivate a modeling of QCD at low energies in terms
of effective chiral Lagrangian theory which we will use to describe
low-energy hadronic/nuclear processes.
\subsubsection{\it The model with $U_A (1)$ anomaly}
\bs
For generality, we consider the flavor group $SU(3)\times SU(3)\times U_A (1)$
appropriate to up, down and strange quark systems. The chiral
symmetry $SU(3)\times SU(3)$ is spontaneously broken down to $SU(3)$, with
the octet Goldstone bosons denoted by $\pi=\frac{\lambda^a}{2}\pi^a$ with
$\lambda$ the Gell-Mann matrices and the $U_A (1)$ symmetry is broken
by anomaly, with the associated massive
scalar denoted
by $\eta^\prime$. Vector mesons can be suitably introduced through hidden
gauge symmetry (HGS) but we will not use them here for simplicity. We do not
yet have a fully satisfactory theory of Cheshire Cat mechanism in the presence
of vector-meson degrees of freedom. We will
develop our strategy using a Lagrangian that contains only the octet
$\pi$ fields
and limit ourselves to the lowest-order derivative order. We will for the
moment consider the chiral limit with all quark masses equal to zero.
The symmetry breaking due to quark masses will be discussed later.

We consider quarks ``confined" within a volume $V$ which we take spherical
for definiteness, surrounded by the octet Goldstone bosons $\pi$
and the singlet $\eta^\prime$
populating the outside space of volume $\tilde{V}$. This is pictorially
represented in Figure I.1.
%
The action is again given by the three terms
\ben
S=S_V+S_{\tilde{V}}+S_{\del V}.
\een
Including gluons, the ``bag" action is given by
\ben
S_V=\int_V d^4x \left(\bar{\psi}i\not\!\!{D}\psi -\frac{1}{2} {\rm tr}\
F_{\mu\nu}F^{\mu\nu}\right)+\cdots
\een
where the trace goes over the color index.
The action for the meson sector must include the usual Goldstone bosons
as well as the massive $\eta^\prime$ taking the form \cite{jschechter},
{\ie},
\ben
S_{\tilde{V}}=\frac{f^2}{4}\int_{\tilde{V}} d^4x \left(\Tr\ \del_\mu U^\dagger
\del^\mu U -\frac{N_f}{4} m^2_{\eta^\prime}({\rm ln}U-{\rm ln}U^\dagger)^2
\right) +\cdots  + S_{WZW}
\een
where $N_f=3$ is the number of flavors and
\ben
U=e^{i\eta^\prime/f_0}e^{2i\pi/f},\\
f_0\equiv \sqrt{N_f/2} f. \nonumber
\een
The WZW term is the five-dimensional analog of the three-dimensional one
we encountered above \cite{witten83},
\ben
S_{WZW}=-N_c \frac{i}{240\pi^2}\int_{D_5[\del D_5=V\times [0,1]]}
\epsilon_{\mu\nu\lambda\rho\sigma} \Tr (L^\mu L^\nu L^\lambda L^\rho L^\sigma)
\een
with $L=g^\dagger (x,s) dg (x,s)$. We will come back to this WZW term later.
Now the surface action is nontrivial. It has the usual term
\ben
\frac{1}{2}\int_{\del V} d\Sigma^\mu (n_\mu \bar{\psi} U^{\gamma_5}\psi)
\label{sa1}
\een
with
\ben
U^{\gamma_5}=e^{i\eta^\prime\gamma_5/f_0}e^{2i\pi\gamma_5/f},
\een
but it has more because of what is called ``color anomaly", analogous to
what we had in two dimensions, which we need to analyze \cite{NRWZ1}.

The trouble with the boundary condition (\ref{sa1}) is that it is not
consistent with quantum anomaly effect and hence breaks color gauge invariance
at quantum level.
To see what happens, first we clarify what the color confinement is
at classical level.
With (\ref{sa1}), the gluons inside the bag are subject to the boundary
conditions
\ben
\hat{n}\cdot \vec{E}^a=0,\ \ \ \hat{n}\times \vec{B}^a=0\label{gluemit}
\een
with $\hat{n}$ the outward normal to the bag surface and $a$ the color
index. These are the celebrated MIT bag boundary conditions.
What these conditions
mean is as follows. A positive helicity fermion near the surface with its
momentum along the direction of the color magnetic field moves freely in the
lowest Landau orbit, because the energy $-|gB|/\omega$ (where $\omega$ is
its total energy and $g$ is the gluon coupling constant) resulting from its
magnetic moment cancels exactly its zero point energy $|gB|/\omega$. Since
the color electric field points along the surface, it cannot force color charge
to cross the bag wall and hence the color charge will be strictly confined
within.

We will now show that quantum mechanical anomaly induces the leakage of
color. The main cause is the boundary condition associated with the
$\eta^\prime$ field
\ben
\left\{i\gamma\cdot\hat{n}+e^{i\gamma_5 \eta (\beta)}\right\}\psi (\beta)
=0 \label{bceta}
\een
where as before $\beta$ is a point on the bag surface $\Sigma=\del V$ and
we define
\ben
\eta\equiv \eta^\prime/f_0.
\een
Consider now the color charge (indexed by the Roman superscript) inside the bag
\ben
Q^a=\int_V d^3x g(\psi^\dagger \frac{\lambda^a}{2}\psi +f^{abc}\vec{A}^b\cdot
\vec{E}^c).
\een
The color current is conserved inside the bag, so integrating over the
spatial volume of the divergence of the current (which is zero), we
have
\ben
\dot{Q}^a=-\oint_{\Sigma=\del V} d\beta \vec{j}^a\cdot \hat{n}\label{charget}
\een
where using temporal gauge, $A_0^a=0$,
\ben
\vec{j}^a=g\bar{\psi}\vec{\gamma}\frac{\lambda^a}{2}\psi+gf^{abc}(\vec{A}^b
\times \vec{B}^c).
\een
As mentioned above, classically there is no color flux leaking out, so
$\dot{Q}^a=0$. However as in the case of (1+1) dimensions encountered above,
the integrand on the right-hand side of
Eq.(\ref{charget}) is not well-defined without regularization. Let us
regularize it by time-splitting it as before. To simplify further our
discussion, we will make the quasi-abelian approximation. After the
calculation is completed, we can extract the non-abelian structure by
inspection. In this case, we can write
\ben
\vec{j}^a (\beta)=g \Tr [\frac{\lambda^a}{2}\vec{\gamma} S^+ (\beta_-;
\beta_+)]
\een
where the fermion propagator $S^+$ is defined as
\ben
S^+ (\beta,\beta^\prime)=\lim_{(x,x^\prime)\rightarrow (\beta,\beta^\prime)}
S(x,x^\prime)
\een
and
\ben
\beta_\pm\equiv \beta\pm \frac{\epsilon}{2}.
\een
Using (\ref{bceta}) for $\beta$ replaced by $\beta-{\epsilon}/{2}$ and
and the hermitian conjugate of it for $\beta+{\epsilon}/{2}$, we have
\ben
\dot{Q}^a=\frac{i}{2} g \lim_{\epsilon\rightarrow 0_+} \epsilon\oint_{\Sigma}
\Tr [\frac{\lambda^a}{2}\gamma_5\vec{\gamma}\cdot\hat{n} S^+
(\beta_-;\beta_+)] \dot{\eta}.
\een
Therefore if we take, for the moment, $\eta$ to
depend on time only (this assumption will be lifted later), we can write
\ben
\frac{dQ^a}{d\eta}=\frac{i}{2} g \lim_{\epsilon\rightarrow 0_+} \epsilon
\oint_{\Sigma}
\Tr [\frac{\lambda^a}{2}\gamma_5\vec{\gamma}\cdot\hat{n} S^+
(\beta_-;\beta_+)].\label{dqeta}
\een
Using the multiple reflection method in four dimensions \cite{goldjaffe},
in Euclidean space,
\ben
S^+ (\beta_-;\beta_+)=\frac{1}{4}(1+\gamma_E\cdot \hat{n} U^5) S(\beta_-;
\beta_+) (3-U^5\gamma_E\cdot\hat{n})\label{mr1}
\een
with $U^5=e^{i\gamma_5\eta}$, $\gamma_E=-i\gamma$, ${A_E}_i^a=-A_i^a$ for
$i=1,2,3$, ${A_E}_0^a=-iA_0^a$ and
\ben
S(\beta_-;\beta_+)\sim \int_{{\cal R}\times V} d^4x S_0 (\beta_--x)
(-ig\gamma_E^\alpha A_E^{b\alpha} (x)\cdot \frac{1}{2}\lambda^b) S_0
(x-\beta_+)\label{mr2}
\een
where ${\cal R}\times V$ is the Euclidean space-time
and $S_0$ is the free field Euclidean propagator
\ben
S_0 (x-x^\prime)=\frac{1}{2\pi^2}\frac{\gamma_E^\alpha (x-x^\prime)^\alpha}
{(x-x^\prime)^4}.
\een
Substituting (\ref{mr1}) and (\ref{mr2}) into (\ref{dqeta}) and returning
to Minkowski space, we arrive at
\ben
\frac{dQ^a}{d\eta}=-\frac{g^2}{4\pi^2}\epsilon^{0\mu\alpha\sigma}
\oint_{\Sigma} d\beta\int_{R\times V} d^4x \delta^4 (\beta-x)\del_\sigma
A_\alpha^a (x) n_\mu.
\een
In the quasi-abelian case that we are considering, we can write
\ben
\frac{dQ^a}{d\eta}=\frac{g^2}{8\pi^2}\int_\Sigma d\beta \vec{B}^a (\beta)\cdot
\hat{n}.\label{leaketa}
\een
Since $Q^a$ is covariant under small gauge transformations that are constant
on the bag surface, Eq.(\ref{leaketa}) equally holds for a nonabelian color
magnetic field. Furthermore, one can generalize (\ref{leaketa}) to a variation
with a space-time dependent $\eta (\vec{x},t)$ by replacing the normal
derivative by a functional derivative with it.

Roughly speaking, what (\ref{leaketa}) means is that color charge ``disappears"
from the bag following the time variation of the $\eta$ field. Coming
from the regularization, this is a quantum effect which signals a
disaster to the model {\it unless} it is cancelled by something else.
A simple remedy to this is to impose, in a complete analogy to the
(1+1)-dimensional Schwinger model considered above,
a boundary condition that cancels the leaking color charge
which can be effected in the Lagrangian
by a surface term of the type
\ben
{\cal L}_{CT}\sim -\frac{g^2 N_F}{8\pi^2}\oint_\Sigma d\beta A_0^a \vec{B}^a
\cdot\hat{n} \eta (\beta)\label{lcounter}
\een
where we have taken into account the number of light flavors involved.
Heavy-quark flavors are irrelevant to the issue \footnote{Why this is so is
easy to understand. When the fermions have a mass $m$, there is a gap between
the positive and negative energy Landau levels. Now if the time variation in
the $\eta$ field is large on the scale of $1/m$, then there is no net flow out
of the Dirac sea. In terms of anomaly, this is seen as follows. In the presence
of heavy fermion masses, the anomaly is formally nonzero. However this is
balanced by the term $2m\langle\bar{\psi}i\gamma_5\psi\rangle$ coming
from the explicit chiral symmetry breaking term in the Lagrangian.}.
Chiral invariance,
covariance and general gauge transformation properties allow us to rewrite
(\ref{lcounter}) in a general form
\ben
{\cal L}_{CT}=i\frac{g^2}{32\pi^2}\oint_\Sigma K_5^\mu n_\mu
(\Tr\ {\rm ln} U^\dagger-\Tr\ {\rm ln} U)\label{cs}
\een
where $K_5^\mu$ is the (properly regularized) ``Chern-Simons current"
\ben
K_5^\mu=\epsilon^{\mu\nu\alpha\beta} (A_\nu^a F_{\alpha\beta}^a-\frac{2}{3}
gf^{abc} A_\nu^a A_\alpha^b A_\beta^c).
\een
Note that (\ref{cs}) is invariant under neither large gauge transformation
(because of the Chern-Simons current) nor small gauge transformation
(because of the surface). Thus at classical level, the Lagrangian is not
gauge invariant. However at quantum level, it is, because of the cancellation
between the anomaly term and the surface term (\ref{cs}). A concise way to
see all this is to note that the surface term (\ref{sa1}), when regularized
in a gauge-invariant way, takes the form
\ben
-\frac{1}{2}\int d^4x \Delta_{\del V}\bar{\psi}(x+\epsilon/2) e^{i\gamma_5
\eta}{\cal P}e^{-ig\int_{x-\epsilon/2}^{x+\epsilon/2} A_\mu d\xi^\mu}
\psi (x-\epsilon/2)
\een
with the $\Delta_{\delta V}$ the surface $\delta$ function and
${\cal P}$ the path ordering operator. Taking the limit
of $\epsilon$ going to zero, one then recovers (\ref{sa1}) suitably
normal-ordered plus the Chern-Simons term (\ref{cs}). It may appear somewhat
bizarre that we have here a theory which violates gauge invariance classically
and becomes gauge invariant upon proper quantization. Usually it is the other
way around. But the point is that
when anomalies are involved as in this hybrid description, only quantum
theory makes sense.

In summary, the consistent Cheshire bag model is given by the action
\ben
S&=& S_V+S_{\tilde{V}}+S_{\del V},\label{cheshire}\\
S_V&=& \int_V d^4x \left(\bar{\psi}i\not\!\!{D}\psi -\frac{1}{2}
{\rm tr}\ F_{\mu\nu}F^{\mu\nu}\right)+\cdots\nonumber\\
S_{\tilde{V}}&=&\frac{f^2}{4}\int_{\tilde{V}} d^4x \left(\Tr\ \del_\mu
U^\dagger
\del^\mu U -\frac{N_f}{4} m^2_{\eta^\prime}({\rm ln}U-{\rm ln}U^\dagger)^2
\right) +\cdots  + S_{WZW},\nonumber\\
S_{\del V}&=&
\frac{1}{2}\int_{\del V} d\Sigma^\mu\left\{(n_\mu \bar{\psi} U^{\gamma_5}\psi)
+i\frac{g^2}{16\pi^2}{K_5}_\mu
(\Tr\ {\rm ln} U^\dagger-\Tr\ {\rm ln} U)\right\}.\nonumber
\een

\subsubsection{\it An application: $\etap$ mass}
\bs
The presence of the Chern-Simons term in the surface action affects at
classical level the boundary conditions satisfied by the gluon fields.
In place of the usual MIT conditions (\ref{gluemit}), we have instead
\ben
\hat{n}\cdot\vec{E}^a&=&-\frac{N_F g^2}{8\pi^2}\hat{n}\cdot\vec{B}^a\ \etap,
\label{newbc1}\\
\hat{n}\times \vec{B}^a &=& \frac{N_F g^2}{8\pi^2}\hat{n}\times
\vec{E}^a\ \etap \label{newbc2}.
\een
The fact that radial color electric fields can exist contrary to the MIT
conditions will most likely
have an important ramification on the so-called ``proton
spin" problem discussed later. But no work has been done so far on
this interesting issue. Here we will discuss how to do a calculation
analogous to
that of the scalar mass in the (1+1)-dimensional Schwinger model to
obtain the mass for the $\etap$ \cite{NRWZ2}.
As there  we shall make use of the $U_A (1)$ (or ABJ) anomaly.

Consider the situation where the $\etap$ excitation is a long wavelength
oscillation of zero frequency and let $V$ be a small volume in space.
The CCP states that it should not matter whether we describe the $\etap$
in $V$ in terms of QCD variables (quarks and gluons) or in terms of effective
variables ({\ie}, mesonic degrees of freedom). Since we are looking at the
static situation, $\dot{Q}_5=0$ where $Q_5$ is the gauge-invariant singlet
axial charge $Q_5=\int_V d^3x j_5^0 (x)$. The $U_A (1)$ (or ABJ) anomaly
translates into the statement that the flux of the singlet axial current
through the bag wall is just the axial anomaly
\ben
\oint_{\del V} \vec{j}_5\cdot\hat{n}
\approx \frac{N_F g^2}{4\pi^2}\int_V d^3 x \vec{E}^a\cdot \vec{B}^a (x).
\label{anom1}
\een
In terms of the bosonized variables, the flux is given by
\ben
\oint_{\del V} \vec{j}_5\cdot\hat{n}=-2\oint_{\del V}f_0 \vec{\nabla}
\etap\cdot\hat{n}=-2\int_V f_0\nabla^2\etap\label{anom2}
\een
where $f_0\equiv \sqrt{\frac{N_f}{2}}f$.
Note that unlike in the previous cases, here the bosonization is done
{\it inside} the volume $V$. Combining (\ref{anom1}) and (\ref{anom2}),
we obtain
\ben
f_0^2\nabla^2\etap (0,\vec{x})\approx -f_0\frac{N_F g^2}{8\pi^2}
\vec{E}^a\cdot\vec{B}^a (0,\vec{x}).\label{massform1}
\een
This is an approximate bosonization condition. The left-hand side of
(\ref{massform1}) is
\ben
\nabla^2\etap\approx m_{\etap}^2 \etap
\een
while the righ-hand side can be evaluated using (\ref{newbc1}) and
(\ref{newbc2})
\ben
\vec{E}^a\cdot\vec{B}^a \approx -\frac{N_F g^2}{8\pi^2} \frac{\etap}{f_0}
\left(1-\left\{\frac{N_F g^2}{8\pi^2}\right\}^{2}\frac{{\etap}^2}{f_0^2}
\right)^{-1}\frac{1}{2}  F^2 .
\een
Expanding to linear order in $\etap$, we obtain from (\ref{massform1})
\ben
m_{\etap}^2 f_0^2\approx 2\left(\frac{N_F g^2}{16\pi^2}\right)^2 \times
\langle F^2 \rangle_V.\label{massform2}
\een
This is the four-dimensional analog of (\ref{schmass}),
the scalar mass in the Schwinger model.
In this expression, the gluon condensate density is averaged over the
bag volume $V$.
A similar result was obtained by Novikov, Shifman, Vainshtein and
Zakharov \cite{novikov} by saturating
the functional integral with only self-dual gauge configurations ({\it i.e.},
instantons).
In obtaining (\ref{massform2}), we have been a bit cavalier
about the definition of the gluon coupling constant. In principle,
it should be a running coupling constant and it makes a difference
whether it is defined on
the surface as in Eqs.(\ref{newbc1}) and (\ref{newbc2}) or in the
volume as in Eq.(\ref{anom1}). We are using both in Eq.(\ref{massform2})
and one should distinguish them. If one puts in rough numerical factors for
the gluon condensate and appropriate running coupling constants
into (\ref{massform2}), we find that the $\etap$ mass comes out to be in the
range (315 MeV--1650 MeV) to be compared with the experimental value
958 MeV. There is a large uncertainty in this prediction for the reason that
it is difficult to pin down the relevant scales involved in the running
of the coupling constant. But the qualitative picture comes out right.
\subsection{Cheshire Cat as a Gauge Symmetry}
\bs
In a nut-shell, the Cheshire cat phenomenon means that the ``bag" in
the sense of the bag model is an artifact that has no strict physical meaning.
An extremely attractive way to view this is to interpret the ``bag" as a gauge
artifact in the sense of
gauge symmetry. Thus a description in terms of a particular ``bag" radius
can be interpreted as a particular gauge fixing. Therefore physics is
strictly equivalent for {\it all} gauge choices (``bag" radii) one may make.
This is a novel way of looking at the classical confinement
embodied by the MIT bag. Let us briefly discuss
this fascinating and elegant approach \cite{damgaard}.
The idea can be best
described in (1+1) dimensions where the Cheshire cat is exact.

Consider the case of massless fermions coupled to external vector
$\V_\mu$ and axial-vector $\A_\mu$ fields, the generating functional of which
(in Minkowski space) is
\be
Z[\V,\A]&=&\int [d\psi][d\bar{\psi}] e^{iS},\label{z1}\\
S&=& \int d^2x \bar{\psi}\gamma^\mu\left(i\del_\mu +\V_\mu +\A_\mu\gamma_5
\right)\psi. \nonumber
\ee
One can get from this functional the massless Thirring model if one adds
a term $\sim \V^\mu \V_\mu$ and integrate over the vector field. One can
get the massive Thirring model if one adds scalar and pseudoscalar sources in
addition. In this sense, the model is quite general. Now do the field
redefinition while {\it enlarging the space}
\be
\psi (x)=e^{i\theta (x)\gamma_5} \chi (x)\ , \ \ \ \bar{\psi} (x)=\bar{\chi}
(x)
e^{i\theta (x)\gamma_5}.\label{localch}
\ee
Since this is just a change of symbols, the generating functional is
unmodified
\be
Z[\V,\A]&=& \int [d\chi][d\bar{\chi}]\, J\, e^{iS^\prime},\label{z2}\\
S^\prime &=& \int d^2x \bar{\chi}\gamma^\mu\left(i\del_\mu +\V_\mu +\A_\mu
\gamma_5 -\del_\mu \theta (x)\gamma_5\right)\chi\nonumber
\ee
where $J$ is the Jacobian of the transformation which can be readily calculated
\be
J={\rm exp}\left\{i\int d^2x \left(\frac{1}{2\pi}\del_\mu \theta \del^\mu
\theta + \frac{1}{\pi} \epsilon^{\mu\nu}\V_\mu\del_\nu \theta -\frac{1}{\pi}
\A_\mu \del^\mu \theta\right)\right\}.
\ee
This nontrivial Jacobian arises due to the fact that the measure is not
invariant under local chiral transformation (\ref{localch})\cite{fujikawa}.
The important thing to note is that since {\it by definition} (\ref{z1})
and (\ref{z2}) are the same, the physics should not depend upon
$\theta (x)$. The latter is redundant.
Therefore modulo an infinite constant which requires a
gauge fixing as described below, we can just do a functional
integral over $\theta (x)$ in (\ref{z2}) without changing physics.
Doing the integral in the path integral amounts to elevating the redundant
$\theta (x)$ to a dynamical variable.
Now if we define a Lagrangian $\L^\prime$ by
\be
Z[\V,\A]&=&\int [d\chi][d\bar{\chi}] e^{i\int d^2x \L^\prime},\\
\L^\prime &=& \bar{\chi}\gamma^\mu \left(i\del_\mu +\V_\mu +\A_\mu\gamma_5
-\del_\mu \theta \gamma_5\right)\chi\nonumber\\
&& +\frac{1}{2\pi}\del_\mu \theta \del^\mu \theta +\frac{1}{\pi}
\epsilon^{\mu\nu} \V_\mu \del_\nu -\frac{1}{\pi} \A_\mu \del^\mu \theta.
\label{z3}
\ee
We note that there is a local gauge symmetry, namely, that (\ref{z3})
is invariant under the transformation
\be
\chi (x)\rightarrow e^{i\alpha (x) \gamma_5} \chi (x)\, ,\ \ \
\bar{\chi} (x)\rightarrow \bar{\chi} (x) e^{i\alpha (x) \gamma_5}\, ,
\ \ \ \theta (x)\rightarrow \theta (x)-\alpha (x)\label{gt}
\ee
provided of course the Jacobian is again taken into account. Thus by
introducing a new dynamical field $\theta$, we have gained a gauge symmetry
at the expense of enlarging the space.

In order to quantize the theory, we have to fix the gauge. This means that
we pick a $\theta$, say, by a gauge-fixing condition
\be
\Phi (\theta)=0.
\ee
Then following the standard Faddeev-Popov method \cite{refield},
we write
\be
Z[V,A]=\int [d\chi][d\bar{\chi}][d\theta] \delta (\Phi[\theta])
|{\rm det}(\frac{\delta \Phi}{\delta \theta})|e^{i\int d^2x \L^\prime}.
\ee
Note that if we choose $\Phi=\theta$, we recover the original generating
functional (\ref{z1}) which describes everything in terms of fermions.
In choosing the gauge condition relevant to the physics of the chiral bag,
we recall that the axial current plays an essential role. Damgaard et al.
\cite{damgaard} choose the following gauge fixing condition
\be
\Phi[\theta,\chi,\bar{\chi}]=\Delta \int_{x_0}^x d\eta^\nu \bar{\chi} (\eta)
\gamma_\nu\gamma_5\chi (\eta) +(1-\Delta)\frac{1}{\pi}\theta (x) =0\label{gf}
\ee
which corresponds to saying that the total axial current consists of the
fraction $\Delta$ of the bosonic piece and the fraction $(1-\Delta)$ of
the fermionic piece. This may be called ``Cheshire Cat gauge".
We will not dwell on the proof which is rather technical but just mention
that the functional integral does not depend upon the path of this
line integral. The Faddeev-Popov determinant corresponding to
(\ref{gf}) is
\be
{\rm det}\left(\frac{\delta\Phi}{\delta\alpha}\right)={\rm det}\left(-\frac{1}
{\pi}\right).\label{fp}
\ee
To see this write the constraint
\be
\delta (\Phi[\theta,\bar{\chi},\chi])&=&\int [db] e^{i\int d^2x\, b\, \Phi}
\nonumber\\
&=& \int [db] e^{i\int d^2x\  \L_{g.f.}}\label{gaugefix}
\ee
defining the gauge-fixing Lagrangian $\L_{g.f.}$.
Now do the (infinitesimal)
gauge transformation (\ref{gt}) on (\ref{gaugefix}). We have
\be
\delta_{\alpha} \L_{g.f.}= -\frac{1}{\pi} \{(1-\Delta) +\Delta\}\alpha
\ee
where the first term in curly bracket comes from the shift in $\theta$
and the second term from the Fujikawa measure.  Now the right-hand side
of this equation is just $b\delta_\alpha \Phi$, from which follows
equation (\ref{fp}).

The Cheshire Cat structure is manifested in the choice of the $\Delta$.
If one takes $\Delta=0$, we have the pure fermion theory as one can see
trivially. If one takes instead $\Delta=1$, one obtains after some nontrivial
calculation (the detail of which is omitted here and can be found
in the paper by Damgaard et al.\cite{damgaard2})
\be
Z[\V,\A]= {\rm const.}\times \int [d\theta] e^{i\int d^2x \L^{\theta}}
\ee
with
\be
\L^{\theta}=\frac{1}{2}\del_\mu \theta \del^\mu \theta -\frac{1}{\sqrt{\pi}}
\epsilon^{\mu\nu} \V_\mu \del_\nu \theta +\frac{1}{\sqrt{\pi}}\A_\mu\del^\mu
\theta.
\ee
This can be recognized as the bosonized form of the fermion theory coupled
to the vector fields. The Cheshire Cat statement is that the theory
is identical for {\it any} arbitrary value of $\Delta$.

One can go further and take $\Delta$ to be a local function. In this case,
one can continuously change representation from one region of space-time
to another, choosing different gauge fixing conditions in different space-time
regions. This leads to a ``smooth Cheshire bag." The standard chiral bag model
\cite{brcom88} corresponds to the gauge choice
\be
\Delta (x)=\Theta ({\bf x} -{\bf z})
\ee
with ${\bf z}=\hat{r}R$. The usual bag boundary conditions employed
in phenomenological studies arise after certain suitable averaging
over part of the space and hence are a specific realization of the
Cheshire Cat scheme. Possible discontinuities or anomalies caused by
sharp boundary conditions often used in actual calculations
should not be considered to be the defects of the
effective model. An interesting conclusion of this point of view is
that one can in principle construct an {\it exact} Cheshire cat model
without knowing the exact bosonized version of the theory considered.

\section{Induced (Berry) Gauge Fields in Hadrons}
\bs
In this section, we introduce another aspect of the Cheshire Cat mechanism
which generalizes the concept to excitations. So far our focus has been on the
ground-state properties. That topological quantities satisfy the CCP exactly
even in (3+1) dimensions (such as the baryon charge) is
not surprising. This is a ground-state property associated with symmetry.
It is a different matter when one wants to establish
an approximate CCP for dynamical properties such as excitations, responses
to external fields etc. To establish that certain response functions can
be formulated in terms of effective variables, it proves to be highly
fruitful to introduce and exploit the notion of induced gauge fields
familiar in other areas of physics \cite{shapere}.
In this section, we discuss
how a hierarchy of ``vector-field" degrees of freedom can be {\it induced} and
how they can lead to
a natural setting for describing {\it excited states and their
dynamic properties}\cite{LNRZannals}. Since the resulting structure is quite
generic, we will spend some time discussing simpler quantum mechanical systems.
When the dust settles, we will see that much of the arguments used for those
systems can be applied with little modifications to hadronic systems.
\subsection{``Magnetic Monopoles" in Flavor Space}
\bs
A useful concept in understanding baryon excitations is the concept of
induced magnetic monopoles and instantons, {\viz}, topological objects, in
order-parameter or in our case flavor space. First we consider a toy
model, the quantum mechanical spin-solenoid
system \`{a} la Stone \cite{stone}
which succinctly illustrates the emergence of a Berry phase.

\subsubsection{\it A Toy Model: (0+1) Dimensional Field Theory}
\bs

Consider a system of a slowly rotating solenoid coupled to a fast
spinning object (call it ``electron") described by the (Euclidean) action
\be
S_E=\int dt \left(\frac{{\cal I}}{2}\dot{\vec{n}}^2+\psi^\dagger
(\del_t-\mu\hat{n}\cdot\vec{\sigma})\psi \right)\label{euaction}
\ee
where $n^a(t)$, $a$=1,2,3, is the rotator with $\vec{n}^2=1$, ${\cal I}$
its  moment of inertia, $\psi$
the spinning object (``electron") and $\mu$ a constant. We will assume
that $\mu$ is large so that we can make an adiabatic approximation
in treating the slow-fast degrees of freedom. We wish to calculate the
partition function
\be
Z=\int [d\vec{n}][d\psi][d\psi^\dagger] \delta (\vec{n}^2-1) e^{-S_E}
\ee
by integrating out the fast degree of freedom $\psi$ and $\psi^\dagger$.
Formally this yields the familiar fermion determinant, the evaluation of which
is tantamount to doing the physics of the system. In adiabatic approximation,
this can be carried out
as follows which brings out the essence of the method useful for handling
the complicated situations which will interest us later.

Imagine that $\vec{n} (t)$ rotates slowly. At each instant $t=\tau$,
we have an instantaneous Hamiltonian $H(\tau)$ which in our case
is just $-\mu \vec{\sigma}\cdot \hat{n} (\tau)$ and the ``snap-shot"
electron state $|\psi_0 (\tau)\rangle$ satisfying
\be
H(\tau)|\psi^0 (\tau)\rangle=\epsilon (\tau)|\psi^0 (\tau)\rangle.
\ee
In terms of these ``snap-shot" wave functions, the solution of the
time-dependent Schr\"{o}dinger equation
\be
i\del_t |\psi (t)\rangle=H(t)|\psi (t)\rangle \label{sequation}
\ee
is
\be
|\psi (t)\rangle=e^{i\gamma(t)-i\int_0^t \epsilon (t^\prime)dt^\prime}
|\psi^0 (t)\rangle.
\ee
Note that this has, in addition to the usual dynamical phase involving the
energy $\epsilon(t)$, a nontrivial phase $\gamma (t)$
which substituted into (\ref{sequation}) is seen to satisfy
\be
i\frac{d\gamma}{dt}+\langle \psi^0|\frac{d}{dt}\psi^0\rangle=0.
\ee
This allows us to do the fermion path integrals to the leading order
in adiabaticity and to obtain (dropping
the trivial dynamical phase involving $\epsilon$)
\be
Z=const\int [d\vec{n}]\delta (\vec{n}^2-1) e^{-S^{eff}}, \label{z}\\
S^{eff} (\vec{n})=\int \L^{eff}=\int [\frac{{\cal I}}{2}
\dot{\vec{n}}^2-i\vec{\A} (\vec{n})\cdot \dot{\vec{n}}] dt \nonumber
\ee
where
\be
\vec{\A} (\vec{n})=\langle \psi^0 (\vec{n})|i\frac{\del}{\del\vec{n}}
\psi^0 (\vec{n})\rangle \label{bpot}
\ee
in terms of which $\gamma$ is
\be
\gamma=\int \vec{\A}\cdot d\vec{n}.\label{berryphase}
\ee
$\A$ so defined is known as Berry potential or connection and $\gamma$
is known as Berry phase~\cite{berry}. That $\A$ is a gauge field
with coordinates defined by $\vec{n}$ can be seen
as follows. Under the transformation
\be
\psi^0 &\rightarrow& e^{i\alpha (\vec{n})}\psi^0, \label{gtpsi}
\ee
$\vec{\A}$ transforms as
\be
\vec{\A} &\rightarrow& \vec{\A}-\frac{\del}{\del\vec{n}} \alpha (\vec{n})
\label{gta}
\ee
which is just a gauge transformation.
The theory is gauge-invariant in the sense that under
the transformation (\ref{gta}), the theory (\ref{z}) remains unchanged.
(We are assuming that the surface term can be dropped.)

One should note that the gauge field we have here is first of all
defined in ``order parameter space", not in real space like electroweak field
or gluon field and secondly it is induced when fast degrees of freedom are
integrated out. This is a highly generic feature we will encounter time and
again. Later on we will see that the space on which
the gauge structure emerges is usually the flavor space like isospin or
hypercharge space in various dimensions.

We shall now calculate the explicit form of the potential $\A$. For this
let us use the polar coordinate and parametrize the solenoid as
\be
\vec{n}=(\sin\theta \cos\phi, \sin\theta \sin\phi, \cos\theta)
\ee
with the Euler angles $\theta (t)$ and $\phi (t)$ assumed to be slowly changing
(slow compared with the scale defined by the fermion mass $\mu$)
 as a function of time. Then the relevant Hamiltonian can be written as
\be
\delta H=-\mu \vec{\sigma}\cdot\hat{n} (t)=S(t)\delta H_0 S^{-1} (t),\\
\delta H_0 \equiv -\mu \sigma_3 \nonumber
\ee
with
\be
S(\vec{n} (t))=\left(\matrix{\cos \frac{\theta}{2} &-\sin \frac{\theta}{2}
e^{-i\phi}\cr \sin\frac{\theta}{2} e^{i\phi} &\cos \frac{\theta}{2}\cr}
\right).
\ee
Since the eigenstates of $\delta H_0$ are $\left(\matrix{1\cr 0\cr}\right)$
with eigenvalue $-\mu$ and $\left(\matrix{0\cr 1\cr}\right)$ with
eigenvalue $+\mu$, we can write the ``snap-shot" eigenstate of $H(t)$
as
\be
|\psi^0_{+\uparrow}\rangle=S\left(\matrix{1\cr 0\cr}\right)=
\left(\matrix{\cos \frac{\theta}{2} \cr\sin \frac{\theta}{2} e^{i\phi}}
\right) \label{ws}
\ee
where the arrow in the subscript denotes the ``spin-up" eigenstate
of $\delta H_0$ and $+$ denotes the upper hemisphere to be specified
below. The eigenstate $|\psi^0_{+\downarrow}\rangle$ is similarly defined
with the ``down spin". Now note that for $\theta=\pi$, (\ref{ws}) depends
on $\phi$ which is undefined. This means that (\ref{ws}) is ill-defined
in the lower hemisphere with a string singularity along $\theta=\pi$.
On the other hand, (\ref{ws}) is well-defined for $\theta=0$ and hence
in the upper hemisphere. The meaning of the $+$ in (\ref{ws}) is that
it has meaning only in the upper hemisphere, thus the name ``wave section"
referring to it rather than wave function.

Given (\ref{ws}), we can use the definition (\ref{bpot}) for the Berry
potential to obtain
\be
\vec{\A}_+ (\vec{n})\cdot d\vec{n}= \langle \uparrow|S^{-1}dS|\uparrow\rangle
=\frac{-1}{2}(1-\cos\theta)d\phi\label{monopole}
\ee
written here in one-form. The explicit form of the potential is
\be
\vec{\A}_+=-\frac{1/2}{(1+\cos\theta)} (-\sin\theta \sin\phi, \sin\theta
\cos\phi, 0)
\ee
which is singular at $\theta=\pi$ as mentioned above. This is the well-known
Dirac string singularity. Since we have the gauge freedom, we are allowed to
do a gauge transformation
\be
\psi^0_+\rightarrow e^{-\phi}\psi^0_+ \equiv \psi^0_-
\ee
which is equivalent to defining a gauge potential regular in the
lower hemisphere (denoted with the subscript $-$)
\be
\vec{\A}_{-}\cdot d\vec{n}=\vec{\A}_+ \cdot d\vec{n}+d\phi \label{GT}
\ee
giving
\be
\vec{\A}_-\cdot d\vec{n}=\frac 12 (1+\cos\theta)d\phi. \label{wsl}
\ee
This potential has a singularity at $\theta=0$. Thus we have gauge-transformed
the Dirac string from the lower hemisphere to the upper hemisphere. This
clearly shows that the string is an artifact and is unphysical. In other
words, physics should not be dependent on the string. Indeed the field
strength tensor, given in terms of the wedge symbol and forms,
\be
{\cal F}=d\A=\frac 12 d\theta\wedge d\phi=\frac 12 d(Area) \label{F}
\ee
is perfectly well-defined in both hemispheres and unique. A remarkable
fact here is that the gauge potential or more properly the field tensor
is completely independent of the fermion ``mass" $\mu$. This means that
the potential does not depend upon how fast the fast object is once it
is decoupled adiabatically. This indicates that the result may be valid
even if the fast-slow distinction is not clear-cut. We will come back
to this matter in connection with applications to the excitation
spectra of the light-quark baryons.

We now explain the quantization rule. For this consider a cyclic path.
We will imagine that
the solenoid is rotated from $t=0$ to $t=T$ with large $T$ such that
the parameter $\vec{n}$ satisfies $\vec{n} (0)=\vec{n} (T)$. We are thus
dealing with an evolution, with the trajectory of $\vec{n}$ defining a circle
$C$. The parameter space manifold is two-sphere $S^2$ since $\vec{n}^2=1$.
Call the upper hemisphere $D$ and the lower hemisphere $\bar{D}$ whose boundary
is the circle $C$, \ie, $\del D=C$. Then using Stoke's theorem, we have
from (\ref{berryphase}) for a cyclic evolution $\Gamma$
\be
\gamma (\Gamma)=\int_{C=\del D} \vec{\A}\cdot d\vec{n}
\equiv \int_{\del D=C} \A=-\int_D d\A=-\int_D {\cal F}.
\ee
Since the gauge field in $\bar{D}$ is related to that in $D$ by
a gauge transform, we could equally well write $\gamma$ in terms of
the former. Thus we deduce that
\be
e^{i\int_C \A}=e^{i\int_D {\cal F}}=e^{-i\int_{\bar{D}} {\cal F}}
\ee
which implies
\be
e^{i\int_{D+\bar{D}=S^2}{\cal F}}=1.
\ee
Thus we get the quantization condition
\be
\int_{S^2}{\cal F}=2\pi n \label{dirac}
\ee
with $n$ an integer. This just means that the total ``magnetic flux" going
through the surface is quantized. Since in our case the field strength
is given by (\ref{F}), our system corresponds to $n=1$ corresponding
to a ``monopole charge" $g=1/2$ located at the center of the sphere.
What we learn from the above exercise is that
consistency with quantum mechanics demands that it be a multiple of
1/2. Otherwise, the theory makes no sense. It will turn out later that
real systems in the strong interactions involve nonabelian gauge fields
which do not require such ``charge" quantization, making the consideration
somewhat more delicate.

We should point out another way of looking at the action (\ref{euaction})
which is useful for understanding the emergence of Berry potentials in
more complex systems. Since
\be
\hat{n} (t)\cdot \vec{\sigma}= S(t) \sigma_3 S^{-1} (t)
\ee
we make the field redefinition
\be
\psi\rightarrow \psi^\prime=S^{-1}(t)\psi.
\ee
Then (\ref{euaction}) can be rewritten as
\be
S_E=S_E^0 +\int dt \psi^\dagger (S^{-1} \del_t S)\psi.
\ee
Here the first term contains no coupling between the fast and slow
degrees of freedom. The second term linear in time derivative generates
the Berry structure analyzed above. We will encounter this structure in
complex systems.

\subsubsection{\it Connection to the WZW term}
\bs

The key point which will be found useful later is this: when the fast degree
of freedom (the ``electron") is integrated out, we wind up with a gauge
field as a relic of the fast degree of freedom that is integrated out.
The effective Lagrangian that results has the form (in Minkowski space)
\be
L^{eff}=\frac 12 \I \dot{\vec{n}}^2+{\vec{\A}} (\vec{n})\cdot \label{Leff}
\dot{\vec{n}}.
\ee
We now show that the Berry structure is closely related to what is
called the Wess-Zumino-Witten term defined in two dimensions. For this,
we recall that the gauge field in (\ref{Leff}) is an induced one, coming
out of the solenoid $\vec{n}$. Therefore one should be able to rewrite
the second term of (\ref{Leff}) in terms of $\vec{n}$ alone. However
this cannot be done {\it locally} because of the Dirac singularity
mentioned above but
the corresponding action can be written locally in terms of $\vec{n}$ by
extending (homotopically) to one dimension higher. This is the
(one-dimensional) WZW term.
When written in this way, the gauge structure will be
``hidden" in some sense.

Let us look at the action
\be
S_{WZ}\equiv \int  \vec{\A} (\vec{n})\cdot \dot{\vec{n}} dt. \label{wz}
\ee
We have already shown how to express this in a local form. We repeat here
to bring the point home. The general procedure goes as follows. First
extend the space from the
physical dimension $d$ which is 1 in our case to $d+1$ dimension.
This extension is possible (``no obstruction") if
\be
\pi_d (\M)=0,\ \ \ \pi_{d+1} (\M) \neq 0
\ee
where $\pi$ is the homotopy group and $\M$ is the parameter space manifold.
In our case
\be
\pi_1 (S^2)=0,\ \ \ \pi_2 (S^2)=Z.
\ee
So it is fine. We therefore
extend the space to $\tilde{n} (s)$, $0\leq s \leq 1$ such that
\be
\tilde{n} (s=0)=1,\ \ \ \tilde{n} (s=1)=\vec{n}.
\ee
Now the next step is to construct a winding number density $\tilde{Q}$
for $\pi_{d+1}$ which is then to be integrated over a region of $\M$
($\approx S^2$ here) bounded by d-dimensional fields $n^i$. The winding
number density $\tilde{Q} (x)$ (say $x_1=t$ and $x_2=s$) is
\be
\tilde{Q}=\frac{1}{8\pi}\epsilon^{ij} \epsilon^{abc}\tilde{n}^a
\del_i \tilde{n}^b \del_j \tilde{n}^c
\ee
and the winding number $n$ (which is 1 for $\pi\leq \theta \leq 0$,
$2\pi\leq \phi \leq 0$) \footnote{This can be calculated as follows.
The surface element is $$d\Sigma^i=\left(\frac{\del\vec{n}}{\del s} ds\times
\frac{\del\vec{n}}{\del t} dt\right)^i=\frac 12 \epsilon^{ab}\epsilon^{ijk}
\frac{\del n^j}{\del x^a}\frac{\del n^k}{\del x^b} d^2 x$$ and hence
$$\int_{S^2} \vec{n}\cdot d\vec{\Sigma}=4\pi.$$}
\be
n=\int_{S^2} d^2 x \tilde{Q} (x).
\ee
Comparing with (\ref{dirac}), we deduce
\be
S_{WZ}=4\pi g  \int_{{{\D}}=t\times [0,1]} d^2 x \tilde{Q} (x).\label{wzp}
\ee
This is the familiar form of the WZW  term defined in two dimensions.
As we saw before this has a ``monopole charge" $g=1/2$. Below we will
carry $g$ as an integral multiple of 1/2.
This way of understanding the WZW term complements the approach given
in subsection 2.1.3 and brings out the universal characteristic of Berry
phases.
\subsubsection{\it Quantization}
\bs

There are numerous ways of quantizing the effective action (hereon we will
work in Minkowski space)
\be
S^{eff}=S_0 +S_{WZ}
\ee
where the Wess-Zumino action is given by (\ref{wzp}) and
\be
S_0=\oint dt\, \frac{\I}{2}\, \dot{\vec{n}}^2.
\ee
We will consider the time compactified as defined above, so the
time integral is written as a loop integral. We will choose one way
\cite{rabino} which illustrates other interesting properties.

The manifold has $SO(3)$ invariance corresponding to  $\sum_{i=1}^3 n_i^2=1$.
Consider now the complex doublet $z\in SU(2)$
\be
z(t)=\left(\matrix{z_1\cr z_2\cr}\right),\\
z^\dagger z=|z_1|^2+|z_2|^2=1.
\ee
Then we can write
\be
n_i=z^\dagger\sigma_i z
\ee
with $\sigma_i$ the Pauli matrices. There is a redundant (gauge)
degree of freedom since under the $U(1)$ transformation $z\rightarrow
e^{i\alpha}z$, $n_i$ remains invariant. This is as it should be since
the manifold is topologically $S^2$ and hence corresponds to the coset
$SU(2)/U(1)$. We are going to exploit this $U(1)$ gauge symmetry to quantize
our effective theory.

Define a 2-by-2 matrix $h$
\be
h=\left(\matrix{z_1&-z_2^\star\cr z_2&z_1^\star}\right)
\ee
and
\be
a(t)=\sum_{k=1}^2 \frac i2 (z_k^\star\stackrel{\leftrightarrow}{\del}_k z_k).
\ee
Then it is easy to obtain (setting $\I=1$) that
\be
S_0=\frac 12 \oint\! dt {\rm Tr} \left[\stackrel{\rightarrow}{D}_t^\dagger
h^\dagger h\stackrel{\leftarrow}{D}_t\right]
\ee
where
\be
\stackrel{\rightarrow}{D}_t=\stackrel{\rightarrow}{\del}_t
-ia\sigma_3.
\ee
Let
\be
\tilde{a}_\mu=\frac i2 (\tilde{z}_k^\star\stackrel{\leftrightarrow}{\del}_\mu
\tilde{z}_k)
\ee
where the index $\mu$ runs over the extended coordinate $(s,t)$. As noted
before, there is no topological obstruction to this extension. This is
defined in such a way that $\tilde{z} (s=0)=1$
and $\tilde{z} (s=1)=z$. Then the Wess-Zumino action can be written
in a Chern-Simons form
\be
S_{WZ}=2g\int_{{\D}}d^2x \epsilon^{\mu\nu}\del_\mu\tilde{a}_\nu
=2g\oint\! dt a(t).
\ee
Introducing an auxiliary function $A$, we can write the partition function
\be
Z=\int [dz][dz^\dagger][dA] exp\ i\oint dt\left(\frac 12 \Tr (\stackrel
{\rightarrow}{\del}_t +iA\sigma_3)h^\dagger h (\stackrel{\leftarrow}
{\del}_t-iA\sigma_3)+2gA+\frac 14 (2g)^2\right). \label{ZP}
\ee
The $U(1)$ gauge invariance is manifest in this action. Indeed if
we make the (local) transformation $h\rightarrow e^{i\alpha (t)
\frac{\sigma_3}{2}h}$
and $A\rightarrow A-\frac 12 \del_t \alpha (t)$ with the boundary condition
$\alpha (T)-\alpha (0)=4\pi N$ where $N$ is an integer, then the
action remains invariant. This means that we have to gauge-fix the ``gauge
field" $A$ in the path integral. The natural gauge choice is the
``temporal gauge" $A=0$. The resulting gauge-fixed action is $\oint \L_{gf}$
with
\be
\L_{gf}=\Tr (\del_t h^\dagger \del_t h)+ g^2 \label{lag}.
\ee
Since there is no time derivative of $A$ in (\ref{ZP}), there is a
Gauss' law constraint which is obtained by taking
$\frac{\delta S}{\delta A}|_{A=0}$ from the action (\ref{ZP})
{\it before} gauge fixing:
\be
\frac i2\Tr \left(\sigma_3h^\dagger\del_t h-\del_t h^\dagger h \sigma_3\right)
+2g=0 \nonumber
\ee
which is
\be
i\Tr\left(\del_t h^\dagger h \frac{\sigma_3}{2}\right)=g.
\ee
The left-hand side is identified as the right rotation around third axis
$J_3^R$, so the constraint is that
\be
J_3^R=g.
\ee
Since (\ref{lag}) is invariant under $SU(2)_{L,R}$ multiplication, we have
that
\be
\vec{J}^2=\vec{J}_L^2=\vec{J}_R^2.
\ee
The Hamiltonian is (restoring the moment of inertia $\I$)
\be
H=\frac{1}{2\I} \left(\vec{J}^2-g^2\right)\label{Hmonopole}
\ee
which has the spectrum of a tilted symmetric top. Now adding the energy of
the ``electron", the total energy is
\be
E=\epsilon + \frac{1}{2\I}\left(J(J+1)-g^2\right)
\ee
with the allowed values for $J$
\be
J=|g|, |g|+1, \cdots.
\ee
The rotational spectrum is the well-known Dirac monopole spectrum.
Later we will derive an analogous formula for real systems in four
dimensions. The corresponding wavefunction is given by monopole harmonics.
\subsubsection*{\it Summary}
\bs

When a fast spinning object coupled to slowly rotating object is integrated
out, a Berry potential arises gauge-coupled to the rotor. The effect of this
gauge coupling is to ``tilt" the angular momentum of the rotor in the spectrum
of a symmetric top. It supplies an extra component to the angular
momentum, along the third direction. The gauge field is abelian and has
an abelian (Dirac) monopole structure. The abelian nature is inherited from
one nondegenerate level crossing another nondegenerate
level. When degenerate levels cross, the gauge field can be nonabelian and
this is the generic feature we will encounter in strong interaction physics.
\subsection{Induced Nonabelian Gauge Field}
\subsubsection{\it Diatomic Molecules in Born-Oppenheimer approximation}
\bs
In hadronic systems we will study below, we typically
encounter nonabelian induced gauge potentials. This is because degeneracy
is present. In order to understand this situation, we first
discuss here a case in which a nonabelian
gauge structure arises in a relatively simple quantum mechanical system.
To do so,  we will study a simple toy-model example of the
induced nonabelian gauge fields and Berry phases
in the Born-Oppenheimer approximation\cite{zygel}.
When suitably
implemented, the treatment can be applied to a realistic description
of the spectrum of a diatomic molecule, wherein
this approximation is usually described as a separation of slow (nuclear)
and fast (electronic) degrees of freedom. This separation is motivated by the
fact that the rotation of the nuclei does not cause the  transitions
between the electron levels. In other words, the splittings between the fast
variables are much larger than the splittings between the slow ones.
We will demonstrate how the integrating-out of the fast degrees of freedom
generates in the slow-variable space an induced vector potential of the
Dirac monopole type in certain special situations.

Before we develop the main argument, we should mention one point which may not
be clear at each stage of the development but should be kept in mind. In
the strong interactions at low energy, it is not always easy to delineate
the ``fast" and ``slow" degrees of freedom. In particular in the chiral
(light-quark) systems
that we shall consider later, it is not even clear whether it makes sense
to  make the distinction. Nonetheless
we will see that once the delineation is made, whether it is sharp or not,
the result does not depend on how good the distinction is. This was already
clear from the quantum mechanics of the solenoid-spin system we discussed
above where the final result had no memory of how strong the coupling of the
spin to the solenoid -- namely, the coefficient $\mu$ -- was.
In a later section where we deal with heavy-quark systems, we will see that
there
the concept developed here applies much better than in light-quark systems.

Let us define a generic Hamiltonian describing a system consisting of the
slow (``nuclear") variables $\vec{R} (t)$ (with $\vec{P}$ as conjugate
momenta) and the fast (``electronic") variables $\vec{r}$
(with $\vec{p}$ as conjugate momenta) coupled through a potential
$V(\vec{R},\vec{r})$
\be
H=\frac{\vec{P}^2}{2M} + \frac{\vec{p}^2}{2m} + V(\vec{R},\vec{r})
\ee
where we have reserved the capitals for the slow variables and lower-case
letters
for the fast variables. We expect the electronic levels to be stationary
under the adiabatic (slow) rotation of the nuclei. We split therefore the
Hamiltonian into the slow and fast parts,
\be
H&=& \frac{\vec{P}^2}{2M} + h \nonumber \\
h(\vec{R})&=& \frac{\vec{p}^2}{2m} + V(\vec{r},\vec{R})
\ee
where the fast Hamiltonian $h$ depends {\em parametrically}
on the slow variable $\vec{R}$. The snapshot Hamiltonian (for fixed $\vec{R}$)
leads to the Schr\"{o}dinger equation:
\be
h\phi_n(\vec{r},\vec{R}) = \epsilon_n(\vec{R})\phi_n(\vec{r},\vec{R})
\ee
with electronic states labeled by the set of quantum numbers $n$.
The wave function for the whole system is
\be
\Psi(\vec{r},\vec{R})=\sum_n \Phi_n(\vec{R})\phi_n(\vec{r},\vec{R}).\label{sum}
\ee
Substituting the wave function into the full Hamiltonian and using the equation
for the fast variables we get
\be
\sum_n\left[\frac{\vec{P}^2}{2M}
+\epsilon_n(\vec{R})\right]\Phi_n(\vec{R})\phi_n(\vec{r},\vec{R})=
E\sum_n\Phi_n(\vec{R})\phi_n(\vec{r},\vec{R})
\ee
where $E$ is the energy of the whole system.  Note that the operator of the
kinetic energy of the slow variables acts on {\em both} slow and fast
parts of the wavefunction. We can now ``integrate out"
the fast degrees of freedom.
A bit of algebra leads to the following effective Schr\"{o}dinger equation
\be
\sum_m H_{nm}^{eff} \Phi_m = E \Phi_n
\ee
where the explicit form of the matrix-valued Hamiltonian (with respect to
the fast eigenvectors) is
\be
H_{mn}^{eff} = \frac{1}{2M}\sum_k \vec{\Pi}_{nk}\vec{\Pi}_{km} +
 \epsilon_n\delta_{nm}
\ee
where
\be
\vec{\Pi}_{nm}= \delta_{nm}\vec{P}-
 i<\phi_n(\vec{r},\vec{R})|\vec{\nabla}_R|\phi_m(\vec{r},\vec{R})>\equiv
\delta_{mn}\vec{P}-\vec{A}_{nm}.
\ee
 The above equation is exact. We see that the effect of the fast variables
 is summarized by an effective gauge field $\vec{A}$. The vector part couples
 minimally to the momenta with the
fast eigenvalue acting like a scalar potential. The vector field is in general
nonabelian and corresponds to a nonabelian Berry potential first discussed
by Wilczek and Zee \cite{wilczee}.

In case one can neglect the off-diagonal transition
terms in the induced gauge potentials ({\ie}, if the adiabatic
approximation is valid), then the Hamiltonian simplifies
to
\be
H_n^{eff}= \frac{1}{2M}(\vec{P}-\vec{A}_n)^2 + \epsilon_n\label{diatomH}
\ee
where the diagonal component of the Berry phase $A_{nn}$ is denoted by $A_n$.
Suppose that the electronic eigenvalues are degenerate so that there are
$G_n$ eigenvectors with a degenerate eigenvalue $\epsilon_n$.
Then instead of a single Berry phase, we have
a whole set of the $G_n \times G_n$ Berry phases, forming the
matrix
\be
A_n^{k, k^{'}} = i <n, k|\nabla|n, k^{'}>  \,\,\,\,\,\,\,\,k,k^{'}=1,2,...G_n.
\label{generic}
\ee
The gauge field generated in such a case is nonabelian valued in the
gauge group $U(G_n)$.
In practical calculations, one truncates the infinite sum in (\ref{sum})
to a few
finite terms. Usually the sum is taken over the degenerate subspace
corresponding to the particular eigenvalue $\epsilon_n$. This is so-called
Born-Huang approximation, which we will use in what follows.
(A Berry potential built of the whole space would be a pure gauge type
and would have a vanishing stress tensor, so it would be trivial.)

As stated, the fast variable will be taken to be the motion of the
electron around the internuclear axis.
The slow variables are the vibrations and rotations of the internuclear
axis. This case corresponds to the situation where the energy of the spin-axis
interaction is large compared with the energy splittings between the rotational
levels, so that the adiabaticity condition holds.  This is called
``Hund case a". We follow the standard notation
\cite{landau}: Introducing the unit vector $\vec{N}$
along the internuclear axis, we define then the following quantum numbers
\be
\Lambda&=&\,\,{\rm eigenvalue\,\,\,of\,\, }\vec{N}\cdot\vec{L} \nonumber \\
\Sigma &=&\,\,{\rm eigenvalue\,\,\,of\,\, }\vec{N}\cdot\vec{S} \nonumber \\
\Omega &=&\,\,{\rm eigenvalue\,\,\,of\,\, }\vec{N}\cdot\vec{J}
=|\Lambda+\Sigma|,
\ee
so $\Lambda,\Sigma,\Omega$ are the projections of the orbital momentum,
spin and total angular momentum of the electron on the molecular axis,
respectively. For simplicity we focus on the simple case of $\Sigma=0$,
$\Lambda=0, \pm 1$. The $\Lambda=0$ state is referred to as $\Sigma$ state
and the $\Lambda=\pm 1$ states are called $\pi$, a degenerate doublet.
We are interested in the property of these triplet states, in particular
in the symmetry associated with their energy splittings.

To analyze the model, let us write the Lagrangian corresponding to
the Hamiltonian (\ref{diatomH})
\be
L_{nm}^{eff}= \frac{1}{2}M \dot{\vec{R}}(t)^2\delta_{mn}
 + \vec{A}_{mn}[\vec{R}(t)]\cdot
\dot{\vec{R}}(t)-\epsilon_m \delta_{mn}.\label{LEFF}
\ee
This can be rewritten in non-matrix form (dropping the trivial
electronic energy $\epsilon$)
\ben
{\cal L}=\frac{1}{2}M\dot{\vec{R}}^2 + i\theta^{\dagger}_a(\frac{\partial}
{\partial t}-i\vec{A}^{\alpha}T^{\alpha}_{ab}\cdot\dot{\vec{R}})\theta_b
\label{glagran}
\een
where we have introduced a Grassmannian variable $\theta_a$ as a trick
to avoid using the matrix form of (\ref{LEFF}) \cite{casal}
and ${\bf T}^{\alpha}$ is a matrix representation in
the vector space in which the Berry potential lives satisfying the commutation
rule
\ben
[{\bf T}^{\alpha},{\bf T}^{\beta}]=i f^{\alpha\beta\gamma}{\bf T}^{\gamma}.
\label{talgeb}
\een
To get to (\ref{diatomH}) from this form of Lagrangian, we use the standard
quantization procedure to obtain the following commutation relations,
\ben
[R_i,p_j]=i\delta_{ij}, \ \ \ \ \ \{\theta_a,\theta_b^{\dagger}\}=i\delta_{ab}.
\label{quantiz}
\een
The Hamiltonian
\ben
H=\frac{1}{2M}(\vec{P}-\vec{{\bf A}})^2\label{thetah}
\een
follows with
\ben
\vec{{\bf A}} &=& \vec{A}^{\alpha}I^{\alpha},\nonumber \\
   I^{\alpha} &=& \theta^{\dagger}_a T^{\alpha}_{ab}\theta_b.\label{ialpha}
\een
Using the commutation relations, it can be verified that
\ben
[I^{\alpha},I^{\beta}]=if^{\alpha\beta\gamma}I^{\gamma}.
\een
The Lagrangian, Eq.(\ref{glagran}), is
invariant under the gauge transformation
\ben
\vec{A}^{\alpha} &\rightarrow& \vec{A}^{\alpha} + f^{\alpha\beta\gamma}
\Lambda^{\beta}\vec{A}^{\gamma} - \vec{\nabla}\Lambda^{\alpha}\label{gaugea},
\\         \theta_a &\rightarrow& \theta_a -i\Lambda^{\alpha}T^{\alpha}_{ab}
\theta_b.\label{gaugetr}
\een
It should be noted that  Eq.(\ref{gaugetr})
corresponds to the gauge transformation on $|a\rangle$.
\subsubsection{\it Conserved angular momentum of the nonabelian monopole}
\bs
Before discussing structure of the diatomic molecular system,
we digress here and review the known case
of a particle coupled to an external gauge field of 't Hooft-Polyakov
monopole \cite{thooft} with a coupling constant $g$
to which our nonabelian gauge field will correspond.
Asymptotically the magnetic field involved is of the form
\ben
\vec{\bf B} = -\frac{\hat{r}(\hat{r}\cdot{\bf T})}{gr^2}\label{bmag}
\een
which is obtained from the asymptotic form of the gauge field $\vec{{\bf A}}$
\ben
A^{\alpha}_i &=& \epsilon_{\alpha i j} \frac{r_j}{gr^2},
\label{hooft}\\
\vec{{\bf B}} &=& \vec{\nabla} \times \vec{{\bf A}} -ig[\vec{{\bf A}},
\vec{{\bf A}}].\label{twoform}
\een
The Hamiltonian of
a particle coupled to a 't Hooft-Polyakov monopole can therefore be written as
\footnote{Hereafter we put $g=1$ for close analogy with Eq.(\ref{thetah}).}
\ben
H &=& \frac{1}{2M}(\vec{p}-\vec{\bf A})^2\nonumber\\
  &=& \frac{1}{2M}\vec{{\bf D}}\cdot \vec{{\bf D}}\label{dhamil}
\een
where $\vec{{\bf D}}=\vec{p}-\vec{\bf A}$.

It is obvious that the mechanical angular momentum $\vec{L}_m$ of a particle
\ben
\vec{L}_m=M\vec{r}\times\dot{\vec{r}}=\vec{r}\times\vec{{\bf D}}
\een
does not satisfy the $SU(2)$ algebra after canonical quantization
as in Eq.(\ref{quantiz})
and moreover it cannot be a symmetric operator that commutes with the
Hamiltonian.
The conventional angular momentum, $\vec{L}_o =\vec{r}\times\vec{p}$,
satisfies the usual angular momentum commutation rule.
However it does not commute with the
Hamiltonian and hence cannot be a conserved angular momentum of the
system.  This observation shows us that the construction of a conserved angular
momentum of a system coupled to a topologically nontrivial gauge field is not
a trivial matter.

The conserved angular momentum can be constructed by modifying $\vec{L}_m$ to
\be
\vec{L} = \vec{L}_m + \vec{{\bf Q}},\label{rxd}
\ee
where $ \vec{{\bf Q}}=\vec{Q}^{\alpha}I^{\alpha}$ is to be obtained as follows.
The methods to determine $\vec{{\bf Q}}$ are standard
\cite{jackiw}. The first condition required for $\vec{{\bf Q}}$ is the
consistency condition that $\vec{L}$ satisfy the $SU(2)$ algebra
\be
[L_i,L_j]=i\epsilon_{ijk}L_k.\label{lll}
\ee
This leads to an equation for $\vec{{\bf Q}}$,
\be
\vec{r}(\vec{r}\cdot\vec{{\bf B}}) + \vec{r}\vec{{\cal D}}\cdot\vec{{\bf Q}}
-\vec{{\cal D}}(\vec{r}\cdot\vec{{\bf Q}})=0.\label{c1}
\ee
where
\be
\vec{{\cal D}}=\vec{\nabla} - i[\vec{{\bf A}},\ \ \ ].\label{covd}
\ee
The second equation is obtained by requiring that $\vec{L}$ commute with H,
\be
[\vec{L},H]=0.\label{c2}
\ee
Eq.(\ref{c2}) can be replaced by a stronger condition
\be
 [L_i, {\bf D}_j]=i\epsilon_{ijk}{\bf D}_k,\label{c3}
\ee
which leads to
\be
{\cal D}_i{\bf Q}_j +\delta_{ij}\vec{r}\cdot\vec{{\bf B}} -
r_i{\bf B}_j=0. \label{c31}
\ee
It is obvious that $\vec{L}$ satisfying Eq.(\ref{c3}) or (\ref{c31}) commutes
with the Hamiltonian Eq.(\ref{dhamil}).

In the case of  the 't Hooft-Polyakov monopole, eqs.(\ref{bmag})
and (\ref{hooft}), it can be shown that
\be
\vec{{\bf Q}} = \hat{r}(\hat{r}\cdot{\bf I})\label{phitm}
\ee
satisfies eqs. (\ref{c1}) and (\ref{c31}). After inserting Eq.(\ref{phitm})
into Eq.(\ref{rxd}), we get
\be
\vec{L} &=& \vec{L}_m + \hat{r}(\hat{r}\cdot{\bf I})\label{ltm0}\\
        &=& \vec{r} \times \vec{p} + \vec{I},\label{ltm}
\ee
where
\be
I_i=\delta_{i\alpha}I^{\alpha}.\label{ii}
\ee
Equations (\ref{ltm}, \ref{ii}) show clearly how the isospin-spin transmutation
occurs in a system where a particle is coupled to a nonabelian monopole
\cite{jackiwrebbi}.

This analysis can be applied to the abelian $U(1)$ monopole
just by replacing $\hat{r}\cdot\vec{I}$ by -1 in Eqs.(\ref{phitm}) and
(\ref{ltm0})\footnote{We are considering a Dirac monopole with $e=g=1$.}. Then
\be
\vec{Q} &=& \hat{r},\label{phidm}\\
\vec{L}    &=& m\vec{r}\times\dot{\vec{r}} - \hat{r}.\label{ldm0}
\ee
This can be rewritten as
\be
\vec{L} = \vec{r}\times\vec{p} - \vec{\Sigma},\label{ldm}
\ee
where
\be
\vec{\Sigma} =\left(\frac{(1-\cos\theta)}{\sin\theta}\cos\phi,\,
\frac{(1-\cos\theta)}{\sin\theta}\sin\phi,\, 1\right)
\ee
which is the form frequently seen in the literature.
\subsubsection{\it Symmetry and spectrum of the diatomic molecule}
\bs
We now turn to the problem of constructing conserved angular momentum
in a diatomic
molecule in which a Berry potential couples to the dynamics of slow degrees of
freedom, corresponding to the nuclear coordinate $\vec{R}$,
a system that has been studied extensively \cite{zygel}.
The procedure is in complete parallel to that reviewed above.

As mentioned before,
the Berry potential is defined on the space spanned by the electronic
states $\pi(|\Lambda|=1)$ and $\Sigma(\Lambda=0)$, where $\Lambda$'s are
eigenvalues of the third component of the orbital angular momentum of
the electronic
states.  The electronic states responding to slow rotation of $\vec{R}$
described by $U(\vec{R})$,
\be
U(\vec{R})={\rm exp}(-i \phi L_z){\rm exp}(i \theta L_y){\rm exp}(i \phi L_z),
\label{uzg}
\ee
induce a Berry potential of the form
\be
\vec{{\bf A}} &=& i\langle \Lambda_a|U(\vec{R})\vec{\nabla}U(\vec{R})^{\dagger}
|\Lambda_b\rangle\label{zga0}\\
              &=& \frac{{\bf A}_{\theta}}{R} \hat{{\bf \theta}} +
\frac{{\bf A}_{\phi}}{R \sin\theta}\hat{{\bf \phi}},
\ee
where
\be
{\bf A}_{\theta}&=&\kappa(R)({\bf T}_y \cos\phi - {\bf T}_x \sin\phi),
\nonumber\\
{\bf A}_{\phi}&=& {\bf T}_z(\cos\theta - 1) - \kappa(R)\sin\theta
({\bf T}_x \cos\phi + {\bf T}_y \sin\phi). \label{aphi}
\ee
$\vec{{\bf T}}'$s are spin-1 representations of the orbital angular momentum
$\vec{L}$ and $\kappa$ measures the transition amplitude between the
$\Sigma$ and $\pi$ states
\be
\kappa(R)=\frac{1}{\sqrt{2}}|\langle\Sigma|L_x-iL_y|\pi\rangle|.\label{kappa}
\ee
The nonvanishing field strength tensor is given by
\be
\vec{{\bf B}}=\frac{F_{\theta\phi}}{R^2 \sin\theta}=-\frac{(1-\kappa^2)}
{R^2}T_z\hat{R}.\label{zgb0}
\ee
 Following the procedure described above, introducing a Grassmannian
variable for each electronic state and replacing ${\bf T}$ by {\bf I} defined
in Eq.(\ref{ialpha}) and quantizing the corresponding Lagrangian, we obtain
the Hamiltonian
\be
H=\frac{1}{2M} (\vec{P} - \vec{{\bf A}})^2 \label{hzg0},
\ee
where $\vec{{\bf A}} = \vec{A}^{\alpha} I^{\alpha}$.
It should be noted that the presence of
the constant $\kappa$ which is not quantized that appears in the
Berry potential is a generic feature of nontrivial nonabelian Berry potentials
as can be seen in many examples \cite{shapere}.

To find a solution of Eq.(\ref{c1}) and
Eq.(\ref{c31}), we can exploit the gauge freedom and rewrite
the Hamiltonian in the  most
symmetric form. This can be done by a gauge transform of the form
\be
\vec{{\bf A}}'&=& V^{\dagger}\vec{{\bf A}}V + iV^{\dagger}\vec{\nabla}V
\label{zgas}\\
        {\bf F}' &=& V^{\dagger}{\bf F}V\label{zgbs}
\ee
where $V$ is an inverse operation of $U$ in Eq.(\ref{uzg}), {\it i.e.},
$V = U^{\dagger}$.  Then
\be
{\bf A}_{\theta}' &=& (1-\kappa)(I_x \sin\phi-I_y \cos\phi),\nonumber\\
{\bf A}_{\phi}' &=& (1-\kappa)\{-I_z \sin^2 \theta +
\cos\theta \sin\theta (I_x \cos\phi + I_y \sin\phi )\},\label{zgasf}
\ee
or more compactly
$$\vec{{\bf A}}' = (1-\kappa)\frac{\hat{R} \times \vec{I}}{R},$$
and
\be
\vec{{\bf B}}'=-(1-\kappa^2)\frac{\hat{R}(\hat{R}\cdot{\bf I})}
{R^2}.\label{zgbsf}
\ee
It is remarkable that the above Berry potential has the same structure as
the 't Hooft-Polyakov monopole, Eq.(\ref{bmag}) and Eq.(\ref{hooft}), except
for the different constant factors
$(1-\kappa)$ for vector potential and $(1-\kappa^2)$ for magnetic field.
Because of these two different factors, however, one cannot simply
take Eq.(\ref{phitm}) as a solution of (\ref{c31}) for the case of
nonabelian Berry potentials.
Using the following identities derived from Eq. (\ref{zgasf}),
\be
\vec{R}\cdot\vec{{\bf A}}' &=& 0,\label{da}\\
\vec{R} \times \vec{{\bf A}}' &=& -(1-\kappa)\{\vec{I} -
(\vec{I}\cdot\hat{R})\hat{R}\},\label{ca}
\ee
the Hamiltonian, Eq.(\ref{hzg0}), can be written as
\be
H = -\frac{1}{2M R^2}\frac{\partial}{\partial R}R^2\frac{\partial}
{\partial R}
+ \frac{1}{2M R^2}(\vec{L}_o + (1-\kappa)\vec{I})^2
- \frac{1}{2M R^2}(1-\kappa)^2(\vec{I}\cdot\hat{R})^2.\label{hzg1}
\ee
Now one can show that the conserved angular momentum $\vec{L}$ is
\be
\vec{L} &=&  \vec{L}_o + \vec{I},\label{zglf}\\
        &=&  M\vec{R}\times\dot{\vec{R}} + \vec{{\bf Q}},\label{zglmf}
\ee
with
\be
\vec{{\bf Q}} = \kappa \vec{I} + (1-\kappa)\hat{R}(\hat{R}\cdot\vec{I}).
\label{zgqf}
\ee
Hence, in terms of the conserved angular momentum $\vec{L}$,
the Hamiltonian becomes
\be
H = -\frac{1}{2M R^2}\frac{\partial}{\partial R}R^2\frac{\partial}
{\partial R}
+ \frac{1}{2M R^2}(\vec{L} - \kappa\vec{I})^2
- \frac{1}{2M R^2}(1-\kappa)^2\label{hzg2}
\ee
where  $(\vec{I}\cdot\hat{R})^2 = 1$ has been used.

It is interesting to see what happens in the two extreme cases of $\kappa = 0$
and $1$. For $\kappa = 1$, the degenerate $\Sigma$
 and $\pi$ states form a representation of the rotation group and hence the
Berry potential (and its field tensor) vanishes or becomes a pure gauge.
Then $\vec{{\bf Q}} = \vec{I}$ and $\vec{L} = M\vec{R}\times\dot{\vec{R}} +
\vec{I} $.  Now $\vec{I}$ can be understood as the angular momentum of the
electronic system which is
decoupled from the spectrum. One can also understand what happens here
as the restoration
of rotational symmetry in the electronic system. Physically $\kappa \rightarrow
1$ as $R\rightarrow \infty$.

For $\kappa=0$, the
$\Sigma$ and $\pi$ states are completely decoupled and only the
$U(1)$ monopole field can be developed on the $\pi$
states \cite{moody}. Equation (\ref{zgqf}) becomes identical
to Eq.(\ref{phidm}) as
$\kappa$ goes to zero and the Hamiltonian can be written as
\be
H = -\frac{1}{2M R^2}\frac{\partial}{\partial R}R^2\frac{\partial}
{\partial R}
+ \frac{1}{2M R^2}(\vec{L}\cdot\vec{L} - 1)\label{hzgu1}
\ee
which is a generic form for a system coupled to an $U(1)$ monopole field.
Physically this corresponds to small internuclear distance at which
the $\Sigma$ and $\pi$ states decouple.  Therefore, a truly
nonabelian Berry potential can be obtained
only for a  $\kappa$ which is not equal to zero or one.
We will encounter later an analogous situation in heavy-quark
baryons.

\subsection{Berry Phases in the Chiral Bag Model}
\subsubsection{\it Induced gauge fields}
\bs
We now turn to hadronic systems in (3+1) dimensions and use
the intuition we gained in the quantum mechanical cases, exploiting largely the
generic structure of the resulting expressions. Consider the generating
functional for the chiral bag in $SU(2)$ flavor space
\ben
 Z = \int d[{\bf \pi}] \,  d[\psi][\bar{\psi}]e^{ iS}
\label{CBZ}
\een
with the (Minkowski) action  S for the chiral bag given by
\be
S = \int_V\,\overline{\psi}
i\gamma^{\mu}\partial_{\mu}\psi
-\frac 12\int \Delta_s \,\,
\overline{\psi} \,S\,e^{i\gamma_5\vec{\tau}\cdot\hat{r} F(r)}\,S^{\dag}\,
\psi
+ S_M(SU_0 S^{\dag})\label{gfa1}
\ee
where $F$ is the chiral angle appearing in the ``hedgehog" configuration
$U_0=e^{i\vec{\tau}\cdot\hat{r}
F(r)}$, $\Delta_s$ is a surface delta function and the space rotation
has been traded in for an isospin rotation ($S(t)$)
due to the hedgehog symmetry considered here\footnote{The role of hedgehog and
``grand spin" etc. will be clarified in discussing the skyrmion structure in
the following lecture. In the present context, it suffices to consider them
as classical objects to be suitably quantized.}. The quarks are assumed to be
confined in the spatial volume $V$ and
the purely mesonic terms outside the bag are described by $S_M$.

Suppose that we adiabatically rotate the bag in space. As we will see later,
the ``rotation" is required to quantize the zero modes to obtain the
quantum states of the physical baryons. In this model, the rotation
implies transforming the hedgehog configuration living outside the bag as
\be
U_0\rightarrow S(t) U_0 S^\dagger (t)
\ee
to which the quarks in the bag respond in a self-consistent way. This leads to
computing the path integral in (\ref{CBZ}) with (\ref{gfa1}).
Because of
the degeneracy of the Dirac spectrum, mixing between quark levels
is expected no matter how small the rotation is. This mixing takes
place in each quark ``grand spin" $K$-band (where $K$ is the vector sum of spin
and isospin that labels hedgehog quark states) and leads to a nonabelian
Berry or gauge field. Instead of the boundary term rotating as a function
of time, it is more convenient to work with a static boundary condition
by ``unwinding" the boundary with
the redefinition $\psi\rightarrow S\psi$, leading to
\be
S = S_{S={\bf 1}} +\int\,{\psi}^{\dag}\,{S}^{\dag}i\partial_t S\,\psi\,.
\label{Ss}
\ee
The effect of the rotation on the fermions inside the bag is the
same as a time dependent gauge potential. This is the origin of the induced
Berry potential analogous to the solenoid-electron system described above.
Now as the meson field outside rotates, the quarks inside the bag could
either remain in their ground state which is in the lowest-energy
$K=0^+$ orbit or excited to higher orbits. In the former case, we
have a ground-state baryon and no Berry phases remain active: There is
no Wess-Zumino term in the SU(2) case. In the latter case,
as sketched in Fig. I.2),  there are, broadly, two classes of quark
excitations that can be distinguished.
\bitem
\item Class A: One or more quarks get excited to $K\neq 0^+$ levels, changing
no flavors;
\item Class B: One or more quarks get excited to flavor Q levels through a
flavor change where Q could be strange, charm or bottom quark.
\eitem
In what follows, we will address the problem using Class A but
the discussion applies equally well to Class B as we will detail later.

To expose the physics behind the second term of (\ref{Ss}) which is the
principal element of our argument, we  expand
the fermionic fields in the complete set of states $\psi_{KM}$
with energies $\epsilon_{K}$ in the unrotating bag corresponding to
the action $S_{S=1}$ in (\ref{Ss}), and $M$ labels
$2K+1$ projections of the grand spin $K$. Generically,

\be
\psi (t,x) =\sum_{K,M} c_{KM}(t) \psi_{KM} (x)
\ee
where the $c$'s are Grassmannians, so that

\be
S = \sum_{KM} \int dt
\,\, c^{\dag}_{KM} (i\partial_t -\epsilon_K)c_{KM} +
\sum_{KMK^{'}N} \int dt\,\, c^{\dag}_{KM} A^{KK^{'}}_{MN}
                c_{K^{'}N}
\label{Action}
\ee
where
\be
A_{MN}^{KK^{'}}= \int_V d^3x\, \psi^{\dag}_{KM} S^{\dag}i\partial_t S
\psi_{K^{'}N}\,.\label{gaugef}
\ee
No approximation has been made up to this point. If the $A$ of (\ref{gaugef})
were defined in the whole $K$ space, then $A$ takes the form of a pure gauge
and the field strength tensor would be identically
zero \footnote{As we saw in the case of the diatomic molecule, the vanishing
of the field tensor does not imply that there is no effect. It describes
the restoration of certain symmetry. See later a similar phenomenon in
heavy-quark baryons.}. However we are forced to truncate the space.
As in the preceding
cases, we can use now the adiabatic approximation and neglect the
off-diagonal
terms in $K$, {\it i.e.}, ignore the effect of adiabatic rotations that
can cause the jumps between the energy levels of the fast quarks.
Still, for every $K \neq 0$ the adiabatic rotation mixes $2K+1$ degenerate
levels corresponding to particular fast eigenenergy $\epsilon_K$.
In this form we clearly see that the rotation induces a hierarchy
of Berry potentials in each K-band, on the generic form  identical to
Eq.(\ref{generic}).  This field is truly a gauge field. Indeed,
any local rotation of the $\psi_{KM} \rightarrow D^K_{MN} \psi_{KN}$
where $D^K$ is a $2K+1$ dimensional matrix spanning the representation of
rotation in the $K$-space,
can be compensated by a gauge transformation of the Berry potential
\be
A^K \rightarrow D^K (\partial_t + A^K ) D^{K \dag}
\ee
leaving $S_S$ invariant. A potential which has this
transformation property is called ``invariant gauge potential"
\cite{jackiwprl}.

The structure of the Berry potential depends on the choice of the
parametrization of the isorotation $S$ (which is related to
the gauge freedom). For the
parametrization $S =a_4 +i\vec{a}\cdot\vec{\tau}$ with the unitary
constraint $a\cdot a =1$ (unitary gauge), we have
\be
A^K = T_K^a\, A_K^a = T_K^a\,
\left( \,g_K\,\,\frac{\eta_{\mu\nu}^a\,a_{\mu}da_{\nu}}{1+a^2}\right)
\label{berrypot}
\ee
where $\eta$ is the 't Hooft symbol and $g_K$ the induced coupling
to be specified below.  The $T$'s refer to the K-representation
of $SU(2)$, the group of isorotations. In the unitary gauge the Berry
potential has the algebraic structure of a unit size instanton in isospace,
{\ie}, the space of the slow variables. It is not the Yang-Mills instanton,
since the above configuration is not self-dual due to the unitary gauge
constraint.
This configuration is a non-abelian generalization of the monopole-like
solution present in the diatomic molecular case.

To make our analogy more quantitative, let us refer to the
Grassmannians $c$ in the valence states by $\alpha$'s and those in the Dirac
sea by $\beta$'s. Clearly (\ref{Action})
can be trivially rewritten in the form
\be
S &=& \sum_{KMN} \int dt
\,\, {\alpha}^{\dag}_{KM} \left[(i\partial_t -\epsilon_K) {\bf 1}_{MN} +
                (A^K)_{MN}\,\right] \alpha_{KN} \nonumber\\
     &+& \sum_{KMN} \int dt\,\,
 {\beta}^{\dag}_{KM} \left[(i\partial_t -\epsilon_K) {\bf 1}_{MN} +
                (A^K)_{MN}\,\right] \beta_{KN}\,.
\ee
Integrating over the Dirac sea {\em in the presence of valence
quarks} yields the effective action
\be
S =&& \sum_{KMN} \int dt \,\,
 {\alpha}^{\dag}_{KM}\left[ (i\partial_t -\epsilon_K) {\bf 1}_{MN} +
                (A^K)_{MN}\,\right] \alpha_{KN} \nonumber\\
    + && i{\rm Tr}\,{\rm ln}\,
          \left( (i\partial_t -\epsilon_K) {\bf 1}_{MN} +
                (A^K)_{MN}\, \right)
\label{full}
\ee
where the Trace is over the Dirac sea states. The latter can be
Taylor expanded in the isospin velocities $\dot{a}_\mu$ in the
adiabatic limit,
\be
i{\rm Tr}\,{\rm ln}\,((i\partial_t-\epsilon_k){\bf 1}_{MN} - ({\cal
A}_{\mu}^K)_{MN} \dot{a}_{\mu})= \int\, dt \, \frac {{\I}_q}2
\dot{a}_{\mu}\dot{a}_{\mu} + \cdots
\label{sea}
\ee
where ${\I}_q$ is the moment of inertia contributed by the quark sector.
We have exposed the velocity dependence by rewriting the form
$A^K_{MN}= ({\cal
A}_{\mu}^K)_{MN} \dot{a}_{\mu}$.
Linear terms in the velocity are absent since the Berry phases
in the sea cancel pairwise in the SU(2) isospin case under
consideration. (For SU(3) they do not and are at the origin of
the WZW term.) The ellipsis in (\ref{sea}) refers to higher derivative
terms. We should
point out that Eq.(\ref{sea}) includes implicitly the valence quark effect,
 because the
levels of the Dirac sea are modified due to the presence  of the valence
quarks.

To clarify the general motivation for studying the excited states in terms of
Berry phases,
let us consider the case of the bag containing one valence quark in the $K=1$
state.
The action for the adiabatic motion of this quark is obtained from the
above formulae and yields (sum over repeated indices is understood)
\be
S = \int dt\,\,[i{\alpha}^{\dag}_{1M}\dot{\alpha}_{1M}-\epsilon_1
\alpha^{\dag}_{1M}\alpha_{1M} +\frac{1}{2}{\I}_q
\dot{a}_{\mu}\dot{a}_{\mu} + \dot{a}_{\mu}({\cal A}_{\mu}^1)_{MN}
\alpha_{1M}^{\dag}\alpha_{1N}]\,.\label{SA1}
\ee

As we will see below, when canonically quantized,
the generic structure of the resulting Hamiltonian is identical to
(\ref{hzg1}). This illustrates the universal character of the Berry phases.

\subsubsection{\it Excitation spectrum}
\bs

To quantize the system, we note that (\ref{SA1}) describes the motion of a
particle with nonabelian charges coupled to an instanton-like
gauge field on $S^3$. Since $S^3$ is isomorphic to the
group manifold of $SU(2)$, it is convenient to use the left or right
Maurer-Cartan forms as a basis for the vielbeins (one-form notation understood)

\be
S^\dagger idS = - \omega_a \tau_a = -v_a^c(\theta) d\theta^c \tau^a\label{s3}
\ee
where we expressed  the ``velocity" forms $\omega$ in the basis  of the
vielbeins $v_a^c$, and $\theta$ denotes some arbitrary parametrization
of the $SU(2)$, {\it e.g.} Euler angles.
In terms of vielbeins, the induced Berry potential simplifies to

\be
{\cal A}^c=g_K A_K^c = -\, g_K\, v_a^c(\theta) T^a\label{gkd}
\ee
where $T$ are generators of the induced Berry structure in the $K$
representation  and $g_K$ is the corresponding charge \cite{LNRZannals},
the explicit form of which is not needed.

Note that we could have equally well
used the right-invariant Maurer-Cartan form instead of the left-invariant
Maurer-Cartan form (\ref{s3}). The field strength can be
written  in terms of $A$ defined in Eq.(\ref{gkd})

\be
{\cal F}_K= dA_K-ig_KA_K\wedge A_K =-(1-g_K/2)\epsilon^{mij}T_K^m\,v^i\wedge
v^j.
\ee
${\cal F}_K$ vanishes for $g_K=2$, $i.e.$ the Berry potential becomes a
pure gauge.
The vielbeins, and hence $A$ and ${\cal F}$ are frame-dependent, but
to quantize the system no specific choice of framing is  needed.
The canonical momenta are
$p_a= {\partial L}/{\partial \dot{\theta}_a}$.
Our system
lives on $S^3$ and is invariant under $SO(4) \sim SU(2) \times SU(2)$.
Right and left generators are  defined as
\be
R_a &=& u_a^c\,p_c \nonumber \\
L_a &=& D_{ab}(S)\,R_b\label{rgenerator}
\ee
where $u^a_i\,v^i_c = \delta_c^a$ and $D(S)$ spans the adjoint
representation
of $SU(2)$. Following the standard procedure we get
our Hamiltonian in terms of the generators
\be
H^* = \epsilon_K \, {\bf 1}     + \frac{1}{8\I}\,
  \left( R_j -g_K\,T_{Kj}\right)
   \left(R_j -g_K\,T_{Kj} \right). \label{hstar}
\ee
Here ${\I}$ is the total moment of inertia, the sum of the quark contribution
and the mesonic contribution.
As a result the Hamiltonian for a single excited quark\footnote{If we were
to add a second quark to this band (doubly excited state), then we would no
 longer have irreducible representation of $T_{K}$ but a reducible
representation instead. This case will not be pursued here. In the
next lecture, we shall introduce the notion of quasiparticle to describe
the excitations of more than one quark in heavy baryons. }
 takes the simple form
\be
H^* =  \epsilon_1 {\bf 1} +\frac{1}{8\I}\left(\vec{R}^2 -2g_K\vec{R}
\cdot\vec{T}_K
        +g_K^2\vec{T}_K^2\right) \label{hstarf}
\ee
The spectrum  can be readily constructed if we notice that (\ref{hstarf})
can be rewritten solely in terms of the independent Casimirs
\be
H^* =  \epsilon_K {\bf 1}+ \frac{1}{2\I}\left[
+\frac{g_K}{2}\vec{J}_K^2 + (1 - \frac{g_K}{2}) \vec{I}^2
-\frac{g_K}{2}(1-\frac{g_K}{2})\vec{T}^2 \right]\label{hexcit}
\ee
where $\vec{J}_K = {\vec{R}}/{2}-\vec{T}_K$ and $\vec{I}=\vec{L}/2$.
We recall that $\vec{L}^2 = \vec{R}^2$ on $S^3$. The set
$J_K^2$, $I^2$, $J_{K3}$ and $I_3$ form a complete set of independent
generators that commute with $H^*$. They can be used to span the Hilbert
space of a single quark in the $K$ band. We will identify $J_K$ with the
angular momentum and $I$ with the isospin.

The spectrum associated with (\ref{hexcit}) will be analyzed both in
the light-quark and heavy-quark sectors when we have discussed the
skyrmion structure. Let us briefly mention here that in the light-quark
system, the limit $g_K=0$ for which the induced angular momentum $-\vec{T}_K$
decouples from the system (which would correspond to $(1-\kappa)=0$
in the diatomic molecule) is never reached. Neither is the abelian limit
that will be attained for $g_K=2$ physically relevant. On the other hand,
as we will explicitly show in Lecture II, the decoupling limit
is attained in heavy-quark baryons because of the presence of heavy-quark
symmetry that is present in QCD in the limit the heavy-quark mass becomes
infinite.

\part{Skyrmions}
\renewcommand{\theequation}{II.\arabic{equation}}
\setcounter{equation}{0}
\setcounter{section}{0}
\section{Ground-State Baryons}
\bs
In this lecture, we shrink the bag to a point using the Cheshire
Cat strategy and consider it to be
a gauge-fixed version of the chiral bag. We assume that the resulting system
is a skyrmion \cite{skyrme} and describes the physics of the baryons.
First we describe the ground-state
systems, the nucleon and the $\Delta$. Excitations both in the light-flavor
sector and in the heavy-flavor sector will then be discussed
in terms of Berry potentials.
\subsection{Large $N_c$ Consideration}
\bs
The notion of large number of colors ($N_c$) plays an essential role in
deriving or guessing
effective Lagrangians of QCD applicable at low energy in terms of
finite number of long wavelength meson excitations.  We have used
it (albeit implicitly) for the Cheshire Cat mechanism in four dimensions.
The outcome of large $N_c$ considerations is a Lagrangian that contains
weakly coupled meson fields only.
If such a Lagrangian
is the {\it entire} content of QCD at low energy, then where are the
ground-state baryons? We have shown in the previous section that even
when only Goldstone bosons are excited, there is a certain sense in which
baryons could emerge through the approximate CCP. Here we discuss how
in large-$N_c$ QCD baryons can emerge as solitons in mesonic effective
theories. Note that the CCP argument did not rely on large $N_c$ explicitly.
But the argument based on large $N_c$ implies that the CCP is a better
approximation the larger the $N_c$.

That baryons must appear as solitons in meson field theory \cite{ncbaryon}
is plausible but not fully established. The reason is that even
in the limit $N_c=\infty$, the exact solution of QCD is not available.
In any event, in the large $N_c$ limit, QCD
is a weakly interacting meson field theory, with the meson mass going
as $O(1)$ in $1/N_c$ and the $n$-Goldstone boson function going as
$\sim N_c^\lambda$, $\lambda=(2-n)/2$. The meson scattering amplitude
is described by the tree graphs of this effective Lagrangian with the
coupling constant proportional to $1/\sqrt{N_c}$, so goes to zero as
$N_c$ tends to infinity. Now this theory has no fermions and hence no
{\it explicit} baryons.
In quark picture, one needs $N_c$ quarks to make up a baryon,
so the mass of a baryon must go like
\be
m_B\sim N_c=\frac{1}{1/N_c}.
\ee
Such an object can occur from mesons only if there is a soliton in the
meson field
configuration. Such a configuration is known to exist in less than four
dimensions and well understood. In four dimensions, we have the magnetic
monopole which has not yet been seen experimentally. The soliton mass in
weakly coupled meson field theory does indeed go like $1/\alpha$ where
$\alpha$ is the coupling constant of the meson fields. Thus it must be that
a baryon in QCD is a soliton with a large mass $\sim N_c$ with a size of
$O(1)$. In the same vein, a nucleus made of $A$ nucleons must be a soliton
of mass $\sim A N_c$ of the same mesonic Lagrangian.

In treating baryons as solitons, the basic assumption is that the bulk of
baryon physics is dictated by the leading $N_c$ contributions and next
$1/N_c$ corrections are in some systematic fashion calculable. It is not
obvious that such corrections are necessarily small. For instance,
nuclear binding is of $O(1/N_c)$ and hence nuclei exist in the subleading
order. So how useful is the concept of nucleons and nuclei as solitons?
This is the question addressed below.

\subsection{Soliton}
\subsubsection{\it Stability of the soliton}
\bs
The large $N_c$ consideration {\it suggests} that baryons, if at all,
can only arise from mesonic effective theories as solitons. Let us first see
whether a stable soliton can indeed be obtained from simple meson Lagrangians.

The longest wavelength chiral Lagrangian is the leading nonlinear
$\sigma$-model
Lagrangian
\ben
{\L}_2=\frac{f^2}{4}\Tr (\del_\mu U \del^\mu U^\dagger).
\een
Does this Lagrangian by itself give a stable soliton?
One can answer this question quite simply.

Before proceeding to answer this question, we have to introduce a basic notion
of a soliton in the way that we are using here \footnote{For a more
precise discussion, A.P. Balachandran, Ref.\cite{balachandran}.}.
(The notion of soliton as used
in particle and nuclear physics is different from other fields where the term
soliton is applied to localized solutions which maintain their form even in
scattering process.) Since we are going to shrink the bag to a point,
we will consider only topological solitons, not
the nontopological variety one finds in the literature. The topology involves
space of fields of $U$. A sufficient condition for the existence of a soliton
is that
\ben
U(\vec{x},t)\rightarrow 1 \ \ \ \ {\rm as}\ \
|\vec{x}|=r\rightarrow \infty
\een
with the rate of approach of $U$ to $1$
sufficiently fast so that the energy is finite, say, $E<\infty$. Consider the
usual hedgehog ansatz made by Skyrme a long time ago \cite{skyrme}
\ben
U_0 (\vec{x})=e^{i\hat{x}\cdot \vec{\tau} F(r)}.
\een
Then the static energy functional is
\ben
E&=& 2\pi f^2\int_\epsilon^\infty r^2 dr\left[(dF/dr)^2+\frac{2}{r^2}
\sin^2F\right]\nonumber\\
&=& 2\epsilon\pi f^2\int_1^\infty y^2 dy \left[(dF/dy)^2+\frac{2}{y^2}
\sin^2F\right]\label{energyf}
\een
where the second equality is obtained with the change of variable
$y\equiv r/\epsilon$. We have put $\epsilon$ for the lower limit for the reason
that will be apparent soon. If we let $\epsilon=0$, then the energy of the
system is zero. Since the integral is positive definite, this is the lowest
energy. Now $f$ has a dimension of mass and since the only other scale
parameter possible is the size of the system ($R$), the energy must be
proportional to $f^2 R$ and hence the lowest energy solution corresponds
to the zero size, namely, a collapsed system. This is the well-known
result of ``Hobart-Derrick theorem" \cite{derrick}.
Since there is no known massless point-like hadron
with fermion quantum number, this must be an artifact of the defects of the
theory.

There is one simple way \cite{schechter1} to avoid the Hobart-Derrick collapse
without introducing any further terms in the Lagrangian or any other degrees
of freedom, which is interesting and in a way, physically relevant. It is to
simulate short-distance physics by ``drilling" a hole
into the skyrmion or equivalently by taking a nonzero $\epsilon$ in the
energy functional (\ref{energyf}). This is not as academic an issue as it
seems since a hole of that type arises naturally in the chiral bag model
discussed before. Let us for the
moment ignore what is inside that hole and see what that hole
does to the system. Since the
detail of the analysis that establishes the existence of a stable soliton
for $\epsilon \neq 0$ is not very relevant to us as there is a natural way
of assuring a stable system using a realistic quark bag, it suffices to
just summarize what transpires from the analysis: For {\it any} nonzero value
of $\epsilon$, there is always a solution which is classically stable
in the sense of escaping the Hobart-Derrick theorem and is stable against
small fluctuation $\eta$ of the solution $F$
\ben
F(y)\rightarrow F(y)+\eta (y).
\een
Since the system definitely collapses for $\epsilon=0$, it is clearly
a singular point. This singular behavior is probably
related to the presence or absence of a topological quantum number
in the chiral bag \cite{baacke}.
Indeed a chiral bag can be shrunk to a point without losing
its topological structure as long as a proper chiral boundary condition is
imposed. Without
the boundary condition, the topological quantum number is not conserved and
the bag decays into a vacuum. Thus a nontrivial topological structure
requires a finite $\epsilon$ to allow the boundary condition to be imposed at
the bag boundary and if the bag is shrunk, then at the center.
A relevant physical question is how to determine $\epsilon$. One practical
approach suggested by Balakrishna {\it et al} \cite{schechter1} is to
elevate it to a quantum
variable whose expectation value is related to the ``bag radius." From this
point of view, a bag is a necessary physical ingredient. On the other hand,
although this is highly suggestive, the discussion
on the Cheshire cat phenomenon given above clearly indicates that
the picture of replacing a quark bag by a quantized cut-off variable
can only be an academic one.

Another way of stabilizing the system for $\epsilon=0$ might be to
introduce quantum fluctuations before solving for the minimum configuration.
It is plausible that
quantum effects could stabilize the system by themselves just as the
collapse of the S-wave state of the classical hydrogen atom can be prevented
by quantum fluctuations. It turns out that just quantizing a few collective
degrees of freedom such as scale variable does not work without a certain
``confinement" artifact of the sort mentioned above. So far nobody has been
able to stabilize the skyrmion of (\ref{energyf}) with $\epsilon=0$
in a fully convincing way. This of course does not mean that such a scheme
is ruled out. There are many other collective variables
than the scale variable to quantize and
it is possible that by suitable quantization of appropriate variables, the
system could be stabilized.
We want to stress here however that
all these are really academic issues. We are working with effective Lagrangians
that are supposed to model QCD. Hence there are many other degrees of freedom
in nature,
be they in quark variables as in the chiral bag or in meson variables such as
the strong vector mesons. We saw above that a quark-bag does trivially
avoid the Hobart-Derrick collapse. As one eliminates the quark bag,
higher derivative terms in the chiral field arise through various induced
gauge fields as discussed above, among which we can have the quartic term
\ben
\frac{1}{32g^2}\Tr [U^\dagger\del_\mu U,U^\dagger\del_\nu U]^2
\een
the so-called ``Skyrme quartic term" which Skyrme introduced (in a somewhat
{\it ad hoc} way) to stabilize the soliton. We know now that such a term can
arise in some particular way from the $\rho$ meson when the latter is
integrated out.
We can readily verify that this term is enough to stabilize the soliton.
To see this,
let us scale $U(x)\rightarrow U(\lambda x)$. Then the energy of the system
changes
\ben
E_\lambda=\lambda^{2-d} E_{(2)} +\lambda^{4-d} E_{(4)}
\een
where the first term is the contribution from the quadratic term and
the second from the quartic term, with subscripts labeling them appropriately
and $d$ is the space dimension (equal to 3 in nature).
The minimum condition implies
\ben
dE_\lambda/d\lambda=0\rightarrow E_{(2)}/E_{(4)}=-(d-4)/(d-2)
\een
and the stability implies
\ben
d^2 E_\lambda/d\lambda^2 >0 \rightarrow 2(d-2) E_{(2)}>0.
\een
Thus for stability, we need $d>2$ which is fine with $d=3$ for which
we have $E_{(2)}=E_{(4)}$, the virial theorem. What this implies is then
that once one has more massive strong interaction degrees of freedom, the
stability issue is no longer a real issue. Indeed there have been considerable
controversy as to what vector mesons stabilize the soliton,
in particular, the question as to whether
the $\rho$ meson can do the job by itself. It is now understood that while
the Skyrme quartic term which can be traced to the $\rho$ meson in origin
stabilizes the system, the finite-range $\rho$ meson cannot and it is
instead the $\omega$ meson which plays the role of stabilization. There is
no lack of candidates for the role if some do not do the job.
Our attitude will be that the skyrmion is safe from the Hobart-Derrick collapse
and so we can simply proceed to the real stuff, namely physics of
the skyrmion. Those readers who want to know more about the technicality
are referred to the literature.
\subsubsection{\it Structure of the soliton}
\bs
The original Skyrme model captures the essence of the soliton structure, so
we will focus on the Lagrangian
\ben
{\L}_{sk}=\frac{f^2}{4}\Tr (\del_\mu U \del^\mu U^\dagger)
+\frac{1}{32g^2}\Tr [U^\dagger\del_\mu U,U^\dagger\del_\nu U]^2.
\label{skmodel}
\een
A quantitatively more precise prediction can be made with a Lagrangian
that contains the strong vector mesons in consistency with hidden gauge
symmetry \cite{bando}. If the vector mesons are
introduced disregarding HGS, one has to make sure that the
chiral counting comes out correctly, which is not always transparent
but the physics is not qualitatively modified by it.
The static energy of the skyrmion given by (\ref{skmodel}) reads
\ben
E=\int d^3 x \left\{-\frac{f^2}{4}\Tr L_i^2 -\frac{1}{32 g^2}\Tr [L_i,L_j]^2
\right\}
\een
with $L_i=U^\dagger\del_i U$. It is convenient to scale the length by
$1/gf$ by setting $x\rightarrow (1/gf)x$ and give the energy in units of
$f/2g$. In these units, the energy reads
\ben
E=\int d^3x \left\{-\frac{1}{2} \Tr L_i^2 -\frac{1}{16} \Tr[L_i,L_j]^2
\right\}.\label{statenergy}
\een
One can rewrite this
\ben
E &=& \int d^3x \Tr \left\{-\frac{1}{2} L_i \pm \frac{1}{4} \epsilon_{ijk}
[L_j,L_k]\right\}^2\nonumber\\
&\pm& 12\pi^2 \int d^3x \frac{1}{24\pi^2} \epsilon_{ijk} \Tr (L_i L_j L_k).
\een
The first term is non-negative and the second integral measures the winding
number which can be identified with the baryon number $B$
\ben
B=\frac{1}{12\pi^2}\int d^3x \epsilon_{ijk} \Tr (L_i L_j L_k)\label{baryonno}
\een
and so we have the topological bound on the energy
\ben
E\geq 12\pi^2 |B|.
\een
This bound is known as Faddeev bound \cite{faddeev}.
This bound would be saturated if we had the equality
\ben
L_i+\frac{1}{4}\epsilon_{ijk} [L_j,L_k]=0.\label{boundeq}
\een
But it turns out that the only solution to this equation is the trivial
one, namely, $U=constant$ and so $L_i=0$ \cite{manton}.
It turns out also that in curved space,
this bound {\it can} actually be satisfied and plays an interesting role
in connection with chiral symmetry restoration.

To proceed further, we shall now take an ansatz for the chiral field
$U$. In particular we assume the hedgehog form
\ben
U({\bf x})=e^{i F(r)\hat{x}\cdot {\bf \tau}}\label{hedgehog}.
\een
With this, the energy (\ref{statenergy}) is
\ben
E &=& 4\pi \int_0^\infty \, dr \left\{r^2[(dF/dr) + 2r^{-2} \sin^2 F]
\sin^2 F[r^{-2}\sin^2 F + 2(dF/dr)^2]\right\}\nonumber\\
&=& 4\pi \int_0^\infty \, dr[r^2 (dF/dr+ r^{-2} \sin^2 F)^2+2\sin^2 F
(dF/dr+1)^2]\nonumber\\
& & -24\pi\int_0^\infty \, dr \sin^2 F \frac{dF}{dr}.\label{statenergy2}
\een
For the energy to be finite, $U$, (\ref{hedgehog}), must approach a
constant at infinity which can be chosen to be the unit matrix. This means that
the space is compactified to three-sphere ($S^3$) and $U$ represents a
map from $S^3$ to to $SU(2)$ which is topologically $S^3$, the winding number
of which was given by the right-hand side of (\ref{baryonno}). The implication
of this on Eq. (\ref{statenergy2}) is that since the ``profile function"
$F$ at the origin and at infinity must be an integral multiple of $\pi$, the
second term of (\ref{statenergy2}) must correspond to $12\pi [F(0)-F(\infty)]
=12\pi^2 B$, the Faddeev bound. As mentioned above, the first integral of
(\ref{statenergy2}) cannot vanish for a nontrivial $U$ in flat space, so
the energy is always greater than $12\pi^2 B$.

Minimizing the energy with respect to the profile function $F$ gives
us the classical equation of motion. The exact solution with
appropriate boundary conditions at the origin and at infinity is known
\cite{jacksonrho,atkins} but only numerically.
For the present qualitative discussion,
we do not need a precise form for $F$. Indeed one can infer a simple
analytic form by the following argument\cite{manton}.
First of all, it turns out numerically that
the the first integral in (\ref{statenergy2}) is small so that the
soliton energy is only slightly greater than the Faddeev bound.
The equation of motion shows that $F(r)$ goes like
$B\pi-\alpha (r/r_0)$ as $r\rightarrow 0$
and like $\beta (r_0/r)^2$ as $r\rightarrow \infty$ with some constant $r_0$.
Let us therefore take the form for $B=1$
\ben
F &=& \pi -\alpha (r/r_0), \ \ \ \ r\leq r_0,\\
&=& \beta (r_0/r)^2, \ \ \ \ \ \ \ \ r\geq r_0.\label{profile1}
\een
Demanding the continuity of $F$ and its first derivative with respect to
$r$ at $r=r_0$, one determines
\ben
\alpha=2\pi/3, \ \ \ \ \beta=\pi/3.\label{profile2}
\een
Substituting (\ref{profile1}) with (\ref{profile2}) into (\ref{statenergy2}),
one finds
\ben
E\approx 4\pi (4.76 r_0 +7.55/r_0)
\een
where the coefficients of $r_0$ and $1/r_0$ are estimated numerically.
Minimizing with respect to $r_0$, one gets $r_0=1.26$ (in units of $(gf)^{-1}$)
and $E=151$ (in units of $f/2g$) which is 1.03 times the exact energy evaluated
numerically. The ansatz (\ref{profile1}) is compared in Fig. II.1 with
the exact numerical solution.

It is mentioned above that the true energy of the soliton is only slightly
greater than the Faddeev bound $12\pi^2$. Numerically it turns out to be
about 23 \% above. An interesting question is: When can the Faddeev bound be
saturated? As stated, this can never happen in flat space, which is the case
in nature since the space we are dealing with is $R^3$ or $S^3 (\infty)$, the
three-sphere with an infinite radius. Consider instead
the space $S^3 (L)$ with $L$ finite. Let its (inverse) metric be $g^{ij}$
and denote the dreibein $e_m^j$ so that
\ben
g^{ij}=e^i_m e^j_m.
\een
If we define
\ben
X_m\equiv e^i_m \del_i U U^\dagger
\een
then the static energy of the system is (for $n_B=1$)
\ben
E=\int d^3x \sqrt{g} [-\frac{1}{2} (X_m +\frac{1}{4}
\epsilon_{mnp}[X_n,X_p])^2] +12\pi^2.
\een
The topological term, $12\pi^2$, is independent of the metric as expected. Now
the bound is saturated provided
\ben
X_m+\frac{1}{4} \epsilon_{mnp}[X_i,X_j]=0.
\een
When the space is flat ({\ie}, $L=\infty$), this equation reduces to Eq.
(\ref{boundeq}) which has no nontrivial solution as already mentioned above.
In curved space, however, it has a solution when $L=1$.\footnote{A simple
way of showing
this is to use a geometrical technique patterned after nonlinear elasticity
theory. See N.S. Manton, Commun. Math. Phys. {\bf 111}, 469 (1987).}
This is an identity map, inducing no distortion and corresponds to isometry.
One can perhaps associate this particular solution with a state of symmetry
different from the spontaneously broken chiral symmetry for which the bound
can never be satisfied.

\subsubsection{\it Quantization of the soliton}
\bs
To give a physical meaning, the soliton has to be quantized. As it is, the
hedgehog solution is not invariant under separate rotation in
configuration space or in isospace\footnote{It is localized in space also,
so breaks translational invariance. For the moment, we will ignore this
problem.}. It is invariant under the simultaneous
rotation of the two, with the conserved quantum number being the ``grand spin"
which is the vector sum of the spin and the isospin
\ben
\vec{K}=\vec{J}+\vec{I}.
\een

The hedgehog configuration breaks the rotational symmetry as mentioned,
as a consequence of which there are three zero modes corresponding to
the triplet of the isospin direction. For $S\in SU(2)$,
the equation of motion of the hedgehog is invariant under the transformation
\ben
U (\vec{x})\rightarrow S U(\vec{x}) S^\dagger
\een
so the states obtained by rotating the hedgehog are degenerate
with the unrotated hedgehog. To restore the symmetry, one has then to elevate
$S$ to a quantum variable by endowing it a time dependence and
quantize the zero modes. This is essentially equivalent to rotating the
hedgehog. It is also equivalent to projecting out good angular momentum and
isospin states. Note that in discussing induced gauge fields above, we have
already encountered the notion of an adiabatic rotation. Since this rotation
leads to next order in $1/N_c$ in $N_c$ counting in physical quantities,
we are implicitly making the adiabatic approximation.

Substituting $U(\vec{x},t)=S^\dagger (t) U(\vec{x}) S (t)$ into the Skyrme
Lagrangian (\ref{skmodel}), we get
\ben
L_{sk}=-{\cal M} + {\cal I}\Tr (\dot{S} \dot{S}^\dagger)
\een
with
\ben
{\cal I}=\frac{4\pi}{3} \frac{1}{g^3f}\int_0^\infty \, s^2 ds \sin^2 F
[1+{F^\prime}^2 + \frac{\sin^2 F}{s^2}]\label{mominertia}
\een
where $s=fg r$ is the dimensionless radial coordinate
and ${\cal M}=E$, the soliton energy. Note that in terms of $N_c$, both
${\cal M}$ and ${\cal I}$ are of order $N_c$ while $\dot{S}$ is
of order $1/N_c$. The theory can now be quantized canonically in a standard
way. To do so, introduce the collective coordinates $(a_0, \vec{a})$
by
\ben
S=a_0+i\vec{\tau}\cdot \vec{a}
\een
with
\ben
S^\dagger S=1=a_0^2 +\vec{a}^2.
\een
The momentum conjugate to $a_k$ is
\ben
\pi_k=\frac{\del L_k}{\del \dot{a}_k}=4{\cal I}\dot{a}_k.
\een
The Hamiltonian is
\ben
H=\sum_{k=0}^3 \pi_k \dot{a}_k -L_{sk}={\cal M}+\frac{1}{8{\cal I}}
\sum_{k=0}^3 \pi_k^2.
\een
Quantization is effected by taking $\pi_k=-i (\del/\del a_k)$; thus
\ben
H={\cal M} +\frac{1}{8{\cal I}} \sum_{k=0}^3\left(-i\frac{\del}{\del a_k}
\right)^2.\label{topH}
\een
The second term is just the Laplacian on $S^3$ whose eigenstates are the Jacobi
polynomials in $a$'s. Using the Noether method, the spin $J$ and isospin $I$
can be calculated
\ben
J_k &=& \frac{i}{2} \left(a_k\frac{\del}{a_0}-a_0\frac{\del}{\del a_k}-
\epsilon_{klm} a_l \frac{\del}{\del a_m}\right), \\
I_k &=& \frac{i}{2} \left(a_0\frac{\del}{\del a_k}- a_k \frac{\del}{\del a_0}
-\epsilon_{klm} a_l \frac{\del}{\del a_m}\right)
\een
with
\ben
\vec{J}^2=\vec{I}^2=\frac{1}{4}\sum_{k=0}^3 \left(-\frac{\del^2}{\del
a_k^2}\right).
\een
Therefore the Hamiltonian describes a spherical top
\ben
H={\cal M} +\frac{1}{2{\cal I}} \vec{J}^2.\label{top}
\een
The energy eigenstates of this Hamiltonian can be explicitly constructed
as polynomials in $a$'s. Whether they are even or odd polynomials depends
upon whether the object satisfies bosonic or fermionic statistics. A detailed
discussion on this matter will be given in the next subsection. If one
considers fermions, then for the ground state $J=I=1/2$ and
its wavefunction will then be a monomial in $a$'s. Specifically\cite{atkins},
\ben
|p\uparrow\rangle=\frac{1}{\pi} (a_1+ia_2), \ \ \ \
|p\downarrow\rangle=-\frac{i}{\pi} (a_0-ia_3),\\
|n\uparrow\rangle=\frac{1}{\pi} (a_0+ia_3), \ \ \ \
|n\downarrow\rangle=-\frac{1}{\pi} (a_1-ia_2).
\een
These can be identified with the physical proton with spin up or down and
neutron with spin up or down.
A similar construction can be made for the excited states with $J=I=3/2$ etc.
A more general construction will be given later when we include the
strangeness in the scheme.

\subsection{Light-Quark Skyrmions}
\bs
We have thus far motivated within the Cheshire
cat picture that the object we have, namely the skyrmion, is a baryon.
In what follows, we will supply further evidences that we are indeed dealing
with what amounts to baryons expected in QCD. We will start with
phenomenological aspects with $SU(2)$ skyrmions and later more mathematical
aspects associated with the WZW term. For more rigorous mathematical
exposition, we refer to the article by Fr\"{o}lich and
Marchetti \cite{frohlich}.

Some people have difficulty accepting the skyrmion as a bona-fide baryon
consistent with QCD. They ask where the quark-gluon imprints are in the
skyrmion description. They even attempt to find in skyrmions what should be
there in QCD. In other words, their suspicion is that the skyrmion ``misses"
something basic of QCD. Throughout what follows, we will argue that this
attitude is unfounded.

\subsubsection{\it $SU(2)$ skyrmions}
\subsubsection*{\it Masses}
\bs
Let us assume that for $N_c$ odd, we have a fermion and when quantized
as sketched above, it is a baryon. For $N_c=3$, we have two states
$J=I=1/2$ and $J=I=3/2$. The first is the nucleon, the ground state and
the second the excited state $\Delta$. In terms of $N_c$, their masses
are of the form
\ben
M_J= M_1 +M_0 +M^J_{-1} +O(N_c^{-2})\label{mass}
\een
where the subscript stands for $n$ in $N_c^n$. The $O(N_c)$ term corresponds to
the soliton mass ${\cal M}$ of Eq. (\ref{top}) and the $O(N_c^{-1})$ to the
``hyperfine structure" term, {\ie}, the second term of (\ref{top}).
The ``fine structure" term $M_0$ has not yet been encountered. It is
the first ($O(\hbar)$) quantum correction to the mass known as Casimir
energy which describes the shift in energy of the vacuum produced
by the presence of the soliton. Note that the two terms of $O(N_c)$
and $O(1)$ define the ground state of the baryons and the dependence on quantum
numbers appears at $O(1/N_c)$. When heavy flavors are introduced, there
will be an {\it additional} overall shift at $O(1)$ for each flavor. An
important point is that the first two terms of (\ref{mass}) is common to
all baryons, be they light-quark or heavy-quark baryons.

We will later present an argument that the term $M_0$ is attractive
and can be of order of $\sim -0.5$ GeV. For the moment, let us treat it
as a parameter independent of quantum numbers and estimate how much it
could be from phenomenology. Let us consider the nucleon and $\Delta$.
For definiteness, we will consider the Skyrme Lagrangian ${\cal L}_{sk}$
(\ref{skmodel}). There are two constants in the model, $f$ and $g$.
We fix the constant $f$ to the pion decay constant $f_\pi=93$ MeV.
The constant $g$ can be obtained from a more general Lagrangian containing
HGS vector mesons but here we shall fix it from imposing that the $g_A$
of the neutron decay come out to be the experimental value 1.25
(see below how $g_A$ is calculated)\cite{jacksonrho}.
The resulting value is $g\approx 4.76$.
Restoring the energy unit $f/2g$, the soliton mass is
\ben
M_1={\cal M}=1.23\times 12 \pi^2 \left(\frac{f}{2g}\right)\approx 1423
\ {\rm MeV}.
\een
Similarly from (\ref{mominertia}), we get numerically
\ben
{{\cal I}^{-1}}\approx 193 \ {\rm{MeV}}.
\een
We can now {\it predict} the mass difference between the $\Delta$ and $N$
\ben
\Delta M=\frac{3}{2} {\cal I}^{-1}\approx 290 \ \rm{MeV}
\een
to be compared with the experiment  296 MeV.
The Casimir energy is a difficult quantity to calculate accurately and
since our knowledge on it is quite uncertain, we cannot make a prediction
for the absolute value of the ground state baryon ({\ie}, nucleon) mass.
We can turn the procedure around and get some idea on the magnitude of
the Casimir energy {\it required} for consistency with nature.
Taking the nucleon mass formula with
$J=1/2$, we have $M_1+\frac{3}{8}{\cal I}^{-1}\approx 1495$ MeV and hence
\ben
M_0\approx (940 -1495) \ \rm{MeV}= -555 \ \rm{MeV}.\label{mzero}
\een
This may look large but it is in the range expected theoretically
as discussed below~\cite{casimir} and
also consistent with the expectation based on the $N_c$ counting that
\ben
M_1 >|M_0| > M_{-1}.
\een
There is a caveat to this discussion which would modify the numerics
somewhat. We have not taken into account the translational mode; the
center-of-mass correction, which is of $O(N_c^{-1})$, should increase
the centroid energy. Therefore the value of $M_0$ may even be higher
in magnitude than (\ref{mzero}).
\subsubsection*{\it Static properties}
\bs
As an example of static quantities, consider the axial-vector coupling constant
$g_A$. The nucleon matrix element of the axial current is
\ben
\langle N(p_\prime)|A_\mu^a|N(p)\rangle=\bar{U} (p^\prime)\frac{\tau^a}{2}
\left[g_A(q^2)\gamma_\mu\gamma_5 +h_A (q^2)q_\mu\gamma_5\right] U(p)
\een
with $q_\mu=(p-p^\prime)_\mu$ and $U(p)$ the Dirac spinor for the nucleon.
We will continue working with the chiral limit,
so the pseudoscalar form factor $h$ has the pion pole. Thus in the limit
that $q_\mu\rightarrow 0$, we have
\ben
\lim_{q\rightarrow 0}\langle N(p^\prime)|A_i^a|N(p)\rangle&=&\lim_{q\rightarrow
0} g_A (0) (\delta_{ij}-\hat{q}_i\hat{q}_j)
\langle N|\frac{\tau^a}{2}\sigma_j|N\rangle \nonumber\\
&=& \frac{2}{3} g_A(0) \langle N|\frac{\tau^a}{2}\sigma_i|N\rangle
\een
where we have expressed the matrix element in the nucleon rest frame.
We now obtain the relevant matrix element in the present model. Using the
axial current suitably ``rotated", we have
\ben
\int d\vec{x}A_i^a (\vec{x})=\frac{1}{2}\bar{g} \Tr (\tau_i S^\dagger\tau_a S),
\een
with
\ben
\bar{g}=\frac{4\pi}{3g^2} \int_0^\infty s^2ds\left(F^\prime +\frac{\sin 2F}
{s}(1+{F^\prime}^2) +2F^\prime \frac{\sin^2 F}{s^2}
+\frac{\sin^2 F}{s^3} \sin 2F\right)
\een
where we have used the dimensionless radial coordinate $s=(fg) r$. Note that
given the profile function, $\bar{g}$ depends only on the constant $g$.
The relevant matrix element is
\ben
\lim_{q\rightarrow 0}\int d\vec{x} e^{i{\vec{q}}\cdot {\vec{x}}}
\langle N|A_i^a (\vec{x})|N\rangle=\frac{1}{2}\bar{g}\langle N|\Tr (\tau_i
S^\dagger\tau_a S)|N\rangle=\frac{2}{3}\bar{g}\langle N|\frac{\tau^a}{2}
\sigma_i|N\rangle.
\een
We are therefore led to identify
\ben
g_A(0)=\bar{g}.
\een
Given $g_A$ from experiment, one can then determine
the constant $g$ from this relation.
The value $g\approx 4.76$ used in the mass formula was determined in this way.

Other static quantities like magnetic moments and various radii are calculated
in a similar way and we will not go into details here, referring to standard
review articles\cite{review}.

\subsubsection*{\it The Casimir energy}
\bs
The calculation of the $O(N_c^0)$ Casimir energy is in practice
a very difficult affair although in principle it is a well-defined problem
given an effective Lagrangian. The reason is
that there are ultraviolet divergences that have to be cancelled by counter
terms and since the effective theory is nonrenormalizable in the conventional
sense, increasing number of counter terms intervene as one makes higher-order
computations. In principle, given that the finite counter terms
can be completely determined from experiments, the nonrenormalizability is
not a serious obstacle
to computing the Casimir contribution. The main problem is that chiral
perturbation theory applicable in the Goldstone boson sector cannot
be naively applied because of the baryon mass which is not small compared
with the chiral scale $\Lambda_\chi\sim 4\pi f_\pi$. We will describe later
how this problem is avoided in chiral expansion when baryon fields are
explicitly introduced but it is not clear how to do so in the skyrmion
structure.

The basic idea of computing the Casimir energy of $O(N_c^0)$ is as follows:
Let us consider specifically the Skyrme Lagrangian (\ref{skmodel}). The
argument applies to any effective Lagrangian with more or less complications.
Let $L_\mu\equiv i{\bf \tau}\cdot {\bf L}_\mu$. Then (\ref{skmodel}) takes
the form
\ben
{\cal L}_{sk}=\frac{f^2}{2}\left( {\bf L}_\mu \cdot {\bf L}^\mu +\frac{1}
{2\lambda^2} [({\bf L}_\mu \cdot {\bf L}^\mu)^2 - ({\bf L}_\mu \cdot
{\bf L}_\nu)^2]\right)\label{skmodelp}
\een
with $\lambda=gf$. We denote by $\tilde{\bf L}_\mu$ the static solution of
the Euler-Lagrange equation of motion
\ben
\del^\mu \left(\tilde{L}_\mu^a + \frac{1}{\lambda^2} (\tilde{L}_\mu^a
\tilde{\bf L}_\nu \cdot \tilde{\bf L}_\nu -\tilde{\bf L}_\mu \cdot
\tilde{\bf L}_\nu \tilde{\bf L}_\nu)\right)=0.\label{el}
\een
In terms of the chiral field $U$, the solution is of course the
hedgehog we have seen above which we will denote by $U_0 (\vec{x})$.
We introduce fluctuation around $U_0$ by
\ben
U (x)= U_0 e^{i{\bf \tau}\cdot{\bf \phi}}
\een
and $L_\mu$ as
\ben
{\bf L}_\mu=\tilde{\bf L}_\mu +\del_\mu {\bf \phi} +\del_\mu {\bf \phi}
\wedge {\bf \phi} \cdots
\een
Substitution of this expansion into (\ref{skmodel}) gives rise to a term
zeroth order in ${\bf \phi}$ which is what we studied above for the nucleon
and $\Delta$, a term quadratic in ${\bf \phi}$ and higher orders.
There is no term linear in $\phi$ because its coefficient is just
the equation of motion (\ref{el}). To next to the leading
order, the relevant Lagrangian is the quadratic term which takes the form
\ben
\Delta {\cal L}=\phi^a\left(H^{ab}+\cdots \right)\phi^b,\\
H^{ab}=-\nabla_\mu^2\delta^{ab}-2\epsilon^{abc} \tilde{L}_\mu^c \nabla_\mu
\een
where the ellipsis stands for the term coming from fluctuations around
the Skyrme quartic term and higher derivative terms (and/or other matter
fields in a generalized Lagrangian). Ignoring terms higher order
in $\phi$'s, the effective potential $V^{eff}$ we wish to calculate is
\ben
e^{-i\int d^4x V^{eff} (\tilde{L})}=\int d[\phi] e^{i\int d^4 x \Delta
{\cal L}}.
\een
The fluctuation contains the zero modes corresponding to three translational
modes and three (isospin) rotational modes in addition to other
``vibrational modes".
The three rotational modes were quantized to give rise to the correct quantum
numbers of the baryons. Quantizing the translational zero modes is to restore
the translational invariance to the system. We are here interested in the other
modes which are orthogonal to these zero modes. This requires constraints
or ``gauge fixing". Now doing the integral over the modes minus the six
zero modes, we get formally
\ben
\int d^4x V^{eff} (\tilde{L})= -\frac{1}{2} \Tr^\prime \ln (H+\cdots)
\label{veff}
\een
where the prime on trace means that zero energy modes of $H$ be omitted in the
sum. This generates one-loop corrections to the effective potential.
The $O(N_c^0)$ Casimir energy, to one-loop order, is then given by
\ben
M_0=\frac{1}{T}\int d^4x \left(V^{eff} (\tilde{\bf L})-V^{eff} (0)\right)
\een
where $T$ is the time interval.
This can be evaluated by chiral perturbation theory. In doing so, one can
ignore the quartic and higher derivative terms in (\ref{veff}) in
the one-loop graphs since it leads to higher chiral order terms
relative to the one-loop terms evaluated with the quadratic
term $H$. In accordance with general strategy of chiral perturbation theory,
the ultraviolet divergences are to be eliminated by counter terms involving
four derivatives, with finite constants fixed by experiments. Those constants
are available from analyses on $\pi \pi$ scattering. Using this strategy,
several people have estimated this Casimir contribution to
the nucleon mass. Because of the dubious validity in using the derivative
expansion for skyrmions mentioned above, it is difficult to pin down
the magnitude precisely: The magnitude
will depend on the dynamical details of the Lagrangian and chiral loop
corrections. The calculations, however, agree on its sign which is negative.
Numerically it is found to range \cite{casimir}
from $\sim -200$ MeV to $\sim -1$ GeV. The recent more reliable
calculations \cite{holz}
give $\sim -500-- -600$ MeV, quite consistent with what seems to be
needed for the ground-state baryon mass.

\subsubsection{\it $SU(3)$ skyrmions: eight-fold way}
\bs
We now study the system of three flavors u,d and s in the highly
unrealistic situation where all three quarks are taken
massless\cite{su3skyrme}. The idea that
one can perhaps treat the three flavors on an equal footing, with the
mass of the strange quark treated as a ``small" perturbation turns out
to be not good at all, but it provides a nice theoretical framework
to study the role of the anomalous (Wess-Zumino) term which pervades
even in realistic treatments given below. Putting the chiral field
in $SU(3)$ space
\ben
U=e^{i\lambda^a \pi^a/f}
\een
where $a$ runs over octet Goldstone bosons, we can generalize (\ref{skmodel})
to apply to $SU(3)$
\ben
{\cal L}_n=-\frac{f^2}{4} \Tr L_\mu L^\mu +\frac{1}{32g^2}\Tr [L_\mu,L_\nu]^2
\label{lnormal}
\een
and
\ben
S_n=\int d^4x {\cal L}_n
\een
with $L_\mu$ defined with $U\in SU(3)$. While this Lagrangian was sufficient
in $SU(2)$ space, it is not in $SU(3)$ space as mentioned before.
Here we give a more physical argument for its ``raison d'\^{e}tre"
following Witten \cite{witten83}.

Consider two discrete symmetries $P_1$ and $P_2$
\ben
P_1:&& \ \ U(\vec{x},t)\rightarrow U(-\vec{x},t),\\
P_2:&& \ \ U(\vec{x},t)\rightarrow U^\dagger (\vec{x},t).
\een
It is easy to see that (\ref{lnormal}) is invariant under both $P_1$
and $P_2$ and of course under parity
\ben
P=P_1 P_2
\een
which transforms
\ben
U(\vec{x},t)\rightarrow U^\dagger (-\vec{x},t).
\een
QCD is of course invariant under parity but does not possess the symmetries
$P_1$ and $P_2$ separately. For instance the well-known five pseudoscalar
process $K^+ K^-\rightarrow \pi^+ \pi^- \pi^0$ is allowed by QCD but is not
by the Lagrangian (\ref{lnormal}). Thus there must be an additional term
that breaks $P_1$ and $P_2$ separately while preserving $P$. This information
is encoded in the WZW term which we have encountered before in lower
dimensions. In (3+1) dimensions, it is given by
\ben
S_{WZ}=-N_c\frac{i}{240\pi^2}\int_D d^5x \epsilon^{ijklm}\Tr
(\bar{L}_i \bar{L}_j \bar{L}_k \bar{L}_l \bar{L}_m)
\een
with $\bar{U} (\vec{x},t,s=0)=1$ and $\bar{U} (\vec{x},t,s=1)= U(\vec{x},t)$.
Here $D$ is a five-dimensional disk whose boundary is the space time. The
presence of the factor $N_c$ will be clarified later.
It is easy to see that this has the right symmetry structure. It also carries
other essential ingredients to render the CCP operative in baryon structure.

The action we will focus on is then the sum
\ben
S_{sk}= S_n +S_{WZ}.\label{sksu3}
\een
As in the nonstrange hadrons, a marked improvement in quantitative agreement
with experiments can be gained by introducing vector and axial-vector fields
but the qualitative structure is not modified by those fields, so we will
confine our discussion to the action (\ref{sksu3}) which we will
call Skyrme action although the WZW term is a new addition.

The $SU(3)$ skyrmion is constructed by first taking the ansatz that embeds
the hedgehog in $SU(3)$ as
\be
U_c=\left(\matrix {e^{i\vec{\tau}\cdot \hat{r} F(r)} & 0 \cr 0 & 1 \cr}\right).
\label{su3hedge}
\ee
This is one of two possible spherically symmetric ans\"{a}tze that one can
make. Instead of embedding the $SU(2)$ group, one could embed the $SO(3)$
group. However the latter turns out to give baryon-number even objects,
perhaps related to dibaryons. We shall not pursue this latter embedding
since it is not clear that such an embedding is physically relevant for
the time being.

Substitution of (\ref{su3hedge}) into (\ref{sksu3}) gives us the equations for
$SU(2)$ soliton with a vanishing WZW term as
the hedgehog is sitting only in that space. To generate
excitations in the isospin as well as in the strangeness direction
we rotate the hedgehog with the $SU(3)$ matrix $S$
\be
U=S(t) U_c S^\dagger (t).
\ee
The resulting Lagrangian from (\ref{sksu3}), after some standard though
tedious algebra, is
\be
L_{sk}=\int d^3x \L_{sk}=-\M -\frac{a(U_c)}{8}[\Tr \lambda_a S^\dagger
\dot{S}]^2
-\frac{b(U_c)}{8} [\Tr \lambda_{\mbox{\tiny A}} S^\dagger\dot{S}]^2
+L_{\mbox{\tiny WZ}}\label{sklag}
\ee
where $\M$ is the static soliton energy and
\be
a &=& \frac{16\pi}{3f g^3}\int_0^\infty s^2 ds\, \sin^2 F\left(1 +\frac{1}{2}[
{F^\prime}^2 +\sin^2 F/s^2]\right),\nonumber\\
b &=& \frac{4\pi}{f g^2}\int_0^\infty s^2 ds\,  (1-\cos F) \left(1+\frac{1}{8}[
{F^\prime}^2 +2\sin F/s^2]\right)\nonumber
\ee
with $\lambda_\alpha$ are the Gell-Mann matrices for $a=$ 1, 2, 3 and
$A=$ 4, 5, 6, 7 and
\be
L_{\mbox{\tiny WZ}}=i\frac{N_c}{2} B(U_c) \Tr [Y S^\dagger \dot{S}]
\ee is the WZW term\footnote{Putting the WZW term in this form
requires quite a bit of work. If the reader is diligent, this is an exercise.
Otherwise, consult A.P. Balachandran's lecture note, Ref.\cite{balachandran}}.
Here $B(U_c)$ is the baryon number corresponding to the configuration
$U_c$ and $Y$ is the hypercharge $Y=\frac{1}{\sqrt{3}} \lambda_8$.
\subsubsection*{\it The WZW term and $U(1)$ gauge symmetry}
\bs
The WZW term can now be shown to have some remarkable roles
in making the skyrmion correspond precisely to the physical baryon.
The Lagrangian (\ref{sklag}) has an invariance under right $U(1)$
local (time-dependent) transformation
\be
S(t)\rightarrow S(t) e^{iY\alpha (t)}.\label{u1gt}
\ee
This is simply because under this transformation $U=S U_c S^\dagger$
is invariant since $U_c$ commutes with $e^{iY\alpha (t)}$. This can be seen
explicitly on $L_n$ in (\ref{sklag}) as
\be
\Tr[\lambda_\alpha S^\dagger \dot{S}]\rightarrow \Tr[e^{iY\alpha (t)}
\lambda_\alpha e^{-iY\alpha (t)} S^\dagger\dot{S}] +i\dot{\alpha} \Tr
[\lambda_\alpha Y]
\ee
and $\Tr [\lambda_\alpha Y]=\frac{2}{\sqrt{3}}\delta_{\alpha 8}$.
Now the Wess-Zumino term is also invariant modulo a total derivative
term
\be
L_{\mbox{\tiny WZ}}\rightarrow L_{\mbox{\tiny WZ}} +\frac{N_c B}{3}
\dot{\alpha}.\label{noether}
\ee
The last term being a total derivative does not affect the equation of
motion. Normally a group of symmetries implies conservation laws but this
is not the case with a {\it time-dependent} gauge symmetry. It imposes
instead a constraint. To see this, let $\hat{Y}$ be the operator that generates
this $U(1)$ transformation which we will call right hypercharge generator.
By Noether's theorem,
\be
\hat{Y}_{\mbox{\tiny R}}=\frac{N_c B}{3}
\ee
and hence the allowed quantum state must satisfy the eigenvalue
equation
\be
\hat{Y}_{\mbox{\tiny R}}\psi=\frac{N_c B}{3}\psi.
\ee
We will see later that this is the condition to be imposed on the
wavefunction that we will construct.
\subsubsection*{\it A proof that the skyrmion is a fermion}
\bs
We can now show explicitly that the skyrmion is a fermion for three
colors ({\ie}, $N_c=3$). To do so, we note that the flavor $SU(3)_f$
and the spin $SU(2)_s$ are related to the left and right transformations,
respectively
\be
S &\rightarrow& BS\, \ \ \ B\in SU_L (3),\nonumber\\
S &\rightarrow& SC^\dagger\, \ \ \ C\in SU(2) \subset SU_R (3).
\ee
The proof is simple. Under $SU(3)_f$ transformation,
\be
S U_c S^\dagger \rightarrow B(S U_c S^\dagger)B^\dagger=(BS) U_c (BS)^\dagger
\label{flavortr}
\ee
so it is effectuated by a left multiplication on $S$ while under the space
rotation
$e^{i\alpha_i \hat{L}_i}$
with $\hat{L}_i$ ($i=$ 1, 2, 3), the angular
momentum operator,
\be
e^{i\alpha_i \hat{L}_i} (S U_c S^\dagger) e^{-i\alpha_i \hat{L}_i} =
(S e^{-i\alpha_i \lambda_i/2}) U_c (S e^{-i\alpha_i \lambda_i/2})^\dagger
\label{spacetr}
\ee
where we have used the fact that $S$ commutes with $\hat{L}_i$ and
$L_i +\lambda_i/2 =0$ with the hedgehog $U_c$. This shows that the spatial
rotation is a right multiplication.

Now let us see what happens to the wavefunction of the system when we make
a space rotation by angle $2\pi$ about the z axis for which we have
\be
C=\left(\matrix{-1 & 0 & 0 \cr 0 & -1 & 0\cr 0 & 0 &1 \cr}\right).
\ee
This induces the transformation on $S$
\be
S\rightarrow S C^\dagger= S e^{3\pi i Y}.
\ee
Therefore the wavefunction is rotated to
\be
\psi \rightarrow \psi e^{i \pi N_c B}
\ee
and hence for $N_c=3$ and $B=1$, the object is a fermion.
\subsubsection*{\it Canonical quantization}
\bs
To quantize the theory described by (\ref{sklag}),
we let\footnote{One can formulate what follows in a much more general
and elegant way by using vielbeins. There is of course no gain
in physics but it brings out the generic feature of the treatment.
Write (using one-form notation)
\be
S^\dagger dS=i v_\alpha^c (\theta) d\theta^c \lambda_\alpha.\nonumber
\ee
Here $v_\alpha^c$ are the vielbeins, $\theta^c$ denotes
some arbitrary parametrization
of $SU(3)$ with $\lambda_\alpha$ being the corresponding generators.
The canonical momenta are $p_c=\del \L_{sk}/\del \theta_c$.
If one defines $u_c^\beta$ as $u_c^\beta v_\alpha^c=\delta_\alpha^\beta$, then
the right generators are $R_\alpha=u_\alpha^c p_c$. The rest of the procedure
for quantization is identical.}
\be
S^\dagger \del_0 S=-i\lambda_\alpha \dot{q}_\alpha
\ee and reexpress the Lagrangian (\ref{sklag})
\be
L_{sk}= -\M +\frac{a}{2} (\dot{q}_a)^2 +\frac{b}{2}
(\dot{q}_{\mbox{\tiny A}})^2+\frac{N_c}{\sqrt{3}} B \dot{q}_8.
\ee
The momenta conjugate to $q_\alpha$ are
\be
\Pi_\alpha=\frac{\del L_{sk}}{\del \dot{q}_\alpha}= a \dot{q}_a \delta_{a
\alpha} + b \dot{q}_{\mbox{\tiny A}} \delta_{{\mbox{\tiny A}} \alpha}
+\frac{N_c}{\sqrt{3}} B \delta_{8\alpha}.
\ee
The Hamiltonian is
\be
H=\Pi_\alpha \dot{q}_\alpha -L_{sk}=\M+ \frac{1}{2} a (\dot{q}_a)^2
+\frac{1}{2}
b (\dot{q}_{\mbox{\tiny A}})^2.
\ee
The right generators are
\be
R_\alpha=\Pi_\alpha=a\dot{q}_\alpha \delta_{a\alpha}+b \dot{q}_{\mbox{\tiny A}}
\delta_{{\mbox{\tiny A}}\alpha} +\frac{N_cB}{\sqrt{3}} \delta_{8\alpha}
\ee
in terms of which the Hamiltonian takes the form
\be
H=\M +\frac{R_a^2}{2a}+\frac{R_{\mbox{\tiny A}}^2}{2b}+q_8 \chi
\ee
where we have incorporated the right hypercharge
constraint into the Hamiltonian as a (``primary") constraint (\`{a} la Dirac)
\be
\chi=R_8 +\frac{N_c B}{\sqrt{3}}\approx 0.
\ee
Expressed in terms of the $SU(2)$ Casimir operator $\sum_{\alpha=1}^3
R_\alpha^2= 4\hat{J}^2$ and the $SU(3)$ Casimir operator $\sum_{\alpha=1}^8
R_\alpha^2=4\hat{C}_2^2$, the Hamiltonian can be rewritten
\be
H=\M +\frac{1}{2} \left(a^{-1}-b^{-1}\right)4\hat{J}^2 +\frac{1}{2}
b^{-1} \left(4\hat{C}_2^2 -3\hat{Y}_{\mbox{\tiny R}}^2\right)\label{hamilt}
\ee
\subsubsection*{\it Wave functions}
\bs
To write down the spectrum for the Hamiltonian (\ref{hamilt}), consider
the irreducible representation (IR) of the $SU(3)$ group. The IR's are
characterized by two integers (p,q). The element  $S\in SU(3)$ is represented
by the operator $D^{(p,q)} (S)$, with the basis vector on which it acts denoted
by $|I,I_3,Y\rangle$. If we denote the generators of the $SU(2)$ subgroup of
$SU(3)$ by $\I_a$ ($a$=1,2,3), with the commutation relations
$[\I_a,\I_b]=i\epsilon_{abc}\I_c$, and the hypercharge operator by
$\hat{Y}$, we have
\be
\I_a^2 |I,I_3,Y\rangle &=& I(I+1)|I,I_3,Y\rangle, \nonumber\\
\I_3|I,I_3,Y\rangle &=& I_3|I,I_3,Y\rangle,\nonumber\\
\hat{Y}|I,I_3,Y\rangle &=& Y|I,I_3,Y\rangle.
\ee
The basis for the corresponding Hilbert space is
\be
D^{(p,q)}_{(I,I_3,Y),(I^\prime,I_3^\prime,Y^\prime)}\equiv
\langle I,I_3,Y|D^{(p,q)} (S)|I^\prime,I_3^\prime,Y^\prime\rangle.
\label{basis}
\ee
The right hypercharge constraint then requires that $Y^\prime=N_c B/3$
which is 1 for $N_c=3$ and baryon number 1. We will write the
wavefunction as
\be
\psi^{(p,q)}_{ab}=\sqrt{{\rm dim} (p,q)} \langle a|D^{(p,q)} (S)|b\rangle.
\ee

To determine the quantum numbers involved, let us see how the wavefunctions
transform under relevant operations. Under flavor $SU(3)$, $S$ gets
transformed by left multiplication, namely, $S\rightarrow B S$.
Under this transformation,
\be
D^{(p,q)}_{(I,I_3,Y),(I^\prime,I_3^\prime,Y^\prime)} (S)\rightarrow
D^{(p,q)}_{(I,I_3,Y),(I^",I_3^",Y^")} (B)
\times D^{(p,q)}_{(I^",I_3^",Y^"),(I^\prime,I_3^\prime,Y^\prime)} (S)
\ee
which is just the consequence of the group property. From this one sees that
the flavor is carried by the left group $(I,I_3,Y)$. Thus $I$ and $I_3$
are the isopin and its projection respectively and $Y$ is the hypercharge.
Under space rotation, $S$ is multiplied on the right $S\rightarrow
S C^\dagger$ with $C^\dagger$ a $2I^\prime +1$ dimensional IR of $SU(2)$.
Therefore we have
\be
D^{(p,q)}_{(I,I_3,Y),(I^\prime,I_3^\prime,Y^\prime)} (S)\rightarrow
D^{(p,q)}_{(I,I_3,Y),(I^",I_3^",Y^\prime)} (S)\times
D^{(p,q)}_{(I^",I_3^",Y^\prime),(I^\prime,I_3^\prime,Y^\prime)} (C^\dagger).
\ee
This means that the space rotation is characterized by the right group
$(I^\prime,I_3^\prime)$. Thus $I^\prime_a$ is the angular momentum $J_a$,
$I^\prime_3$ its projection $-J_3$ and $Y^\prime$ the right hypercharge.

The system we are interested in has $N_c=3$, $B=1$, and $Y^\prime=1$
for which the lowest allowed IR's are $(p=1, q=1)$ corresponding to
the octet with $J=1/2$ and $(p=3,q=0)$ corresponding to the decuplet
with $J=3/2$. In this wavefunction representation, the Casimir operators
have the eigenvalues
\be
\hat{J}^2 \psi_{..J..}^{(p,q)}&=& J(J+1)\psi_{..J..}^{(p,q)},\nonumber\\
\hat{C}_2^2\psi_{ab}^{(p,q)} &=& \frac{1}{3}[p^2+q^2+3(p+q)+pq]
\psi_{ab}^{(p,q)}.\nonumber
\ee
With this, we can write down the mass formula valid in the chiral limit
for the Hamiltonian (\ref{hamilt})\cite{su3skyrme}
\be
M=\M +2(a^{-1}-b^{-1})J(J+1)+2b^{-1}\{\frac{1}{3}[p^2 +q^2 +3(p+q)
+pq]-\frac{3}{4}\}.\label{massu3}
\ee

Calculation of the matrix elements relevant to  static
properties and other observables involves standard (generalized)
angular momentum algebra and is straightforward, so
we will not go into details here.
\subsubsection*{\it Elegant disaster}
\bs
The mass formula (\ref{massu3}) is valid for the chiral limit
at which all three quarks have zero mass. The u and d quarks are
practically massless but in reality, the s-quark mass is not.
Given the scheme sketched above, it is tempting to retain
the elegance of the $SU(3)$ collective rotation by assuming that the
mass difference can be treated as perturbation. This does not sound so
bad in view of the fact that the Gell-Mann-Okubo mass formula which
is proven to be so successful
is obtained by taking the quark mass terms as a perturbation and that
current algebra works fairly well even with s-quark hadrons.
The procedure to implement the strange quark mass then is
that one takes the $SU(3)$ wavefunctions constructed
above and takes into account the symmetry breaking to first order
with
\be
\L_{sb}=a\Tr\, [U(M-1)+{\rm h.c.}] \label{lsb}
\ee
where $M$ is the mass matrix which we can choose to be diagonal (since we
are focusing on the strong interaction sector only), $M={\rm diag}\,
(m_u,m_d,m_s)$. This is the standard form of symmetry breaking transforming
$(3,\bar{3})+(\bar{3},3)$. The spectrum then will be given by the
mass formula (\ref{massu3}) plus the lowest-order perturbative
contribution from (\ref{lsb}). The matrix
elements of physical operators will however be given by the unperturbed
wavefunctions constructed above.
We will not say much anymore since this elegant
scheme {\it does not work!}. The phenomenology is a simple disaster
\cite{prasz} not only for the spectrum
but also for other observable matrix elements, showing
that considering the s-quark mass to be light in the present context
is incorrect. From this point of
view, it seems that the s-quark cannot be classed in the chiral family.
\subsubsection*{\it Yabu-Ando method}
\bs
That the eight-fold way of treating collective variables fails completely
seems at odds with the relative success one has with current algebras with
strangeness. Why is it that current algebras work with kaons to $\sim$ 20
\% accuracy if the strange quark mass is too heavy to make chiral symmetry
meaningless ?

A way out of this contradiction was suggested by the method of
Yabu and Ando \cite{yabuando}. The idea is to preserve all seven (eight minus
one right hypercharge constraint)
zero modes as relevant dynamical variables but treat the symmetry breaking
due to the s quark nonperturbatively. The result indicates that in consistency
with three-flavor current algebras, the s quark {\it can still be} classed in
the chiral family. This together with what we find below (the
``heavy-quark" model of Callan and Klebanov)
presents a puzzling duality that the s quark seems at the same time {\it
light} and {\it heavy}. Testing experimentally
which description is closer to Nature will be addressed later.

We start with the eight-fold way Hamiltonian (\ref{hamilt}) which we will
call $H_0$ plus a symmetry breaking term
\be
H_{\mbox{\tiny SB}}=\frac{1}{2}\gamma (1-D_{88})\label{sb}
\ee
with
\be
D_{ab} (S)\equiv \frac{1}{2} {\rm Tr} (\lambda_a S \lambda_b S^\dagger).
\ee
The coefficient $\gamma$ measures the strength of the symmetry breaking
that gives the mass difference between the pion and the kaon.
Now if we adopt the parametrization used by Yabu and Ando
\be
S=R(\alpha,\beta,\gamma)e^{-i\nu\lambda_4}
R(\alpha^\prime,\beta^\prime,\gamma^\prime)e^{-i\frac{\rho}{\sqrt{3}}\lambda_8}
\ee
where $R(\alpha,\beta,\gamma)$ is the $SU(2)$ isospin rotation matrix with
the Euler angles $\alpha,\beta,\gamma$ and the eighth right generator $R_8$ and
$\rho$ are the conjugate variables, then
\be
H_{\mbox{\tiny SB}}=\frac{3}{4}\gamma \sin^2 \nu.
\ee
The idea then is to diagonalize exactly the total
Hamiltonian $(H_0 +H_{\mbox{\tiny SB}})$
in the basis of the eight-fold way wavefunction. Specifically
write the wavefunction as
\be
\Psi_{YII_3,JJ_3}=(-1)^{J-J_3}\sum_{M_{\mbox{\tiny L}},M_{\mbox{\tiny R}}}
D_{I_3,M_{\mbox{\tiny L}}}^{(I)*} (\alpha,\beta,\gamma) f_{M_{\mbox{\tiny L}}
M_{\mbox{\tiny R}}} (\nu) e^{i\rho}D_{M_{\mbox{\tiny R}},-J_3}^{(J)*}
(\alpha^\prime,\beta^\prime,\gamma^\prime).
\ee
Then the eigenvalue equation
\be
[\hat{C}_2^2+\omega^2 \sin^2 \nu]\Psi=\epsilon_{\mbox{\tiny SB}}\Psi
\ee
leads to a set of coupled differential equations for the functions
$f_{M_{\mbox{\tiny L}}M_{\mbox{\tiny R}}} (\nu)$. Here
$\omega^2=\frac{3}{2}\gamma b$ with $b$ the moment of inertia in the
``strange" direction appearing in (\ref{hamilt}).
When the parameter $\gamma$ or $\omega^2$ is zero, the wavefunction is
just the pure $SU(3)$ symmetric one corresponding to the {\it eightfold
way} given above, namely the matrix element of an irreducible representation of
$SU(3)$ containing states that satisfy $Y=1$ and $I=J$.

The mass spectrum is still very simple. It is given by the $SU(3)$
symmetric formula (\ref{massu3})) supplemented by the eigenvalue
$\epsilon_{\mbox{\tiny SB}}$ of the $H_{\mbox{\tiny SB}}$
\be
M &=& \M +2(a^{-1}-b^{-1})J(J+1)+2b^{-1}\{\frac{1}{3}[p^2 +q^2 +3(p+q)
+pq]-\frac{3}{4}\}\nonumber\\
&+& 2b^{-1}\epsilon_{\mbox{\tiny SB}} .\label{massya}
\ee
The wave functions are also quite simple. A simple way of understanding
the structure of the resulting wave function is to write it in perturbation
theory. For instance, since the symmetry-breaking Hamiltonian mixes
the ${\bf 8}$ with $\overline{{\bf 10}}$ and ${\bf 27}$ for spin $1/2$
baryons, the proton wave function would look like \cite{yabuhkl}
\be
|p\rangle=|p;{\bf 8}\rangle +\frac{\omega^2}{9\sqrt{5}}|p;
\overline{{\bf 10}}
\rangle+  \frac{\sqrt{6}\omega^2}{75}|p;{\bf 27}\rangle +\cdots
\ee
\subsubsection*{\it Phenomenology with Yabu-Ando method}
\bs
Treating the s-quark mass term exactly improves considerably the baryon
spectrum as well as their static properties. Once the ground state is
suitably shifted, the excitation spectra are satisfactory for physical
values of the coupling constants $f$, $g$ etc. Other static observables
such as magnetic moments, radii, weak matrix elements etc. are
also considerably improved\cite{parkweigel}.
The problem of high ground-state energy gets worse,
however, for $SU(3)$ if one takes physical values of
the coupling constants but as mentioned above, this is expected as the Casimir
energy is proportional to the number of
zero modes and the Casimir attraction will be $7/3$ times greater than the
two-flavor case. Again a careful Casimir calculation will be needed to
confront Nature in a quantitative way.

A surprising thing is that the results obtained in the
Yabu-Ando $SU(3)$ description are rather similar to the
Callan-Klebanov method described below, though the basic assumption is
quite different. In the latter, while the vacuum is assumed to be $SU(3)$
symmetric, the excitation treats the strangeness direction completely
asymmetrically with respect to the chiral direction. This seems to indicate
that
there is an intriguing duality that the s quark can be considered
both light and heavy.

An important caveat with the exact treatment of the symmetry breaking is that
it is not consistent with chiral perturbation theory. The point is that
the symmetry-breaking Hamiltonian $H_{SB}$ is a linear term in
quark-mass matrix $M$ and in principle
one can have terms higher order in $M$ in combination with derivatives.
Treating the linear mass term exactly while ignoring higher order
mass terms is certainly not consistent. On the other hand, as mentioned
before, the naive chiral expansion does not make sense in the baryon sector.
Since the Yabu-Ando procedure seems to work rather satisfactorily, this
may indicate that while the higher order chiral expansion is meaningful in
the meson sector, it may not be in the baryon sector. This is somewhat
like the six-derivative term in the skyrmion structure which is formally
higher order in derivative but is more essential for the baryon
than higher orders in
quark mass term. This suggests that chiral perturbation theory powerful
in certain aspect (see Lecture III) is not very useful in {\it baryon
structure.}
\subsection{On the ``Proton Spin" Problem}
\bs
We have here an ``ingenious" model, full of clever ideas. But such a model
would be useless if it could not explain existing experimental data {\it and}
could not make new predictions. Up to date, there is no evidence
against the model but neither is there an unambiguous support for it.
As an illustration, we show what the skyrmion description says about the
controversial ``proton spin" problem.
\subsubsection{\it Flavor singlet axial current matrix element of the
proton}
\bs
The operator that is widely believed to be carrying
information on the strangeness content as well as spin content
of the nucleon  is the current
\be
\bar{q}\Gamma_\mu q
\ee
where
$q$ is the quark field $u$, $d$ and $s$ and $\Gamma_\mu=\gamma_\mu$,
$\gamma_\mu \gamma_5$. The axial current $\gamma_\mu \gamma_5$
will be considered in this subsection.

The most intersting quantity is the flavor-singlet axial
current (FSAC in short)
\be
J_{5\mu}^0=\frac 12 \left(\bar{u}\gamma_\mu\gamma_5 u +\bar{d}\gamma_\mu
\gamma_5 d +\bar{s}\gamma_\mu\gamma_5 s \right)\label{FSA}
\ee
and its matrix element of the proton. The matrix element at zero
momentum transfer measures the flavor-singlet axial charge $g_A^0$
which can be extracted from deep inelastic experiments on spin-dependent
structure functions. By its intrinsic
structure, it can also carry information of the strangeness content of
the spin of the proton. This quantity touches on various aspects of
QCD proper and therefore merits an extensive discussion. Here we shall
address this issue in terms of chiral effective theories to the extent
that it is possible.

In order to treat the flavor singlet channel of (\ref{FSA}), we have to include
the nonet of pseudoscalar mesons rather than the usual octet so far treated.
Thus we write the chiral field as
\be
\tilde{U}=e^{i\lambda_0 \eta^\prime/f} U, \ \ \ \ \pi\equiv \frac{\lambda}{2}
\cdot \pi
\ee
where $U=e^{2i\pi/f}$ with ${\rm det} U=1$
is the octet chiral field. Here $\lambda_0=\sqrt{\frac 23}
\vec{1}$. Since the singlet
$\eta^\prime$ has a mass due to $U_A (1)$ anomaly, we need
to take into account the anomaly in the effective Lagrangian as discussed
in Lecture I. We do so in the simplest possible way following Schechter
\cite{schechtspin} whose discussion captures the essence of the problem
without too much complication.
For this we introduce the pseudoscalar glueball field $G=\del_\mu K^\mu$
with
\be
\del_\mu K^\mu=\frac{N_f \alpha_s}{4\pi} \Tr\left (F_{\mu\nu}
{\tilde{F}}^{\mu\nu} \right)
\ee
(in the literature $K^\mu=\frac{N_f \alpha_s}{2\pi}\epsilon^{\mu\nu\alpha\beta}
\left(A^a_\nu F^a_{\alpha\beta}-\frac 23 g f^{abc}A^a_\nu
A^b_\alpha A^c_\beta\right)$ is called Chern-Simons current and
$\tilde{F}$ is the dual to the field tensor $F$)
as given in Lecture I and write the chiral Lagrangian as
\be
\L=\frac{f^2}{4} \Tr (\del_\mu \tilde{U} \del^\mu \tilde{U}^\dagger)
+\frac{1}{12 f^2 m_{\eta^\prime}^2} G^2 +\frac{i}{12} G \ln \left(
\frac{{\rm det}\tilde{U}}{{\rm det} \tilde{U}^\dagger}\right) +\cdots
\label{schechter}
\ee
where the ellipsis denotes higher derivative terms. The field $G$ has
no kinetic energy term, so it does not propagate and hence the
equation of motion just relates $G$ and $\eta^\prime$
\be
G=\sqrt{6} f m_{\eta^\prime}^2\ \eta^\prime.\label{Gnaive}
\ee
{}From the Lagrangian density one obtains the flavor-singlet axial current
(FSAC) in the form
\be
J_{5\mu}^0=\sqrt{6} f \del_\mu \eta^\prime\label{FSAC}
\ee
and hence using the equation of motion for the $\eta^\prime$ field
\be
\del^\mu J_{5\mu}^0= G \label{u1anomaly}
\ee
which expresses the $U_A (1)$ anomaly. Consider now the matrix element of
(\ref{FSAC}) for the proton
\be
\langle p^\prime|J_{5\mu}^0|p\rangle=\frac 12\overline{U} (p^\prime)\left[
g_A^0 \gamma_\mu +g_P^0 \frac{q_\mu}{m_N}\right]\gamma_5 U(p)
\ee
where $U(p)$ is the Dirac spinor for the nucleon with four-momentum $p$.
At zero momentum transfer $q_\mu\equiv p^\prime-p=0$, the second term
(pseudoscalar term)
vanishes since there is no zero mass pole due to $U_A (1)$ anomaly.
On the other hand
from the right-hand side of (\ref{FSAC}), we have that
\be
g_A^0=0.\label{zerofsac}
\ee
{\em This is the prediction of the skyrmion picture in its simplest version.}
This can be easily understood as follows. Since the axial current operator
(\ref{FSAC}) is a gradient operator, its matrix element
is proportional to the momentum
transfer $q_\mu=(p^\prime -p)_\mu$ times a form factor, so its divergence
goes to zero as $q^2\rightarrow 0$ as long as there is no zero mass pole in
the form factor. Indeed in the flavor-singlet channel, there is no zero
mass pole due to the anomaly.

A vanishing flavor-singlet axial charge {\it naively}
implies that the proton contains
strange-quark axial charge. To see this point, define
\be
\langle p|\int d^3x \bar{q}\gamma_\mu\gamma_5 q|p\rangle\equiv
\Delta q S_\mu
\ee
where $q=$ $u$, $d$, $s$ is the quark field and $S_\mu$ the spin polarization
vector. One can relate $\Delta q$ to the parton distribution function
\be
\Delta q=\int_0^1 dx [q_+ (x) +\bar{q}_+ (x) -q_- (x) -\bar{q}_- (x)]
\ee
where $q_\pm$ $(\bar{q}_{\pm})$ are the densities of parton quarks
(antiquarks) with helicities $\pm \frac 12$ in the proton. Thus
\be
\langle p|J_{5\mu}^0|p\rangle\equiv \Delta \Sigma S_\mu
\ee
with
\be
\Delta \Sigma=\Delta u +\Delta d +\Delta s.
\ee
The result (\ref{zerofsac}) implies a startling consequence on polarized
quark moments, namely that
\be
\Delta \Sigma=0.\label{delsigma0}
\ee
The simplest Skyrme model we used above (without vector
mesons or higher derivatives or quark bag) would predict this.
Experimentally with the help of Bjorken sum rule and hyperon decays, the
data give\cite{elliskarliner}
\be
\Delta \Sigma^{exp}=0.22\pm 0.10\label{delsigexp}
\ee
so one may be inclined to conclude as often done in the literature
that ``the valence quarks in the proton carry negligible spin," generating
a ``proton spin crisis." This interpretation comes about in the following
way. In the nucleon, the helicity sum rule
says that the proton spin $1/2$ consists of
\be
\frac 12= \frac 12 \Delta \Sigma +\Delta g +\Delta L_z\label{h-sumrule}
\ee
where
$\Delta g$ is a possible contribution from gluons (see later) and
$\Delta L_z$ the orbital contribution. In the usual quark model such as
non-relativistic quark model and the MIT bag model, $\Delta g=0$
and $\Delta L_z\approx 0$, so the spin is lodged mainly in the valence
quarks, so $\Delta \Sigma \approx 1$ (the relativistic effect reduces this
by about 25\%).  The observed value (\ref{delsigexp}) is in a violent
disagreement with this quark-model description.

How does one understand the skyrmion prediction $\Delta \Sigma
\approx 0$ ? The pure skyrmion without glueball degrees of freedom and
without vector mesons or quark bags has $\Delta g=0$. Therefore from
the helicity sum rule (\ref{h-sumrule}), the spin is entirely lodged
in the orbital part. Is this unreasonable?

We will later show that none of these is a correct picture.
It turns out that the skyrmion picture starts with $g_A^0\approx 0$
whereas usual quark models start with $g_A^0\approx 1$ with the corrections
moving upwards in the former and downwards in the latter.

Let us assume for the moment that (\ref{delsigma0}) holds and see what
that means on the ``strangeness in the proton."
Now from neutron beta decay (using Bjorken sum rule), we have
\be
\Delta u-\Delta d=g_A=1.25
\ee
and from the hyperon decay
\be
\Delta u +\Delta d -2\Delta s= 0.67
\ee
Using $\Delta \Sigma=0$, we obtain
\be
\Delta u \approx 0.74, \ \ \ \Delta d \approx -0.51, \ \ \
\Delta s \approx -0.23.
\ee
Naively then one would be led to the conclusion that there is large strangeness
content in the proton current matrix element as already indicated in the matrix
element of the scalar condensate $\bar{s}s$.

It turns out that this interpretation is too naive because of the subtle
role that the $U_A(1)$ anomaly plays. The point is that the gauge-invariant
quantity $\Delta q$ is not simply measuring the $q$ quark content. It contains
information on gluons. Furthermore the Lagrangian (\ref{schechter})
itself is not the whole story when baryons are around. We have to consider
the baryons as solitons of the theory and take into account
the soliton coupling to the glueball
field $G$ as well as the quarkish scalar field $\eta^\prime$. This one
could do in a self-consistent way starting with (\ref{schechter}).
It is however much simpler, following Schechter \cite{schechtspin}
to introduce the baryon fields directly and
couple them to the $\tilde{U}$ and $G$ fields while treating
(\ref{schechter}) as consisting of fluctuating chiral fields.
We will denote this mesonic Lagrangian $\L_{M}$.
In fact in the next chapter, we will be forced to take this approach in
treating nuclei and nuclear matter.

We wish to preserve the anomaly equation
(\ref{u1anomaly}) (ignoring quark masses)
\be
\del^\mu J_{5\mu}^0=G=\del^\mu K_\mu=\frac{N_f\alpha_s}{4\pi}
\Tr (F_{\mu\nu} \tilde{F}^{\mu\nu}).\label{uanomaly}
\ee
Since the last term of (\ref{schechter}) embodies fully the anomaly,
we need to introduce only the couplings that are invariant under chiral
$U(3)\times U(3)$ so that the anomaly relation is not spoiled.
Such terms involving $\eta^\prime$ and $G$ can be written down
immediately
\be
\Delta \L=-\Tr (\overline{B}\gamma_\mu\gamma_5 B) (C_{\eta^\prime}
\del^\mu \eta^\prime +C_{\mbox{\tiny G}} \del^\mu G).\label{deltal}
\ee
For later purpose, it is useful to define a $GNN$ coupling
\be
C_{\mbox{\tiny G}} &=& g_{\mbox{\tiny GNN}}/2m_B
\ee
where $m_B$ is a baryon (say, nucleon) mass and $B$ is the $SU(3)$
matrix-valued baryon field.  The physical $\eta^\prime NN$ coupling
will be defined and written later as $g_{\eta^\prime \mbox{\tiny NN}}$.
It is clear that the
coupling with the $G$ field contributes neither to the anomaly nor
to the FSAC $J_{5\mu}^0$. While the coupling with $\eta^\prime$ leaves the
anomaly relation intact it will however contribute to the FSAC
since $\eta^\prime$ does transform under $U_A(1)$.  We use the Noether
method to construct the FSAC from the Lagrangian $\L_{M}+\Delta \L$
\be
J_{5\mu}^0=\sqrt{6} f\del_\mu \eta^\prime +\sqrt{6}f C_{\eta^\prime}
\Tr(\overline{B} \gamma_\mu\gamma_5 B).\label{FSACP}
\ee
The second term is the new entry owing to the fact that the baryon matter is
present. Such a term would be present whenever other degrees of
freedom than Goldstone bosons enter into the baryon structure.
Note that it is not a gradient of something. Now
calculating the matrix element between the proton state at zero
momentum transfer, we get zero from the first term as explained above
but there is a (possibly non-vanishing) contribution from the second term
\be
g_A^0=\sqrt{6}f C_{\eta^\prime}.\label{fsacharge}
\ee
Although there is nothing which says that this cannot be zero,
there is no compelling reason why $g_A^0$ should vanish either.
Since $C_{\eta^\prime}$ cannot measure directly the $\eta^\prime NN$
coupling (see below), even if $g_A^0\approx 0$ experimentally, it would not
be possible
to conclude that the $\eta^\prime$ coupling to the nucleon goes to zero.

The divergence of the FSAC (\ref{FSACP}) is still given by (\ref{uanomaly})
since $\Delta\L$ is $U_A(1)$ invariant. However now
$G$ is modified from the previous equation (\ref{Gnaive}). The equation
of motion for $G$ is
\be
G=\sqrt{6}f m_{\eta^\prime}^2 \eta^\prime -6f^2 m_{\eta^\prime}^2
C_{\mbox{\tiny
G}} \del^\mu \Tr (\overline{B}\gamma_\mu\gamma_5 B).
\ee
Substituting this into $\L_M +\Delta \L$ (or ``integrating out" the $G$ field
in field theory jargon) and putting the baryons on-shell, we get
\be
\L_{\eta^\prime B} &=&\frac{f^2}{4}\Tr(\del_\mu U \del^\mu U^\dagger) +\frac 12
(\del_\mu \eta^\prime)^2 -\frac 12 m_{\eta^\prime}^2 \, {\eta^\prime}^2
\nonumber\\
& & +\left(2m_B C_{\eta^\prime}+\sqrt{6}f g_{\mbox{\tiny GNN}}
m_{\eta^\prime}^2\right) \,\eta^\prime \Tr(\overline{B}\gamma_5 B)\nonumber\\
& & -3 f^2 m_{\eta^\prime}^2 g_{\mbox{\tiny GNN}}^2
[\Tr (\overline{B}\gamma_5 B)]^2.\label{LetaB}
\ee
{}From this Lagrangian one can read off the effective $\eta^\prime NN$ coupling
\be
g_{\eta^\prime {\mbox{\tiny NN}}}=2m_B C_{\eta^\prime}
+\sqrt{6} f m_{\eta^\prime}^2 g_{\mbox{\tiny GNN}}.\label{grelation}
\ee
Eliminating $C_{\eta^\prime}$ from Eqs.(\ref{fsacharge}) and (\ref{grelation}),
we obtain what is known as ``$U_A(1)$ Goldberger-Treiman relation"
\be
m_B g_A^0=f_{\eta^\prime} (g_{\eta^\prime{\mbox{\tiny NN}}}-2f_{\eta^\prime}
m_{\eta^\prime}^2 g_{\mbox{\tiny GNN}})
\ee
with $f_{\eta^\prime}=\frac{\sqrt{6}}{2} f$.
Shore and Veneziano showed that this relation follows rigorously from
$U_A(1)$ Ward identities \cite{shore}.
The derivation given here, while less rigorous,
is certainly much simpler and conveys the essential idea that the flavor
singlet axial charge has two components: One matter component involving
the $\eta^\prime$ field and the other gluon component involving
pseudoscalar glueball $G$. Smallness of $g_A^0$ could arise from
either small $\eta^\prime$ contribution {\it and} small glueball
contribution or a cancellation of the two components regardless of
their individual magnitudes. Thus the issue of strangeness in the proton
spin may just be an artifact of a particular way of looking at the
spin polarization moments.

While the $U_A(1)$ Goldberger-Treiman
relation involves gauge invariant quantities, it is not obvious that
all the quantities that appear there are ``measurable." For instance
one can think of ``measuring" the gluon coupling $g_{\mbox{\tiny GNN}}$
in lattice QCD. However it is nor clear what the operator is since
there is no gauge invariant dimension-3 operator gluon operator.
Furthermore as one can surmise from the last term of (\ref{LetaB}),
the $G$ is either a massive object above the chiral scale or else
unphysical. It is interesting to note that a term quartic in baryon
field is naturally generated in the Lagrangian (\ref{LetaB}), contributing
a contact term to baryon-baryon interactions. There will thus be two terms
representing $\etap$ exchange in nucleon-nucleon potential, this contact term
plus a finite range $\etap$ exchange.

\section{Excited Baryons}
\bs
So far we have treated what we might classify as ``ground-state" baryons
and their rotational excitations. Thus in the light u- and d-quark sector,
we encountered only the states with $J=I$ and in the $SU(3)$ sector,
strange baryons were rotational excitations in the strange direction on top
of the $J=I$ nonstrange baryons. But in nature there are excitations that
cannot be generated in this way. In the $SU(2)$ sector, for instance,
there are excitations with $J\neq I$. To describe this class of excitations,
one requires ``vibrational" modes and Berry potentials associated with these
modes.
\subsection{Light-Quark Excitations}
\bs
In the next subsection, we shall consider the strangeness as a ``vibration"
with its frequency $\omega_K$ much ``faster" than the rotational
frequency discussed above and describe strange hyperons in terms of nonabelian
Berry potentials induced when the vibrational mode is integrated out.
Since the description is better the faster the fast variable is, the
description is expected to be more reliable the heavier the heavy quark
is. The strange flavor is quite marginal in this sense while the c and b
quarks are certainly of that type.

Here let us simply assume that we can apply the same adiabaticity relation to
the vibrational excitations in the light u- and d-sector. Again using the
chiral-bag picture, we can think of the vibration in the meson sector to
be the fluctuating pion field and in the quark sector to be the excitation
of one of the quarks sitting in $K=0^+$ level to higher $K$ levels, the
lowest of which is the $K=1^+$ level. The Hamiltonian describing this
excitation was given in Lecture I, {\ie}, Eq. (\ref{hexcit}),
\be
H^* =  \epsilon_K {\bf 1}+ \frac{1}{2\I}\left[
+\frac{g_K}{2}\vec{J_K}^2 + (1 - \frac{g_K}{2}) \vec{I}^2
-\frac{g_K}{2}(1-\frac{g_K}{2})\vec{T}^2 \right]\label{hexcit2}
\ee
This describes only one quark excited from the $K=0^+$ level to a
$K$ level. One could generalize this to excitations of more than one quark
with some simplifying assumptions ({\ie}, {\it quasiparticles})
but this may not give a realistic
description in the light-quark sector whereas in the heavy-quark sector
considered in the next chapter, it is found to be quite accurate.
Here we will not consider multi-quark excitations.

If one confines to one-quark excitations, then there are three parameters
involved. They are all calculable in a specific model as described in the
chiral bag model \cite{LNRZpl} but given the
drastic approximations, one should not expect that the calculation would
yield reasonable values. On the other hand, the generic structure of the
mass formula suggests that given a set of parameters, the mass formula
could make useful predictions. Suppose then that we write mass formulas
eliminating the parameters $\epsilon_K$, $\I$ and $g_K$.
We give a few of the resulting mass formulas and see how they work.
For instance, in the Roper channel, it follows from (\ref{hexcit2})
that
\be
P11-N=P33-\Delta
\ee
where here and in what follows we use particle symbols for their masses.
Empirically the left-hand side is 502 MeV and the right-hand side is
688 MeV. In the odd-parity channel,
\be
D13-D35+\Delta-N=-\frac 14 (D35-S31).
\ee
{}From the data, we get 116 MeV for the left side and 76 MeV for the right.
Also
\be
S31-S11=\frac 52 (\Delta-N)-\frac 32 (D35-D13).
\ee
Empirically the left-hand side is 85 MeV and the right-hand side gives
125 MeV.

The agreement of the mass formulas is reasonable. But it is not very good
either. This is not surprising as already warned above. Specifically
the following should be noted. In the pure skyrmion description, we know that
the vector mesons $\rho$, $\omega$ and possibly $a_1$ play an important role
in describing some of the channels involved. For instance, to understand
the phase shifts of the $S31$ and $S11$ channels, it is found that an
anomalous parity (Wess-Zumino) term involving vector mesons consistent with
hidden gauge symmetry is essential \cite{masak}.
Now in our framework what this
means is that the quark excitations corresponding to the vector mesons
$\rho$ and $\omega$ etc. must be identified. This problem has not yet been
fully understood in terms of induced gauge structure.
\subsection{Heavy-Quark Baryons as Excitations}
\bs
As announced, we will describe baryons with one or more heavy quarks as
{\it excitations} of the Class B defined in subsection 2.3.1, Lecture I.
Before doing this, we describe the picture
of strange hyperons as {\it bound soliton-kaon complex} suggested by Callan and
Klebanov \cite{ck}.

\subsubsection{\it Strangeness as a heavy flavor: Callan-Klebanov model}
\bs
Since the s-quark is not ``light," let us imagine that it is very heavy.
In a strict sense, this does not sound right. After all the s-quark
mass in the range of 150 to 190 MeV is comparable to the QCD scale parameter
$\Lambda_{\mbox{\tiny QCD}}$, not much larger as would be required to be in the
heavy-quark category. As we will see later, heavy-quark hadrons exhibit
a heavy-quark symmetry which is not explicitly visible but known to be
present in the QCD Lagrangian that applies to the quarks $Q$
whose mass satisfies
$m_{\mbox{\tiny Q}} >> \Lambda_{\mbox{\tiny QCD}}$. The b quark belongs
to this category and to some extent so does the charm (c) quark. In this
subsection, we will simply assume that the s quark can be treated as
``heavy." This leads to the successful Callan-Klebanov model
\cite{ck,snnr}.
Instead of working with a complicated model, we will
consider a simplified model based on the following intuitive idea.
Imagine that the s quark is very heavy. Instead of ``rotating" into the
s-flavor direction as we did above, we treat it as an independent
degree of freedom.
The effective field carrying the s-quark flavor is the kaon which can be put
into isospin doublet corresponding to the quark configurations $s\bar{u}$ and
$s\bar{d}$. Let $K$ represent the doublet $K^T= (K^{-} \bar{K^0})$.
(In anticipation of what is to come, let us mention that other ``heavy"
quark flavors $Q$ will also be put into independent doublets, the
$Q$ replacing the s quark.)
Suppose one puts a kaon into the $SU(2)$ soliton field. One can
then ask what happens to the soliton-kaon system. The Lagrangian density
takes the form
\be
\L_{toy}=\L_{\mbox{\tiny SU(2)}} +\L_{\mbox{\tiny K-SU(2)}}.\label{toyL}
\ee
Here the first term describes the chiral (u- and d--quark) sector
\be
\L_{\mbox{\tiny SU(2)}}=\frac{f^2}{4}\Tr (\del_\mu U \del^\mu U^\dagger)
+\cdots\label{toyL1}
\ee
where $U=e^{i\vec{\tau}\cdot\vec{\pi}/f}$ is the $SU(2)$ chiral field
and the ellipsis stands for terms of higher derivative, mass term etc.
For simplicity, we will consider a Lagrangian density constructed enitrely
of the pion field since more realistic Lagrangians do not modify the
discussion qualitatively. To write the second term of (\ref{toyL}),
introduce the ``induced vector field" $v_\mu$ and axial field $a_\mu$
\be
v_\mu &=& \frac{1}{2}(\del_\mu \xi \xi^\dagger +\del_\mu \xi^\dagger \xi),\\
a_\mu &=& \frac{1}{2i}(\del_\mu \xi \xi^\dagger -\del_\mu \xi^\dagger \xi)
\ee
with $\xi=\sqrt{U}$. Under chiral transformation $\xi\rightarrow
L\xi h^\dagger= h\xi R^\dagger$, $v_\mu$ and $a_\mu$ transform
\be
v_\mu &\rightarrow& h(\del_\mu +v_\mu) h^\dagger,\\
a_\mu &\rightarrow& h a_\mu h^\dagger.
\ee
Note that $h$ is a complicated local function of $L$, $R$ and $\pi$.
We assume that under chiral transformation, $K$ transforms
\be
K\rightarrow hK.\label{chiralK}
\ee
(Later on we will implement
this with heavy-quark spin transformation when the quark mass becomes heavy
compared with the chiral symmetry scale as for instance in the case of $D$ and
$B$ mesons containing a c quark and b quark respectively. Since $K^*$ is still
much higher in mass, we need not worry about this transformation here.)
Therefore
in $\L_{\mbox{\tiny K-SU(2)}}$, we must have the ``covariant" kinetic energy
term $(D_\mu K)^\dagger D^\mu K$ with $D_\mu K=(\del_\mu +v_\mu)K$
and mass term $m_K^2 K^\dagger K$.
We must also have a potential term $K^\dagger V K$ with $V$ transforming
$V\rightarrow h V h^\dagger$ in which we could have terms like
$a_\mu^2$ etc. If the kaon is very massive, we can ignore terms higher
order than quadratic in the kaon field. In accordance with the spirit
of chiral perturbation theory, we ignore higher derivative terms involving
pion fields. There is one term of higher derivatives however that
we cannot ignore,
namely the term involving the baryon current $B_\mu$ which contains
three derivatives but while higher order in the chiral counting
in the meson sector, is of $O(1)$ in the baryon sector.
Therefore we also have a term of the form
$(K^\dagger D^\mu K B_\mu +{\rm h.c.})$. Putting these together, the minimal
Lagrangian density we will consider is
\be
\L_{toy} &=& \frac{f^2}{4}\Tr (\del_\mu U \del^\mu U^\dagger)
+\cdots\nonumber\\
&& +(D_\mu K)^\dagger D^\mu K -K^\dagger(m_K^2 +V)K +
\kappa (K^\dagger D^\mu K B_\mu +{\rm h.c.})\nonumber\\
&& +\cdots \label{toyL2}
\ee
with $\kappa$ a dimensionful constant to be fixed later.
\subsubsection*{\it Isospin-spin transmutation}
\bs
We will now show that in the field of the soliton
the isospin of the kaon in (\ref{toyL})
transmutes to a spin, the latter behaving effectively as an s quark.
This is in analogy to the transmutation of a scalar doublet into
a fermion in the presence of a magnetic monopole\cite{JRHT}
except that in the present case, the scalar is ``wrapped" by the soliton while
in the other case, the scalar is ``pierced" by the monopole.
To see the isospin-spin transmutation in our soliton-kaon complex, we first
note that the Lagrangian density (\ref{toyL})
with (\ref{toyL1}) and (\ref{toyL2}) is invariant under the (global) rotation
$S\in SU(2)$
\be
U &\rightarrow& S U S^\dagger,\nonumber\\
K &\rightarrow& S K
\ee
which implies $D_\mu K\rightarrow S(D_\mu K)$. Thus we again have three
zero modes. These zero modes have to be quantized to get the quantum numbers.
Thus we are invited to write
\be
K(\vec{x},t)=S(t)\tilde{K} (\vec{x},t).
\ee
Here $\tilde{K}$ is the kaon field defined in the soliton rotating frame.
Under an isospin rotation $B$ which is a subgroup of the left
transformation $L\in SU(3)$,
\be
S\rightarrow BS, \ \ \ \ \ \tilde{K}\rightarrow \tilde{K}.
\ee
This shows that the kaon field in the rotating frame carries no isospin.
Under a spatial rotation,
\be
\tilde{K}\rightarrow e^{i\vec{\alpha}\cdot\vec{L}}\tilde{K}
=e^{-i\vec{\alpha}\cdot \vec{I}}e^{i\vec{\alpha}\cdot\vec{T}}\tilde{K}
\ee
where $\vec{T}$ is the grand spin $\vec{J}+\vec{I}$ under which the
hedgehog $U_c$ is invariant.  Under a spatial rotation
\be
S\rightarrow Se^{-i\vec{\alpha}\cdot\vec{I}},
\ee
so we deduce
\be
\tilde{K}\rightarrow e^{i\vec{\alpha}\cdot\vec{T}}\tilde{K}
\ee
which means that the grand spin operator
$\vec{T}$ is the {\it angular momentum operator} on $\tilde{K}$.
For the kaon the grand spin $\vec{T}$ is the vector sum $\vec{L}+\vec{I}$
with $I=1/2$
and since, as we will see later, the lowest kaon mode has $L=0$, the
kaon seen from the soliton rotating frame carries spin 1/2.
This shows that in the soliton rotating frame, the isospin lost from the
kaon field is transmuted to a spin. This is the analog to the transmutation
of the isospin of the scalar doublet to a spin in the presence of a
magnetic monopole. There is one difference however. While the kaon acquires
spin in this way, it retains its bosonic statistics while in the case of
the monopole-scalar, the boson changes into a fermion. Since the soliton-kaon
complex must be a baryon (fermion), the soliton which will be quantized
with {\it integer} spin will carry {\it fermion} statistics. With one kaon
in the system, the soliton carries $J=I=0, 1, \cdots$

Suppose we have two kaons in the soliton rotating frame. They now have
integer spins $T=0, 1$, so the soliton carries $J=I=1/2, 3/2, \cdots$.
For three kaons in the system, the kaons carry $T=1/2, 3/2, \cdots$
with the soliton carrying $J=I=0, 1, \cdots$ and so on. One can show that
this pattern of quantization follows naturally from the CK theory
\cite{ck,sw}. The resulting quantum numbers and particle identification
are given in Table II.1. It is remarkable that they are identical to what one
expects from
quark models. Thus {\it effectively}, the kaon behaves like an s quark
in all respect except for its statistics. By field redefinition one can indeed
make it behave {\it precisely} like an s quark,
thus making the picture identical to the quark-model picture.

\begin{center}
{\bf Table II.1}
\end{center}
\noindent
Even-parity $J={1 \over 2}$ and ${3 \over 2}$ baryons containing
strange and charm quarks. $S$ and $C$ stand for the strangeness and
charm numbers, respectively. $R$ is the rotor angular momentum
which is equal to the isospin of the rotor $I$.
\begin{center}
$$
\begin{array}{ccccccccc}\hline\hline
{\rm particle}&I&J&S&C&R&J_1&J_2&J_m\\ \hline
\Lambda&0&{1 \over 2}&-1&0&0&{1 \over 2}&0&{1 \over 2}\\
\Sigma&1&{1 \over 2}&-1&0&1&{1 \over 2}&0&{1 \over 2}\\
\Sigma^*&1&{3 \over 2}&-1&0&1&{1 \over 2}&0&{1 \over 2}\\
\Xi&{1 \over 2}&{1 \over 2}&-2&0&{1 \over 2}&1&0&1\\
\Xi^*&{1 \over 2}&{3 \over 2}&-2&0&{1 \over 2}&1&0&1\\
\Omega&0&{3 \over 2}&-3&0&0&{3 \over 2}&0&{3 \over 2}\\ \hline
\Lambda_c&0&{1 \over 2}&0&1&0&0&{1 \over 2}&{1 \over 2}\\
\Sigma_c&1&{1 \over 2}&0&1&1&0&{1 \over 2}&{1 \over 2}\\
\Sigma_c^*&1&{3 \over 2}&0&1&1&0&{1 \over 2}&{1 \over 2}\\
\Xi_{cc}&{1 \over 2}&{1 \over 2}&0&2&{1 \over 2}&0&1&1\\
\Xi_{cc}^*&{1 \over 2}&{3 \over 2}&0&2&{1 \over 2}&0&1&1\\
\Omega_{ccc}&0&{3 \over 2}&0&3&0&0&{3 \over 2}&{3 \over 2}\\
\Xi_c&{1 \over 2}&{1 \over 2}&-1&1&{1 \over 2}&{1 \over 2}&{1 \over 2}&1,0\\
\Xi_c^\prime&{1 \over 2}&{1 \over 2}&-1&1&{1 \over 2}&{1 \over 2}
&{1 \over 2}&1,0\\
\Xi_c^*&{1 \over 2}&{3 \over 2}&-1&1&{1 \over 2}&{1 \over 2}&{1 \over 2}&1\\
\Omega_c&0&{1 \over 2}&-2&1&0&1&{1 \over 2}&{1 \over 2}\\
\Omega_c^*&0&{3 \over 2}&-2&1&0&1&{1 \over 2}&{3 \over 2}\\
\Omega_{cc}&0&{1 \over 2}&-1&2&0&{1 \over 2}&1&{1 \over 2}\\
\Omega_{cc}^*&0&{3 \over 2}&-1&2&0&{1 \over 2}&1&{3 \over 2}\\ \hline\hline
\end{array}
$$
\end{center}

\subsubsection*{\it Heavy baryons from chiral symmetry}
\bs
In discussing dynamical structure of the baryons in the CK approach,
we have to specify the quantities that enter into the Lagrangian
(\ref{toyL2}). Suppose that we want to arrive at (\ref{toyL2}) starting
from a chiral Lagrangian as Callan and Klebanov did. Then the basic
assumption to be made is that the vacuum is $SU(3)$-symmetric while
the quark ``particle-hole" excitation may not be. This means that
we can start with a Lagrangian that contains a piece that is
$SU(3)$ symmetric with the $U$ field in $SU(3)$ plus a symmetry-breaking
term. Thus the $f$ that appears in (\ref{toyL2}) is directly related
to the pion decay constant $f_\pi$ and the $V$ and $\kappa$ will be
specified. In particular the coefficient $\kappa$ will be fixed by
the topological Wess-Zumino term to
\be
\kappa=N_c/4f^2.
\ee

The structure of the resulting baryons can be simply described as follows.
In large $N_c$ limit, the $SU(2)$ soliton contributes a mass ${\cal M}$
which is of $O(N_c)$. The kaon field which will be bound by the Wess-Zumino
term of (\ref{toyL2}) (with some additional contribution from the potential
term $V$) contributes a fine-structure splitting $n\omega$ of $O(N_c^0)$
where $n$ is the number of bound kaons and $\omega$ is the eigenenergy
of the bound kaon (so $< m_K$). The quantization of the collective
rotation of the soliton brings in the rotational energy of the
soliton as well as a hyperfine-structure splitting of $O(1/N_c)$.

Let us now show how this structure arises in detail. We write the
energy of the baryons as before
\be
E_{JIs}={\cal M}+M_0 +M_{-1}.\label{MASS}
\ee
We know how ${\cal M}$ is obtained. Now $M_0$ contains the Casimir term
discussed above and a term which depends on heavy flavor. Let us call the
latter as $\delta M_0$.
To obtain $\delta M_0$, we may write from (\ref{toyL2}) the equation
of motion satisfied by the kaon field and solve it to $O(N_c^0)$. Alternatively
we calculate the Lagrangian (\ref{toyL2}) to $O(N_c^0)$ by taking the hedgehog
$U=U_c=e^{iF (r)\tau\cdot\hat{r}}$. To do this, we note that the hedgehog
is symmetric under the ``grand spin" rotation $\vec{T}=\vec{L}+\vec{I}$
(we are changing the notation for the ``grand spin" to $T$), so we may write
the kaon field as
\be
K=k(r)Y_{TLT_z}.
\ee
The Lagrangian then is (using the length scale $1/gf$ as before, so that
$s=(gf) r$ is dimensionless)
\be
L_{\mbox{\tiny K}}=\frac{\pi}{2}\int \, ds s^2 \left[\dot{k}^\dagger \dot{k}
+i\lambda (s) (k^\dagger \dot{k}-\dot{k}^\dagger k)-\frac{d}{ds}k^\dagger
\frac{d}{ds} k -k^\dagger k (m_{\mbox{\tiny K}}^2 +V (s;T,L))\right]\label{KL}
\ee
where
\be
\lambda (s)=-\frac{2N_c g^2}{\pi^2 s^2} F^\prime \sin^2 F
\ee
is the contribution from the Wess-Zumino term. This expression differs
from what one sees in the literature \cite{cargese}
in that the kinetic energy
terms are not multiplied by nonlinear functions of $F$ which come from
the Skyrme term involving kaon fields. They are not essential, so we will
not include them. The Euler-Lagrange equation of motion obtained from
(\ref{KL})  is
\be
\frac{d^2}{dt^2}k -2i\lambda (s)\frac{d}{dt}k +{\cal O}k=0
\ee
where
\be
{\cal O}= -\frac{1}{s^2}\frac{d}{ds} s^2 \frac{d}{ds} +m_{\mbox{\tiny K}}^2
+V(s;T,L).
\ee
Expanded in its eigenmodes, the field $k$ takes the form
\be
k(s,t)=\sum_{n>0}\left(\tilde{k} (s) e^{i\tilde{\omega}_n t} b_n^\dagger
+ k_n(s) e^{-i\omega_n t}a_n\right)
\ee
with $\omega_n\geq 0$ and $\tilde{\omega}_n\geq 0$. The Euler-Lagrange
equation gives the eigenmode equations
\be
\left(\omega_n^2 +2\lambda (s)\omega_n +{\cal O}\right)k_n=0,\\
\left(\tilde{\omega}_n^2-2\lambda (s)\tilde{\omega}_n +{\cal O}\right)
\tilde{k}_n=0.
\ee
An important point to note here is the sign difference in the term linear
in frequency associated with the Wess-Zumino term, which is of course
due to the linear time derivative. We shall discuss its consequence shortly.
Now recognizing that ${\cal O}$ is hermitian, one can derive
the orthonormality condition
\be
4\pi\int\, ds s^2k_n^* k_m\left((\omega_n +\omega_m) +2\lambda (s)\right)
&=& \delta_{nm},\\
4\pi\int\, ds s^2 \tilde{k}_n^*\tilde{k}_m \left((\tilde{\omega}_n+
\tilde{\omega}_m) -2\lambda (s)\right) &=& \delta_{nm},\\
4\pi\int\, ds s^2 k_n^* \tilde{k}_m \left((\omega_n-\tilde{\omega}_m)
+2\lambda (s)\right)&=& 0.
\ee
The creation and annihilation operators of the modes satisfy the usual bosonic
commutation rules
\be
[a_n,a_m^\dagger]=\delta_{nm}, \ \ \ \ \ \ [b_n,b_m^\dagger]=\delta_{nm}
\ee
with the rest of the commutators vanishing. The diagonalized
Hamiltonian so quantized takes the form
\be
H=\sum_{n>0}\left(\omega_n a_n^\dagger a_n +\tilde{\omega}_n b_n^\dagger b_n
\right)
\ee
and the strangeness charge is given by
\be
S=\sum_{n>0} \left(b_n^\dagger b_n -a_n^\dagger a_n\right).
\ee
It follows that the mode $k_n$ carrying $S=-1$ receives
an attractive contribution
from the Wess-Zumino term while the mode $\tilde{k}_n$ carrying $S=+1$
gets a repulsive contribution. The potential $V$ is numerically small compared
with the strength of the Wess-Zumino term, so the $S=-1$ state is bound
whereas the $S=+1$ state is not. The former has the strangeness quantum number
of the hyperons seen experimentally
and the latter the quantum number of what is called
``exotics" in quark models.

To $O(N_c^0)$, at which the kaon binding takes place, no quantum numbers
of the bound system other than the strangeness are defined. It is at
$O(N_c^{-1})$ that the proper quantum numbers are recovered.
What we have at this point is
a bound object in which the kaon has $T=1/2$ and $L=1$ with $S=-1$.
We shall see shortly that the bound object, when collective-quantized,
has the right quantum numbers to describe $\Lambda$, $\Sigma$ and $\Sigma^*$.
The corresponding frequency $\omega_n$ constitutes the ``fine-structure"
contribution $\delta M_0$ to Eq. (\ref{MASS}).

To get the ``hyperfine-structure" term $M_{-1}$
and the quantum numbers, we have
to quantize the three zero modes encountered above.
Let
\be
U(\vec{r},t)=S(t) U_0 (\vec{r}) S^\dagger (t), \ \ \ \ \
\tilde{K}(\vec{r},t)=S(t) K(\vec{r},t).\label{rotation}
\ee
As defined, $K$ is the kaon field ``seen" in the soliton rotating frame
and $\tilde{K}$ is the kaon field defined in its rest frame.
As before, we let
\be
S^\dagger \del_0 S=-i\lambda_\alpha \dot{q}_\alpha\equiv
\lambda_\alpha \Omega_\alpha
\ee
defining the rotational velocities $\Omega_\alpha$
where $\alpha$ here runs over 1 to 3. When (\ref{rotation}) is substituted
into (\ref{toyL2}) and integrated over space, one obtains the Hamiltonian
at $O(N_c^{-1})$ of the following form
\be
H_{-1}=2\I \vec{\Omega}^2 -2 \vec{\Omega}\cdot\vec{Q}.\label{Hrot}
\ee
Here ${\I}$ is the moment of inertia of the rotating $SU(2)$ soliton
which can be determined by the $\Delta N$ mass difference as discussed
above.  Equation (\ref{Hrot}) has
a generic form of the Hamiltonian with a linear time derivative
encountered before.
The quadratic velocity-dependent term arises from the kinetic energy term
of the soliton rotation.
The linear velocity-dependent term was seen above as arising from
integrating out certain fast degrees of freedom.
After some straightforward and tedious algebra, one can get an
explicit form for $\vec{Q}$ from (\ref{toyL2})
\be
Q_i=\int\, d^3r \chi_i
\ee
with
\be
\chi_i&=& i \dot{K}^\dagger M_i K+ \lambda K^\dagger M_i K-
2i\lambda\epsilon_{ijk} r_j K^\dagger D_k K \nonumber\\
&+& 2i \dot{K}^\dagger D_j K {\rm Tr} (P_i A_j) -6i \dot{K}^\dagger
[P_i,A_j]D_j K +h.c.
\ee
where
\be
M_i &=& \frac{1}{2} (\xi^\dagger \tau_i \xi +\xi\tau_i \xi^\dagger),\\
P_i &=& \frac{1}{2} (\xi^\dagger \tau_i \xi -\xi\tau_i \xi^\dagger)
\ee
and
\be
A_i\equiv \frac{1}{i} a_i
=\frac{1}{2} (\xi^\dagger \del_i \xi-\xi\del_i \xi^\dagger)
\ee
with $\xi =\sqrt{U_0}$. Canonically quantizing (\ref{Hrot}), we obtain
\be
H_{-1}= \frac{1}{2\I}\left( R_i-\int\, d^3r \chi_i\right)^2\label{hf}
\ee
where $R_i$ is the right generator we encountered above which in the present
case is the rotor angular momentum which we will denote $\vec{J}_l$
(the subscript $l$ stands for ``light" quark making up the soliton).
Explicitly evaluating with the known solutions for the soliton and
the bound kaon, one finds that
\be
\int\, d^3r \chi_i=c J_{\mbox{\tiny K}}^i\label{hfff}
\ee
where $J^i_{\mbox{\tiny K}}$ stored in the bound kaon identified with the
grand spin $T^i$ and
\be
c=2\int\, r^2 dr\omega\, k^* k h(r;\lambda,F)
\ee
where
\be
h=1-\frac{3}{4} \cos^2\frac{F}{2} +\frac{1}{2r}F^\prime \sin F
+\frac{1}{2} \frac{d^2F}{dr^2}\sin F +\frac{1}{3r^2}
\sin^2 F \cos^2 \frac{F}{2}.
\ee
The Hamiltonian (\ref{hf}) which can be rewritten in a form
encountered in Lecture I
\be
H_1= \frac{1}{2\I}\left(\vec{J}_l +c\vec{J}_K\right)^2\label{CKhyperfine}
\ee
gives the hyperfine splitting spectrum
\be
E_{-1}=\frac{1}{2\I} \left(cJ(J+1) +(1-c)J_l (J_l+1) +c(c-1) J_K (J_K +1)
\right)\label{hfs}
\ee
where $\vec{J}=\vec{J}_l +\vec{J_K}$ is the total angular momentum  of the
bound system.

The quantity $c$, as we will see shortly, can be interpreted in terms of
a Berry charge quite analogous to the charge $(1-\kappa)$ that figures
in the spectrum of the diatomic molecule.
The hyperfine
spectrum can be used to relate the $c$ coefficient to the masses of
the hyperons. It takes the form (using the particle symbol for the masses)
\be
c=\frac{2(\Sigma^*-\Sigma)}{2\Sigma^* +\Sigma -3\Lambda}
\approx 0.62\label{cemp}
\ee
where the last equality comes from experiments. Note that this is close to
unity. Were it precisely equal to 1, then the hyperfine spectrum would have
been that of a symmetric top depending only on the total angular momentum
$J$. Deviation from 1 will become accentuated as the mass of the
heavy meson increases. We will find that as the heavy meson mass becomes
infinite, the charge $c$ will go to zero.
\subsubsection{\it Hyperon spectrum}
\bs
So far, we have considered one strange quark in the system. How can one
describe the baryons with one or more s quarks such as $\Xi$ and
$\Omega$? For this, we need to make an assumption on the interaction
between ``heavy" mesons. For simplicity, we will assume that mesons interact
weakly so that higher order terms in meson field can be ignored. This was
of course the basic premise under which the Lagrangian (\ref{toyL2})
was written to start with. We expect that this assumption gets better
the larger the mass of the meson. Now two or more heavy mesons, each of
which undergoes spin-isospin transmutation, can then be combined to give
the states of total flavor quantum number and angular momentum with the
$SU(2)$ soliton. It has been shown that the quantum numbers of
the hyperon octet and decuplet
do come out correctly in agreement with experiments and with quark models
\cite{sw}. The quantum number identification of the strange and charmed
hyperons is given in Table II.1.
A straightforward calculation gives the fine- and hyperfine-structure
mass formula
$$M(I,J,n_1, n_2, J_1, J_2, J_m) -M_{\rm sol}- M_0
= n_1 \omega_1 + n_2 \omega_2+$$
$${1 \over 2 {\cal I}} \, \Bigl\{ I (I + 1) + (c_1 - c_2) [c_1 J_1
(J_1 + 1) - c_2 J_2
(J_2 + 1)] + c_1 c_2 J_m (J_m + 1) +$$
\ben
[J(J + 1) - J_m (J_m + 1) - I (I + 1) ] \Bigl[ {c_1 + c_2 \over 2} +
{c_1 - c_2 \over 2} \, {J_1 (J_1 + 1) - J_2 (J_2 + 1) \over J_m (J_m + 1)}
\Bigr] \Bigr\}. \label{massform}
\een
Here the subscript $i$ in $J_i$ and $c_i$ represents the heavy-meson flavor
and $J_m$ stands for all possible vector addition of the two angular momenta
$\vec{J}_1$ and $\vec{J}_2$ labeling different states.


\subsubsection{\it Connection to Berry structure}
\bs
In Lecture I, we introduced the notion that a nonabelian Berry potential
emerges when a ``particle-hole" excitation corresponding to a $Q\bar{q}$
configuration is eliminated in the presence of a slow rotational field,
{\ie}, the class B case. We now show that the corresponding excitation
spectrum is identical to what we obtained above, (\ref{CKhyperfine}),
when a pseudoscalar
meson $K$ is bound to the soliton as in the Callan-Klebanov scheme.
For this it suffices to notice that the angular momentum lodged in the
rotating soliton is $J_l$ and that lodged in the Berry potential is
$J_K$. The coefficient $c$ is just the Berry charge $(1-\kappa)$
in the diatomic-molecular case and $g_K$ in the case of the chiral bag.
Although Eq.(\ref{CKhyperfine}) is derived in the Callan-Klebanov framework,
the resulting spectrum is quite generic with a general interpretation in terms
of a Berry potential.
Now given an effective Lagrangian, one can calculate the meson frequency
$\omega_i$ and the hyperfine coefficients $c_i$ and use Eq.(\ref{massform})
to predict the spectrum. For strange hyperons, this is known to work quite
well \cite{snnr}.
Given the generic form, it is tempting to extend the
mass formula (\ref{massform}) to heavier-quark systems such as charmed and
bottom baryons by taking the $\omega_i$'s and $c_i$'s as parameters and
by assuming one or more bound mesons as {\it quasiparticles} ({\ie},
independent
particles) which allows us to use the formula (\ref{massform}). Figure II.2
shows how well strange hyperons are described when $\omega_K$ and $c_K$
are taken as parameters fit to $\Lambda$ and $\Sigma$. (The ground-state
parameters $\I$, $f_\pi$ and $g$ -- in the case of the Skyrme term --  are
known from that sector.) This is a remarkable agreement with experiment.
The parameters so determined are rather close to the Callan-Klebanov
prediction made with the Skyrme Lagrangian with a suitable symmetry
breaking term implemented.

In Figure II.3 is given the spectrum of charmed baryons containing one or
more charm quarks and strange quarks. The only parameters of the model,
$\omega_D$ and $c_D$, are again fixed by fitting $\Lambda_c$ and $\Sigma_c$.
The prediction is compared with quark-model predictions. The agreement with
quark models are quite impressive, much better than one would have the right
to expect.
\subsubsection{\it Heavy-quark symmetry and skyrmions}
\subsubsection*{\it Hyperfine splitting}
\bs
One obvious feature in Figures II.2 and II.3 is that the hyperfine coefficients
$c$ decrease as the heavy quark mass (or heavy-meson mass) increases.
To be specific, let us consider the structure of baryons with one heavy quark
$Q=s, s, b$ (we are considering the s quark to be heavy as mentioned above).
{}From (\ref{hfs}), we see that
%
\be
\Sigma_{\mbox{\tiny Q}}-\Lambda_{\mbox{\tiny Q}}=\frac{1}{\cal I}
(1-c_{\mbox{\tiny Q}})
\simeq 195 {\rm MeV} (1-c_{\mbox{\tiny Q}}) \label{finesplit}
\ee
where the last equality is obtained from the $\Delta N$ mass splitting
which fixes $1/{\I}$.
This implies a simple relation for heavy-quark flavors $Q$ and $Q^\prime$
\be
\frac{\Sigma_Q-\Lambda_Q}{\Sigma_{Q^\prime}-\Lambda_{Q^\prime}}\simeq
\frac{1-c_{\mbox{\tiny Q}}}{1-c_{\mbox{\tiny Q}^\prime}}.
\ee
With the experimental value $\Sigma_c-\Lambda_c\approx 168 {\rm MeV}$ for
the charmed baryons, we get $c_{c}\simeq 0.14$ as given in Figure II.3.
In the heavy-quark limit, we expect the hyperfine splitting to go to
zero \cite{isgur}, so let us assume that
\be
c_\Phi \approx a m_\Phi^{-1}
\ee
where $\Phi$ stands for the heavy meson in which the heavy quark
$Q$ is lodged and $a$ is a constant of mass dimension of $O(N_c^0)$. Taking
$m_{\mbox{\tiny D}}=1869 {\rm MeV}$ in the charmed sector, we have
$a\simeq 262$ MeV. So
\footnote{It may be coincidental but it is surprising that this formula
works satisfactorily even for the kaon for which one predicts
$c_s \simeq 0.53$ to be compared with the empirical value 0.62.}
\be
c_{\mbox{\tiny Q}}\simeq 262 {\rm MeV}/m_\Phi.\label{ccoeff}
\ee
Now for b-quark baryons, using $m_{\mbox{\tiny B}}=5279$ MeV, we find
$c_b\simeq 0.05$ which with (\ref{finesplit}) predicts
\be
\Sigma_b-\Lambda_b\approx 185 {\rm MeV}.
\ee
This agrees well with the quark-model prediction. Furthermore
the $\Sigma^*-\Sigma$
splitting comes out correctly also. For instance, it is predicted that
\be
\frac{\Sigma^*_b-\Sigma_b}{\Sigma^*_c-\Sigma_c}\simeq \frac{c_b}{c_c}\simeq
\frac{m_{\mbox{\tiny D}}}{m_{\mbox{\tiny B}}}\approx 0.35
\ee
to be compared with the quark-model prediction $\sim 0.32$.
If one assumes that the heavy mesons $\Phi$'s are weakly interacting, then
we can put more than one heavy mesons in the soliton and obtain the spectra
for $\Xi$'s and $\Omega$'s given in Figs. II.2 and II.3\cite{RRS}.
\subsubsection{\it Berry potentials in heavy-quark limit}
\bs
The analysis made above suggests that the hyperfine splitting goes
to zero in heavy quark limit as $O(N_c^{-1}m_\Phi^{-1})$. Here
we will show that in the limit heavy-quark symmetry is restored,
the Berry potential obtained in the last lecture vanishes, namely that
$A=F=0$. To do this, we first construct the Lagrangian that possesses
chiral symmetry and heavy quark symmetry, following Ref.\cite{wisetal}.

Suppose that a meson is made up of a heavy quark
$Q$ and a light antiquark $\bar{q}$. We denote the spin singlet
$P$, the pseudoscalar used in the previous subsection and the spin
triplet $P^*$. When the quark is light, then the $P^*$ is much heavier
than $P$. For instance, the mass ratio $K^*/K\approx 0.5$. We understand this
in terms of one-gluon exchange that pushes $P^*$ up above $P$
proportional to $(m_q m_{\mbox{\tiny Q}})^{-1}$. However as
$m_{\mbox{\tiny Q}}\rightarrow \infty$, the $P$ and $P^*$ become
degenerate. As a consequence, the heavy-quark spin that decouples from
the spectrum becomes a good quantum number. This is the heavy-quark
spin symmetry. Thus for a baryon containing a heavy quark $Q$, we need
both $P$ and $P^*$ to construct a skyrmion in the Callan-Klebanov scheme
if one wants to assure correct symmetry in heavy baryons.
We will see that this degeneracy plays a key role in giving the correct
Berry potential. An elegant way of incorporating the symmetry of $P$
and $P^*$ is to write the meson field as\cite{georgilecture}
\be
H = \frac{ ( 1 + \not\!{v})}{2} \left[ {P^*}_{\mu} \gamma^\mu -
P \gamma^5 \right] \label{Ha}
\ee
where $v_\mu$ is the four-velocity of the heavy quark. This field
transforms as follows. Under chiral transformation $h$ ({\ie}, $\xi\rightarrow
L\xi h^\dagger=h\xi R^\dagger$), it transforms
\be
H\rightarrow Hh
\ee
which is equivalent to (\ref{chiralK}) generalized to the $H$ field
and under heavy-quark spin symmetry $S$, it transforms as
\be
H\rightarrow SH.
\ee
This is the symmetry we did {\it not} impose when we were discussing the
Callan-Klebanov model.
Writing in derivative expansion the Lagrangian density invariant simultaneously
under the two symmetry transformations, we have
\be
\L^{\mbox{\tiny HQS}} &=&\frac{f^2}{4}\Tr (\del_\mu U \del^\mu U^\dagger)
-i \Tr \overline{H}_a v^\mu \del_\mu H_a
 + i \Tr \overline{H}_a H_b v^\mu \left(V_\mu \right)_{ba} \nonumber \\
& & + ig \Tr \overline{H}_a H_b \gamma^\mu \gamma_5
\left( A_\mu \right)_{ba} +  \cdots \ \ ,  \label{LH}
\ee
where the subscript $a$ labels the light quark flavor and the ``induced"
vector and axial vector currents are given by
\be
V_\mu &=& \frac{1}{2}(\del_\mu \xi \xi^\dagger +\del_\mu \xi^\dagger \xi),\\
A_\mu &=& \frac{1}{2i}(\del_\mu \xi \xi^\dagger -\del_\mu \xi^\dagger \xi)
\ee
with $\xi=\sqrt{U}$. The $g$ is an unknown axial vector coupling constant
to be determined from experiments.
The ellipsis stands for higher derivative terms
and terms that break both chiral and heavy-quark symmetries.
As written, the $H$ field has dimension 3/2 since a factor of $\sqrt{m_\Phi}$
has been incorporated into the component fields $P$ and $P^*$. The
heavy meson mass term is cancelled away by the redefinition of the meson field
$H\rightarrow e^{iv\cdot x m_H} H$ where $v_\mu$ is the velocity four
vector.

We will now show that the vanishing of $c$ is tantamount to the disappearance
of the associated nonabelian Berry potential that emerges from the Lagrangian
(\ref{LH}). What happens is that
the Berry phases generated from $P$ and $P^*$ when they are integrated out
{\it cancel exactly} in the IW limit \cite{nrz93}.
This is the analog of the vanishing Berry
potential in the diatomic molecule when the electronic rotational symmetry
is restored.

For the present purpose, it is convenient to rewrite (\ref{LH}) as
\be
{\cal L}_H &=& -i \Tr \overline{H}_a v^\mu \partial_\mu H_a
 + i \Tr \overline{H}_a H_b v^\mu \left(U^{\dagger}\partial_{\mu}
U \right)_{ba} \nonumber \\
& & + ig \Tr \overline{H}_a H_b \gamma^\mu \gamma^5
\left(U^{\dagger}\partial_{\mu}U \right)_{ba} +
\frac{f^2}{4}\Tr (\del_\mu U \del^\mu U^\dagger)+ \cdots   \label{hq}
\ee
This is obtained from (\ref{LH}) by ``dressing" $H$ with pion field
as $H\rightarrow H \xi$. To $O(m_Q^0\cdot N_c^0)$, we get the
fine-structure spectrum by taking the hedgehog configuration $U_c$ in
(\ref{hq}). In the infinite heavy-quark mass limit,
the heavy meson is bound at the center of
the soliton. Now we go to $O(m_Q^0\cdot N_c^{-1})$ by slowly rotating
the hedgehog
\be
U=S(t)U_c S^\dagger (t).
\ee
As we did in the case of the chiral bag, we let the rotation operator
act on the $H$ field, thereby unwinding the soliton
\be
H\rightarrow HS^\dagger,\ \ \ \ \overline{H}\rightarrow S\overline{H}.
\ee
The Lagrangian (\ref{hq}) becomes
\be
{\cal L}_H &=& -i \Tr \overline{H} \partial_t H
    -i \Tr H \partial_t S^{\dagger} S\overline{H}
   + \frac{{\cal I}}{4}\Tr(S^{\dagger} \partial_t  S)^2 \nonumber\\
   & & + i \Tr \overline{H}_a H_b  \left(S^{\dagger}(U^{\dagger}\partial_t
          U) S \right)_{ba}
    + ig \Tr \overline{H}_a H_b \gamma^o \gamma^5
\left(S^{\dagger} (U^{\dagger}\partial_t U)S \right)_{ba}\nonumber\\
 & & + ig \Tr \overline{H}_a H_b \gamma^i \gamma^5
\left(U_c^{\dagger}\partial_iU_c \right)_{ba}+ \cdots \ \ .  \label{la}
\ee
Now one expects that
the heavy meson is bound at the origin of the soliton; therefore
the fourth and fifth terms of (\ref{la}) vanish in the heavy-meson limit,
so we are left with
\be
{\cal L}_H &=& -i \Tr \overline{H} \partial_t H
    + \frac{{\cal I}}{4}\Tr(S^{\dagger} \partial_t  S)^2 +
    -i \Tr H \partial_t S^{\dagger} S\overline{H}\nonumber \\
   & & + ig \Tr \overline{H}_a H_b \gamma^i \gamma^5
\left(U_c^{\dagger}\partial_iU_c \right)_{ba}+ \cdots \ \ . \label{laa}
\ee
In complete analogy with the diatomic molecule and the chiral bag,
we can identify the second term of (\ref{laa}) with the Berry potential.
As in the chiral bag case, $\dot{S}^\dagger S$ lives in the isospin space
of the light-quark sector and hence operates on the space of the light
antiquark $\bar{q}$. When the $H$ is bound to the soliton, the light
antiquark is characterized by its grand spin $T=0$ where
$\vec{T}=\vec{I}_{\bar{q}} +\vec{J}_{\bar{q}}$. This can be understood simply
in the chiral bag model. In the chiral bag model, the lowest light-quark
orbit is given by the hedgehog $T=0$. The heavy meson $Q\bar{q}$ is therefore
made of a ``hole" in the hedgehog and a (heavy)
particle in the flavor $Q$ orbit.
Thus the configuration of the light antiquark in $H$ takes the form
\be
|T=0\rangle=\frac{1}{\sqrt{2}}\left(|\bar{d}\downarrow\rangle -|\bar{u}
\uparrow\rangle \right).\label{hh}
\ee
Now the heavy particle decouples in the infinite mass limit, so behaves
as a spectator and the role of fast variable is taken up entirely by
the light antiquark, making it completely analogous to the quark sitting
in $T=0$ orbit in the chiral bag model. We know from Lecture I that
the operator $\dot{S}^\dagger S$ sandwiched between
the $T=0$ states vanishes as one can readily verify with (\ref{hh}) acted on
by an isospin operator. In other words, the Berry potential vanishes.

\subsubsection{\it Comparison with large $N_c$ QCD}
\bs
The general characteristics of the skyrmions, both the ground state
(light-quark
baryons) and the
excited states (heavy-quarks baryons), can be understood in terms
of large $N_c$ QCD.   In large $N_c$ limit, the consistency in $N_c$ behavior
as well as unitarity requires that the $O(1/N_c)$ correction to
the axial coupling constant $g_A$ or equivalently to the $\pi$-baryon coupling
constant $g_{\pi NN}$ should vanish \cite{dashenmanohar}\footnote{
We will refer to the constraints discussed by these authors as Dashen-Manohar
constraints.}.
The leading nonvanishing correction comes at order $1/N_c^2$.
This can happen only if there is an infinite tower of degenerate states
contributing in the intermediate state in the limit of large $N_c$. The
degeneracy is lifted at the order
$1/N_c$. This can be seen by observing that the Born term for baryon-pion
scattering violates both large $N_c$ constraint and unitarity, and hence
must be cancelled -- among the tower of intermediate states -- to $O(1/N_c)$
or alternatively by
assuring that one loop chiral perturbation  corrections to pion-baryon
scattering do not violate the large $N_c$ constraint and unitarity.

A consequence of the Dashen-Manohar QCD constraints in light-quark
baryons is found to be that baryon splittings must be
proportional to $1/N_c$ and $\vec{J}^2$ where $J$ is the angular momentum
of the baryon\cite{jenkins1}. This is precisely the structure of the
light-quark spectrum
we obtained in the skyrmion picture. Now if the baryons contain one heavy quark
$Q$ plus two light quarks, heavy-quark symmetry requires that the hyperfine
splitting in the heavy baryons must be inversely proportional to the
heavy-quark mass $m_Q$. Thus large $N_c$ QCD predicts the splitting to be
$\sim (N_c m_Q)^{-1}$. Furthermore from the Dashen-Manohar constraints,
it follows \cite{jenkins2}
that the hyperfine splitting for, say, $\Sigma^*_Q-\Sigma_Q$ must be of
the form,
$\sim \vec{J}_l\cdot \vec{J}_Q$, where $J_l$ is the light-quark soliton angular
momentum and $J_Q$ the heavy-quark angular momentum. This is identical
to what we found in the heavy baryon spectrum (see (\ref{CKhyperfine}))
\be
\Delta H\sim \frac{1}{\I} c_ \Phi \vec{J}_l\cdot \vec{J}_\Phi
\ee
where $J_\Phi$ is the angular momentum lodged in the Berry potential generated
from the heavy meson $\Phi$ containing the heavy quark $Q$ when the latter is
integrated out. (Recall that $c$ is of the form, Eq. (\ref{ccoeff}).)

The same argument applies to non-relativistic quark models as one expects
from the large $N_c$ equivalence between the skyrmion model and
non-relativistic quark model\cite{manohar84}.
\vskip 1cm
\part{Chiral Symmetry in Nuclei}
\renewcommand{\theequation}{III.\arabic{equation}}
\setcounter{equation}{0}
\setcounter{section}{0}
\section{Introduction}
\bs
In the preceding two lectures we discussed the structure of hadrons
and their interactions
in an isolated environment with a focus on chiral structure of the vacuum and
light and heavy excitations on it. In this lecture, we turn to nuclei
and nuclear matter and
their interactions. We will focus on many-body systems with light-quark
(up, down and strange \footnote{In the context used in this chapter, the
strange quark can be considered as light, in contrast to the last chapter
where it ``behaved" more properly as ``heavy".}) hadrons only.
We will not discuss the behavior of heavy baryons in medium.
Broadly we can classify two classes of issues involved here.
One is the change of the {\it environment}
when there are many but finite number
of baryons present in the system, thus addressing the
issue of {\it vacuum change} in the medium and consequently modifications in
the intrinsic property of the particles (both baryons and mesons) such as
their masses and coupling constants. The other is the interaction of the
particles among themselves, which is
intrinsically many-body in nature. The question then is: Given an effective
Lagrangian as discussed in Lectures I and II, with low-excitation mesons and
baryons relevant at the energy scale below the chiral symmetry scale
$\Lambda_{\chi}$, how do we describe nuclei and their interactions?

There are basically two ways of addressing this question. We will discuss
both. The first described in the next section is to start with an effective
Lagrangian given in large $N_c$ expansion as we have gotten in the previous
lectures and develop many-body dynamics along the line that we have pursued for
a baryon, that is to look at the multibaryon sector of the skyrmion picture.
Unfortunately the mathematics of multibaryon structure
is yet obscure and beyond
baryon number 2, we have no clear understanding of the basic structure. There
is a large number of literature on the infinite baryon number system, that is,
nuclear matter, but again apart from a general global property, little is
understood at the moment. In this chapter, we will treat the two-nucleon
system, {\ie}, the deuteron and then nuclear force for which there is
a promising new development that deserves to be recognized.

The second approach is to start with an effective Lagrangian that consists
of mesons and baryons and guided by chiral symmetry and other symmetries of
the strong interactions, develop chiral perturbation theory for nuclear
dynamics. There is an inherent limitation to such an approach that
is associated with low momentum/energy expansion. Nonetheless it
has certain predictive powers as we shall show later.
The main task then would be to understand nuclear forces from the point of
view of chiral symmetry and then develop nuclear response functions
in the same scheme. This approach will be discussed with applications
to both finite nuclei and infinite nuclear matter, possible phase
transitions under extreme conditions of density etc.
\section{Nuclei As Skyrmions}
\bs
Given a realistic effective Lagrangian of meson fields, the topological
soliton sector should in principle
give rise, not only to the baryons we considered above,
but also to arbitrary winding-number objects. In the strong interactions, the
winding number $W$ is the baryon number $n_B$
and hence we are to obtain nuclei with
the baryon number $n_B=A$ where $A$ is the mass number of the nucleus when
properly quantized. Unfortunately we have at present very little
understanding of the structure of multi-skyrmion configurations.
Given the approximations that
one is forced to make in solving the problem, the question is how close the
multi-winding-number solution is to real nuclei. We will discuss later an
alternative approach of the skyrmion phenomenology that sidesteps solving
this difficult problem. There the idea is to deduce an effective
nucleon-nucleon
potential using the skyrmion structure of the $n_B=2$ soliton and let the
ensemble of $n_B=1$ solitons interact through the
potential. This is closer to the conventional approach to the nucleus.
In this Section, we discuss how much of nuclear properties we can understand
solely from a skyrmion point of view. We discuss the simplest nucleus,
the deuteron and then nucleon-nucleon potential. There are studies of
$A\geq 3$ nuclei in terms of multi-winding-number skyrmions but
the mathematics involved has not yet been worked out to the point
as to be able to gauge whether such a description is realistic.
We shall not pursue this approach any further \footnote{See for references,
V.G. Makhanov, Y.P. Rybakov and V.I. Sanyuk \cite{skyrme}}.
Also instead of working with a general,
vector-meson-implemented Lagrangian, we will consider the simplest one,
the original Skyrme Lagrangian discussed in Lecture II,
\ben
\L_{skyrme}=\frac{f_\pi^2}{4}\Tr\ [\del_\mu U \del^\mu U^\dagger]+
\frac{1}{32g^2}\Tr\ [U^\dagger\del_\mu U, U^\dagger \del_\nu U]^2
+ r \Tr\ (\M U + h.c.)\label{lskyrme}
\een
with $f_\pi$ the pion decay constant and $g$ the Skyrme term constant
related to vector meson gauge coupling discussed in Lecture II.
The last term in (\ref{lskyrme}) is the leading chiral symmetry-breaking
term of $O(\del^2)$ in the ``chiral counting" defined more precisely later.
In dealing with
nonstrange nuclear systems, $U$ will be the chiral field in $SU(2)$, so
the Wess-Zumino term is missing from (\ref{lskyrme}). Later we will consider
the chiral field valued in $SU(3)$ in dealing with kaon-nuclear interactions.
In what follows, one should keep in mind that in some sense, this is a
Lagrangian that results from a large vector-meson mass
limit of a Lagrangian with explicit vector mesons with the so-called
Skyrme quartic term representing in some average sense the effect of all
phenomenology. More sophisticated
Lagrangians with vector mesons and baryon resonances
bring about quantitative improvements but no qualitative changes.
This was quite clear in hyperon phenomenology discussed in the
preceding chapter.

\subsection{Deuteron}
\bs
Consider the simplest nucleus, the deuteron, which
is a loosely bound proton-neutron system with a large quadrupole
moment: a binding energy of 2.2 MeV and a root-mean-square radius of
2.095 fm.  Since the two nucleons are on the average widely separated,
we could start with the product {\it ansatz} for the chiral field
with winding number 2. Let us
put the centers of the two skyrmions symmetrically about the origin along the
x axis. When the two nucleons are widely separated, a reasonable
configuration for the system is the product form,
\ben
U_1 (\vec{r}+s\hat{x})SU_1(\vec{r}-s\hat{x})S^\dagger
\een
where
$U_1$ is the hedgehog with baryon number 1 and $S$ is a constant (global)
$SU(2)$ matrix describing the relative isospin orientation of the two
skyrmions \footnote{In the literature, one frequently uses $A$ for the
global $SU(2)$ rotation. Here as in Lecture II, we reserve $A$ for the induced
vector field.}. The two skyrmions are separated along the $x$ axis by
the distance of $2s$. It turns out from the early numerical studies\cite{JJP}
that the most attractive
configuration is obtained when the isospin of one skyrmion is rotated
by an angle of 180$^\circ$ about an axis perpendicular to the $x$ axis
({\ie}, the axis of separation).
Choosing the z axis for the rotation,
\ben
A=e^{i\pi \tau_3/2}=i\tau_3
\een
the $n_B=2$ chiral field takes the form
\ben
U_s (x,y,z)=U_1(x+s,y,z)\tau_3 U_1(x-s,y,z)\tau_3.
\een
Let us see what symmetries this {\it ansatz} has \cite{braatencarson}.
With the explicit form
for the single hedgehog $U_1=\exp (i\vec{\tau}\cdot \hat{r} F(r))$,
it is easy to verify that it satisfies the
following discrete symmetry relations
\ben
U_s (-x,y,z)&=& \tau_2 U_s^\dagger (x,y,z)\tau_2,\\
U_s (x,-y,z)&=& \tau_1 U_{-s}^\dagger (x,y,z)\tau_1,\\
U_s (x,y,-z)&=& U_{-s}^\dagger (x,y,z).
\een
It also has the symmetry under the parity transformation
\ben
U_s^\dagger (-x,-y,-z)=\tau_3 U_s (x,y,z)\tau_3.
\een
These discrete symmetries can be summarized by one nontrivial one
\ben
U_s (x,-y,-z)=\tau_1 U_s (x,y,z)\tau_1. \label{sym1}
\een
What this says is that a spatial rotation about the axis of separation
by angle of $\pi$ is equivalent to an isospin rotation about the $x$ axis by
$\pi$. This symmetry forbids the quantum number $I=J=0$ where $I$ is the
isospin and $J$ the angular momentum of the state. This is
consistent with Pauli exclusion principle. However
the product {\it ansatz} is not a solution of the equation of motion and
in addition there is nothing to prevent another deuteron-like state
(say, $d^\prime$) with $I=0$ and $J=1$
to appear nearly degenerate with the deuteron. Indeed a simple analysis
does show that a $d^\prime$ is present when the product {\it ansatz} is used.
In Nature
there is only one bound state, so while the product {\it ansatz} describes
the deuteron property (and more generally static two-nucleon properties
\cite{vinhmau}) in qualitative agreement with experiments,
it does not have enough symmetry of Nature. In addition to the symmetry
(\ref{sym1})
\ben
U_2 (x,-y,-z)=\tau_1 U_2 (x,y,z)\tau_1, \label{sym1p}
\een
other discrete symmetries not present in the product {\it ansatz}
but needed to eliminate the spurious $d^\prime$ are
\ben
U_2 (-x,y,-z)&=& \tau_1 U_2 (x,y,z)\tau_1,\label{sym2}\\
U_2 (-x,-y,z)&=& U_2 (x,y,z).\label{sym3}
\een
(We are using the subscript 2 to indicate the winding number-2 configuration.)
In fact the last symmetry (\ref{sym3}) is a special case ({\ie}, $\alpha=\pi$)
of the continuous cylindrical symmetry
\ben
U_2 (\rho, \phi+\alpha,z)=e^{-i\alpha\tau_3} U_2 (\rho,\phi,z)e^{i\alpha
\tau_3}\label{symtoroid}
\een
where the $z$ axis is taken to be the symmetry axis.
This has a toroidal geometry. A $n_B=2$ skyrmion with the symmetries
(\ref{sym1p}), (\ref{sym2}), (\ref{sym3}) and (\ref{symtoroid}) is found to
be the lowest energy solution of the skyrmion equation\cite{verbar}.
With this configuration, there is no bound $d^\prime$ state as it is pushed up
by rotation.  We will refer to
this as toroidal skyrmion. To be more quantitative, let us see what the
classical energies of various configurations come out to be\cite{manton}.
For this we put the length in units of $(ef)^{-1}$ and the energy in units of
$f/2e$. In these units,
the energy of two infinitely separated skyrmions is $1.23\times 24 \pi^2$
whereas the toroidal structure with coincident skyrmions comes slightly
lower, at
$1.18\times 24\pi^2$. The latter corresponds to the lowest energy solution of
the $n_B=2$ skyrmion. The $n_B=2$ spherical hedgehog configuration has much
higher energy, $1.83\times 24 \pi^2$.

The question is: What is the deuteron seen in Nature?
While lowest in energy, the toroidal configuration cannot be the dominant
component of the deuteron. For with that configuration, the size -- and
hence the quadrupole moment -- of the deuteron would be much too
small\cite{braatenetal},
the calculated size being $\sqrt{\langle r^2\rangle_d}\approx 0.92$ fm while
experimentally it is $2.095$ fm and the calculated quadrupole moment
being $Q\approx 0.082$ fm$^2$ while experimentally it is $0.2859$ fm$^2$.
This compactness of the structure is easily understandable in terms of much
too large binding energy -- about an order of magnitude too large --
associated with this configuration. Clearly then this configuration
must constitute only a small component of the deuteron wave function.
Despite the problem alluded above, the size and quadrupole
moment indicate that the widely separated configuration should be
closer to Nature.
Even so, the deuteron seems to share the torus symmetry in a variety of ways
as observed in its static electromagnetic properties and other moments.

Even though reducing the problem
from a field theory involving solitons to a quantum
mechanics in ``moduli space" brings in enormous simplification,
the reality must still be rather complex in terms of skyrmion configurations.
A realistic treatment\cite{manton,atiyahmanton}
must encompass the unstable spherical hedgehog with
winding number 2 \cite{jacksonrho},
the manifold of which is twelve dimensional including
six unstable modes, the infinitely separated two skyrmion configuration --
which is also twelve dimensional -- and the eight-dimensional toroidal
configuration \cite{verbar}. This is a very difficult problem
and it is very unlikely that we will
see analytical solutions in the near future. It is however expected that
numerical simulations will provide accurate solutions within a near
future. Indeed an initial ``experiment" of that
type has been recently
performed. While the result is only preliminary and
semiquantitative at best, it is nonetheless
a remarkable development. Since it is most
likely to be soon superseded by more refined results, we will not go into
the details of the present calculation. Let it suffice to
summarize what has been achieved so far\cite{crutch}.
In this analysis, initially two widely separated skyrmions (which are
not a solution on a finite lattice which they are using) are prepared
by relaxation. The continuum solution is imposed on the lattice. As the
skyrmions relax, they approach each other through the dissipation of
their orbital energy. At the point of closest approach, the two skyrmions
merge into the optimal toroidal configuration and then scatter at
angle $\pi/2$, a feature generic of two solitons scattering
at low energy, such as monopole-monopole, skyrmion-skyrmion and
vortex-vortex  scattering. This process is found
to repeat, with the skyrmions falling into the toroidal form and scattering at
right angles. At each cycle, energy is transferred from the orbital motion to
a radial excitation of the individual skyrmion. Quantizing the nearly periodic
motion, it is found that for large $g^2$ (for which the Skyrme quartic term
emerges from the $\rho$ meson kinetic energy term in a more realistic
hidden gauge symmetry Lagrangian\cite{bando}), there is only one bound
state with $J=1$ and $I=0$, with an unbound $J=0$ and $I=1$ state nearby.
(There are instead many bound states for weak $g^2$. But for a weak $g^2$,
there is no reason why the Skyrme Lagrangian would be a good approximation.)
The bound wavefunction describes a state peaked at a configuration
with two nucleons widely separated, hence elongated, with the size about
twice that of the static toroidal structure
and a quadrupole moment about 4 times. These are all in semiquantitative
agreement with experiments.
\subsection{Nuclear Forces}
\bs
The static minimum-energy configuration is known to be
tetrahedron for $n_B=3$, octahedron for $n_B=4$ etc.\cite{braatenetal}.
It would undoubtedly be most exciting to see what happens when one
extends the sort of analysis made on the torus configuration
of the deuteron to $^3$He, $^4$He etc. Unfortunately this will take a
tour-de-force computational effort and extending beyond would probably be
impossible except for qualitative features. As in the case of the torus,
these configurations will make up only a small component of the wavefunctions
even though symmetries may be correctly described by them. The alternative to
this -- which is definitely more economical and more predictive-- is to
construct a nucleon-nucleon potential
from skyrmion structure and treat the nuclei in the conventional way.
Furthermore, constrained static solutions that are required for constructing
NN potentials are much easier to work out than the time-dependent problem
mentioned above.
The question that is properly posed is: Can one derive a nucleon-nucleon
potential which is close to, say, the Paris potential\cite{parispot}~?

The $n_B=2$ baryon has twelve collective degrees of freedom: three for
translation (or center-of-mass position), three for orientations in space and
isospin space, three for the spatial separation and three for the relative
orientation in isospin. For the asymptotic configuration where two
skyrmions are widely separated, the twelve degrees of freedom correspond
to the coordinate $\vec{r}_i$ and the isospin orientation $S_i$ for each
skyrmion $i=1,2$. When quantized, they represent two free nucleon states.
The product {\it ansatz} possessing twelve collective degrees of freedom
is close to two free nucleons when widely separated but lacks the correct
symmetry at short distance. When two skyrmions are on top of each other,
as mentioned above,
the lowest energy configuration is that of a torus (with axial symmetry)
with eight collective degrees of freedom. Thus four additional collective
degrees of freedom are needed for describing general $n_B=2$ configurations.
This constitutes the moduli space that is involved for the problem.

For general $n_B=2$ configurations, global collective coordinates do not
affect the energy, so for considering a potential, we may confine
ourselves to the separation $r$ of the two skyrmions and the relative isospace
orientation which can be expressed by three Euler angles $\alpha$, $\beta$
and $\gamma$
\ben
C=S_1^\dagger S_2=e^{i\tau_3 \alpha/2}e^{i\tau_2 \beta/2}e^{i\tau_3 \gamma/2}.
\een
When global collective coordinates are quantized, the resulting Hamiltonian
involves a potential which will then depend only on the separation $r$
which we choose here to be along the $z$ axis and the three Euler angles. The
potential
therefore must be expressible in terms of the Wigner $\D$-functions
as\cite{wambach}
\ben
V(r,C)=\sum_{j=0}^\infty \sum_{m=-j}^j \sum_{n=-j}^j V_{jmn} (r)
{\cal D}^{(j)}_{m,n}(C).
\een
This is the most general form since the ${\cal D}$ functions form a
complete set over $SU(2)$ for
integer $j$. This can be further reduced by using the symmetry associated
with rotation about the $z$ axis -- therefore reducing the Euler angles
to two, namely, $\beta$ and $(\alpha-\gamma)$ --and
the relation $V(r,C)=V(r,C^\dagger)$ due to the $SU(2)$ symmetry, to
the form
\ben
V(r,C)=\sum_{j=0}^\infty \sum_{m=0}^j V_{jm}\frac{1}{2}\left[
{\cal D}^{(j)}_{m,m} (C) +{\cal D}^{(j)}_{-m,-m} (C)\right].\label{pot}
\een
The nucleon-nucleon potential can then be defined by taking the matrix element
with the asymptotic two-nucleon wavefunction
\ben
\Psi (1,2)\sim {\cal D}^{(1/2)}_{i_1,-s_1} (S_1) {\cal D}^{(1/2)}_{i_2,-s_2}
(S_2).\label{wf}
\een
The potential so defined is guaranteed to give the correct asymptotic behavior
shared by the product {\it ansatz} and the correct short-distance behavior
shared by the toroidal configuration.
Although there is no rigorous proof, it appears that only the partial waves
$j\leq 1$ contribute significantly\cite{wambach}. Thus ignoring the higher
partial waves $j\geq 2$, we have
\ben
V(r,C)=V_{00} (r) {\cal D}^{(0)}_{0,0} (C) +V_{10} {\cal D}^{(1)}_{0,0} (C)
+V_{11} (r) [{\cal D}^{(1)}_{1,1} (C) +{\cal D}^{(1)}_{-1,-1} (C)]
\een
which when sandwiched with the wavefunction (\ref{wf}) gives
\ben
V(1,2)=V_C (r) +V_{SS} (r) \vec{\tau}_1\cdot\vec{\tau}_2 \vec{\sigma}_1
\cdot\vec{\sigma}_2 +V_T (r) \vec{\tau}_1\cdot\vec{\tau}_2 (3 \vec{\sigma}_1
\cdot\hat{r} \vec{\sigma}_2\cdot\hat{r}-\vec{\sigma}_1\cdot\vec{\sigma}_2)
\een
with
\ben
V_C=V_{00}, \ \ \ V_{SS}=25(V_{10}+V_{11})/247, \ \ \ V_T=25(2V_{10}-
V_{11})/486
\een
with potential functions $V_i$ given by the skyrmion solution
of the classical equation of  motion. The $V$'s are obtained numerically.

There is considerable uncertainty in defining the separation $r$ between
two skyrmions because of their extended structure. A convenient definition is
\cite{wambach}
\ben
(\frac{r}{2})^2=\frac{1}{n_B}\int d^3 x \vec{x}^2 {\cal B}^0 (\vec{x})
\een
where ${\cal B}^0$ is the baryon number density. This reduces to the
usual separation $r=|\vec{r}_1-\vec{r}_2|$ if one takes the form
$${\cal B}^0 (\vec{x})=\delta^3 (\vec{x}+\frac{1}{2}r\hat{e}_3)
+\delta^3 (\vec{x}-\frac{1}{2} r\hat{e}_3).$$
The resulting central $V_c^T$ potentials obtained
by Walhout and Wambach \cite{wambach}
and Walet and Amado \cite{walet1} are given in Fig.III.1
and compared with a realistic phenomenological potential.
The intermediate-range attraction in the central potential has been the
most difficult one to obtain in the skyrmion picture. (Other components
of the force have been better understood.) One sees that
the difficulty is now largely eliminated. A considerable improvement
is obtained by introducing state-mixing between NN,
N$\Delta$ and $\Delta \Delta$ intermediate states that involves $1/N_c$
corrections\cite{walet1}.

In summary, the skyrmion picture is found\cite{walet2}
to provide a generally satisfactory account of
long-range pion exchange, shor-distance repulsion, tensor forces
and spin-orbit forces. Considering the enormous simplicity of the effective
Lagrangian, this is a truly remarkable success of the model.

\section{Chiral Perturbation Theory}
\setcounter{equation}{0}
\subsection{Foreword}

\bs Chiral symmetry has played
indirectly an extremely important and powerful role
in nuclear physics since many years \cite{brcom81}; yet it is only
recently that
the nuclear physics community started to pay attention to this, particularly
in connection with the structure of the nucleon and associated issues.
Even the role of mesons in nuclei -- a problem which has been around
for more than four decades -- has only recently been elucidated, thanks
largely to a better understanding of the workings of chiral symmetry
in strongly interacting many-body systems. Though
chiral symmetry addresses specifically to pions, it turns out to play
a crucial role
where other degrees of freedom than pions intervene. The reason for this is
that {\it pions dominate whenever they can and if they do not, then there are
reasons for it, which have something to do with chiral symmetry.}

In this part of the lecture, we discuss to what extent
nuclear dynamics -- both under
normal and extreme conditions -- are dictated
by chiral symmetry. There are several layers of problems dealing with
nuclei. The first is the property of the basic building block of the nucleus,
which is of course the nucleon and the nucleon is made of quarks and gluons:
We have to understand what the nucleon is to understand
how two or more nucleons interact. Now given our understanding of the nucleon,
we have to understand how they interact -- following what symmetries and
dynamics -- and how the individual property of the nucleon is modified
by the change of environment. In all these, we would like that the same
guiding principles apply for both the basic structure of the nucleon
and the residual interaction between them.

We shall discuss from the vantage point of chiral symmetry what
the state of hadronic matter may be at densities near that of
nuclear matter $\rho_0\approx 0.16 fm^{-3}$ and ultimately higher including the
regime $\rho\gg \rho_0$ where phase transitions from normal hadronic
matter to ``abnormal" matters are supposed to take place.

\subsection{Chiral Lagrangians for Baryons and Goldstone Bosons}
\bs
In Lecture II, we dealt with chiral Lagrangians with mesons
only. We discussed there how baryons could emerge from meson fields
as solitons (skyrmions). We saw that this picture of the baryons gets
sharpened for large $N_c$. In principle,
given such mesonic Lagrangians, we could expect to calculate properties of
an A-skyrmion
system and suitably quantized, to describe a nucleus with A nucleons. A brief
discussion on this matter was given in the preceding section. One
day it may be possible to describe, say, $^{208}$Pb with a Lagrangian
whose parameters are completely fixed in the zero-baryon sector. In fact we
might be  able to expect that the Pb nucleus is understood {\it because}
the nucleon is understood, through a concept of ``tumbling" as
stressed by Nambu\cite{nambu}, the idea being that it is a
hierarchy of symmetries which
tumble down from high energy to low energy with the strength of the condensates
setting the scale. The key picture of such a hierarchy
is a $\sigma$-model structure, which applies to the Higgs sector, to the
QCD sector (nucleon), to the complex nuclear sector and to superconductivity.
In Lecture I, we discussed the possibility
that a hierarchy of induced
gauge potentials might provide the link.

In this lecture, we approach the problem from a different vantage point.
We shall assume that it is possible to map the soliton-meson Lagrangian that
results from the soliton quantization and  fluctuations around it to an
effective Lagrangian of baryons and mesons. In large $N_c$ limit, the baryons
are infinitely massive, so one can do a simultaneous expansion in
$1/m_{\boxB}$ (where $m_{\boxB}$ is the baryon mass), $\del/\Lambda_\chi$
and ${\cal M}/\Lambda_\chi$ (where ${\cal M}$ is the quark mass matrix).
The baryon here is put in a way quite analogous to the heavy quark in
the chiral symmetric heavy quark Lagrangian considered in the preceding
lecture.

We will construct an appropriate
Lagrangian in the spirit of chiral expansions. In so doing, we
will {\it postulate} that mesons and baryons that figure in nuclei satisfy
chiral symmetry and other flavor symmetries.   We will simply write
it with local fields. The underlying assumption is that Weinberg's
``theorem" \cite{wein79} holds equally in the case that baryons are
present explicitly. In this framework, the effective Lagrangian approach is
as general as any field theory.

Although for the moment we will be mostly concerned with the up- and
down-quark ($SU_f (2)$) sector, we shall write the Lagrangian for the
$SU(3)$ flavor as it will be needed for kaon-nuclear interactions
and kaon condensation in the later part of the lecture.
As explained in the previous lectures, the $U_A(1)$ symmetry
is broken by anomaly, so the associated ninth pseudoscalar meson
$\eta^\prime$ is massive.
It will not concern us here. The scale invariance is broken by
the trace anomaly with the associated scalar $\chi$ playing an important
role for scale properties of the hadrons.
This feature will be incorporated later in the scheme.
Let the octet pseudoscalar field $\pi=\pi^a T^a$ be denoted by
\ben
\sqrt{2}\pi=\left(\begin{array}{ccc}
\frac{1}{\sqrt{2}}\pi^0+\frac{1}{\sqrt{6}}\eta & \pi^+ & K^+ \\
\pi^- & -\frac{1}{\sqrt{2}}\pi^0 +\frac{1}{\sqrt{6}}\eta & K^0 \\
K^- & \bar{K}^0 & -\frac{2}{\sqrt{6}}\eta
\end{array}\right)
\een
and the octet baryon field $B=B^a T^a$ be given by
\ben
B=\left(\begin{array}{ccc}
\frac{1}{\sqrt{2}}\Sigma^0 +\frac{1}{\sqrt{6}}\Lambda & \Sigma^+ & p \\
\Sigma^- & -\frac{1}{\sqrt{2}}\Sigma^0 + \frac{1}{\sqrt{6}} & n \\
\Xi^- & \Xi^0 & -\frac{2}{\sqrt{6}}\Lambda
\end{array}\right).
\een
Define as before the ``induced" vector current $V_\mu$ and
axial-vector vector current $A_\mu$
\ben
\left(\begin{array}{c} V^\mu \\ -i A^\mu \end{array}\right)=\frac{1}{2}
\left(\xi^\dagger\del^\mu \xi \pm \xi \del^\mu \xi^\dagger\right)
\een
where $U=\xi^2={\rm exp} (2i\pi^a T^a/f)$ with $\Tr T^a T^b=\frac{1}{2}
\delta_{ab}$ ({\ie}, $T^a\equiv \lambda^a/2$ where $\lambda^a$ are the usual
Gell-Mann matrices). Recall:
Since $U$ transforms $U\rightarrow LUR^\dagger$ under
chiral transformation with $L\in SU_L (3)$ and $R\in SU_R (3)$,
$\xi$ transforms $\xi\rightarrow L\xi g^\dagger (x)=
g(x) \xi R^\dagger$ with $g\in SU(3)$ a local, nonlinear function of
$L$, $R$ and $\pi$. Under the local transformation $g(x)$, $V_\mu$ transforms
as a gauge field
\ben
V_\mu\rightarrow g(V_\mu+\del_\mu)g^\dagger
\een
while $A_\mu$ transforms homogeneously (or covariantly)
\ben
A_\mu\rightarrow g A_\mu g^\dagger.
\een

The next thing to do is to decide how the baryon field transforms under
chiral $SU_L(3)\times SU_R(3)$ symmetry. Now the baryon field can be considered
as a matter field just as the vector meson fields $\rho$ are. The general
theory of matter fields interacting with Goldstone boson fields was formulated
by Callan, Coleman, Wess and Zumino many years ago \cite{CCWZ}.
Since matter fields transform as definite representations {\it only} under the
unbroken diagonal subgroup $SU_{L+R} (3)$, one may choose any representation
that reduces to this subgroup. Different choices give rise to  the same
S-matrix. A convenient choice turns out to have the baryon field transform
\ben
B\rightarrow g B g^\dagger.\label{hgauge}
\een
Under vector transformation $L=R=V$, we have $g=V$ so that Eq.(\ref{hgauge})
reduces to the desired vector (octet) transformation $B\rightarrow
VBV^\dagger$. With this representation, the pion field has only derivative
coupling to the baryons which facilitates the chiral counting developed below.
\cite{georgibook}. With the ``gauge" field $V_\mu$, we write the
covariant derivative
\ben
D^\mu B=\del^\mu +[V^\mu, B]
\een
which transforms $D^\mu B\rightarrow g D^\mu B g^\dagger$.
The Lagrangian containing the lowest derivative involving baryon fields is
then
\ben
\L_B &=& i \Tr (\bar{B}\gamma^\mu D_\mu B) - m_B\Tr(\bar{B} B) \nonumber\\
&+& D \Tr (\bar{B} \gamma_\mu \gamma_5 \{A^\mu, B\})
+F \Tr (\bar{B} \gamma_\mu \gamma_5 [A^\mu, B])+ O(\del^2) \label{lbaryon}
\een
where $D$ and $F$ are constants to be determined from experiments as we will
see later.
The mesonic sector is of course described by Eq. (\ref{lskyrme}), the chirally
invariant part of which is, in terms of the chiral field $U$ alone,
\ben
\L^{U}=\frac{f_\pi^2}{4} \Tr (\del_\mu U
\del^\mu U^\dagger) +O(\del^4). \label{lmeson}
\een
In Eqs.(\ref{lbaryon}) and (\ref{lmeson}), truncation is made at the
indicated order of derivatives.

Next we need to introduce chiral symmetry breaking terms since the symmetry
is indeed broken by the quark mass terms.
We assume as suggested by QCD Lagrangian that
the quark mass matrix $\M$ transforms as $(3,\bar{3}) + (\bar{3},3)$,
namely, as
\ben
\M \rightarrow L \M R^\dagger.
\een
In our discussion, the mass matrix will be taken to be in the diagonal form
\ben
\M={\rm diag}\  (m_u, m_d, m_s).
\een
Writing terms involving $\M$ that transform invariantly under
chiral transformation, we have
\ben
\L_{\chi SB} &=& r\Tr\ (\M U+h.c.)\nonumber\\
&+& a_1\Tr \barB\left(\xi \M \xi+ h.c.\right) B +a_2\Tr \barB B
\left(\xi \M\xi
+ h.c.\right) +a_3\Tr \barB B \Tr\left(\M U + h.c.\right).\nonumber\\
\label{sbp}
\een
In writing this only $CP$ even terms are kept. In general, without restriction
on $CP$, the $h.c.$ term would be multiplied by complex conjugation of the
coefficient which would in general be complex. As detailed later, the mass
matrix $\M$ is formally $O(\del^2)$, so to the order involved in (\ref{sbp}),
we have to implement with a Lagrangian that contains two derivatives.
We will call it $\delta {\L_B}$. It will be specified below.


As given, the baryon Lagrangian (\ref{lbaryon}) does not lend itself
to a straightforward chiral expansion. This is because the baryon mass
$m_B$ is not small compared with the chiral scale $\Lambda_\chi$: As noted
below, while the space derivative acting on the baryon field can be considered
to be small, the time derivative cannot be considered to be so
since it picks up the baryon mass
term. It is therefore more appropriate to consider the baryon to be ``heavy"
and make a field redefinition so as to eliminate the mass term from
the Lagrangian. To do this we let
\cite{heavyfermion}
\ben
B\rightarrow e^{im_B v\cdot x} \frac{1+\gamma\cdot v}{2} B
\een
where $v_\mu$ is the velocity four-vector of the baryon with $v^2=1$.
This rescaling eliminates the centroid mass dependence $m_B$ from the
baryon Lagrangian ${\L}_B$ and banish negative-energy Dirac components
$\frac 12 (1+\gamma\cdot v) B$ to $1/m_B$ corrections
\ben
\L_B = \Tr \barB iv\cdot D B
+2D\Tr \barB S^\mu \{A_\mu, B\}+2F\Tr \barB [A_\mu, B]
+\cdots \label{lbaryonp}
\een
and simplifies $\delta {\L_B}$ which is somewhat awesome looking to
\ben
\delta \L_B &=&
c_1 \Tr \barB D^2 B +c_2\Tr \barB (v\cdot D)^2 + d_1\Tr \barB A^2 B
+d_2 \Tr \barB (v\cdot A)^2 B + \cdots \nonumber\\
&+& f_1 \Tr \barB (v\cdot D)(S\cdot A)B +f_2\Tr \barB (S\cdot D)(v\cdot A)B
+f_3\Tr \barB[S^\alpha,S^\beta]A_\alpha A_\beta B + \cdots\nonumber\\
\label{deltalb}
\een
where the ellipsis stands for other terms of the same class as well as $1/m_B$
corrections which are of the same order as the
symmetry-breaking term (\ref{sbp}), that is, $O(\del^2)$.
Here $S_\mu$ is the spin operator defined as $S_\mu=\frac{1}{4}\gamma_5
[\not\!\!{v},\gamma_\mu]$ constrained to $v\cdot S=0$ and the constants
$a_i$, $c_i$, $d_i$, $f_i$ and $r$ are parameters to be fixed from
experiments.

In many-body systems like nuclei, we have to include also terms higher order in
baryon field, such as\cite{wein90}
\ben
\L_{4}=\sum_i \Tr \left(\barB \Gamma_i B\right)^2 +\cdots\label{L4}
\een
where $\Gamma_i$ are matrices constrained by flavor and Lorentz symmetries
which we assume do not contain derivatives. The ellipsis stands for
four-fermion interactions containing derivatives and also for higher-fermion
interaction terms. Such terms with quartic fermion fields can be thought of
as arising when meson exchanges with the meson mass greater than the chiral
scale are eliminated. We have already encountered such a term in  Lecture
II in connection with the $\etap$ exchange between baryons.
We can also have derivatives multiplying $\Gamma_i$. We will encounter
such derivative-dependent four-fermion
terms later on -- as counter terms -- when we consider nuclear responses
to electroweak interactions
at which time we shall write down the explicit forms.

In the form given above for $\L_B$ and $\delta \L_B$,
derivatives on the baryon field pick up small residual
four-momentum $k_\mu$
\ben
p^\mu=m_B v^\mu +k^\mu.
\een
In this form, a derivative
on the baryon field is of the same order as a derivative on
a pseudoscalar meson field or the mass matrix $\sqrt{\M}$.

The Lagrangian we will work with then is the sum of (\ref{lmeson}),
(\ref{lbaryonp}), (\ref{deltalb}) and (\ref{L4})
\ben
\L=\L^U +\L_B +\delta \L_B +\L_4.\label{Thelag}
\een
Now given this Lagrangian for three-flavor
system, it is easy to reduce it to one for two flavors -- which is what we need
for most of nuclear physics applications. The pseudoscalar field is then
a $2\times 2$ matrix
\ben
\sqrt{2}\pi=\left(\begin{array}{cc}
\frac{1}{\sqrt{2}}\pi^0 & \pi^+ \\
\pi^- & -\frac{1}{\sqrt{2}}\pi^0 \end{array}\right)
\een
and the baryon field becomes a simple doublet
\ben
N=\left(\begin{array}{c} p\\ n \end{array}\right)
\een
with the resulting Lagrangian
\ben
\L &=& N^\dagger i\del_0 N+iN^\dagger V_0 N
-g_A N^\dagger\vec{\sigma}\cdot \vec{A} N \nonumber\\
&+&b N^\dagger\left(\xi \M \xi + h.c.\right) N +
h^\prime N^\dagger N \Tr \left(\M U + h.c.\right) +\cdots.\nonumber\\
&+& \sum_i (N^\dagger\Gamma_i N)^2 + \delta \L_B \cdots\label{lsu2}
\een
where we have replaced $(F+D)$ by $g_A$, relevant for neutron beta decay,
a procedure which is justified at the tree order. This relation no longer
holds, however,  when loop corrections are included.
This Lagrangian (\ref{lsu2}) will be used in nuclear physics applications
in what follows.

So far, we have written down Lagrangian terms that are lowest in derivatives
and in the quark mass matrix $\M$. Going further in the expansion can be done
in two ways: one, higher derivatives in pseudoscalar fields (and higher orders
in $\M$); two, introduce vector fields (and higher orders in $\M$).
For instance,
we can take the vector and axial-vector fields to be $O(\del)$ and if they are
coupled
to the lowest derivatives on pseudoscalars consistent with chiral
transformation, then we wind up with a Lagrangian valid up to $O(\del^4)$,
{\it e.g.}, the vector
meson field tensor which involves $(\del_\mu V_\nu)^2$. In all applications we
will discuss, we will follow the second route which is simpler at
low orders we use.
\subsection{Chiral Power Counting}
\bs
In nuclear physics, we are concerned with an amplitude that involves
$E_N$ nucleon lines (sum of incoming and outgoing lines), $E_\pi$ pion lines
and one electroweak current. Chiral Lagrangian approach is relevant at low
energy and low momentum, so we are primarily concerned with slowly varying
electroweak fields. In dealing with chiral power counting, the pion field
and its coupling are simple because chiral symmetry requires derivative
coupling. The chiral field $U$ can be counted as $O(1)$ in chiral counting,
a derivative acting on the pion field as $O(Q)$ where $Q$ is a characteristic
pion four-momentum involved in the process, assumed to be of $O(m_\pi)$
or less -- and in any case much less than the
chiral scale of order 1 GeV -- and $m_q\sim m_\pi^2\sim O(Q^2)$.
As remarked above, the situation with the baryon as given in (\ref{lbaryon})
is somewhat more complicated because the baryon mass
is of $O(1)$ relative to the chiral scale. Indeed, one should count
$m_B\sim O(1)$, the generic baryon field
$\psi\sim O(1)$ and hence $\del^\mu \psi\sim O(1)$. This means that
\ben
(i\gamma_\mu \del^\mu-m_B)\psi=iv_\mu \del^\mu N \sim O(Q).
\een
However the three-momentum of the baryon -- and hence the space derivative
of the baryon field -- should be taken to be $O(Q)$ while the time derivative
of it should be taken to be $O(1)$. The heavy-fermion formalism takes care
of this problem of the heavy baryon mass. We will illustrate how this
works out in specific examples we will treat later.

In deducing the chiral counting rule, we must distinguish two types of
graphs: one, {\it reducible} graphs and the other, {\it irreducible}
graphs. Chiral expansion is useful for the second class of diagrams.
An irreducible graph is a graph that cannot be made into disconnected graphs
by cutting any intermediate state containing $E_N/2$ nucleon lines
and either all the initial pions or all the final pions. Reducible graphs
involve energy denominators involving the difference in nucleon kinetic
energy which is much smaller than the pion mass whereas irreducible graphs
involve energy denominators of $O(Q)$ or the pion mass. We must apply
chiral perturbation theory to the irreducible graphs, but not to the
reducible ones. The reducible ones involve infrared divergences and are
responsible for such nonperturbative effect as binding or pairing transitions.
In the language of effective field theories \cite{polchinski}
it is the effect of reducible graphs that transform ``irrelevant terms"
to ``marginal" leading to interesting phenomena such as superconductivity
as discussed in Polchinski's lecture and elsewhere. In the present context,
the reducible graphs are included when one solves Lippman-Schwinger equation
or Schr\"{o}dinger equation. We will therefore confine ourselves to
the irreducible graphs only.

Furthermore the counting rule is much more clearly defined
without vector mesons than with vector mesons. The vector mesons with
their mass comparable to the chiral scale bring in additional scale to the
problem.

In what follows, using heavy-fermion formalism (HFF),
we rederive and generalize
somewhat Weinberg's counting
rule applicable to {\it irreducible} graphs \cite{wein79,wein90,wein92}.
Although we shall not consider explicitly
the vector-meson degrees of freedom, we include them here in addition
to pions and nucleons. Much of what we obtain later turn out to be valid in
the presence of vector mesons. Now in dealing with them, their
masses will be regarded as comparable to the
characteristic mass scale of QCD $\sim 4\pi f_\pi$ which can be considered
heavy compared to $Q$ -- say, scale of external three momenta or $m_\pi$
\footnote{One subtlety to note here: In considering the vector limit in the
later part of this chapter, we will be taking the limit $m_\rho\rightarrow 0$.
For this it would be dangerous to consider the vector mass to be $O(1)$
as we do here. It is an open problem how to organize an effective theory
for this situation that involves eventually a phase transition. It must be
analogous to instabilities in Landau Fermi liquid theory. See Polchinski's
lecture.}.
\\ \indent
In establishing the counting rule, we make the following key assumptions:
Every intermediate meson (whether heavy or light) carries a four-momentum
of order of $Q$ while every intermediate nucleon carries a four-momentum
of order of $m_N$, namely, of the QCD scale. In addition we assume that
for any loop, the effective cut-off in the loop integration
is of order of $Q$. We will be more precise as to what this means physically
when we discuss specific processes, for this clarifies the limitation of
the chiral expansion scheme.

An arbitrary Feynman graph can be characterized by the number $E_N (E_H)$
of external -- both incoming and outgoing -- nucleon (vector-meson)
lines, the number $L$ of loops, the number
$I_N (I_\pi,\ I_H)$ of internal nucleon (pion, vector-meson) lines.
Each vertex can in turn be  characterized by
the number $d_i$ of derivatives and/or of $m_\pi$ factors and
the number $n_i$ $(h_i)$ of nucleon (vector-meson) lines attached
to the vertex. Now
for a nucleon intermediate state of  momentum $p^\mu=mv^\mu + k^\mu$ where
$k^\mu = {\cal O}(Q)$, we acquire a factor $Q^{-1}$ since
\be
S_F(mv+k) = \frac{1}{v\cdot k} = {\cal O}(Q^{-1}).
\ee
An internal pion line contributes a factor $Q^{-2}$ since
\be
\Delta(q^2;m_\pi^2)=\frac{1}{q^2 - m_\pi^2} = {\cal O}(Q^{-2})
\nonumber
\ee \noindent
while a vector-meson intermediate state contributes
$Q^0$ $(\sim O(1))$ as one can see from its propagator
\be
\Delta_F (q^2;m_V^2) = \frac{1}{q^2 - m_V^2} \simeq \frac{1}{-m_V^2}
= {\cal O}(Q^0)
\ee \noindent
where $m_V$ represents a generic mass of vector mesons.
Finally a loop contributes a factor $Q^4$ because
its effective cut-off is assumed to be of order of $Q$.
We thus arrive at the counting rule that an arbitrary graph is characterized
by the factor $Q^\nu$ (times a slowly varying function $f(Q/\mu)$ where
$\mu\sim \Lambda_\chi$) with
\be
\nu = - I_N - 2 I_\pi + 4 L + \sum_i d_i
\label{piN0}\ee \noindent
where the sum over $i$  runs over all vertices of the graph.
Using the identities,
$I_\pi + I_H + I_N = L + V -1$,
$I_H = \frac12 \sum_i h_i - \frac{E_H}{2}$ and
$I_N = \frac12 \sum_i n_i - \frac{E_N}{2}$,
we can rewrite the counting rule
\be
\nu = 2 - \frac{E_N + 2 E_H}{2} + 2 L + \sum_i \nu_i,\ \ \ \ \
\nu_i \equiv d_i + \frac{n_i+ 2 h_i}{2} - 2 .\label{count1}
\ee
We recover the counting rule derived by Weinberg \cite{wein90,wein92}
if we set $E_H=h_i=0$.

The situation is different depending upon whether or not
there is external gauge field ({\it i.e.}, electroweak field) present
in the process. In its absence (as in nuclear forces),
$\nu_i$ is non-negative
\be
d_i + \frac{n_i + 2 h_i}{2} - 2 \geq 0.
\ee \noindent
This is guaranteed by chiral symmetry. This means that the
leading order effect comes from graphs with vertices satisfying
\be
d_i + \frac{n_i + 2 h_i}{2} - 2 = 0\,.
\ee \noindent
Examples of vertices of this kind are:
$\pi NN\, (d_i=1,\ n_i=2,\ h_i=0)$,
$h N N\, (d_i=0,\ n_i=2,\ h_i=1)$, $(\Nbar \Gamma N)^2\,
(d_i=0,\ n_i=4,\ h_i=0)$, $\pi\pi NN \, (d_i=1,\ n_i=2,\ h_i=0)$,
$\rho\pi\pi\,(d_i=1,\ n_i=0,\ h_i=1)$, etc.
In $NN$ scattering or in nuclear forces, $E_N=4$ and $E_H=0$, and so we have
$\nu \geq 0$. The
leading order contribution corresponds to $\nu=0$,
coming from three classes of diagrams; one-pion-exchange,
one-vector-meson-exchange and four-fermi contact graphs.
In $\pi N$ scattering, $E_N=2$ and $E_H=0$, we have
$\nu \geq 1 $ and the  leading order comes from
nucleon Born graphs, seagull graphs and one-vector-meson-exchange
graphs.\footnote{We note here that scalar glueball fields $\chi$ play only
a minor role in $\pi N$ scattering because the $\chi \pi\pi$
vertex ($d_i=2,\ n_i=0,\ h_i=1$) acquires an additional $Q$ power.}

In the presence of external fields, the condition becomes \cite{mr91,pmr93}
\be
\left( d_i + \frac{n_i + 2 h_i}{2}-2\right) \geq -1 \,.\label{exchcond}
\ee \noindent
The difference from the previous case comes from the fact that a derivative
is replaced by gauge fields. The equality holds only when $h_i=0$. We will
later show that this is related to what is called
the ``chiral filter" phenomenon.
The condition (\ref{exchcond}) plays an important role in determining
exchange currents in nuclei.
Apart from the usual nucleon Born terms which are in the class of
``reducible" graphs and hence do not enter into our consideration,
we have two graphs that contribute in the leading order to the exchange
current: the ``seagull" graphs and ``pion-pole" graphs,
both of which involve
a vertex with $\nu_i=-1$. On the other hand, a vector-meson-exchange graph
involves a $\nu_i= +1$ vertex. This is because $d_i=1,\ h_i = 2$
at the $J_\mu hh$ vertex. Therefore vector-exchange graphs are suppressed
by power of $Q^2$. This counting rule is the basis for chiral filtering.

The key point of chiral perturbation theory is that for small enough $Q$
with which the
system is probed, physical quantities should be dominantly given by the
low values of $\nu$. The leading term is therefore given by tree
diagrams with the next-to-the leading order being given by one-loop
digrams with vertices that involve lowest derivative terms.
The divergences present in one-loop graphs are then to
be cancelled by counter terms in the next order derivative terms. The finite
counter terms of $O(1)$ are to be determined from empirical data.
This can be continued to an arbitrary order in a well-defined manner.
``Weinberg theorem" states that this should correspond to QCD at
long wavelength.
A systematic application of this strategy to $\pi \pi$ scattering has been
made with an impressive success \cite{gasseretal}.

\subsection{Effective Theories in Nuclear Medium}
\bs
Before we apply chiral perturbation theory (ChPT)
to many-body nuclear systems, we need to extend the
concept of effective theories somewhat further. For this we restate what
the basic premise of writing down the chiral Lagrangian (\ref{Thelag})
was \footnote{See Polchinski lectures for further discussions.}.
Given a set of fields denoted by $\phi$ which can be decomposed into
a high-energy component $\phi_H$ with $\omega > \Lambda$ in which {\it we are
not interested} and a low-energy component $\phi_L$ with $\omega <\Lambda$
in which {\it we are interested}. Here $\Lambda$ is some cut-off.
Now integrating out the uninteresting ``fast" component and leaving only
the interesting ``slow" component
\be
\int [d\phi_L] [d\phi_H] e^{iS(\phi_L,\phi_H)}=\int [d\phi_L]e^{iS_\Lambda (
\phi_L)}
\ee
where
\be
e^{iS_\Lambda (\phi_L)}\equiv\int [d\phi_H] e^{iS(\phi_L,\phi_H)}.
\ee
Now the effective action $S_\Lambda$ is given by the sum of all the terms
allowed by symmetries of the fundamental theory
\be
S_\Lambda=\sum_i g_i O_i\label{expansion}
\ee
where $g_i$ are some constants and $O_i$ are operators which can in general
be nonlocal. The chiral expansion we sketched above corresponds to making
a local expansion in terms of derivatives. In Lecture I, we have discussed
how this operation could be done in terms of the chiral bag: It led to
among others the non-trivial emergence of the Wess-Zumino term (or Berry
phases). There $\Lambda$ was identified with the chiral scale $\Lambda_\chi$.

In going to nuclear systems, we make further reductions. We do this in
recognition of the existence of the Fermi surface characterized by the
Fermi momentum $k_F$. Since we will be interested in excitations near
the Fermi surface, we would like to integrate out further the sector
$\frac{\Lambda}{s} <\omega \leq \Lambda$ where $s>1$ which we assume
depends on $k_F$
\be
\int [d\phi_L^{<}][d\phi_L^>] e^{iS_\Lambda (\phi_L^<,\phi_L^>)}
=\int [d\phi_L^<] e^{iS_L^* (\phi_L^<)}
\ee
where the superscript $<$ ($>$) represents the sector $\omega <\Lambda/s$
($\omega >\Lambda/s$).
The successive reduction satisfies renormalization group equation
and forms the usual program proven to be a powerful technique in condensed
matter physics \cite{shankar}.
It is clear that we will have in place of (\ref{expansion})
\be
S_L^*=\sum_i g_i^* O_i^*\label{expansionp}
\ee
where the star indicates the dependence on $k_F$ through $s$.
The key point here is that
(\ref{expansionp}) is of the same structure as (\ref{expansion}) dictated
by the symmetries of the original action. This is quite analogous to Nambu's
proposition. The assumption is that
there is no phase transition so that symmetries are not modified as the cut-off
$\Lambda^*=\Lambda/s$ is decreased with an increasing density $k_F$.
As discussed by Polchinski and Shankar\cite{polchinski,shankar},
Landau's Fermi liquid theory can be derived using
the notion of effective field theories. In our discussion, we are
essentially transposing this same argument to
nuclear systems in a way analogous to Migdal's formulation of
finite Fermi systems \cite{migdal}.
\subsubsection{\it In-medium effective chiral Lagrangian}
\bs
As a concrete application of the above idea, we will construct an
effective Lagrangian applicable in medium based on chiral symmetry
and (broken) scale invariance of QCD~\cite{scalinglag,brscaling}.

Let us denote the baryon matter density that we will be concerned with
by $\rho$.
For a given density, we suppose that we can define a ``vacuum", which
corresponds
to the ground state with vacuum quantum number. We choose to characterize it by
the vacuum expectation value (VEV)
of a generic scalar field denoted as $\Omega$.
In QCD, $\Omega$ can be either the quark scalar density
$\bar{q}q$ or the gluon scalar density $G_{\mu\nu}G^{\mu\nu}$ or a linear
combination. A reasonable working hypothesis is
that the condensates $\langle\Omega\rangle^*$ (where the star indicates medium)
get modified as
the density is changed. This should be verifiable by lattice QCD calculations.
A baryon or meson in that system could then be described as
a {\it quasiparticle} excitation on top of that
vacuum $|0^*\rangle$, propagating with a mass $m^*$ and interacting with matter
with coupling constants $g^*$. As mentioned, our key assumption is
that that {\it the effective Lagrangian
relevant in the system is given by one with known symmetries of
QCD which in our case will be taken to be approximate chiral and
scale (or conformal) invariances}. We wish to establish the scaling of
the relevant masses and coupling constants consistent with the renormalization
group equations and with general properties of
chiral and scale invariances, so that used at the tree level, it captures
the essence of QCD at that length scale \footnote{It should be emphasized
that the mass discussed here is the world scalar mess, not to be confused
with the ``mass" calculated in mean-field theories which contains vector
contribution. The latter is to be computed explicitly with the Lagrangian
constructed below.}. This would define a Lagrangian for a given
background density. Given this Lagrangian for $\rho$, we are to
calculate physical amplitudes involving mesons, baryons and electroweak
currents {\it in the medium}
in terms of the chiral expansion developed above.
We shall illustrate in what follows how this strategy can be confronted
with experimental data.
\subsubsection{\it Construction of the Lagrangian}
\bs
Since we are in a phase in which chiral symmetry is spontaneously broken,
we have nonzero quark condensates
$$\langle\bar{q}q\rangle^*\equiv \langle 0^*|\bar{q}q| 0^* \rangle$$
as a consequence of which the quark mass is dynamically generated.
Since the vacuum is modified, it is different from the matter-free vacuum.
It can in principle be calculated but we will take it as a parameter here.
Baryons and mesons (other than Goldstone bosons)
--bound states of the ``constituent" quarks-- are believed to
pick up their dynamical masses in this way.
(We will first ignore
current quark masses and discuss the chiral limit. The explicit symmetry
breaking will be brought in later.) Thus a scale is generated through
spontaneous symmetry breaking (SSB).
A scale is also generated through the conformal (trace) anomaly, {\ie},
\ben
\theta^\mu_\mu=-\frac{\beta}{g} \Tr \ F_{\mu\nu}F^{\mu\nu};\ \ \
(m_q\rightarrow 0)\label{an1}
\een
where $\theta_{\mu\nu}$ is the energy momentum tensor,
$\beta$ is the renormalization group $\beta$ function which at one loop
order is
\ben
\beta(g)/g=-\frac{\alpha_s}{4\pi} (11- \frac{2}{3} N_f),
\een
$g$ the color gauge coupling constant ($\alpha_s=\frac{g^2}{4\pi}$)
and $G$ the gluon field strength tensor. That the trace of the
energy-momentum tensor is non-vanishing is just equivalent to
saying that the divergence of
the dilatation current is nonvanishing,
\ben
\del^\mu D_\mu\neq 0
\een
with the dilatation current defined by
\ben
D_\mu \equiv x^\nu \theta_{\mu\nu}.
\een
This means that the scale invariance --
which is the symmetry of the QCD Lagrangian in the chiral limit -- is broken.
The breaking is a quantum mechanical effect, associated with
the dimensional transmutation occurring in QCD. We are also
interested in the change
of the ``vacuum" expectation value $\langle\theta^\mu_\mu\rangle^*
\equiv \langle 0^*| \theta^\mu_\mu|0^* \rangle$ as the density is increased.
Since the gluon field has
canonical scale dimension --1, $\theta^\mu_\mu$ has scale dimension --4,
so it is convenient to define~\cite{scalinglag}
a scalar glueball field $\chi$ of scale dimension --1 and write
\ben
\theta^\mu_\mu\equiv \chi^4\label{an2}
\een
and express the gluon condensate as $\langle 0^*|\chi^4|0^* \rangle$.
This scalar glueball field plays an important role in implementing
trace anomaly to effective chiral Lagrangians.

Before proceeding, we give a heuristic derivation of (\ref{an1})
and (\ref{an2}).
Consider the Yang-Mills Lagrangian for the gluons,
\ben
\L_{YM}=-\frac{1}{4} \vec{F}_{\mu\nu}\cdot\vec{F}^{\mu\nu} \label{yangmills}
\een
where the vector product is in the color space. The Lagrangians for the
quarks, gauge-fixing and ghosts are not written down. To get the
energy-momentum tensor, we take the action
\ben
S=\int d^Dx \sqrt{-g} g^{\mu\sigma}g^{\nu\lambda} \left(-\frac{1}{4}
\vec{F}_{\mu\nu}\cdot\vec{F}_{\sigma\lambda}\right)
\een
with $g=\det g$ and $D$ the space-time dimension (=4 in Nature).
The metric is introduced to define the energy-momentum tensor. In
reality, the metric is flat, so $-g=1$ with $g_{\mu\nu}=\eta_{\mu\nu}
={\rm diag} (1,-1,-1,-1)$.
The energy-momentum tensor is obtained as
\ben
\theta_{\mu\nu}=-\frac{2}{\sqrt{-g}}\frac{\delta S(g)}{\delta g^{\mu\nu}}.
\een
Since
\ben
\delta \sqrt{-g}=\frac{1}{2}\sqrt{-g}g_{\mu\nu}\delta g^{\mu\nu}
\een
and
\ben
\delta\left(g^{\mu\sigma} g^{\nu\lambda} F_{\mu\nu} F_{\sigma\lambda}\right)
=2F_{\mu\sigma} F_\nu^\sigma \delta g^{\mu\nu},
\nonumber
\een
we have
\ben
\theta_{\mu\nu}=-g_{\mu\nu}\L_{YM}- F_{\mu\sigma} F_\nu^\sigma + \cdots
\een
Thus
\ben
\theta_\mu^\mu= -(D-4)\L_{YM}+\cdots =\epsilon \L_{YM}+\cdots
\een
where $\epsilon=4-D$. Since $D=4$ in reality, the trace of the energy-momentum
tensor would vanish {\it classically}. Quantum effects change this in a basic
way. With dimensional regularization, we reexpress
(\ref{yangmills}) in terms of the regularized field as
\ben
\L_{YM}=\frac{1}{\epsilon} \frac{2\beta (g)}{g} \left(-\frac{1}{4}
F_{\mu\nu}F^{\mu\nu}\right)_R +\cdots
\een
where the subscript R stands for renormalized quantity.
Therefore we finally obtain
\ben
\theta_\mu^\mu=-\frac{\beta(g)}{2g} \left(F_{\mu\nu} F^{\mu\nu}\right)_R.
\een

The objective of an effective Lagrangian is that the quantum anomaly is
to be reproduced by the Lagrangian calculated at the tree level. To implement
this feature, we first note that under the scale transformation
\ben
x\rightarrow ^\lambda x=\lambda^{-1} x, \ \ \lambda >0
\een
a field $\phi$ of scale dimension $d_\phi$ transforms as
\ben
\delta\phi (x)=\delta \epsilon (d_\phi +x\cdot\del )\phi (x)
\een
with $\delta \epsilon=\ln \lambda$. This can be phrased in terms of
the dilatation generator $D$
\ben
D=\int d^3x D_0 (x),
\een
where $D_0$ is the time component of the dilatation current $D_\mu$
defined above. We have
\ben
[D, \phi (x)]=i (d_\phi +x.\del )\phi (x).
\een
Under this transformation, the action $S=\int d^4x \L \{\phi \}$ transforms
\ben
\delta S=\delta \epsilon \int d^4x \del_\mu D^\mu (x).
\een
Now let us consider our effective Lagrangian in the chiral limit
\ben
\L^{eff}=\L_4 + V(\chi)
\een
where the first term is of scale dimension $-4$ by construction and
\ben
V(\chi)\sim \chi^4 \ln \chi.
\een
Under scale transformation, $\delta \int d^4x \L_4 =0$ and the $\delta
V(\chi)$ gives
\ben
[-4+\chi \frac{\del}{\del\chi}]V(\chi)=\chi^4\equiv -\frac{\beta
(g)}{2g} \vec{F}\cdot\vec{F}
\een
which follows from $\chi (^\lambda x)=\lambda \chi (x)$
This is the desired result.

In what follows, we will argue in terms of only the scalar field $\chi$ and
the chiral field $U=e^{2i\pi/f_\pi}$ but in actual applications, we will
generalize the consideration
to vector-meson and baryon fields using arguments based, respectively,
on hidden gauge symmetry \cite{bando}
and skyrmion description. For simplicity,
we will restrict to the $SU(2)$ flavor. The generalization to the $SU(3)$
flavor is straightforward.

First consider the chiral limit. Chiral invariance requires that only the
derivative of the chiral field, $\del_\mu U$, appear in building up an
effective Lagrangian. We thus have a quadratic current algebra term
\ben
\Tr (\del_\mu U \del^\mu U^\dagger),
\een
a quartic (Skyrme) term,
\ben
\Tr [U^\dagger \del_\mu U, U^\dagger\del_\nu U]^2
\een
and so on. These terms, not involving
glueball fields, should not be expected to contribute to the trace of the
energy momentum tensor. Therefore we would like them to be scale invariant.
Since the chiral field $U$ has scale dimension 0,
one can see that the quartic term has scale dimension --4, so it is
scale-invariant. But the
quadratic term has scale dimension --2 and hence does not have the right
dimension. We can make it scale correctly by multiplying it by an object of
scale dimension --2, {\ie}, by $\chi^2$. We can continue doing this to {\it
all}
the chiral derivative terms by simply using various powers of the scalar
glueball field. Now having assured that the chiral fields do not contribute
to $\theta_\mu^\mu$, we need to have a term made up of the glueball fields that
will supply the contribution $\sim \chi^4$. We may do this by adding
the potential term that as shown above has a correct scale transformation,
\ben
V(\chi) \sim \chi^4 \ln\chi.
\een

Now what about the explicit chiral symmetry breaking which introduces an
additional scale parameter? This is an intricate story which introduces
a certain element of uncertainty in constructing effective Lagrangians.
The reason is that the quark mass gives an additional term to $\theta_\mu^\mu$
with an anomalous dimension contribution $\gamma$ (because of the
chiral symmetry breaking by the mass term)
\ben
\theta_\mu^\mu = \sum_q	 (1+\gamma_q) m_q \bar{q}q -\frac{\beta}{g}
\Tr\ [F_{\mu\nu}F^{\mu\nu}].
\een
Apart from the anomalous dimension contribution, the mass term brings in
a dimension --3 term. How does one represent such a term in terms of
the chiral fields? Here we will take the most obvious possibility which is to
multiply by a scale dimension-3 field $\chi^3$ to the mass term
$\Tr (\M U +h.c.)$ -- which is of scale dimension 0 --
where $\M$ is the quark mass matrix.
Just as the anomalous dimension modifies the scale dimension of
the quark mass term in the QCD Lagrangian, we would expect that quantum
effects {\it with a given effective Lagrangian} would bring in terms of
other scale dimensions. While the anomalous dimension in QCD proper may
be ignorable, it is possible that loop effects in effective theories may not
be small {\it for Goldstone bosons}. This problem has not yet been fully
clarified. In view of theoretical uncertainties, only experiments will
be able to provide the answer.

Collecting all the relevant terms together, our effective Lagrangian is of
the form
\ben
\L^{eff}&=& \frac{f_\pi^2}{4}(\frac{\chi}{\chi_0})^2 \Tr (\del_\mu U
\del^\mu U^\dagger) + \frac{1}{32e^2} \Tr [U^\dagger\del_\mu U, U^\dagger
\del_\nu U]^2+\cdots \nonumber\\
&+& c\left(\frac{\chi}{\chi_0}\right)^3 \Tr (\M U+h.c.) + \cdots \nonumber\\
&+& \frac{1}{2} \del_\mu \chi \del^\mu \chi +V(\chi), \label{leff0}
\een
where the ellipsis stands for higher derivative and mass matrix terms
that can contribute. We have added the scale-invariant kinetic
energy term for the scalar field. The quantity $\chi_0$ is a constant of
mass dimension which will be identified with the vacuum expectation value
of the $\chi$ field in medium-free space, $\chi_0\equiv \langle 0|\chi|
0\rangle$.

The potential $V(\chi)$ is manufactured such that it gives the trace anomaly
correctly~\cite{scalinglag}.
One can always add scale-invariant and chiral-invariant quantities
to it. It can be quite complicated in reality and perhaps quantum effects
may play a role in the sense of Coleman-Weinberg mechanism \cite{coleman}.
Its structure in medium is at present completely unknown. In the total
ignorance, we will simply assume that for a given density $\rho$,
the potential is minimized at a vacuum expectation of the $\chi$ field,
\ben
\chi_*\equiv \langle 0^*|\chi|0^*\rangle.
\een
This suggests to expand the Lagrangian at a given density
around the vacuum value $\chi_*$ by shifting
\ben
\chi=\chi_* + \chi^\prime
\een
where $\chi^\prime$ is the fluctuating scalar field. Defining
a {\it parameter}
\ben
f_\pi^*\equiv f_\pi \frac{\chi_*}{\chi_0}
\een
we may rewrite the Lagrangian (\ref{leff0}) as
\ben
\L^{eff}&=& \frac{{f^*_\pi}^2}{4} \Tr (\del_\mu U
\del^\mu U^\dagger) + \frac{1}{32e^2} \Tr [U^\dagger\del_\mu U, U^\dagger
\del_\nu U]^2+\cdots \nonumber\\
&+& c\left(\frac{f^*_\pi}{f_\pi}\right)^3 \Tr (\M U+h.c.) + \cdots \nonumber\\
&+& \frac{1}{2} \del_\mu \chi \del^\mu \chi -\frac{1}{2} {m_\chi^*}^2
\chi^2+ \cdots, \label{leff1}
\een
with
\ben
{m_\chi^*}^2={m_\chi}^2 \left(\frac{\chi_*}{\chi_0}\right)^2
\approx m_\chi^2 \left(\frac{f_\pi^*}{f_\pi}\right)^2.
\een
The last ellipsis in (\ref{leff1}) stands for all possible $\chi$ field
couplings to other fields that enter into the Lagrangian. Clearly in the
medium-free
vacuum where $\chi_*=\chi_0$, we recover the familiar chiral Lagrangian
with an additional $\chi$-$U$ coupling.
\subsubsection{\it Scaled parameters}
\bs
We can now read off from (\ref{leff1}) the consequences of this procedure
for the {\it tree level} masses and coupling constants. This procedure is
equivalent to the familiar
Ginzburg-Landau approximation used in statistical mechanics.
We invoke the result obtained in
Lecture II, namely, that baryons emerge as skyrmions of this Lagrangian
and hence their tree-level masses scale as
\ben
m_B^* \sim f_\pi^*/e \sim f_\pi^* \sqrt{g_A^*}
\een
from which we obtain
\ben
\frac{m_B^*}{m_B} &\approx& \frac{f_\pi^*}{f_\pi}\equiv \Phi (\rho).
\label{baryon}
\een
It follows also that
\be
\frac{g_A^*}{g_A} &\approx& 1. \label{ga}
\ee
That at the tree level $g_A$ does not get renormalized by the density
can perhaps be understood by the fact that in large $N_c$ limit, $g_A$
gets corrections only at $O(1/N_c^2)$. We will see later that the in-medium
renormalization of $g_A$ cannot be understood within the framework of low-order
chiral perturbation theory.
The in-medium Goldberger-Treiman relation which states that
\ben
m_N^* g_A^*=g_{\pi NN}^* f_\pi^*
\een
implies that the $\pi NN$ coupling is also unscaled at tree order,
\ben
\frac{g_{\pi NN}^*}{g_{\pi NN}}\approx 1.
\een
If one assumes that vector mesons emerge as hidden gauge bosons
from the chiral Lagrangian as suggested by Bando et al. \cite{bando}
then the mass formula
\ben
{m_V^*}^2=2 {f_\pi^*}^2 g^2
\een
leads to the scaling \footnote{We will show in a subsection below that
the constant
$g$ undergoes density-dependent loop corrections and becomes $g^*$. In
mean-field approximation, it remains unscaled by density.}
\ben
\frac{m_V^*}{m_V}\approx \Phi (\rho). \label{mvector}
\een
To leading order in $N_c$, this scaling applies to the $\rho$, $\omega$,
$A_1$ and other vectors. We do not have explicitly in our effective Lagrangian
the quarkish scalar meson $\sigma$ that figures in nuclear physics:
It is actually an interpolating field for
a two-pion correlation. We expect that it will couple with the scalar glueball
and scale as
\ben
\frac{m_\sigma^*}{m_\sigma}\approx \frac{m_\chi^*}{m_\chi}\approx \Phi.
\label{msigma}
\een

So far we have a ``universal scaling" characterized by one parameter, $\Phi$.
We expect this scaling factor to be calculable from the fundamental theory,
QCD. In our application, we will however take it as a parameter.
It should be stressed that the scaling established so far
applies only to those hadrons whose masses are generated by
spontaneous symmetry breaking. In the chiral limit,
Goldstone-boson masses are zero
and remain zero at all density protected by chiral invariance.
Therefore the behavior of the pion or kaon masses must reflect directly
on the way the chiral symmetry is {\it explicitly} broken at the electroweak
scale. This implies that how their masses change in density
(or temperature) must be quite intricate.

Within the framework thus far developed, the effective Lagrangian
(\ref{leff1}) predicts at
tree level that the pion mass scales
\ben
\frac{m_\pi^*}{m_\pi}\approx \sqrt{\Phi}. \label{pion0}
\een
But this is probably not the entire story.
The reason is simply that in the case of pions,
there are other terms which are equally important.
For instance, Pauli-blocking effects, appearing at one-loop level may be enough
to compensate the scaling of (\ref{pion0}).
Consequently it will not be safe to take (\ref{pion0}) without further
corrections in applying the formalism to nuclei. In fact, there is no
indication in nuclear processes so far examined that the pion mass changes
as a function of density. While awaiting experimental determination, we will
assume that up to nuclear matter density, $m_\pi^*\approx
m_\pi$.
\subsection{Applications}
\subsubsection{\it Nuclear forces}
\subsubsection*{\it Two-body forces}
\bs
As a first application, we discuss Weinberg's approach to nuclear
forces \cite{wein90,wein92},  in particular
the consequences on n-body nuclear
forces (for n=2,3,4) of the leading order chiral expansion with (\ref{lsu2}).
{}From Eq.(\ref{count1}), one can see that we want $L=0$ (or tree
approximation)
and for nuclear forces,
\ben
d_i+\frac{1}{2}n_i-2\geq 0,
\een
hence we want $d_i+\frac{1}{2}n_i-2= 0$ which can be satisfied by two
classes of vertices:
the vertices with $d_i=1$ and $n_i=2$ corresponding to the pseudovector
$\pi NN$ coupling and the vertices with $d_i=0$ and $n_i=4$. The former
gives rise to the standard one-pion exchange force and the latter, coming
from the
four-nucleon interaction in (\ref{lsu2}), gives rise to a short-ranged
$\delta$-function force. For two-body forces, we have
\ben
V_2 = &-&\left(\frac{g_A^*}{2{f_\pi^*}}\right)^2
(\vec{\tau}_1\cdot\vec{\tau}_2)
(\vec{\sigma}_1\cdot\vec{\nabla}_1)(\vec{\sigma}_2\cdot\vec{\nabla}_2)
Y(r_{12})- (1\leftrightarrow 2) \nonumber\\
&+& 2(C_S+C_T \vec{\sigma}_1\cdot\vec{\sigma}_2)\delta (\vec{r}_{12})
\cdots \label{potential}
\een
where $Y(r)\equiv e^{-m_\pi r}/4\pi r$ and
$C_{S,T}$ are unknown constants.
The $\delta$-function force summarizes effects
of all the massive degrees of freedom consistent with chiral invariance
that have been integrated out from the effective Lagrangian.
It will be shown below that in medium, we have $g_A^*/f_\pi^*\approx
g_A/f_\pi$ and $m_\pi^*\approx m_\pi$ which will be used for applications
in nuclei. An important point to note here is that as long as the vector mesons
are not integrated out, they will contribute to nuclear forces at the same
chiral order as the pion as
one can see in the counting rule (\ref{count1}). This is in
agreement with the observation that the $\rho$ contribution to the
tensor force is as important as the $\pi$ contribution \footnote{This point
was emphasized to me by G.E. Brown.}.
\subsubsection*{\it Many-body forces}
\bs
Next we consider three-body forces. The graph given in Figure III.2 is
a Feynman diagram which is genuinely irreducible and hence is a bona-fide
three-body force as defined above. However it involves the middle vertex
involving $NN\pi\pi$  of the form
$N^\dagger V_0 N \sim N^\dagger\vec{\tau}\cdot \vec{\pi}\times \dot{\vec{\pi}}
N\sim O(\frac{Q^2}{m_B})$. So it does not contribute to the leading
order as it is suppressed to $O(Q^3/m_B^3)$.
Next consider the Feynman diagrams Fig. III.3. In terms of time-ordered graphs,
some graphs are reducible in the sense defined above, so we should separate
them out to obtain the contribution to the potential. A simple argument shows
that when the reducible graphs are separated out in a way consistent with the
chiral counting, {\it nothing} remains to the ordered considered, so the
contribution vanishes provided static approximation in the one-pion exchange
is adopted. This is easy to see as follows. In the Feynman diagram, each pion
propagator of three-momentum $\vec{q}$ and energy $q_0$ is of the form
$(q_0^2-\vec{q}^2-m_\pi^2)^{-1}$. The energy is conserved at the $\pi NN$
vertex, so $q_0$ is just the difference of kinetic energies of the nucleon
$p^2/2m_N -{p^\prime}^2/2m_N$ which is taken to be small compared with the
three-momentum of the pion $|\vec{q}|$. Therefore to the leading order,
$q_0$ should be dropped in evaluating the graphs of Fig.III.3. This means that
as long as the intermediate baryon is a nucleon, it is included in the
iteration of the static two-body potential. Bona-fide three-body potentials
arise only to higher orders, say, $O(Q^3/m_N^3)$ \footnote{This simple
argument is based on S. Weinberg \cite{wein92}.
For higher-order consideration, see C. Ord\'{o}\~{n}ez and U. van Kolck
\cite{vankolck}.}. What this means is that if one does the static
approximation as is customarily done in nuclear physics calculations, there
are no many-body forces to the leading chiral order. {\it This result justifies
the standard approximation in nuclear physics from the point of view of chiral
symmetry and, in an indirect sense, of QCD.}
%

We should remark at this point that if we were to include baryon resonances
(in particular the $\Delta$'s) in the chiral Lagrangian as we  will do later,
there could be many-body contributions at the level of Fig. III.3. The graph
Fig. III.4
with a $\Delta$ in the intermediate state is a genuine three-body force
in the space of nucleons and such graphs have been considered in the literature
\cite{deltathree}.
\subsubsection*{\it Problems}
\bs
The problem with nuclear forces
\`{a} la  chiral symmetry is the number of unknown constants that appear
if one goes beyond the tree order we have focused on. To one loop and beyond,
many counter terms are needed to renormalize the theory and to make a
consistent
chiral expansion. The contact four-fermion interactions will receive further
contributions from the loop graphs and its number will increase as derivatives
and chiral symmetry breaking mass terms are included. The contact terms are
$\delta$ functions in coordinate space and describe high-energy degrees of
freedom that are ``integrated out" to describe low-energy sectors.
To handle the zero-range interactions, we need to understand short-range
correlations which we know from phenomenology play an important role in
nuclear dynamics. The scale involved here is not the small scale $Q$ of order
of pion mass but vastly greater.
In order to have an idea as to whether -- and in what sense -- the
chiral expansion is meaningful, we need to calculate graphs
beyond the tree order. However
all these indicate that chiral perturbation theory beyond tree is a difficult,
if not impossible task and it is not at all clear whether there will be
any predictive power in this scheme. In view of this, the pragmatic point
of view one can take is that chiral symmetry can {\it constrain}
phenomenological approaches in nuclear-force problem, but not really
{\it predict} any new phenomena.

One of the crucial questions in connection with the in-medium Lagrangian
discussed above is: In view of the scaling nucleon mass in medium, at what
point does the heavy-fermion approximation break down? The point is that the
smallness of many-body forces depends crucially on the largeness of the
nucleon mass. Will the dropping mass not invalidate the low-order
chiral expansion and thereby increase the importance of many-body forces in
heavy nuclei? A similar problem arises with the mass of the vector mesons
which scale in dense medium. Considering the vectors as ``light" requires
the implementation of hidden gauge symmetry. This will be discussed later.

We now turn to the case where the situation is markedly
different, where chiral symmetry can have a real predictive power.

\subsubsection{\it Exchange currents}
\bs
When a nucleus is probed by an external field, the transition is described
by an operator that consists of one-body, two-body and many-body ones.
It is usually dominated by a one-body operator and in some situations
two-body operators play an important role. It has been known since some time
that under certain kinematic conditions, certain two-body operators can even
dominate \cite{chemrho,froismathiot,kdr78}. The well-known example is the
threshold
electrodisintegration of the deuteron at large momentum transfers. Another case
is the axial-charge transition in nuclei. In this section, we describe how
one can calculate many-body currents from chiral symmetry point of view.

First we have to introduce external fields into the effective Lagrangian.
Consider the vector and axial-vector fields
\bea
\V_\mu=\frac{\tau^i}{2}\V_\mu^i,\ \ \ \A_\mu=\frac{\tau^i}{2}\A_\mu^i.
\eea
Then the coupling can be introduced by gauging the Lagrangian (\ref{lbaryonp})
and (\ref{deltalb}). This is simply effectuated by the replacement
\bea
D_\mu &\rightarrow& \D_\mu\equiv D_\mu -\frac{i}{2}\xi^\dagger\le(\V_\mu+\A_\mu
\ri)\xi -\frac{i}{2}\xi\le(\V_\mu-\A_\mu\ri)\xi^\dagger,\nonumber\\
A_\mu &\rightarrow& \Delta_\mu\equiv A_\mu +\frac 12 \xi^\dagger\le(\V_\mu
+\A_\mu\ri)\xi -\frac 12\xi\le(\V_\mu -\A_\mu\ri)\xi^\dagger
\eea
The corresponding vector and axial-vector currents denoted respectively
by $J^{i\mu}$ and $J_5^{i\mu}$ can be read off
\bea
{\vec J}_{5\mu} &=&
- f_\pi \left[\dmu\pivec -\frac{2}{3f_\pi^2}\left(\dmu\pivec
\,\pivec^2 - \pivec\,\pivec\cdot\dmu\pivec\right)\right]
\nonumber \\
&&+\ \frac12 \Bbar \left\{2 g_A S^\mu
\left[\tauvec + \frac{1}{2f_\pi^2}\left(\pivec\,\tauvec\cdot\pivec
-\tauvec\,\pivec^2\right)\right]
+ v^\mu \left[\frac{1}{f_\pi}\tauvec\times\pivec - \frac{1}{6f_\pi^3}
\tauvec\times\pivec\,\pivec^2\right] \right\}B + \cdots,\nonumber\\
\\
{\vec J}_\mu &=& \left[ \pivec\times\dmu\pivec -\frac{1}{3f_\pi^2}
\pivec\times\dmu\pivec\,\pivec^2\right]
\nonumber \\
&&+\ \frac12 \Bbar \left\{v^\mu
\left[\tauvec + \frac{1}{2f_\pi^2}\left(\pivec\,\tauvec\cdot\pivec
-\tauvec\,\pivec^2\right)\right]
+2 g_A S^\mu
\left[\frac{1}{f_\pi}\tauvec\times\pivec - \frac{1}{6f_\pi^3}
\tauvec\times\pivec\,\pivec^2\right]\right\} B + \cdots.\nonumber\\
\eea
Here $B$ is the ``heavy" nucleon field defined before.

It is now easy to see what the chiral counting tells us \cite{mr91,pmr93}.
The following is known as ``chiral filter
phenomenon" in nuclei. Consider an amplitude involving
four external (two in-going and two outgoing) nucleon fields interacting
with a slowly-varying electroweak current. It can be characterized
just as for nuclear forces with the
exponent $\nu$, (\ref{count1}), with one difference: the vertices involving
electroweak currents satisfy instead
\ben
d_i+\frac{1}{2}n_i-2\geq -1.
\een
Thus the leading order graphs are given by $L=0$ with a single vertex probed by
an electroweak current for which $d_i+\frac{1}{2}n_i-2 = -1$ and $\pi NN$
vertices satisfying $d_i+\frac{1}{2}n_i-2 =0$. One can easily see that such
graphs are given only by one-pion exchange terms in which the current is
coupled to a $\pi NN$ vertex or hooks onto a pion line and that to this order,
{\it there is no contribution from the four-nucleon contact interaction
or from vector-meson exchanges}.
Effectively then, there are only two soft-pion graphs for electromagnetic
field ({\ie}, the ``pair" and ``pionic" graphs in the jargon used in
the literature \cite{chemrho})
and only one graph (``pair") for weak currents. These are given in Fig.III.5.
We should emphasize that
there are no heavy-meson or multi-pion exchanges at this order, a situation
that renders
the calculation for currents markedly simpler than for forces.
What this means is that when massive degrees of freedom are explicitly
included, they can contribute only at $O(Q^2)$ relative to the tree order.
Put more precisely, if one includes one soft-pion ($A_\pi$),
two or more pion ($A_{\pi \pi}$) and vector meson ($A_V$) exchanges,
excited baryons ($A_{N^*}$) and appropriate form factors ($A_{FF}$)
as one customarily does in
exchange current calculations, consistency with chiral
symmetry requires that even though each contribution could be large, the sum
must satisfy\cite{mr91}
\ben
A_\pi+A_{\pi \pi}+A_V+A_{N^*}+A_{FF}\approx A_\pi (1+O(Q^2)).
\label{chiralfilter}
\een
Later we will give a simple justification of this result by one-loop chiral
perturbation theory and give a numerical value for the $O(Q^2)$ term.
The surprising thing is that the relation (\ref{chiralfilter}) holds
remarkably well for  the famous threshold
electrodisintegration of the deuteron as well as in the magnetic form factors
of $^3$He and $^3$H to large momentum transfers \cite{froismathiot}.
In the case of nuclear axial-charge
transitions mediated by the time component of the isovector axial current
which will be discussed below, one can construct a phenomenological model
that fits nucleon-nucleon interactions which satisfies this condition well
with the $O(Q^2)$ less than 10\% of the dominant soft-pion contribution.


\subsubsection{\it Axial-charge transitions in nuclei}
\bs
We now illustrate how our formalism works by applying it to
what is known as axial-charge transition in nuclei. It involves the time
component of the isovector axial current $\vec{J}_5^{\mu\pm}$ inducing the
transition $0^+\leftrightarrow 0^-$ with $\Delta I=1$. This transition
in heavy nuclei is quite sensitive to the soft-pion exchange as well
as to the scaling effect\cite{kr91}
and illustrates nicely the power of chiral symmetry in
understanding subtle nuclear phenomena.
\subsubsection*{\it Tree order}
\bs
The axial-charge transition is described by a one-body (``impulse")
axial-charge operator plus a two-body
(``exchange") axial-charge operator described
above. Three-body and higher many-body operators do not contribute
at small energy-momentum transfer we will focus on. In a work done some time
ago\cite{kdr78}, these operators were calculated without the scaling
effect. Now from the foregoing discussion, it is obvious that we will get
the operators of exactly the same form except that the masses $m_i$'s are
replaced by $m_i^*$'s and the coupling constants $g_i$'s are replaced by
$g_i^*$'s. This means that in the results obtained previously, we are to make
the following replacements
\ben
m_N \rightarrow m_N \Phi (\rho), \ \ \ \
f_\pi \rightarrow f_\pi \Phi, \ \ \ \ m_\pi\rightarrow m_\pi\sqrt{\Phi}
\een
with $\Phi$ a density-dependent constant to be determined empirically.
(This quantity is probably calculable from a microscopic Hamiltonian,
but for the moment we should simply take it as a parameter.)
At the mean-field level, the coupling constants $g_{\pi NN}$ and $g_A$
are not affected by density as shown above (see Eq.(\ref{ga}).
We shall come back to loop effects later.
Evaluating the diagrams of Fig. III.4, we have the axial-charge density
one-body and two-body operators
\ben
J_5^{0\pm} &=& J_5^{(1)\pm} + J_5^{(2)\pm},\label{axialch}\\
J_5^{(1)\pm} &=& -g_A\sum_i \tau_i^{\pm} (\vec{\sigma}_i\cdot
\vec{p}_i /m_N^\star) \delta (\vec{x}-\vec{x}_i),\\
\label{singlep}
J_5^{(2)\pm} &=& \frac{{m_\pi^\star}^2}{8\pi f_\pi^\star}
\left(\frac{g_A^\star}{f_\pi^\star}\right)
\sum_{i<j} (\tau_i\times \tau_j)^\pm [\vec{\sigma}_i\cdot \hat{r}
\delta (\vec{x}-\vec{x}_j) + \vec{\sigma}_j\cdot\hat{r}\delta (\vec{x}
-\vec{x}_i)] Z(r_{ij}) \label{softpion}
\een
with
\ben
Z(r)=\le(1+\frac{1}{{m_\pi^\star} r}\ri) e^{-{m_\pi^\star} r}/{m_\pi^\star} r
\nonumber
\een
where $r_{ij}=\vec{x}_i-\vec{x}_j$ and $\vec{p}$ the initial momentum of
the nucleon making the transition. We have set the current momentum equal
to zero. Note that the star is put on all the quantities that scale within
the framework given above. However in applying the result to actual
transitions in nuclei, there is a subtlety due to the fact that the
Gamow-Teller coupling constant $g_A$ scales in medium due to  short-range
interactions between baryons as described later. As a consequence,
{\it effectively}, at least up to nuclear matter density, the constant
$g_A^\star/f_\pi^\star$ associated with a pion exchange remains unscaled.
Furthermore consideration based on QCD indicates that the properties
of Goldstone bosons are not affected by the changes in the quark condensates.
Thus we should take in Eq. (\ref{softpion}) $m_\pi^\star\approx m_\pi$ and
$\frac{g_A^\star}{f_\pi^\star}\approx \frac{g_A}{f_\pi}$. The $g_A$ that
figures
in the single-particle axial charge operator in (\ref{singlep}) remains
unscaled since it is {\it not} associated with a Gamow-Teller operator.

Let us denote the one-body matrix
element calculated with the scaling by ${\cal M}_1^*$ and the corresponding
two-body exchange current matrix element by ${\cal M}_2^*$
\ben
{\cal M}^*_a\equiv \langle f|J_5^{(a)\pm}|i\rangle, \ \ \ \ a=1,2.\label{me}
\een
Let ${\cal M}_a$
denote the quantities calculated with the operators in which no scaling
is taken into account. For comparison with experiments, it is a common practice
to consider the ratio
\ben
\epsilon_{MEC}\equiv \frac{{\cal M}^{observed}}{{\cal M}_1}.
\een
This is a ratio of an experimental quantity over a theoretical quantity,
describing  the deviation of the impulse approximation from the observation.
The prediction that the present theory makes is remarkably simple. It is
immediate from (\ref{axialch}) and the argument given above that\footnote{
The result in \cite{kr91} differs from this in that in \cite{kr91} the
non-scaling $g_A^*/f_\pi^*\approx g_A/f_\pi$ was not taken into account.
As a consequence, a factor $\Phi^{-1}$ was multiplied into $R$. This I
think is not correct.}
\ben
\epsilon_{MEC}^{th}=
\frac{{\cal M}_1^*+{\cal M}_2^*}{{\cal M}_1}
=\Phi^{-1} (1+R)
\een
where $R={\cal M}_2/{\cal M}_1$.
\subsubsection*{\it One-loop corrections}
\bs
There are two ways of computing the $O(Q^2)$ correction in
(\ref{chiralfilter}). One is to introduce vector mesons and saturate
the correction by tree graphs and the other is to compute one loop graphs
with pions. Clearly the former is easier. However tree graphs with vector
mesons are not complete as we know from $O(\del^4)$ in $\pi\pi$ scattering.
There are nonanalytic terms coming from pion loops in addition to the
counter terms saturated by the vectors. In this subsection, we shall compute
the $O(Q^2)$ correction to $R$ using pions and nucleons only. The $\Delta$
resonance does not contribute and the effect of the vector mesons $\rho$ and
$a_1$ are effectively included in the treatment through the loop effect.
These are discussed in the work of Park {\it et al} \cite{pmr93}.

Let the corrected ratio be
\ben
R\rightarrow (1+\Delta)R
\een
where $\Delta$ is given by the loops according to the counting rule given
in (\ref{count1}).
$\Delta$ can come from two sources. One is the loop correction to
the soft-pion term of Fig. III.5, essentially given by one-loop corrections
to the vertex $A_0 NN\pi$ since there is no one-loop correction in HFF to the
$\pi NN$ vertex other than renormalizing the $\pi NN$ coupling constant
$g_{\pi NN}$. Now the one-loop correction to the $A_0 NN\pi$ is nothing
but the one-loop correction to the isovector Dirac form factor $F_1^V$, so
it is completely determined by experiments. We will call the
corresponding correction $\Delta_{1\pi}$. The other class of diagrams
contributing to $\Delta_{2\pi}$ is two-pion exchange graphs given in
Figure III.6. There are many other graphs one can draw but they do not
contribute to the order considered when the HFF is used. In particular
the four-fermion contact interactions that appear in nuclear forces
do not play any role in the currents. In evaluating these graphs, Fig. III.6,
counter terms (there are effectively two of them)
involving four-fermion field   with derivatives are needed
to remove divergences. They would also contribute to nuclear forces
at the same chiral order. Now the finite counter terms that appear are the
so-called ``irrelevant terms" in effective theory (as they are down
by the inverse power of the chiral scale parameter) and can only be probed at
very short distance. Indeed the currents that come from such terms are $\delta$
functions and derivatives of the $\delta$ functions, so they should be
suppressed if
the wave functions have short-range correlations. Therefore the counter terms
that appear here can be ignored. This can be also understood by the fact that
such terms do not figure in phenomenological potentials; this also means that
one cannot extract them from nucleon-nucleon scattering data. They could
perhaps be seen if one measures small effects at very short distances.
Thus {\em embedded in nuclear medium, the unknown counter terms do not
play any role. This is the part of ``chiral filtering" by nuclear matter of
short-wavelength degrees of freedom}. A corollary to this ``theorem" is that
if the dominant soft-pion contribution is suppressed by kinematics or by
selection rules, then the next order terms including $1/m_N$ corrections
must enter. This is the other side of the same coin as predicted a long
ago \cite{kdr78}. For instance, s-wave pion production in
\ben
pp \rightarrow pp\pi^0
\een
is closely related to the matrix element of the axial charge operator.
But because of an isospin selection rule, the soft-pion exchange current
cannot contribute. Since the single-particle matrix element is kinematically
suppressed,  the contribution from higher chiral
corrections will be substantial. This turns out to be the case
\cite{leeriska}. The case with the space component of the axial current,
namely, the coupling constant $g_A$ in nuclei, is discussed below.

Apart from these counter terms which are rendered ineffective inside
nuclei, there are no unknown parameters once the
masses of the nucleon and the pion and constants such as $g_A$, $f_\pi$ are
taken from free space. We quote the results, omitting details.
The quantity we want is the ratio of nuclear matrix elements (denoted by
$<$ $>$)
\ben
\Delta_{1\pi} &=& \langle \delta J_5^{(2)} (1\pi)\rangle /\langle
J_5^{(2)}(soft)\rangle,
\\
\Delta_{2\pi} &=& \langle \delta J_5^{(2)} (2\pi)\rangle /\langle J_5^{(2)}
(soft)\rangle
\een
where $J_5^{(2)} (soft)$ is given by the soft-pion term (\ref{softpion})
\ben
J_5^{(2)}(soft) = {\cal T}^{(1)} \frac{d}{dr}\left[
- \frac{1}{4 \pi r} e^{- m_\pi r}\right],
\label{msoft}
\een
with the matrix element to be evaluated in the sense of (\ref{me}) and
\ben
\delta J_5^{(2)} (1\pi) &=& \frac{m_\pi^2}{6}\langle r^2\rangle_1^V
\frac{m_\pi^2}{f_\pi^2}
\,J_5^{(2)}(soft)\nonumber \\\
&+&\frac{{\cal T}^{(1)}}{16\pi^2 f_\pi^2}
\frac{d}{dr} \left\{
 - \frac{1+3 g_A^2}{2}\left[K_0(r)- {\tilde K_0}(r)\right]
 + (2+4g_A^2) \left[K_2(r) - {\tilde K_2}(r) \right]\right\},\nonumber\\
\label{onepiloop}\\
\delta J_5^{(2)} (2\pi) &=& \frac{1}{16\pi^2 f_\pi^2} \frac{d}{dr}\left\{
-\left[\frac{3 g_A^2-2}{4} K_0(r) +\frac12 g_A^2 K_1(r)\right] {\cal T}^{(1)}
+ 2\gB^2 K_0(r){\cal T}^{(2)}\right\},\nonumber\\
\label{twopiloop}
\een
where $n\pi$ in the argument of the $\delta J_5$'s stands for $n\pi$ exchange
in the one-loop graphs and
$\langle r^2\rangle_1^V$ is the isovector charge radius of the nucleon.
We have used the notations
\ben
{\cal T}^{(1)} &\equiv& \tauvec_1 \times \tauvec_2 \, {\hat r}\cdot
({\vec \sigma_1} + {\vec \sigma_2}),\\
{\cal T}^{(2)} &\equiv&  (\tauvec_1 + \tauvec_2) \, {\hat r}\cdot
{\vec \sigma_1} \times {\vec \sigma_2}
\een
and the functions $K_i$ and $\tilde{K}_i$ defined by
\bea
K_0(r) &=& - \frac{1}{4\pi r}\, \int_0^1 dx \,
\frac{2 x^2}{1-x^2}\, E^2 e^{-  E r},
\nonumber \\
K_2(r) &=& - \frac{1}{4\pi r}\, \int_0^1 dx \,
\frac{2 x^2}{1-x^2}\, \left(\frac{1}{4}-\frac{x^2}{12}\right)\,
E^2 e^{-  E r}
\eea
and
\bea
{\tilde K}_0(r) &=& \frac{1}{4\pi r}\, \int_0^1 dx \,
\frac{2 x^2}{1-x^2}\, \frac{m_\pi^2}{m_\pi^2- E^2}\,\left(
m_\pi^2 e^{- m_\pi r} - E^2 e^{-  E r}\right),
\nonumber \\
{\tilde K}_2(r) &=& \frac{1}{4\pi r}\, \int_0^1 dx \,
\frac{2 x^2}{1-x^2}\, \left(\frac14 -\frac{x^2}{12}\right)\,
\frac{m_\pi^2}{m_\pi^2- E^2}\,\left( m_\pi^2 e^{- m_\pi r} -
E^2 e^{-  E r}\right).
\eea
Here $E(x)=2m_\pi/\sqrt{1-x^2}$. These are Fourier-transforms of McDonald
functions and their derivatives with $\delta$ function terms excised
on account of short-range correlations.

%

For a first estimate, we shall calculate the relevant matrix elements
in fermi-gas model with a short-range correlation function in the
simple form (in coordinate space)
\be
g(r)=\theta (r-d),\ \ \ r=|\vec{r}_1 -\vec{r}_2|.
\ee
Figure III.7 summarizes the result for various short-range cut-off $d$.
The relevant $d$ commonly used in the literature is 0.7 fm. It is difficult
to pin down the cut-off accurately but this is roughly what realistic
calculations give.

Numerically although $\Delta_{1\pi}$ and $\Delta_{2\pi}$ can be non-negligible
individually, there is a large cancellation, so the total is small,
$\Delta < 0.1$. This is the chiral filter mechanism mentioned before.
This cancellation mechanism which is also operative in nuclear electromagnetic
processes mediated by magnetic operators as in the electrodisintegration
of the deuteron and the magnetic form factors of the tri-nucleon systems
is characteristic of the strong constraints coming from chiral symmetry,
analogous to the pair suppression in $\pi$N scattering.
\subsubsection*{\it Absence of multi-body currents}
\bs
For small energy transfer (say, much less than $m_\pi$), three-body and
other multi-body forces do not contribute to $O(Q^2)$ relative to the
soft-pion term. This can be seen in the following way.

Consider attaching an axial-charge operator to the vertices in the
graphs for three-body forces, Figure III.8. We need not consider those
graphs where the operator acts on any of the nucleon line. Now the
argument that the sum of all such irreducible graphs cannot contribute
to the order considered as long as the static pion propagator is used
for two-body soft-pion exchange currents parallels exactly the argument
used for the forces.

The situation is quite different, however, if the energy transfer is larger
than
say the pion mass as in the case where a $\Delta$ is excited by an external
electroweak probe. In such cases, there is no reason why three-body
and higher-body currents should be small compared with one- and two-body
currents. This phenomenon should be measurable by high intensity electron
machines at multi-GeV region.
\subsubsection*{\it Comparison with experiments}
\bs
We shall estimate the ratio $R$ in fermi-gas model.
This will give us a global estimate.
It turns out that the ratio $R$ depends little on nuclear model and on
mass number. In fermi-gas model, it is given by
\ben
R=\frac{g_{\pi NN}^2}{4\pi^2 g_A^2}\frac{p_F}{m_N}\left[1+\lambda-\lambda
(1+\frac{\lambda}{2})\ln (1+\frac{2}{\lambda})\right]
\een
with $\lambda\equiv \frac{m_\pi^2}{2p_F^2}$, where $p_F$ is the Fermi
momentum (which is $(\frac{3\pi^2}{2}\rho)^{1/3}$ for nuclear matter).
Numerically this ranges from $R\approx 0.44$ for $\rho=\frac{1}{2}
\rho_0$ to $R\approx 0.62$ for $\rho=\rho_0$.
Now in the same model with a
reasonable short-range correlation, $\Delta$ comes out to be about
0.1, pretty much independently of nuclear density\cite{pmr93},
so that we can take
\ben
R(1+\Delta)\approx 1.1 R. \label{R}
\een

The only undetermined parameter in this theory is the scaling factor
$\Phi$. Ideally one should determine it from experiments. However
it is a mean-field quantity which makes sense only for large
nuclei and has not been pinned down yet. Once it is fixed in heavy nuclei,
one could then use local-density approximation to describe transitions in
light nuclei. At the moment it is known only from  QCD sum-rule calculations
of hadrons in medium, {\eg}, the vector-meson mass
\cite{hatsuda}. To have a qualitative idea, we shall take
\ben
\Phi \approx 1-0.18\frac{\rho}{\rho_0}. \label{Phi}
\een
With eqs. (\ref{R}) and (\ref{Phi}), we predict
\ben
\epsilon_{MEC} &\approx&  1.63, \, 2.05\, \ \ {\rm for} \
\rho=\rho_0/2,\,\rho_0.
\een
In view of the crudeness in
(\ref{R}) we have an uncertainty of $\pm 0.2$ in these values.
This agrees with the result (experimentally $\approx 1.6$) in light nuclei
as well as with the results (experimentally $\sim 2$)
in heavy nuclei\cite{warburton}.
The empirical value in the Pb region is
\ben
\epsilon_{MEC}^{exp}=2.01\pm0.05.
\een
For a meaningful test of the theory, a finite-nucleus calculation will
be required.

\subsubsection{\it $g_A^*$ in Nuclei}
\bs As already discussed,
at the tree level of the effective Lagrangian, the axial-vector coupling
constant $g_A$ is not renormalized in medium. Here we illustrate how
short-distance {\it in-medium} effects which cannot be included by low-order
chiral perturbation theory can induce an important density
dependence for the Gamow-Teller vertex.
Such a correction is actually required by the observation that
the value of $g_A$ measured in the neutron beta
decay \cite{gA}
\ben
g_A=1.262\pm 0.005
\een
is quenched to about 1 in medium and heavy nuclei\cite{denys}.
This explains why we must take $g_A^\star/f_\pi^\star\approx g_A/f_\pi$ for the
$\pi NN$ vertex contributing to the one-pion exchange operators, at least
up to nuclear matter density.  The important point here is that
this quenching is not to be attributed to an
influence of a modified vacuum, in other words, a ``fundamental quenching"
as often thought in the past. This is obviated by the fact that
in large $N_c$ QCD, it is the $O(1/N_c^2)$ correction that is affected by
nuclear medium. Of course at an asymptotic density,
$g_A$ should reach unity as a consequence of the melting of the quark
condensate.

In nuclei, $g_A^*$ is measured in Gamow-Teller transitions, namely the
space component of the axial current. For this component, the exchange
current given by the soft-pion theorem discussed above is suppressed. Therefore
according the corollary to the ``chiral filter" argument,
corrections to $g_A^*$ must come from loop terms and/or higher $Q$ Lagrangian
pieces or many-body currents involving short-distance physics~\footnote{
For the Gamow-Teller operator, a
three-body current where one of the nucleons contains $A_\mu\pi\pi NN$
vertex has no suppression factor as the axial-charge operator does, so a
tree graph of this type may be non-negligible. This graph has not been
evaluated so far, so its importance is unknown at present.}.
It is known that the dominant correction comes from the excitation
of $\Delta$-hole states as given in Fig.III.9.
The introduction of the $\Delta$ requires extending the effective
Lagrangian to decuplet (spin-3/2) baryons in the case of
three flavors. This can be done in the following way\cite{jenkins91}.

Introduce a Rarita-Schwinger field $\Delta^\mu$ obeying the constraint
$\gamma_\mu \Delta^\mu=0$. This field carries $SU(3)$ tensor indices
$abc$ that are completely symmetrized to form the decuplet.
The chirally symmetric Lagrangian involving the decuplet with its coupling
to octet baryon and mesons equivalent to (\ref{lbaryon}) is
\ben
\L^{\Delta}=i\bar{\Delta^\mu}\gamma_\nu {\cal D}^\nu \Delta_\mu -
m_{\Delta}\bar{\Delta}^\mu \Delta_\mu +\alpha \left(\bar{\Delta}^\mu
A_\mu B +h.c. \right)+\beta\bar{\Delta}^\mu \gamma_\nu
\gamma_5 A^\nu \Delta_\mu \label{decuplet},
\een
where $\alpha$ and $\beta$ are unknown constants,
$B$ stands for the spin-1/2 octet baryons, $\Delta^\mu$ the spin-3/2
decuplets
$\Delta$, $\Sigma^*$, $\Xi^*$ and $\Omega$ and $A^\mu$ is the ``induced"
axial vector $A^\mu=\frac{i}{2}(\xi \del^\mu \xi^\dagger -\xi^\dagger\del^\mu
\xi)$. Sum over all $SU(3)$ indices (not explicitly written down here)
is understood. The covariant derivative
acting on the Rarita-Schwinger field is defined by
\ben
{\cal D}^\nu \Delta^\mu_{abc}=\del^\nu \Delta^\mu_{abc} + (V^\nu)^d_a
\Delta^\mu_{dbc}+(V^\mu)^d_b \Delta^\mu_{adc} +(V^\mu)^d_c \Delta^\mu_{abd}
\een
with $V^\mu=\frac{1}{2}(\xi \del^\mu \xi^\dagger +\xi^\dagger\del^\mu \xi)$.
Under chiral $SU_L(3)\times SU_R(3)$ transformation, the Rarita-Schwinger
field transforms as
\ben
\Delta^\mu_{abc}\rightarrow g^d_a g^e_b g^f_c \Delta^\mu_{def}.
\een
As before, we have terms quadratic in baryon fields which we may write
\ben
\L^{\Delta}_4 =\gamma \left(\bar{\Delta}^\mu {\cal C}_{\mu\nu} B\bar{B}
{\cal F}^\nu B
+h.c.\right)+ (\Delta\Delta BB)+ (\Delta B\Delta B)+ (\Delta\Delta\Delta
\Delta)+\cdots.
\een
where ${\cal C}$ and ${\cal F}$ are appropriate tensors allowed by
Lorentz and isospin invariance. We have indicated explicitly only the
$\Delta BBB$ components as they play the essential role for $g_A^*$.

As with the nucleon, we can use the heavy-baryon formalism for the
$\Delta$. Evaluating the diagram in Fig.III.9 requires information on
the axial-vector coupling constant $g_A^\Delta$ connecting
to $\Delta$ and $N$ and the contact interaction $g_{\Delta}^\prime$
for $\Delta N\leftrightarrow NN$ in the {\it particle-hole} channel.
The latter is analogous to the Landau-Migdal particle-hole interaction
$g_0^\prime \vec{\tau}_1\cdot \vec{\tau}_2 \vec{\sigma}_1 \cdot \vec{\sigma}_2
\delta (r_{12})$ governing spin-isospin interactions in nuclei and can be
written in terms of the transition spin operator $\vec{S}$ and transition
isospin operator $\vec{T}$ as\cite{brownweise}
\ben
g_{\Delta}^\prime \vec{T}_1 \cdot \vec{\tau}_2 \vec{S}_1 \cdot \vec{\sigma}_2
\delta (r_{12}).
\een
Neither ${g_A^\Delta}$ nor $g_{\Delta}^\prime$ is known sufficiently well.
Assuming that the axial-vector current couples to the $\Delta$ and $N$ in the
same way as the pion does, one usually takes
\ben
g_A^\Delta/g_A\approx 2.
\een
We can then roughly estimate in small-density approximation that
\ben
g_A^*/g_A=1-a g_{\Delta}^\prime \rho (m_{\Delta}-m_N)^{-1} +\cdots
\een
with $a\approx (\frac{2}{3}{m_\pi}^{-1})^2$ (where the factor $(2/3)$ comes
from the matrix element of the transition spin and isospin). It turns out that
the acceptable empirical value (coming from pion-nucleus scattering)
for $g_\Delta^\prime$ is about $1/3$.
Taking the $\Delta N$ mass difference $\sim 2.1 m_\pi$, we have
\ben
g_A^*/g_A\approx 0.8 \ \ \ {\rm for} \ \ \rho=\rho_0.
\een
This makes $g_A^*$ approximately 1 in nuclear matter.

There are many other diagrams contributing at the same order but none of them
is as important as Fig.V.9 except possibly for core polarizations involving
particle-hole excitations through tensor interactions \footnote{The strength of
tensor interactions is believed to be considerably weaker in dense nuclear
matter than what has been accepted, as will be mentioned shortly.
If this is true, then the mechanism due to core
polarizations may be much less important in heavy nuclei where density
is higher than the $\Delta$-hole mechanism
described here. This is a controversial issue that has been around for a
long time.}.
What one observes in nuclei is of course the sum of all.
The point of the above exercise is that the $g_A$ is quenched in nuclei due
to higher order chiral effects of short-range character
and not due to soft pions.

\subsection{Nuclei in the ``Swelled World"}
\bs
The case of axial-charge transitions (and the isovector magnetic dipole
process which we did not discuss here) corresponds to small energy and
momentum transfers $q_\mu\approx 0$. The effective Lagrangian (\ref{leff1})
with the parameters scaled as
\ben
\frac{m_B^*}{m_B}\approx \frac{m_V^*}{m_V}\approx \frac{m_\sigma^*}{m_\sigma}
\approx \frac{f_\pi^*}{f_\pi}\equiv \Phi (\rho)\label{brscaling}
\een
can be applied to other nuclear processes of similar nature.
For this relation to be useful, it is necessary that the process studied
should be dominated by tree graphs, with next chiral order corrections
suppressed as in the axial-charge example seen above. In phenomenological
approach in nuclear physics, this is usually the case. Let us therefore
consider cases where the scaling (\ref{brscaling}) makes a marked improvement
in fitting experimental data. Here we discuss
a few of such examples \cite{adami}.
\subsubsection{\it Tensor forces in nuclei}
\bs
If the $\rho$ field is not integrated out assuming that its mass is very high,
then at tree level, there can be two terms contributing to tensor forces in
nuclei. The first is the tensor force coming from one-pion exchange.
As noted before, there is little scaling in the one-pion exchange
potential, so this part will be left untouched by medium. The second
is a contribution from one-$\rho$ exchange and this part will be
scaled. As we saw before, the vector mesons contribute to forces at
the same chiral order as the pion. Effectively the tensor force has the form
\be
V_T (\vec{q},\omega)=&-&\frac{f^2}{m_\pi^2}\left[
\frac{\vec{\sigma}_1\cdot\vec{q}\vec{\sigma}_2\cdot\vec{q}-\frac 13
(\vec{\sigma}_1\cdot\vec{\sigma}_2)q^2}{(q^2+m_\pi^2)-\omega^2}\right]
{\bf \tau}_1\cdot{\bf \tau}_2\nonumber\\
&+&\frac{{f^*_\rho}^2}{{m^*_\rho}^2} \left[
\frac{\vec{\sigma}_1\cdot\vec{q}\vec{\sigma}_2\cdot\vec{q}-\frac 13
(\vec{\sigma}_1\cdot\vec{\sigma}_2)q^2}{(q^2+{m^*_\rho}^2)-\omega^2}\right]
{\bf \tau}_1\cdot{\bf \tau}_2\label{tensor}
\ee
where we have defined $\frac{f^2}{m_\pi^2}=\left(\frac{g_A}{2f_\pi}\right)^2$.
It is assumed in (\ref{tensor}) that
the pion-exchange term is unscaled up to
nuclear matter density as suggested by nature. Using
\be
\frac{{f^*_\rho}^2}{{m^*_\rho}^2}\approx \frac{{f_\rho}^2}{{m_\rho}^2}
\Phi (\rho)^{-2}
\ee
and the empirical observation
\be
\frac{{f_\rho}^2}{{m_\rho}^2}\approx 2 \frac{f^2}{{m_\pi}^2}
\ee
we see that the tensor force would vanish when
\be
\frac 12\Phi^2 \frac{\langle (q^2+{m^*_\rho}^2)\rangle}{\langle (q^2 +m_\pi^2)
\rangle} \approx 1
\ee
where we have set $\omega\approx 0$ for static force. This happens
when $\langle q^2\rangle\approx 7.6 m_\pi^2$
for $\Phi\approx 0.8$ corresponding to nuclear matter density.
This is within the range of momentum a meson carries between nucleons in
nuclear matter. Such effects of a medium-quenched tensor force are observed in
$(e,e^\prime)$ and $(p,p^\prime)$ processes on heavy nuclei
at several hundreds of MeV: The value of $\Phi$ required is
$m^*/m=$ 0.79 -- 0.86 which is consistent with the value needed to explain
the enhanced axial charge transitions seen in Pb nuclei (see Eq.
(\ref{Phi})). This again indicates that the $\rho$ should be considered on
the same footing as the $\pi$ when nucleons are present. This is consistent
with
the picture of the $\rho$ as a light gauge field.

\subsubsection{\it Other processes}
\bs
Other evidences \cite{br89} are:
\begin{itemize}
\item Electroweak probes. In addition to the weak current discussed above,
electromagnetic probes also provide an indirect evidence of the scaling
effect. Specifically, the scaling of the vector meson and nucleon
masses can be used to explain the longitudinal-to-transverse response function
ratios in $(e, e^\prime p)$ processes with such nuclei as
$^3$He, $^4$He and $^{40}$Ca. Experimentally there is quenching in the
longitudinal response function that seems to increase with nuclear density
while the transverse response remains unaffected. This feature can be
economically (but perhaps not uniquely)
explained by the scaling vector meson masses.
\item Hadronic probes. Some long-standing problems in hadron-nucleus
scattering ({\ie}, the kaon-nucleus and
nucleon-nucleus optical potentials) are resolved by a quenched vector-meson
mass. A stiffening of nucleon-nucleus spin-isospin response
is predicted and is observed. These effects are manifested through the forces
mediating the processes and the change in the range of such forces due to
the scaling masses.
\end{itemize}
\section{Hadronic Phase Transitions}
\bs
So far we have been discussing response functions and two-body forces
in terms of chiral Lagrangians. In this section we will turn to the bulk
property of nuclear matter, particularly its structure under extreme
conditions of high density. The question which remains
largely unanswered up to date is whether or not effective Lagrangians
can address the bulk property of hadronic matter. Our discussion given here
is therefore very much tentative and could be even incorrect. Nonetheless
the problem is important for further development in nuclear physics and
our purpose here is more to indicate the issues involved than
to provide any answers.
\subsection{Nuclear Matter}
\bs
Before discussing matter with density $\rho >\rho_0$ (where $\rho_0\approx
\frac 12 m_\pi^3$ is the ordinary nuclear matter density), we discuss briefly
how the chiral Lagrangians that we shall use fare in describing ordinary
nuclear matter.

It has been a disturbing point for chiral symmetry in nuclear physics that
a Lagrangian with chiral symmetry treated at low chiral orders fails to
describe
correctly both nuclear matter and nuclei. Instead the
Walecka model \cite{walecka} with the scalar $\sigma$ and
the vector $\omega$ coupled to
nucleons when treated at mean field is found to work fairly well
both for nuclear matter and nuclei. This model however has no chiral
symmetry. What is more disturbing is that since the pion has no mean field,
the only solidly established low-energy excitations of QCD (as seen in
preceding chapters),  Goldstone bosons, seem to play no role whatsoever
in the ground-state
property and low-energy excitations of nuclear systems. This has led many
people to believe that chiral Lagrangians are useless for nuclear physics.
This is in glaring conflict with the power of low-energy theorems and
effective theories seen in some nuclear {\it response functions}.

This issue is being revived and will presumably be resolved in the
future. The recent attempt (at the time of this writing) is to include
all the known symmetries of QCD, not just chiral symmetry which has
been mainly investigated, in constructing effective Lagrangians. One
of the additional ingredients so far considered
is the broken conformal invariance
associated with the trace anomaly discussed above. Given such a
nonlinear Lagrangian, one could  obtain nuclear matter and finite nuclei
nuclear matter as  a ``chiral liquid" and " chiral liquid drop" as suggested
by Lynn \cite{lynn}. In this description, nuclei are nontopological solitons .
Since the parameters appearing in such Lagrangians
may be fixed by elementary processes involving mesons and baryons that satisfy
low-energy theorems, this
would be a natural way to marry chiral symmetry (and trace anomaly) with
nuclear charateristics. This is an exciting new development in the
field and deserves attention. Indeed, implementing
vector mesons to the chiral Lagrangian that has correct trace
anomaly structure, it has been found \cite{mishustin}
that both nuclear matter and finite nuclei can be described satisfactorily
at mean-field level with effective Lagrangians
provided certain nonlinear couplings of the matter fields are allowed.
The compression modulus K seems to come out somewhat too high but this
is presumably symptomatic of the tree approximation, not of the
Lagrangian structure.

Although the situation is not enirely clear, we believe it to be safe to
assume that there is no
fundamental obstacle to understanding normal nuclear matter in terms of
effective actions. This is similar to the description of Fermi liquids
and instabilities therefrom leading to phase transitions
\cite{polchinski,shankar}
in terms of effective field theories. Here phase transitions are driven
by instabilities associated with ``marginal" terms and some ``irrelevant"
terms kinematically enhanced in the effective Lagrangian absent in normal
Fermi liquids.
We will essentially take the same point of view in treating phase transitions
in nuclear matter associated with meson condensations. In other words,
our somewhat poor understanding of normal nuclear matter and finite nuclei
need not preclude the prediction of possible phase transitions at high density
in terms of effective field theories.

\subsection{Goldstone Boson Condensation}
\subsubsection{\it Kaon condensation with pion condensation}
\bs
At large density or temperature, there can be two basically different
phase transitions seen from the point of view of effective chiral
Lagrangians: Goldstone boson condensation and chiral/deconfinement
phase transitions. In this section, we consider the density effect and
discuss the first. We will focus on
kaon condensation because it is most likely that pion condensation cannot
occur at a density lower than 5-6 times that of nuclear matter and kaon
condensation may occur before that density as suggested first by
Kaplan and Nelson\cite{kaplan}

There are two kinds of driving mechanism as density increases. First of all,
the masses generally drop due to the ``shrinking " of the chiral circle which
is inherently of the vacuum nature. The second mechanism is the tendency of
Nature
to restore symmetry on the chiral circle. Consider infinite (nuclear) matter.
A specific system to imagine is the ``neutron star" matter\footnote{We put
a quotation mark here because we will see later that ``nuclear star"
is a more appropriate name for the system.}. The relevant
Lagrangian we will consider is the sum ${\cal L}^U+{\cal L}_B +\delta
{\cal L}_B$ of Eq. (\ref{Thelag}) in flavor $SU(3)$ sector,
\ben
\L &=& \frac{f^2}{4}\Tr \del_\mu U\del^\mu U^\dagger + c\Tr \left(\Tr \M U
+h.c.\right) \nonumber\\
&+& \Tr B^\dagger i\del_0 B+i\Tr B^\dagger [V_0, B]
-D\Tr B^\dagger\vec{\sigma}\cdot \{\vec{A}, B\}-F\Tr B^\dagger \vec{\sigma}
\cdot [\vec{A}, B]\nonumber\\
&+&b\Tr B^\dagger\left(\xi \M \xi\right) B + d\Tr B^\dagger B \left(\xi \M\xi
+ h.c.\right) +h\Tr B^\dagger B \left(\M U + h.c.\right) +\cdots.\label{lsu3pp}
\een
The ellipsis stands for chiral symmetric terms quadratic in derivatives
(A), terms linear in $\M$ and powers of derivatives (B), terms
higher order in $\M$ (C) etc. The class B and class C terms are not
relevant at tree order that we will be considering here but the class
A term is of the same order in chiral counting as the terms linear
in $\M$ given in (\ref{lsu3pp}). For the moment we will not worry about this
class A terms but return to them later in discussing kaon-nuclear scattering
at the same order.

Taking the Lagrangian (\ref{lsu3pp}) effectively as a Ginzburg-Landau form,
we will treat it in tree order, so the parameters appearing in
(\ref{lsu3pp}) can be fixed directly by experiments in
free space. Weak leptonic pion decay gives $f=f_\pi \approx 93 $ MeV.
Nuclear $\beta$ decay and semileptonic hyperon decay fix $F\approx 0.44$
and $D\approx 0.81$. The kaon and
$\eta$ masses are given by the Gell-Mann-Oakes-Renner relation (at the lowest
order in the mass matrix)
\ben
m_K^2=\frac{3}{4} m_\eta^2=2m_s c/f^2
\een
with $m_s$ the s-quark current quark mass. Empirical mass gives
\ben
m_s c\approx (182 MeV)^4.
\een
Baryon mass splittings are given by
\ben
m_\Sigma-m_N &=& 2dm_s,\\
m_\Lambda-m_N &=& \frac{2}{3} (d-2b) m_s,\\
m_\Xi-m_N &=&2(d-b)m_s.
\een
{}From these we obtain
\ben
bm_s &\approx& -67\ MeV,\label{bms}\\
dm_s &\approx& 134\label{dms}\ MeV.
\een
The coefficient $h$ is related to the $\pi N$ sigma term and has a large
uncertainty. The presently accepted value is
\ben
hm_s\approx -310\label{hms}\ MeV
\een
with an error of order $\pm 50$ MeV.
In fact, this term related to the strangeness
content of the proton could be considerably less than this, making the
following estimate subject to doubt. Indeed by Feynman-Hellman theorem
\ben
m_s\langle p|\bar{s}s|p\rangle=m_s\frac{\del m_N}{\del m_s}=-2(d+h)m_s
\approx 350 \ MeV.
\een
This corresponds to the strange quark contribution ($\sim 350$ MeV)
to the nucleon mass
and it is most certainly too big to be reasonable. A skyrmion estimate
gives a value smaller by as much as an order of magnitude.
One can gain an interesting idea by looking at
the two extreme limits when $m_s\rightarrow 0$ and $m_s\rightarrow
\infty$. \footnote{We follow
closely Ref.\cite{wise}.}
In the former case, the s-quark contribution to the baryon mass is trivially
zero since $\langle p|\bar{s}s|p\rangle$ is smooth in that limit. In the
latter case, the s quark can be integrated out and one gets
\ben
m_s\frac{\del m_N}{\del m_s}|_{m_s\rightarrow \infty}\rightarrow
\frac{2}{9} m_n\approx 65\ MeV.
\een
thus the two limits give a much smaller value than the ``empirical" value.

Let the Hamiltonian obtained from (\ref{lsu3pp}) be denoted as $H$.
To ensure charge neutrality of the matter system, we consider
\ben
\tilde{H}=H+\mu Q
\een
with $Q$ the electric charge and $\mu$ the corresponding Lagrange
multiplier (chemical potential). Let the expectation value of $\tilde{H}$
with respect to the lowest-energy state be denoted by $\tilde{E}$.
Then the charge neutrality condition is
\ben
\frac{\del \tilde{E}}{\del\mu}=\langle Q\rangle=0.
\een
When this condition is satisfied, $\tilde{E}=E$, the quantity we wish to
calculate. We assume the ground-state expectation values
\ben
\langle \pi^- \rangle &=& e^{-i\mu t} e^{i \vec{p}_\pi\cdot\vec{x}}, \\
\langle K^0\rangle &=& e^{i\vec{q}\cdot\vec{x}} v_0,\\
\langle K^+\rangle &=& e^{i\mu t}e^{i\vec{k}\cdot\vec{x}} v_K.
\een
For the neutral kaons, the chemical potential is zero. As for the
charged pions and kaons, the time dependence of the field is precisely given by
the chemical potential with ${p_\pi}_0=k_0=\mu$. In our consideration,
the negatively charged mesons $\pi^-$ and $K^-$ are involved.
The Hamiltonian mixes the neutron state to baryons with unit charge
of strangeness $S=-1$ since we will be looking at the threshold
for kaon condensation and hence limiting to quadratic order in kaon
field, thus allowing $|\Delta S|=1$. The states that can be admixed are
$\Lambda$, $\Sigma^0$, and $\Sigma^-$. The proton will also be mixed through
pion ($\pi^-$) condensation. The lowest such
quasi-particle state can then be occupied up to the Fermi momentum
$p_F=(3\pi^2 \rho)^{1/3}$ to form the ground state whose energy density
to $O(1)$ in $1/m_B$ is then given by
\ben
\tilde{E}/V\equiv \tilde{\epsilon}&=& \tilde{\epsilon} (0)+
({\vec{q}}^2+m_K^2) |v_0|^2 + ({\vec{k}}^2 +m_K^2 -\mu^2)|v_K|^2\nonumber\\
&+& \left({\vec{p}_\pi}^2 +m_\pi^2 +\mu^2\right) |v_\pi|^2 +
\rho \Delta\epsilon
\label{endensity}
\een
where
\ben
\Delta\epsilon &=& \left(\frac{\mu}{2f^2} -\frac{(F+D)^2}{2\mu f^2}
{\vec{p}_\pi}^2\right)|v_\pi|^2 \nonumber\\
&+& \left[(4h+2d+2b)\frac{m_s}{2f^2}-\frac{(3F+D)^2}{8(d-2b)
m_sf^2}{\vec{q}}^2 -\frac{(D-F)^2}{8dm_sf^2}{\vec{q}}^2\right] |v_0|^2
\nonumber\\
&+& \left[-\frac{\mu}{2f^2} +(4h+2d)\frac{m_s}{2f^2}-\frac{(D-F)^2}{(2dm_s
-\mu)2f^2}{\vec{k}}^2\right] |v_K|^2 +\cdots. \label{endensity2}
\een
The $\Delta\epsilon$ given here is an {\it in-medium}
one-loop term and goes formally
up to $O(Q^5)$ in the Weinberg counting (\ref{count1}).
Two-loop terms not included here will go as $O(Q^6)$
in the Weinberg counting rule. The graphs contributing to (\ref{endensity2})
are given in Fig. III.10. It is easy to understand Eq. (\ref{endensity2})
in terms of these graphs. The terms containing the coefficients $b$, $d$
and $h$ come from Fig. III.10a explicitly proportional to the chiral
symmetry breaking quark mass matrix which are of s-wave KN interactions.
The terms proportional to $\vec{q}^2$
and $\vec{k}^2$ come from Fig. III.10b and are intrinsically from p-wave
K-N interactions. The term linear in $\mu$ is from the vector field
term $V_\mu$ coupled to the baryon current $\bar{B}\gamma_0 B$, included
in Fig. III.10a.

Within the one-loop approximation, the p-wave interaction is linear in
the square of the kaon three-momentum. Thus for a density for which
\ben
\left[\frac{(3F+D)^2}{8(d-2b)m_s f^2}+\frac{(D-F)^2}{8dm_s f^2}\right] > 1
\label{ineq}
\een
which corresponds to the density $\rho > \sim 3.2 \rho_0$, increasing
$\vec{q}^2$ would bring in increasing attraction triggering a $K^0$
condensation.  However in reality, there is a form factor associated
with the vertex (which in the chiral counting
would correspond to higher loops) that
would prevent the arbitrary increase in the momentum. It is likely that
the LHS of (\ref{ineq}) is never greater than 1 in which case
one would have $\vec{q}^2=0$. Let us assume this for the moment and ignore
the $K^0$ mode. We now minimize the energy density $\epsilon$ with respect to
the pion momentum and $K^-$ momentum. From $\del \tilde{\epsilon}/\del
\vec{p}_\pi =0$, we get
\ben
\mu=(F+D)^2\frac{\rho}{2f^2}.
\een
We will confine to s-wave KN interactions and set $\vec{k}=0$. Then
the charge neutrality $\del \tilde{\epsilon}/\del \mu =0$ gives
\ben
\vec{p}_\pi^2=\left(\frac{\rho}{2f^2}\right)^2 (F+D)^2 \left\{[2(F+D)^2-1]+
[2(F+D)^2+1]|\frac{v_K}{v_\pi}|^2\right\}. \label{ppi}
\een
We thus have
\ben
\tilde{\epsilon} &=& \epsilon (v_\pi,v_K)= \epsilon (0,0)\nonumber\\
&+& |v_\pi|^2\left\{m_\pi^2-(F+D)^2\left(\frac{\rho}{2f^2}
\right)^2 [(F+D)^2-1]\right\}\nonumber\\
&+& |v_K|^2\left\{m_K^2-(F+D)^2\left(\frac{\rho}{2f^2}\right)^2 +
\left(\frac{\rho}{2f^2}\right)(4h+2d)m_s\right\}\nonumber\\
&+&\cdots.
\een
What is known from an extensive theoretical and experimental development in
nuclear physics, particularly from nuclear beta decay and spin-isospin modes
in complex nuclei is that the main correlation effect which would correspond to
higher order chiral corrections in the framework of
chiral perturbation expansion is
the downward renormalization of the axial-vector coupling constant $g_A$
from its free space value of $1.26$ to about $1$ discussed in the previous
subsection. This means that the main
correction to the above formula would be to replace $(F+D)$ above by
$g_A^*\approx 1$. Including nucleon-nucleon interaction effects generated
through the contact interaction of (\ref{potential}) in the {\it particle-hole
spin-isospin channel} will further reduce the effective $g_A$, as is
well-known in the Landau-Migdal formalism.
This makes the coefficient of $|v_\pi|^2$ always
greater than zero to a much larger density than relevant here. Therefore we
expect that
\ben
v_\pi\approx 0.
\een
The vanishing VEV of the pion field in the presence of a nonvanishing
kaon VEV would imply by (\ref{ppi})
an arbitrarily large pion momentum. Such a situation would be prevented
by form factor effects arising at higher loop order. Since there cannot
be any kaon condensation without pions (condensed or not) (this can be
easily verified by looking at the energy density in the absence of
pion field and showing that the coefficient of $|v_K|^2$ is always
greater than zero), it pays to have a little bit of pion VEV at the
cost of a large pion momentum. {\it What matters is that the pion contribution
to the energy density be as small as possible, not necessarily strictly zero.}

The critical density for kaon condensation $\rho_K$ is
\ben
\rho_K\approx -f^2 m_s (2h+d)\left\{ [1+\frac{2m_K^2}{m_s^2
(2h+d)^2}]^{\frac{1}{2}}-1\right\}
\een
where we have set $g_A^*\approx 1$.
Numerically this gives $\rho_K\approx 2.6 \rho_0$ for $hm_s=-310$ MeV
and $\rho_K\approx 3.7 \rho_0$ for $hm_s=-140$ MeV (which is close to
$hm_s=-125$ MeV corresponding to the extreme case with
$m_s\langle p|\bar{s}s|p\rangle \approx 0$). Even when $(g_A^*)^2=
1/2$, the critical density increases only to about 5.7 times $\rho_0$.
This is of course too high a density for the approximations used to be
valid. Nonetheless
one observes here a remarkable robustness against changes in parameters,
in particular with respect
to the greatest uncertainty in the theory, namely the quantity $h$ and also
$g_A^*$. The insensitivity to the strangeness content of the proton, namely,
the parameter $(h+d)$ makes the prediction surprisingly solid. Furthermore,
while
$(F+D)\rightarrow 1$ banishes the pion condensation, this affects relatively
unimportantly the kaon property. It seems significant that the critical density
for kaon condensation is robust against higher-order corrections while
pion condensation is extremely sensitive to higher-order graphs.

Unfortunately it is highly unlikely that Nature provides the right condition
for pions to trigger this mode of kaon condensation. The trouble is that
once $g_A^*$ falls to near 1 or below, the condensate pion momentum has to be
large and this would be prevented by the form factors that one would have
to append to pion-nucleon vertices. Furthermore if the kaon condensation occurs
at large density, say, at five or six times the matter density, then many of
the approximations used so far (such as the heavy-baryon approximation) would
no longer be justifiable.

\subsubsection{\it Kaon condensation via electron chemical potential}
\bs
In the above discussion, pions played a crucial role in triggering kaon
condensation. Here we present an alternative way for the kaons
to condense in ``neutron star" matter which does not require pion condensation.
The key idea is that in ``neutron star" matter, energetic electrons -- reaching
hundreds of MeV in kinetic energy in dense ``neutron star" matter --
can decay into kaons if the kaon mass falls sufficiently low in dense matter,
as we will show below \cite{BKTR92}, via
\ben
e^-\rightarrow K^- +\nu. \label{ek}
\een
This process can go into chemical equilibrium with the
beta decay
\ben
n\rightarrow p + e^- +\nu_e \label{betadecay}
\een
with the neutrinos diffusing out. This means that the chemical potentials
satisfy
\ben
\mu_{K^-}=\mu_{e^-}=\mu_n - \mu_p \ .
\een
Let $x$ denote the fraction of protons generated by (\ref{ek}) and
(\ref{betadecay}), so
\ben
\rho_n=(1-x)\rho, \ \ \ \ \rho_p=x\rho
\een
and define
\ben
\rho\equiv u\rho_0
\een
with $\rho_0$ being the nuclear matter density.
With the addition of protons, we gain in energy ({\ie}, the symmetry
energy) by the amount
\ben
E_s/V =\epsilon_s [(1-2x)^2 - 1] \rho_0 u =-4x (1-x)\epsilon_s \rho_0 u
\een
where we have assumed, for simplicity,
that the symmetry energy depends on the density
linearly with the constant $\epsilon_s$ determined from nuclei,
\ben
\epsilon_s\approx 32 {\rm MeV}.
\een
More sophisticated formulas for symmetry energy could be used but for
our purpose this form should suffice. In any event the symmetry energy
above nuclear matter density is unknown, so this constitutes one of the major
uncertainties in our discussion.

Instead of (\ref{endensity}), we have a simpler form for the
energy density,
\ben
\tilde{E}/V\equiv \tilde{\epsilon}&=& \tilde{\epsilon} (0)+
(m_K^2 -\mu^2)|v_K|^2+ \rho \Delta\epsilon \label{edensity}
\een
where
\ben
\Delta\epsilon =
-\frac{1}{2f^2}\left[\mu (1-x)+2\mu x -(4h+2d)m_s \right] |v_K|^2
- 4x (1-x) \epsilon_s +\cdots \label{dedens}
\een
The first term in the square bracket of (\ref{dedens}) is the neutron
contribution through vector exchange (namely, the $V_0$ term in
(\ref{lsu3pp})) and the second the corresponding proton contribution.
The factor of 2 for the proton contribution can be easily understood
by the fact that the $K^- p$ interaction with vector meson exchanges
is twice as attractive as the $K^- n$ interaction: This can be readily
verified by looking at the isospin structure of the vector current
$V_\mu$. Defining
\ben
\hat{\mu}=\mu +\frac{\rho_0 u}{4f^2}(1+x)
\een
we can rewrite (\ref{edensity})
\ben
\Delta \epsilon &\equiv& \tilde{\epsilon}-\tilde{\epsilon} (0)\nonumber\\
&=& \left[(m_K^*)^2 -\hat{\mu}^2
+\frac{\rho_0^2 u^2}{16f^4} (1+x)^2\right]|v_K|^2 -4x (1-x)\epsilon_s
\rho_0 u \label{edensp}
\een
where
\ben
(m_K^*)^2=m_K^2+\frac{\rho_0 u}{2f^2}(4h+2d)m_s.
\een
The charge neutrality condition is gotten by balancing the negative
charge of the kaon against the positive charge of the proton, {\ie},
\ben
\hat{\mu}=\mu_n-\mu_p=x \frac{\rho_0 u}{2|v_K|^2}.\label{chneut}
\een
The condition (\ref{chneut}) gives  $x$ as a function of
$|v_K|^2$ since the proton and neutron chemical potentials are
functions of $x$. Since $\mu_n-\mu_p=-\del \epsilon/\del x$, we
have
\ben
\hat{\mu}=4\epsilon_s u (1-2x)=x \frac{\rho_0 u}{2|v_K|^2}.\label{chneutp}
\een
Solving for $x$ for small kaon VEV, we get
\ben
x\approx (8\epsilon_s/\rho_0) |v_K|^2 +\cdots
\een
where the ellipsis stands for higher orders in the kaon VEV.
Substituting this into (\ref{edensp}) and using the numerical values
$dm_s\approx 134$ MeV, $hm_s\approx -150$ MeV, $\epsilon_s\approx 32$ MeV,
$f=93$ MeV, we find the critical density
\ben
\rho_K\approx 3.5\rho_0.
\een
This critical density is comparable to that predicted for the most optimistic
case of the kaon condensation triggered by pion condensation.

So far we have not considered the possibility that the pion decay constant
can scale as density increases as discussed above.
The scaling effect can be easily implemented,
by incorporating the scaling factor $\Phi$ according to the rule developed
above.
The scaling factor
$\Phi$, as mentioned before, is, strictly speaking, unknown beyond
the ordinary matter density. Here we take for illustration the fixed value
$\Phi = 2/3$ -- and the other parameters unmodified --
for which Eq.(\ref{edensp}) predicts the critical density
to be
\ben
\rho_k\approx 2.0\rho_0.
\een
As expected, the ``shrinking" of the chiral circle, natural in the high-density
behavior of the quark condensate, greatly facilitates the kaon
condensation. This shows the qualitative tendency expected of the scaling.

An extremely interesting question to ask for physics of neutron stars and
stellar collapse is what the proton fraction
$x$ comes out to be in neutron star matter given the low critical density
predicted for the kaon condensate formation. This is highly
relevant for neutron star cooling.
It is known that kaon condensation by itself could lead to a rapid
cooling \cite{BKPP}. But even more importantly, as pointed out
recently by Lattimer et al \cite{lattimer},
the direct URCA process can occur in neutron stars if the proton concentration
exceeds some critical value in the range of $(11-15)\%$: The direct URCA
can cool the neutron star considerably faster than any other known mechanism,
be that meson condensation or quark-gluon plasma. In order to evaluate the
proton concentration, we have to include higher order (at least
quartic) terms in kaon field.
In free space, kaon-kaon interactions are are known to be
weak, so we may ignore them.
However medium-dependent kaon-kaon interactions, appearing at the one-loop
level, may not be negligible. This requires calculating density-dependent
loop diagrams. In the absence of explicit computation of the loop
diagrams, it is difficult to say what the proton fraction comes out to be.
Here we attempt a simple calculation of $x$ which may not be accurate but
nevertheless gives some idea how things will go.

Even if one ignores the {\em explicit} loop graphs, our expression for
the energy density contains an arbitrary power of the kaon VEV
through the relation between $x$ and $|v_K|^2$, namely from eqs.(\ref{chneut})
and(\ref{chneutp}),
\ben
x=(8\epsilon_s/\rho_0)|v_K|^2 \left(1+(16\epsilon_s/\rho_0)|v_K|^2\right)^{-1}.
\label{x}
\een
In looking for the critical density, it sufficed to retain only the lowest
order term in the VEV from this relation. For determining the proton fraction
$x$, however, we need higher orders, certainly at least quartic.
It may be that it is not really justified to take into account the higher VEV's
through solely the nonlinearity of the charge neutrality constraint while
ignoring explicit contributions from higher loop graphs but to the extent that
we are using a ``phenomenological" symmetry energy, it may be hoped
that the constraint equation supplies not entirely erroneous higher-order
VEV's. Now it is easy to check that expanding (\ref{x}) to
the quartic order in the VEV is not good enough. The resulting quartic term in
the energy density is much too repulsive for the kaon VEV to develop
significantly. It is not surprising that the expansion cannot be trusted
if truncated at low order since next order terms that are ignored have much
larger coefficients with alternating sign\footnote{The reason why one
wants to expand and retain the lowest relevant term is that they are the only
ones that can be systematically calculated in chiral perturbation theory. Once
we take the phenomenological symmetry energy and impose the relation due to
charge neutrality, then the chiral perturbation strategy seems to lose its
predictivity and uniqueness. To make a further progress, it would be
necessary to sort out what is in the
symmetry energy expression we use in terms of the chiral expansion.}.
Instead of expanding it, therefore,
we should substitute (\ref{x}) into (\ref{edensp})
and minimize the resulting energy density with respect to $|v_K|^2$.
The resulting algebraic equation has been solved exactly for the VEV as a
function of $u$.
The result (obtained without the scaling effect)
is that the critical proton concentration $x_c=0.11$ is reached at the
density $u=4$, increasing gently as density increases, {\ie},
$x=0.12$ at $u=5$, $x=0.17$ at $u=10$ etc. The scaling effect as well as
any further attraction (expected) from
loop contributions will increase further the proton concentration.
It looks that the URCA process is very likely to play an important role.
\subsubsection{\it Constraints from kaon-nuclear interactions}
\bs
Although the effective chiral Lagrangian is consistent with nuclear matter
when scale anomaly is suitably incorporated,
we have not asked whether it is consistent with
on-shell kaon interactions with nucleons, both in free space and in medium.
To answer this question, we extract relevant terms for s-wave kaon-nucleon
scattering from the Lagrangian density (\ref{lsu3pp}). The leading term
is of order $Q^1$ or $\nu=1$ and has the form
\be
\L_{\nu=1} (KN)=\frac{-i}{8f_\pi^2}\left(3(\barN \gamma^0 N)\bar{K}\lrarrow K
+(\barN\vec{\tau}\gamma^0 N)\cdot\bar{K}\vec{\tau}\lrarrow K\right)\label{lnu1}
\ee
with $N^T=(p\ \  n)$, $K^T=(K^+\  K^0)$ and $\barK\lrarrow K\equiv
\barK \rarrow K-\barK\larrow K$. For $\bar{K}N$ scattering (with
$\bar{K}^T=(K^-\ \bar{K}^0)$), due to G-parity, the isoscalar term changes
sign. In terms of an effective Lagrangian that contains vector mesons in
hidden gauge  symmetric way, the first term of (\ref{lnu1}) cam be identified
as the $\omega$ exchange and the second term as the $\rho$ exchange between
the kaon and the nucleon. Thus this leading order term can be thought of as
vector-dominated for the scattering amplitude. Now the next order term
is of $\nu=2$ and can be written for s-wave scattering as
\be
\L_{\nu=2} (KN) &=& \frac{\Sigma_{KN}}{f_\pi^2} (\barN N) \barK K
+\frac{C}{f_\pi^2} (\bar{N}\vec{\tau} N)\cdot (\bar{K}\vec{\tau}K)
\label{lnu2}
\ee
where
\be
\Sigma_{KN}&= &-(\frac 12 b+d +2h) m_s,\\
C &=& -\frac{b m_s}{2},
\ee
With (\ref{lnu1}) and (\ref{lnu2}), the s-wave $KN$ scattering lengths are
\be
a_{I=1}^{KN} &=&\frac{1}{4\pi f_\pi^2(1+m_K/m_B)}\left(-m_K+\Sigma_{KN}+
C\right),\label{a1}\\
a_{I=0}^{KN}&=& \frac{1}{4\pi f_\pi^2(1+m_K/m_B)}\left(\Sigma_{KN}-3C \right).
\label{a0}
\ee
With the values for the constants $b$, $d$ and $h$ given above (\ref{bms})-
(\ref{hms}), that is, $bm_s=-67$ MeV, $dm_s=134$ MeV and $hm_\approx -310$
MeV (which imply $\Sigma_{KN}\approx 520$ MeV), we get $a_{I=1}^{KN}\approx
0.07$ fm and $a_{I=0}^{KN}\approx 0.50$ fm to be compared with the presently
available empirical values $-0.31$ fm and $-0.09$ fm respectively. The
empirical value for $a_{I=1}^{KN}$ is reliable but that for $a_{I=0}^{KN}$
is not. In fact the latter is compatible with zero fm. The experimental
uncertainty notwithstanding, it is clear that the Lagrangian so far used is
not consistent with the s-wave $K^+ N$ scattering. It is true that the
$\Sigma_{KN}$ that large implies a substantial amount of strangeness
in the proton which is not in accord with Nature.
If one lowers the value of $\Sigma_{KN}$ to $\sim 2m_\pi$ which is reasonable
for the strangeness content, the predicted values are: $a_{I=1}^{KN}\approx
0.07$ fm and $a_{I=0}^{KN}\approx 0.50$ fm. These are closer to the
empirical values but the $I=0$ amplitude is much too attractive. The point is
that if the amplitudes for the $K^+N$ channel come out wrong, there is no
way that the amplitudes for the $K^-N$ channel can come out right: The
$K^-N$ interaction is much stronger than the $K^+N$ interaction.

This difficulty can be remedied if one recognizes that the $\nu=2$ piece of
the Lagrangian (\ref{lnu2}) is incomplete. It misses the chiral symmetric
two-derivative terms which for s-wave scattering take the form
\be
\delta\L_{\nu=2} (KN) = \frac{\tilde{D}}{f_\pi^2}
(\barN N)\del_t \barK\del_t K+ \frac{\tilde{D}^\prime}{f_\pi^2}
(\barN \vec{\tau} N)\cdot(\del_t \barK\vec{\tau}\del_t K).
\label{deltalnu2}
\ee
There are two sources to these terms. One is what we will call ``$1/m$"
correction. This comes because in the heavy-baryon formalism, baryon-antibaryon
pair terms appear at the $\nu=2$ order in the form of $p/m_N$ where $p$ is the
spatial momentum carried by the nucleon. The other source has to do with
high-energy degrees of freedom integrated out above the chiral scale
$\Lambda_\chi$, appearing as contact counter terms. The former can be
calculated but the latter, while calculable in  some dynamical models,
are unknown constants that need be fixed by experiments.
Including (\ref{deltalnu2}) modifies the scattering lengths to
\be
a_{I=1}^{KN} &=&\frac{1}{4\pi f_\pi^2(1+m_K/m_B)}\left(-m_K+\Sigma_{KN}+
C+(\bar{D}+\bar{D}^\prime))m_K^2\right),\label{a1p}\\
a_{I=0}^{KN}&=& \frac{1}{4\pi f_\pi^2(1+m_K/m_B)}\left(\Sigma_{KN}-3C
+(\bar{D}-3\bar{D}^\prime)m_K^2 \right).
\label{a0p}
\ee

Let us first ask what values of $\tilD$ and $\tilDp$ are needed for (\ref{a1p})
and (\ref{a0p}) to reproduce the experimental numbers. We find:
\be
\tilde{D}&\approx& 0.33/m_K-\Sigma_{KN}/m_K^2,\label{D}\\
\tilde{D}^\prime &\approx& 0.16/m_K-C/m_K^2 .\label{D'}
\ee
For $\tilD$, we take $\Sigma_{KN}\approx 2m_\pi$. We have $\tilD\approx
-0.24/m_K$ and $\tilDp\approx 0.23/m_K$.

As mentioned, the $1/m$ corrections are calculable. They are given by
\cite{BLRT}
\be
\tilde{D}_{\frac{1}{m}} &\approx& -
\frac{1}{48} \left[(D+3F)^2+9(D-F)^2\right]/m_N
\approx -0.06/m_K,\nonumber\\
\tilde{D}^\prime_{\frac{1}{m}} &\approx& -\frac{1}{48}\left[(D+3F)^2
-3(D-F)^2\right]/m_N\approx -0.04/m_K.
\label{paircon}
\ee
Clearly the $1/m$ corrections are much too small to account for the values of
$\tilD$ and $\tilDp$. The main contributions must therefore come from
the contact counter terms since at order $\nu=2$, there are no loop
corrections which start at $\nu=3$.

Given that the $\tilD$ and $\tilDp$ are very important at threshold,
an immediate question is how they affect the critical density and composition
of the condensed matter. Since these terms are quadratic in derivative,
for the s-wave processes we are considering, they will simply modify
$\omega^2$ coming from the kinetic energy term for the kaon as
\be
\omega^2\rightarrow \left(1+[\frac{\tilD}{f_\pi^2}+(2x-1)
\frac{\tilDp}{f_\pi^2}]u\rho_0\right)\omega^2.
\ee
At threshold, $\omega\rightarrow m_K$. For kaon condensation, as mentioned
above, $\omega=\mu$ ($\mu$ being the chemical potential)
due to the charge conservation. Now since in dense matter
the chemical potential gets quenched, the role of the $\tilD$ and $\tilDp$
terms becomes less important. For this reason, little is modified in
the properties of the condensate by the inclusion of these extra terms.
For instance, with $\Sigma_{KN}\approx 2m_\pi$ and {\it without} the BR
scaling,
the critical density is $u_c\approx 3.04$ without $\tilD$ and $\tilDp$
and $u_c\approx 3.27$ with them. The equation of state is also little
affected, the matter turning rapidly to ``nuclear matter" with
both $x$ and strangeness fraction $S/B$ near 0.5 right above the critical
point. This feature is expected to have a profound influence on the
formation of compact ``nuclear stars" and their cooling. We will not pursue
this
here but refer to the literature \cite{BLRT}.

\subsection{Vector Symmetry}
\bs
What happens when density is so high that there is a phase transition
away from the Goldstone mode? Does it go into quark-gluon phase as
expected at some density by QCD? Or does it make a transition into
something else? Up to date, this question has not received a clear answer: It
remains an open question even after so many years of QCD.
Here we discuss a possible state of matter that is
still hadronic but different from the normal hadron state and also
from meson-condensed states -- a state that realizes the ``vector
symmetry" of Georgi\cite{georgivec}.

\subsubsection{\it Pseudoscalar and scalar Goldstone bosons}
\bs
When one has the matrix element of the axial current
\ben
\langle 0|A^\mu|\pi\rangle=iq^\mu f_\pi
\een
this can mean either of the two possibilities. The conventional way is
to consider $\pi$'s as Goldstone bosons that emerge as a consequence
of spontaneous breaking of chiral $SU(3)\times SU(3)$ symmetry, with the
current
$A^\mu$ associated with the part of the symmetry that is broken ({\ie},
the charge $\int d^3 x A_0 (x)$ being the generator of the broken symmetry).
But there is another way that this relation can hold. If there are
scalar bosons $s$ annihilated by the vector current $V^\mu$
\ben
\langle 0|V^\mu|s\rangle=iq^\mu f_s
\een
with $f_\pi=f_s$, then the symmetry $SU(3)\times SU(3)$ can be unbroken.
This can happen
in the following way. Since $A^\mu$ and $V^\mu$ are part of $(8,1)+(1,8)$
of $SU(3)\times SU(3)$, all we need is that $\pi$'s and $s$'s are part of
$(8,1)+(1,8)$. The $SU(3)\times SU(3)$ symmetry follows if the decay constants
are equal. In Nature, one does not see low-energy scalar octets
corresponding to $s$. Therefore the symmetry must be broken.
This can be simply understood if one imagines that the
scalars are eaten up to make up the longitudinal components of
the massive octet vector mesons. If this is so, then
there must be  a situation where the symmetry is restored in such a way
that the scalars are liberated and become
real particles, with the vectors becoming massless. This is the ``vector
limit" of Georgi. In this limit, there will be 16 massless scalars
and massless octet vector mesons with the symmetry swollen to
$[SU(3)]^4$ which is broken down spontaneously to $[SU(3)]^2$.
In what follows, we discuss how this limit can be reached.

The full vector symmetry is
\ben
SU_L (3)\times SU_R (3)\times SU_{G_L}(3)\times SU_{G_R}(3).
\een
This is a primordial symmetry which is dynamically broken down
to
\ben
SU_{L+G_L}(3)\times SU_{R+G_R}(3).
\een
As is well-known, the nonlinear symmetry
\ben
\frac{SU_L (3)\times SU_R (3)\times SU_{G_L}(3)\times SU_{G_R}(3)}
{SU_{L+G_L}(3)\times SU_{R+G_R}(3)}
\een
is equivalent to the linear realization
\ben
\{[SU_L(3)\times SU_{G_L}(3)]_{global}\times [SU_L (3)]_{local}\}
\times \{L\rightarrow R\}
\een
where the local symmetries are two copies of the hidden gauge symmetry.
One can think of the two sets of hidden gauge bosons as the octet
vectors $\rho_\mu$ and the octet axial vectors $a_\mu$.

In order to supply the longitudinal components of both vector and axial
vector octets of spin-1 fields, we would need 16 Goldstone bosons that
are to be eaten up plus 8 pseudoscalars $\pi$ to appear as physical states.
Here we shall consider a minimal model where only the vector fields $\rho$
are considered. The axials are purged from the picture. We can imagine this
happening as an explicit symmetry breaking
\ben
SU_{G_L}\times SU_{G_R}\rightarrow SU_{G_L+G_R}.
\een
What happens to the axials after symmetry breaking is many-fold. They could
for instance be banished to high mass $\gg m_\rho$.

\subsubsection{\it The Lagrangian}
\bs
To be more specific, we shall construct a model Lagrangian with the given
symmetry. Since we are ignoring the axials, the Goldstone bosons can be
represented with the transformation properties
\ben
\xi_L &\rightarrow& L\xi_L G^\dagger,\\
\xi_R &\rightarrow& R\xi_R G^\dagger
\een
with $\xi_L$ transforming as $(3,1,\bar{3})$ under $SU_L(3)\times SU_R(3)
\times SU_G(3)$ (with $G=G_L+G_R$) and $\xi_R$ as $(1,3,\bar{3})$. $L$,
$R$ and $G$ are the linear transformations of the respective symmetries.
The usual Goldstone bosons of the coset $SU_L(3)\times SU_R(3)/SU_{L+R} (3)$
are
\ben
U=\xi_L \xi_R^\dagger
\een
transforming as $(3,\bar{3},1)$. In terms of the covariant derivatives
\ben
D^\mu \xi_L &=& \del^\mu\xi_L -ig \xi_L \rho^\mu,\\
D^\mu \xi_R &=& \del^\mu\xi_R -ig \xi_R \rho^\mu
\een
the lowest derivative Lagrangian having the (hidden) local $SU(3)_{L+R}$
symmetry is
\ben
{\cal L}=\frac{f^2}{2}\{\Tr\left(D^\mu \xi_L D_\mu \xi_L^\dagger\right)+
(L\rightarrow R)\} +\kappa \frac{f^2}{4}\Tr (\del^\mu U \del_\mu U^\dagger).
\label{georgilag}
\een
A convenient parametrization for
the chiral field is
\ben
\xi_{\mbox{\tiny{L,R}}}\equiv e^{i\sigma (x)/f_s} e^{\pm i\pi (x)/f_\pi}
\een
with $\sigma (x)=\frac 12 \lambda^a \sigma^a (x)$ and $\pi (x)=\frac 12
\lambda^a \pi^a (x)$. (Setting $\sigma (x)=0$ corresponds to unitary gauge
with no Faddeev-Popov ghosts.)
At the tree order with the Lagrangian (\ref{georgilag}), we have
\ben
f_s=f, \ \ \ \ \ f_\pi=\sqrt{1+\kappa} f
\een
and the $\rho\pi\pi$ coupling
\ben
g_{\rho\pi\pi}=\frac{1}{2(1+\kappa)} g.
\een
The vector meson mass is
\ben
m_\rho=fg=2\sqrt{1+\kappa} f_\pi g_{\rho\pi\pi}.\label{vecmass}
\een
Thus $f_s=f_\pi$ when $\kappa=0$, {\ie}, when the mixing $LR$ vanishes.
One can see that the standard hidden gauge symmetry result is obtained
when $\kappa=-1/2$. When $\kappa=0$, we have an unbroken $SU(3)\times SU(3)$,
but the
vectors are still massive. The mass can disappear even if $f_\pi\neq 0$ if
the gauge coupling vanishes. In the limit that $g\rightarrow 0$ which
is presumably attained at asymptotic density as described below,
the vector mesons decouple from the Goldstone bosons, the scalars are
liberated and the symmetry swells to $[SU(3)]^4$ with the Goldstone
bosons transforming
\ben
\xi_L &\rightarrow& L\xi_L G_L^\dagger,\\
\xi_R &\rightarrow& R\xi_R G_R^\dagger.
\een
Note that the local symmetry $G$ gets ``recovered" to a {\it global} symmetry
$G_L\times G_R$. (This is a dynamical symmetry restoration, in some sense.)
\subsubsection{\it Hidden gauge symmetry and the vector limit}
\bs
We shall now consider renormalization-group (RG) structure of the constants
$g$ and $\kappa$ of the Lagrangian in $SU(2)$ flavor which is of the
form (\ref{georgilag}) with the fields valued in $SU(2)$,
\ben
{\cal L}= \frac{f^2}{2}\{\Tr\left(D^\mu \xi_L D_\mu \xi_L^\dagger\right)+
(L\rightarrow R)\} +\kappa \frac{f^2}{4}\Tr (\del^\mu U \del_\mu U^\dagger)
\label{georgiL}.
\een
The kinetic energy term of the $\rho$ field should be added for completeness;
we do not need it for this discussion.
We are interested in what happens to hadrons described by this Lagrangian
as matter density $\rho$ and/or temperature $T$ are increased \footnote{We
will follow Ref.\cite{harada}.}.
For this consider the vector meson mass in dense medium at zero temperature.
It turns out that the mass formula (\ref{vecmass})
is valid to all orders of perturbation theory \cite{harada,harada2}.
{\it  We expect this to be true
in medium}, according to the argument given above. Therefore we need to
consider the scaling behavior of the constants $f_\pi$, $g$ and $\kappa$
in medium. Now we know from above that $f_\pi$ decreases as density is
increased. This is mainly caused by the vacuum property, so we can assume
that the effect of radiative corrections we are considering is not important.
Thus we have to examine how $g$ and $\kappa$ scale. The quantities we are
interested in are the $\beta$ functions for the coupling constants
$g$ and $\kappa$. These can be derived in a standard way with the
Lagrangian (\ref{georgiL}). At one-loop order, we have in dimensional
regularization \cite{harada}
\ben
\beta_g (g_r) &\equiv& \mu \frac{dg_r}{d\mu}=-\frac{87-a_r^2}{12}
\frac{g_r^3}{(4\pi)^2},\label{betag}\\
\beta_a (a_r) &\equiv& \mu \frac{da_r}{d\mu}=3a_r (a_r^2-1)
\frac{g_r^2}{(4\pi)^2}\label{betaa}
\een
where $\mu$ is the length scale involved (which in our case is the matter
density) and we have redefined
\ben
a=1/(1+\kappa).
\een
To see how the factors in
the beta function can be understood, let us discuss the first equation
describing the way the coupling constant $g$ runs. In terms of Feynman
diagrams, the quantity on the right-hand side of (\ref{betag}) can be
decomposed as
\ben
-\frac{1}{(4\pi)^2}\frac{87-a_r^2}{12}= -\frac{1}{(4\pi)^2}\left[
\frac{11n_f}{3} -(\frac 12)^2\frac 13 -(\frac{a_r}{2})^2\frac 13\right]
\label{decomp}
\ee
with the number of flavor $n_f=2$. This is an exact analog to the beta
function of QCD, $\beta_g^{\mbox{\tiny QCD}}=-\frac{g^3}{(4\pi)^2}
\left[\frac{11}{3}N_c-\frac 23 n_f\right]$. The first term on the right-hand
side of
(\ref{decomp}) is the vector meson loop which is an analog to the gluon
contribution in QCD with the number of flavors $n_f$ replacing the number
of colors $N_c$ (the vector meson
$\rho$ is the ``gauge field" here), the second term is the $\sigma$ loop
contribution $g_{\rho\sigma\sigma}^2/3$ with $g_{\rho\sigma\sigma}=g/2$
and the last term the pion loop contribution $g_{\rho\pi\pi}^2/3$
with $g_{\rho\pi\pi}=(a_r/2) g$.

As in QCD, the fact that the beta function is negative signals that
the coupling constant runs down as the momentum (or density) scale
increases and at asymptotic momentum (or density), the coupling would
go to zero, that is, asymptotically free.
Now the beta function is negative here as long as $a_r^2 < 87$ with
the coupling constant running as
\ben
g^2 (\mu)=\frac{8\pi^2}{(87-a_r^2)\ln\frac{\mu}{\Lambda}}
\een
with the HGS scale $\Lambda$ defined as in QCD.
The inequality $a_r^2 < 87$ is clearly satisfied as
one can see from Eq. (\ref{betaa}) which says that there is an ultraviolet
fixed point $a_r=1$, so as the momentum increases the constant $a$
would run toward 1. Of course one would have to solve the equations
(\ref{betag}) and (\ref{betaa}) simultaneously to see the trajectory of
$a_r$ but it seems reasonable to assume that the $a_r$ runs monotonically
from the vacuum value $a_r=2$ (or $\kappa=1$) to the fixed point $a_r=1$
(or $\kappa=0$). The in-medium mass formula
\ben
m_\rho^\star=f^\star g^\star=\frac{1}{\sqrt{1+\kappa^\star}}
f_\pi^\star g^\star\label{vecmasspp}
\een
shows that apart from the condensate effect on $f_\pi^\star$, there
are additional effects due to the decreasing $(1+\kappa^\star)^{-1}$ factor.
The vector mass will go to zero for $f_\pi^\star\neq 0$
as the coupling $g^\star$ vanishes, approaching the ``vector limit"
of Georgi.

\subsubsection{\it Physics under extreme conditions}
\bs
What happens to the hadronic matter governed by the HGS Lagrangian
as density or temperature increases?
At present, there is no direct prediction from QCD that the relevant limits
can be reached by temperature or density. No model calculations purporting
to indicate them are available.
Nonetheless, the vector limit ideas are intuitively appealing and could play
an important role in dense cold matter like in nuclear stars and in hot
dense matter like in heavy-ion collision.

In the $\kappa=0$  limit, the vector mass is
\ben
m_\rho=2f_\pi g_{\rho\pi\pi}
\een
which in terms of physical values of  $f_\pi$ and $g_{\rho\pi\pi}$
is too big by roughly a factor of
$\sqrt{2}$ whereas the current algebra result (KSRF)
is $m_\rho=\sqrt{2}f_\pi g_{\rho\pi\pi}$, which is in good agreement with
the experiment. This shows that at the tree order, the limit is rather far
from reality as applied to the medium-free space.
Loop corrections with the Lagrangian treating the $\kappa$
term as a perturbation does improve the prediction\cite{cho},
bringing it much closer to experiments.
One possible scenario that emerges from the discussion given above
is that the density changes the state of matter from the
phase of normal matter with $\kappa\neq 0$ and $g\neq 0$
to first $\kappa=0$ and then to $g=0$, resulting in a significant
increase of light degrees of freedom below the critical point. In the model
we have been considering, we will have a total of
32 degrees of freedom consisting of 16 spin-0 bosons and 8 massless vectors.
If we add the axial vectors, then the total becomes 48. Now in the quark-gluon
phase with three flavors, one expects 47$\frac{1}{2}$. They are of the same
order and if anything, the transition will be {\it extremely} smooth.
\footnote{For further discussions on this matter, see \cite{adami}.}

Strictly speaking, the above counting is not quite relevant because of
the strange
quark mass. The relevant degrees of freedom are more likely to be less than
that. Nonetheless if one counts all the possible light degrees of
freedom in the hadronic sector, they come out certainly much more than the
customary counting based on pion degrees of freedom often invoked
in the literature.
Thus the vector limit is a possible candidate phase
before the would-be chiral transition to
quark-gluon phase (to be described in QCD variables)
at which we will have $f_\pi=f_s=0$.

\pagebreak

\newpage
\centerline{\bf FIGURE CAPTIONS}
\vskip 0.7cm
\centerline{\bf Figure I.1}
\begin{quotation}
\noindent A spherical chiral bag for ``deriving" a four-dimensional
Cheshire Cat model. $\psi$ represents the doublet quark field of
u and d quarks for $SU(2)$ flavor or the triplet of u, d and s for
$SU(3)$ flavor and $\pi$ the triplet $\pi^+, \pi^-, \pi^0$ for
$SU(2)$ and the octet pseudoscalars $\pi$, $K$, $\bar{K}$ and $\eta$
for $SU(3)$. The axial singlet meson $\eta^\prime$ figures in
axial anomaly.
\end{quotation}
\vskip 0.2 cm
\centerline {\bf Figure I.2}
\begin{quotation}
\noindent Two classes of excitation spectrum described by Berry potentials
as viewed in the chiral bag picture:
Class A for excitations with the light-quark (u and d) space;
Class B for excitations involving the massive quarks $Q=s, c, b\cdots$
\end{quotation}
\vskip 0.2cm
\centerline {\bf Figure II.1}
\begin{quotation}
\noindent An approximate profile function $F(r)$ compared with the
exact numerical solution for the Skyrme model.
\end{quotation}
\vskip 0.2 cm
\centerline{\bf Figure II.2}
\begin{quotation}
\noindent Strange hyperon spectrum in the skyrmion model compared with
experiments
(or equivalently nonrelativistic quark model): The parameter $\omega_K$
fixed by the state $\Lambda$ (marked by the star) and the $c_K$ by the
empirical relation (\ref{cemp}) are $\omega_K=223$
MeV and $c_K\equiv 1-\kappa_s=0.62$.
\end{quotation}
\vskip 0.2 cm
\centerline{\bf Figure II.3}
\begin{quotation}
\noindent Charmed hyperon spectrum in the skyrmion model compared with
quark-model predictions: Two parameters fixed by fitting two lowest
experimental levels (marked by stars) are $\omega_D=1418$ MeV and
$c_D\equiv 1-\kappa_c=0.14$.  The parameters for
the strange-quark sector are fixed
from Figure II.2 ``Quark model 1" is the prediction by L.A. Copley, N. Isgur
and G. Karl, \pr \ {\bf D20}, 768 (1979) and K. Maltman and N. Isgur, \pr
\ {\bf D22}, 1701 (1980) and ``Quark model 2"
is the prediction by A. De R\'{u}jula, H. Georgi and S.L. Glashow,
\pr \ {\bf D12}, 147 (1975).
\end{quotation}
\vskip 0.2 cm
\centerline{\bf Figure III.1}
\begin{quotation}
\noindent Nucleon-nucleon central potentials $V_c^T$ derived
from baryon number 2
skyrmions compared with the results of Reid soft-core potentials.
The Paris potentials contain momentum dependence, so cannot be compared
directly. The solid line represents the nucleon-only (leading $N_c$) result
of \cite{wambach}, the dashed line the full Born-Oppenheimer diagonalization
of the $N \Delta$ state mixing of \cite{walet1} and the dotted line
the Reid soft-core potentials.
\end{quotation}
\vskip 0.2 cm
\centerline{\bf Figure III.2}
\begin{quotation}
\noindent Three-body forces involving a nonlinear coupling that are
suppressed to $O(Q^2)$.
\end{quotation}
\vskip 0.2 cm
\centerline{\bf Figure III.3}
\begin{quotation}
\noindent Feynman diagrams for three-body amplitudes involving pairwise
one-pion exchange that consist of both reducible and irreducible
graphs. Note that the pion propagator is a Feynman propagator which is
represented by a wiggly line.
\end{quotation}
\vskip 0.2 cm
\centerline{\bf Figure III.4}
\begin{quotation}
\noindent Three-body force mediated with a $\Delta$ intermediate that
can contribute. The wiggly line stands for  Feynman propagator for the pion.
\end{quotation}
\vskip 0.2 cm
\centerline{\bf Figure III.5}
\begin{quotation}
\noindent The leading soft-pion graphs contributing to exchange EM
currents ([a] and [b])
and to exchange axial-charge operator ([a]). The circled cross stands for
the vector current $V^\mu$ or the axial-vector current $A^\mu$. The
propagators are as defined in Fig. III.3.
\end{quotation}
\vskip 0.2 cm
\centerline{\bf Figure III.6}
\begin{quotation}
\noindent Two-body exchange graphs of $O(Q^2)$ relative to the leading
soft-pion graphs.
The non-vanishing graphs are [a], [b] and [c]; all the rest, most of
which are not shown here, do not contribute to the order considered.
The dashed line represents the time-order pion propagator.
\end{quotation}
\vskip 0.2 cm
\centerline{\bf Figure III.7}
\begin{quotation}
\noindent The $O(Q^2)$ correction $\Delta_{1\pi}$, $\Delta_{2\pi}$
and $\Delta=\Delta_{1\pi}+\Delta_{2\pi}$ calculated
in Fermi gas model as function of density $\rho/\rho_0$ for
short-range cut-off $d=0.5$ fm and $0.7$ fm. The possible uncertainty
in the value of cut-off is indicated by the shaded area.
\end{quotation}
\vskip 0.2 cm
\centerline{\bf Figure III.8}
\begin{quotation}
\noindent Three-body currents suppressed to $O(Q^2)$ relative to the soft-pion
(two-body) graphs. The mechanism for the suppression is the same as
for three-body forces to the same chiral order.  As before, the wiggly line
represents the Feynman propagator and the dotted line the time-ordered
propagator of the pion.
\end{quotation}
\vskip 0.2 cm
\centerline{\bf Figure III.9}
\begin{quotation}
\noindent The $\Delta$-hole graph involving four-baryon contact interaction
that plays a dominant role in quenching $g_A$ in nuclei. $\vec{A}$
is the space component of the axial current, the thick solid line
stands for the $\Delta$ and $N^{-1}$ for the nucleon hole.
\end{quotation}
\vskip 0.2 cm
\centerline{\bf Figure III.10}
\begin{quotation}
\noindent In-medium one-loop graphs contributing to energy-density of
dense nuclear
matter proportional to $v_K^2$ for kaon condensation. The solid line stands
for baryons and the dotted line for the kaon.
\end{quotation}

\end{document}